\documentclass[a4paper,twoside,11pt,openright]{book}

\usepackage{amsmath}
\usepackage{amssymb}
\usepackage{makeidx}
\usepackage{graphicx}
\usepackage{chngpage}
\usepackage{hyperref}
\usepackage{hvfloat}
\usepackage{rotating}
\usepackage{subfigure}
\usepackage{listings}

\usepackage[small,bf]{caption}

\begin{document}

\frontmatter

\begin{titlepage}
  \begin{adjustwidth}{1cm}{-1cm}
  \begin{center}
    \includegraphics[width=0.6\textwidth]{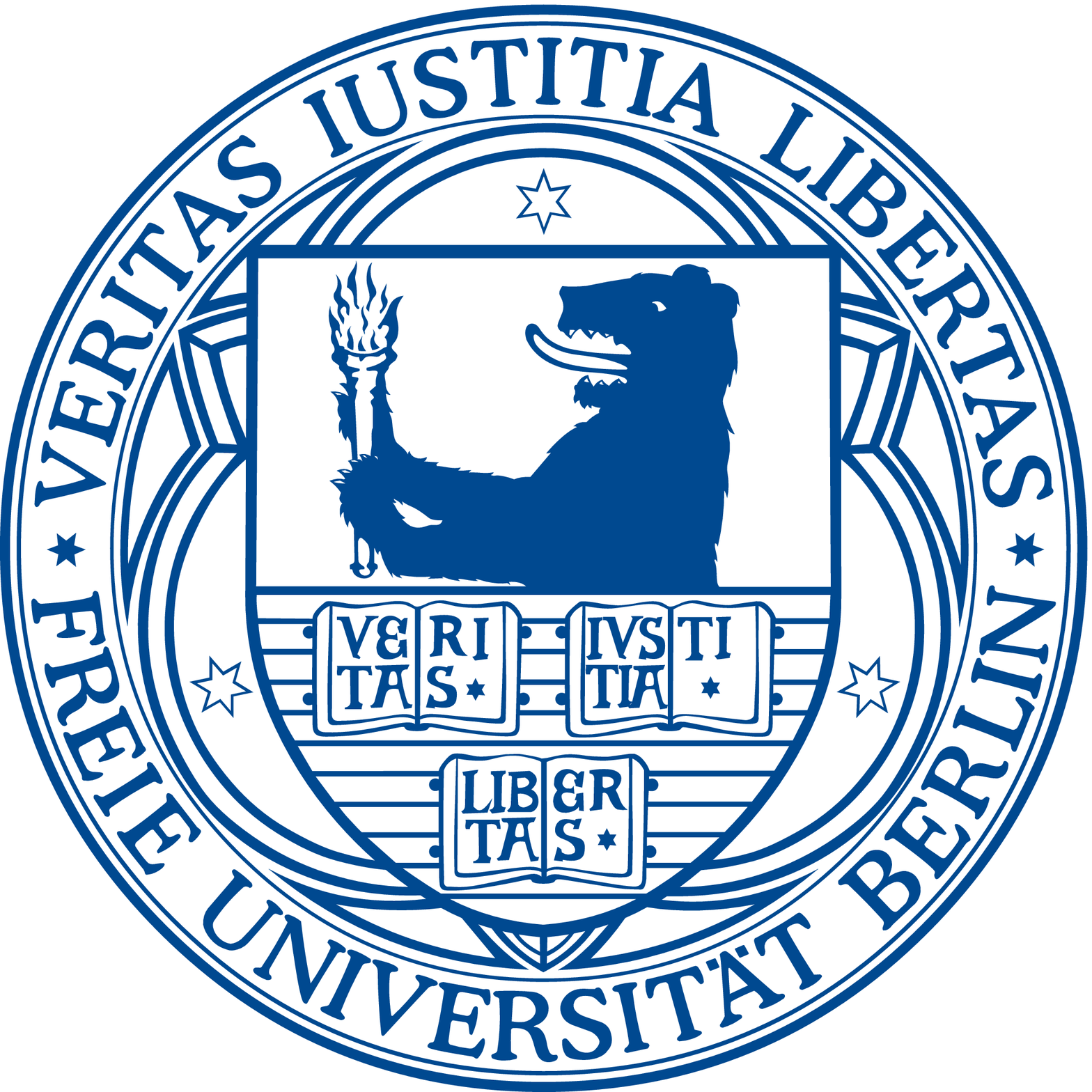}\\[1cm]

    \textsc{\LARGE Freie Universit\"at Berlin\\[.2cm] Department of Physics}\\[1.5cm]
 
    \textsc{\Large DIPLOMA THESIS}\\[1.5cm]

    \hrule
    \vspace{0.4cm}
    { \huge \bfseries Laser-induced Femtosecond Spin Dynamics in Metallic Multilayers}\\[0.4cm]
    \hrule
    \vspace{1.5cm}

    \begin{minipage}{0.4\textwidth}
      \begin{flushleft} \large
        \emph{Author:}\\
        Adrian \textsc{Glaubitz}
        \footnotesize{glaubitz@physik.fu-berlin.de}
      \end{flushleft}
    \end{minipage}
    \begin{minipage}{0.4\textwidth}
      \begin{flushright} \large
        \emph{Supervisor:} \\
        Priv. Doz. Dr. Uwe \textsc{Bovensiepen}
      \end{flushright}
    \end{minipage}

    \vfill
    February 12, 2010 \\
    First revised edition.
  \end{center}
\end{adjustwidth}
\end{titlepage}


\tableofcontents

\cleardoublepage 
\phantomsection  
\addcontentsline{toc}{chapter}{List of Figures} 
\listoffigures

\cleardoublepage 
\phantomsection  
\addcontentsline{toc}{chapter}{List of Tables} 
\listoftables

\mainmatter

\chaptermark{}
\noindent
Diese Arbeit wurde am Fachbereich Physik der Freien Universit\"at Berlin in der Arbeitsgruppe von Prof. Dr. Martin Wolf unter der Anleitung von Priv. Doz. Dr. Uwe Bovensiepen in der Zeit vom 01.~Februar 2009 bis zum 01.~Februar 2010 angefertigt. 

\vspace{3cm}
\noindent
Ich versichere dass die vorliegende Diplomarbeit selbstst\"andig und ohne unerlaubte Hilfe angefertigt wurde.

\vspace{3cm}
\noindent
Datum\hspace{42mm}Unterschrift


\chapter{Introduction}

Magnetism has been known since the ancient times as a phenomena of nature. Lodestones are
natural magnets and are essentially minerals of iron and several of its oxides. They were
already referred to in books dated 400 BC, in China. The name ``magnet'' actually comes
from lodestones found in Magnesia, Greece. A mariner's compass is probably the best-known
and oldest application of magnetic phenomena, it was first mentioned around 1100 AD in
Chinese literature. The easiest construction consists of a magnetized iron needle floating in water.

Despite its long history and scientific interest in magnetic phenomena dating back at least to
William Gilbert in the 1600s, magnetism had not been understood so well like electrostatics.
Its laws were long time unveiled and it was less well understood due to one basic difference
to electrodynamics: \emph{There are no magnetic mono-poles}. It was not until the advent of
electric currents that magnetism could be fully explored. In 1819 \O rsted discovered by accident
that a current-conducting wire deflected the needle of a compass and about one year later,
the two french physicists Jean-Baptiste Biot and Felix Savart deduced the \emph{Biot-Savart-Law}
which relates magnetic fields to electric currents as their sources. At the same time
Michael Faraday conducted experiments to build a device that can be considered to be the first
electric motor. Later he discovered electromagnetic induction and even predicted that
light and electro-magnetism are closely related but he could not provide a mathematical formula
nor prove the relationship with light. It was the mathematician James Clerk Maxwell who
eventually delivered the final mathematical framework now called \emph{Maxwell Equations}.
They describe the behavior of electric and magnetic fields in the presence of electric
charges and currents, also incorporating the laws discovered by Biot and Savart, which can
be deduced from Maxwell's equations.


From electrostatics we know that the basic entity is a single charge (\emph{mono-pole}) whereas in
magneto-statics it is a magnetic \emph{dipole}. Think of it as a small compass needle on the atomic
scale which aligns itself in an external magnetic field. In a solid, the total magnetization
is defined as the amount of magnetic moment per volume, so the more dipoles are aligned
into one direction the stronger the magnetization along this axis will be. To generate such
an external field in order to induce a change of magnetization of the solid, one could simply
drive a current through a wire according to the Biot-Savart-Law and expose the solid to it.
Other ways are mechanical impact or thermal heating, both increase the disorder of all
dipoles and thus reduce magnetization. A rather new and sophisticated method of changing
magnetization is the method of \emph{spin injection} where a change is induced by ``injecting''
electrons into a solid. Due to their intrinsic angular momentum (\emph{spin}) electrons
have a magnetic moment and thus are able to act on the magnetic moments in the solid
(spin-dependent scattering).

Those spin dynamics induced by electronic excitations have potential applications in magnetic
recording with a dramatic speed increase over conventional magnetization techniques. In our
project the electronic excitation is performed optically, with a femtosecond laser in a SHG\footnote{
SHG = Second Harmonic Generation, a method which uses the optical second harmonic signal for surface
sensitive probing.} pump-probe scheme, the excited electrons are called \emph{hot electrons}\footnote{
Hot electrons or hot carriers in general are carriers (electrons or holes) which are not
in thermal equilibrium with the lattice and can move from one region to another without
an externally applied bias.}. Such an optical excitation can occur on a very short time scale which
is limited only by the duration of the laser pulse. Magnetization can thus be changed within
pico- to femtosecond lapses and one is able to overcome the lower barrier of time scale
of $10^{-9}$ of induction with the classical method using an external field. Our model
setup of such a spin injection or spin transfer experiment are layers of gold which
has a large ballistic mean free path and iron as an itinerant ferromagnet.

Spin polarized electrons excited in one ferromagnetic layer travel through a noble metal layer
layer and the scattering in the second ferromagnetic layer results in a spin transfer torque
which eventually changes the net direction of its magnetization. The Fe/Au/Fe-films are grown
epitaxially on transparent, dual-side polished substrates of Magnesium-Oxide (MgO$(001)$) in an
MBE vacuum chamber. The excitation of hot carriers is realized through pumping through the
backside of the transparent substrate, the detection through front-probing at the metal-film surface.

Before the final goal of changing the magnetization over spin transfer torque can be achieved,
we investigate on the proper preparation of our model system, i.e. the samples. Since the
metallic layers are films which are grown epitaxially on the MgO$(001)$ substrate, several
parameters can be tuned during preparation to change the properties of the films and
thus the measurable output signal. For instance, the optimal thicknesses of ferromagnetic
and the gold layer are yet undetermined. The first is important to maximize the output of
spin polarized electrons and the latter for making the gold layer as thick as possible
as long the spin polarized electrons can be detected at the gold surface\footnote{For both
the ``source'' ferromagnetic layer and the gold layer we want to maximize the thickness. A
thick enough source will maximize the output of spin-polarized electrons but keeping it
thin enough that most carriers excited are ballistic. A thick enough gold layer will
decouple the itinerant ferromagnets, assuring that transfer of magnetic
momentum is performed through the ballistic electrons and not some kind of exchange interaction.}
A well established method for determining such thicknesses is by growing the films in wedge-shapes.
The wedge form allows to perform the pump-probe measurements with different film thicknesses by
simply sweeping with the laser spot along the wedge. Growing a gradient shape is
certainly more difficult than simply growing flat films but this way we can perform
all measurements on one sample rather than having to prepare several substrates, all
with different thicknesses. This makes the measurements more consistent, as only the
the sample will be slightly moved with respect to  the laser beam, the rest of the setup
remaining unchanged.

During the experiments it has shown that the roughness and the parameters of the evaporation
process have much more impact on the results obtained in the optical measurements than we
had predicted before. In general, a higher roughness of the film surface may yield higher
SHG signals due to the field enhancement in localized plasmon modes. We therefore
performed experiments to analyze the samples produced to investigate into that direction
as well.


\section{Outline of this thesis}

This thesis is divided into five chapters.

In chapter two we start with the current state of the art in the research field of spin dynamics
of low-dimensional ferromagnetic structures, the preliminary work of this group and the objectives
and goals of the further research and what is covered by this thesis. This includes the use
of a femtosecond laser as an excitation source as well a probe with high time resolution to
detect the magnetization with \emph{Second Harmonic Generation}. To induce a magnetization
in the ferromagnetic films, an electromagnet with dual-nested Helmholtz coils, in which the
sample resides during experiments, was designed and constructed as well as a special sample holder.

Chapter three covers all steps and methods which were involved to produce the samples
used in the measurements. It is explained how the surface quality of the thin films was achieved and
verified\footnote{Impurities and rough film-surfaces impair the excitation transport of the
ballistic electrons through scattering}. Methods for quality analysis include STM-, RHEED-
and AES-measurements\footnote{STM: Scanning tunneling microscope, RHEED: Reflection high-energy
electron diffraction, AES: Auger electron spectroscopy}. The substrates were prepared with ultrasonic
cleansing in organic solvents, degassing and annealing under ultra-vacuum conditions. We also
investigated the influence of the substrate temperature on the sample surface corrugations.
The construction and use of the the shutter for producing the samples with gradient thickness
(\emph{wedge}) is discussed.


The actual measurements and results of the key experiments are shown and discussed in chapter four.

Finally a discussion and conclusion of the whole thesis is given in chapter five. An outlook towards
possible future experiments with the setup are outlined. The future goal of the experiments
is to be able to induce a change of magnetization of a second ferromagnetic (i.e. iron) film on
top of the gold film (\emph{Spin transfer}). For this, the results from our measurements are
hopefully going to be helpful to grow films with the optimum thicknesses to maximize the efficiency of excitation
of the ballistic electrons as well their transfer through the non-ferromagnetic gold film. These
studies may have high impact on the fundamental studies of spin-dependant electron-electron
scattering and transport.






\chapter{Methods \& Materials}

Modern information technologies have gone through a rapid development in the past 30 years.
While the first personal computers started with volatile memory sizes in the order
of a few kilobytes and non-volatile magnetic memory in the order of a few hundreds kilobytes,
concurrent personal computers feature volatile memories with several gigabytes ($10^{9}$ Bytes)
and non-volatile magnetic storage devices with up to terabytes ($10^{12}$ Bytes). This dramatic
increase has been achieved mainly through the boost of data density made possible by smaller
structures in solid-state semiconductors for the volatile and the discovery of the \emph{Giant Magneto
Resistance} and \emph{Tunneling Magneto Resistance} for the non-volatile magnetic devices.

Even though technology slowly shifts from classical magnetic non-volatile storage devices to
flash memory \footnote{``Flash memory'' refers to non-volatile random access memory devices,
which have floating gate transistors within their memory cells. The transistors can be permanently
activated/deactivated by applying a program/erase bias voltage at the gate to fill/empty
electrons from/in the gate. }, the former are still favored over the latter in most personal computers
as they offer a much better cost/data capacity ratio. Furthermore, the overall capacity
of flash memory is limited by a lower data density as compared to magnetic storage.
The research on fast and high-density non-volatile magnetic memory is therefore
still in focus of solid-state physics, more precisely the field of spin
dynamics in low-dimensional ferromagnetic structures \cite{hillebrands_spindynamics1,
hillebrands_spindynamics2,hillebrands_spindynamics3}.

\section{Spin transfer torque}


Current state-of-the-art magnetic storage devices employ several advanced techniques
which have been developed in the past twenty years to increase data density. Two
of them are the already mentioned GMR \cite{gruenberg_gmr, baibich_gmr} and TMR effects
which allow to replace the coils as detectors for surface magnetization by
small solid-state devices to allow measurement of much smaller magnetizations,
the spatial dimensions of the device being reduced as well. To increase the data density
on the storage medium, i.e. the disk, material scientists have developed new
ferromagnetic coatings\footnote{The platters themselves are made from non-magnetic
materials like aluminum or glass. The data is stored on thin magnetic
films, made from iron oxides or cobalt alloys.} which allow smaller magnetic
domains on the disk. A very recent approach to increase data density further is the technique
of \emph{perpendicular recording} \cite{gao_perpendicular} which uses materials with
higher coercivity\footnote{The coercivity of a ferromagnet determines the necessary
strength of an applied field to fully demagnetize it.} with the help of an additional magnetically
soft layer which couples to the field of the write head to increase the effective
field. Higher coercivity of the material then allows to reduce domain sizes
and therefore increasing data density.

However, non of these techniques tackle the problem of faster storage directly.
All speed improvements originate from the fact that data density increases and
thus more data can be read or written within the same time the read/write-head
flies over the spinning disks. A direct approach has to address the method
how magnetization is changed in the ferromagnetic storage layer. Currently
all magnetic and magneto-optic storage devices use coils to generate
magnetic fields to change magnetization. Since the coils are driven
by electronic circuits, the speed of magnetic switching is limited
by the cut-off frequency of the electronic components. Fastest switching
in electronic circuits that can be achieved with current technologies
is in the nanosecond regime. In \cite{tudosa_magnetization}, fastest
magnetic switching with conventional pulses is stated to have
100 ps pulses at best. However, in chapter 15 \cite{stoehr_mangnetism},
St\"ohr elaborates, that faster switching is possible when using
femtosecond lasers.

\emph{Spin transfer torque} is a completely new concept to induce magnetization
changes in ferromagnets \cite{slonczewski_current}, \cite{ralph_spintransfer}.
Like GMR, the fundamental physics behind it is spin-dependent scattering.
In GMR, electrons are filtered by their angular momentum (due to
spin-dependent scattering) passing through one of the ferromagnetic
layers of a \emph{spin valve} structure and tunnel through a non-magnetic interlayer into a
second ferromagnetic layer. Electrons can only pass when both
ferromagnetic layers have parallel magnetization. Now, if
we change the setup such that the second layer is rather magnetically
soft and its magnetization is non-parallel to that one in the first layer,
the electrons from the first layer change the magnetization in the second layer
by transfer of angular momentum. The magnetization in the ferromagnet
exerts a torque on this flow of angular momentum and the flow generates
an opposite torque onto the ferromagnets magnetization so that both
the flowing spins and the magnetization in the ferromagnet finally
reach parallel alignment.

\begin{figure}
  \subfigure[Initial magnetization]{\includegraphics[width=0.5\textwidth]{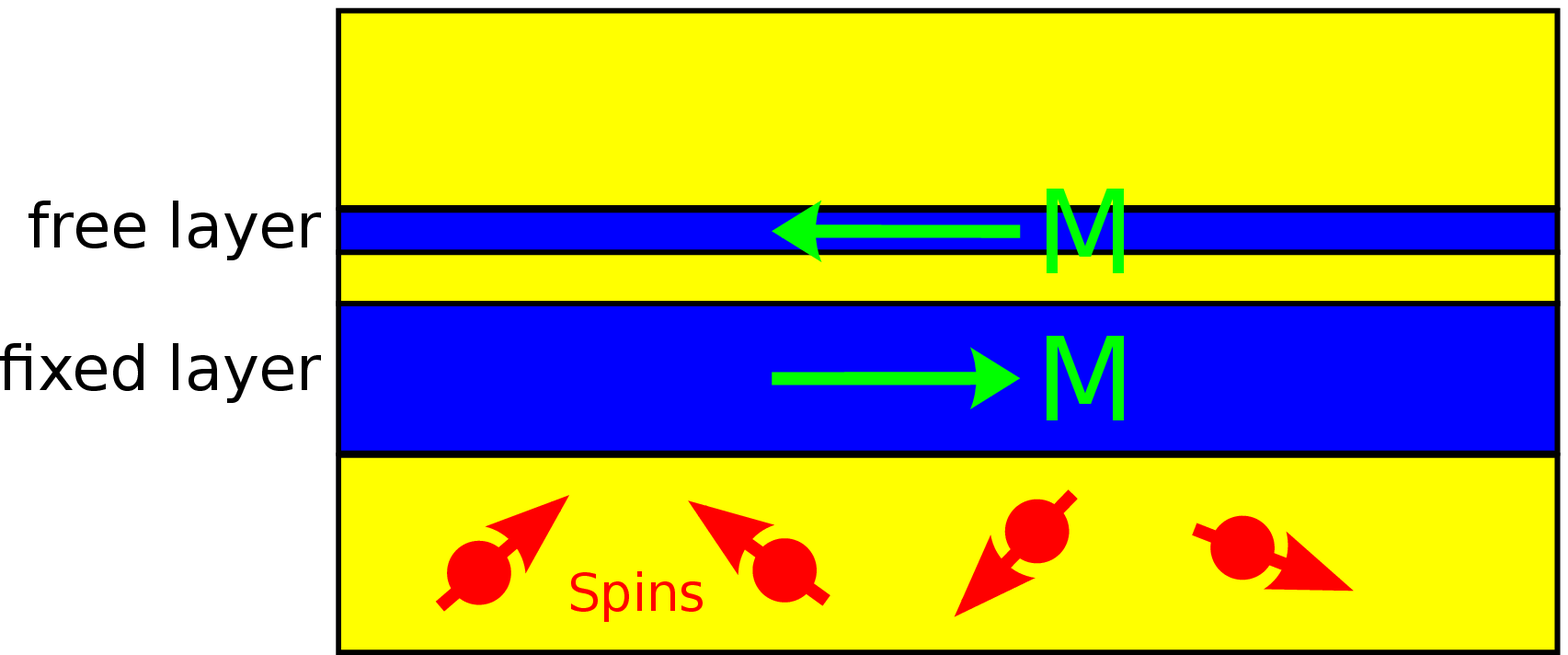}}
  \hfill
  \subfigure[Current is applied and is spin-filtered]{\includegraphics[width=0.5\textwidth]{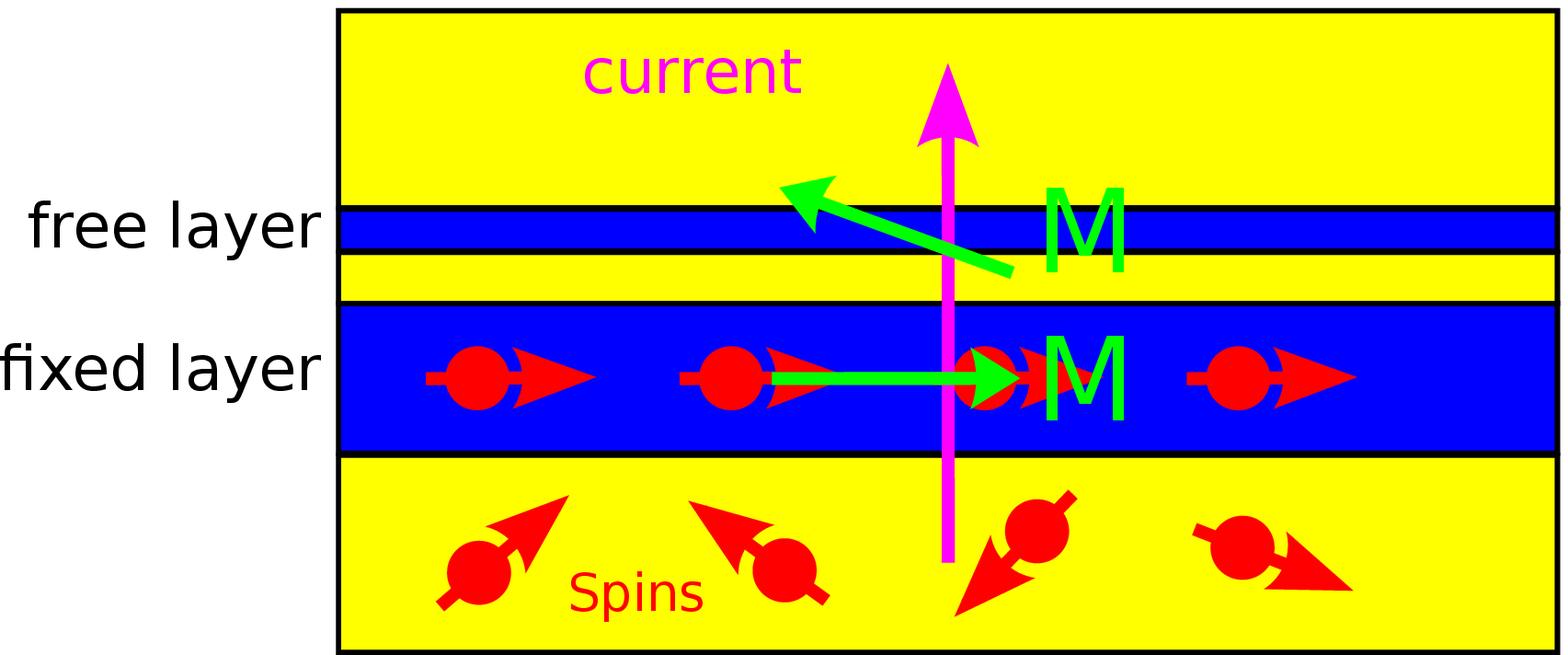}}
  \vfill
  \subfigure[The spin-filtered electrons flip $\vec{M}$]{\includegraphics[width=0.5\textwidth]{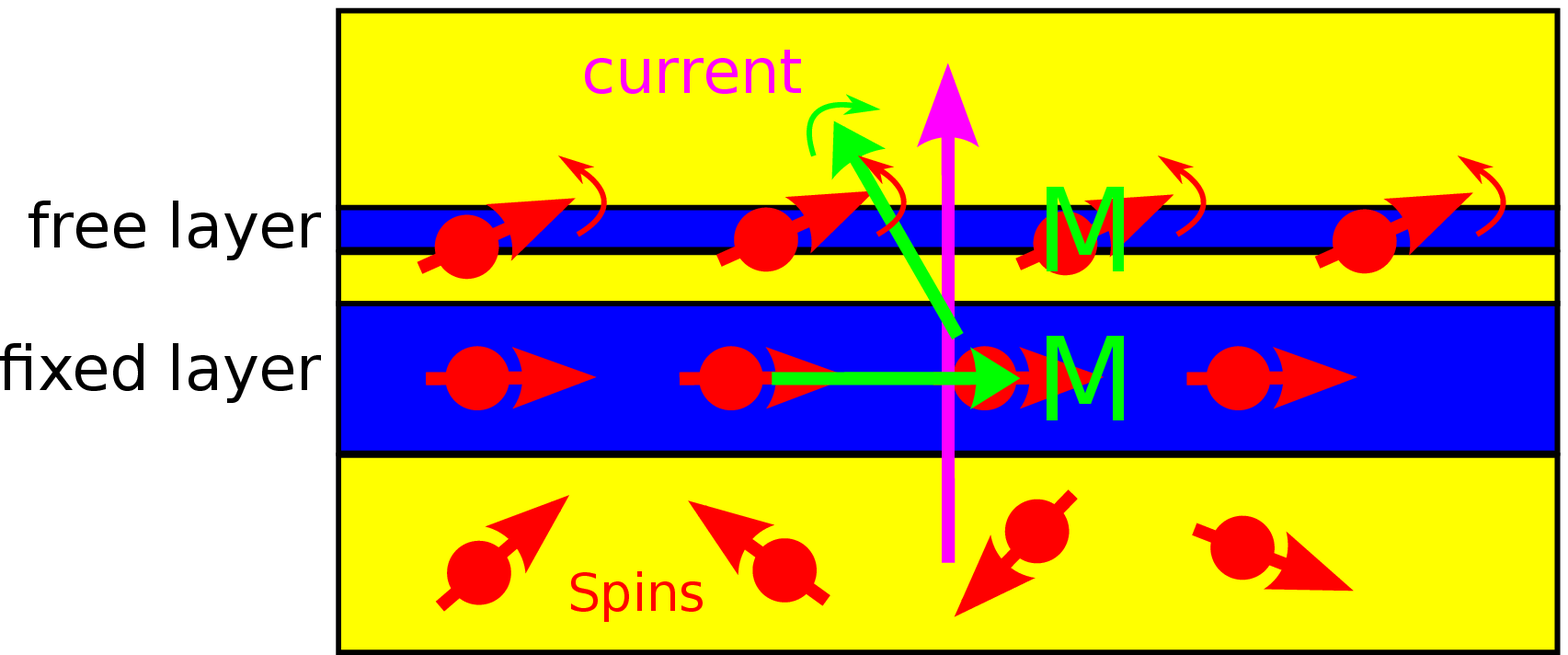}}
  \hfill
  \subfigure[After removing the current]{\includegraphics[width=0.5\textwidth]{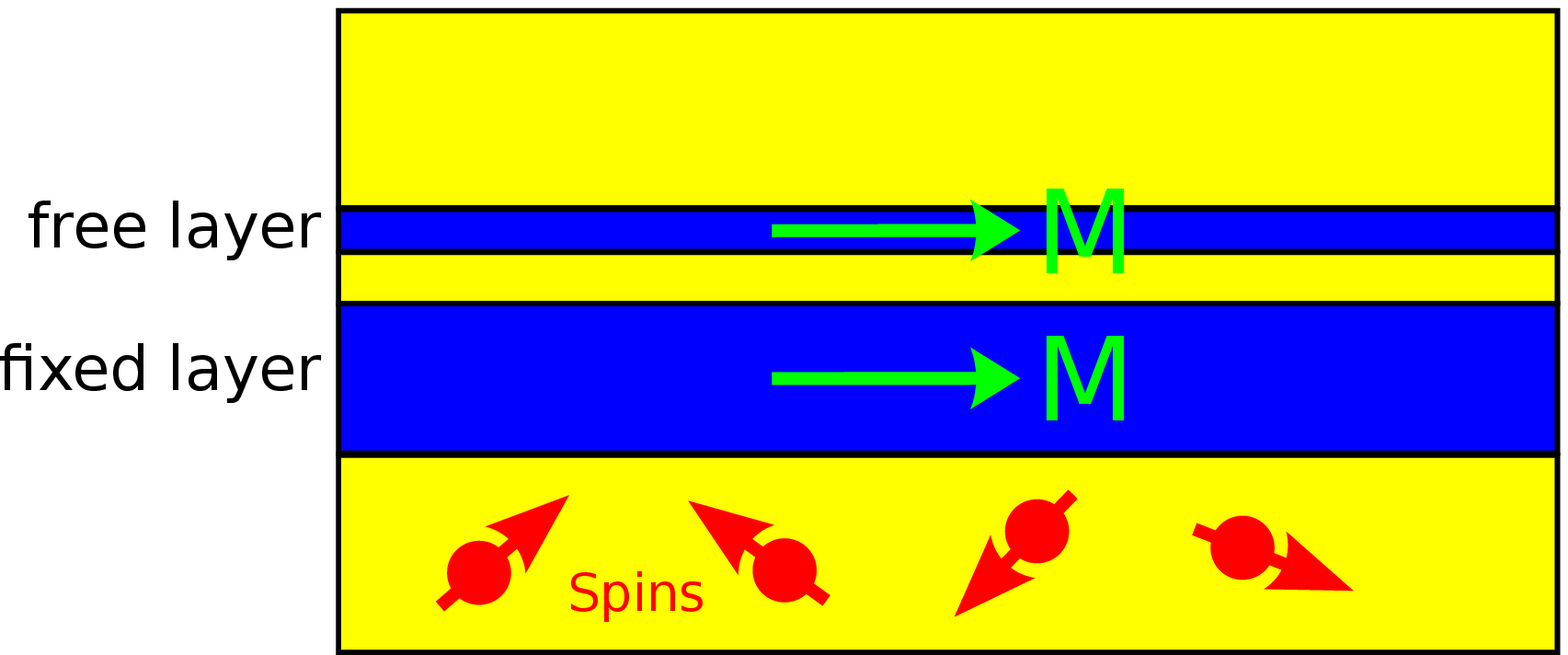}}
 \caption[Electrically induced spin transfer]{A nano pillar multi-layer setup to induce spin transfer
   with an electric current. Electrons from the lower copper layer have initially disordered
   spins, the magnetizations in the two ferromagnetic layers are anti-parallel (a). Due to
   fluctuations the magnetization in the free layer ceases to have an anti-parallel orientation
   with the first magnetic layer. Upon applying an electric current, electrons from the lowest
   copper layer flow through the \emph{fixed layer} where they are spin-filtered to pass
   through another non-magnetic layer which separates the two magnetic ones (b). Once the spin-filtered
   electrons pass through the \emph{free layer}, spin torque is exerted to the non-parallel magnetization
   in that layer resulting which aligns it parallel to the fixed layer (c). After the current is removed,
   magnetization, the free layer can remain in the switched state (d).}
 \label{fig:pointcontact}
\end{figure}

From this rather simple model it is obvious how a device for spin
transfer has to be constructed. It is a spin-valve structure consisting of two
magnetic layers separated by a non-magnetic one. Additional itinerant
conducting layers have to be provided as source and drain for the
electrons. The electrons are spin-filtered by the first magnetic layer, travel
through the non-magnetic layer to finally pass through the second
magnetic layer to exert an angular momentum, finally they are collected
by the drain layer. Figure \ref{fig:pointcontact} illustrates such
a setup which was actually put into practice in \cite{tsoi_excitation,
myers_current}.

\begin{figure}
  \subfigure[Setup for spin transfer experiment]{\includegraphics[width=0.5\textwidth]{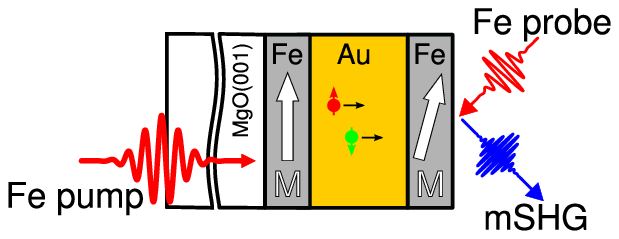}}
  \hfill
  \subfigure[Setup for investigation of spin transport]{\includegraphics[width=0.5\textwidth]{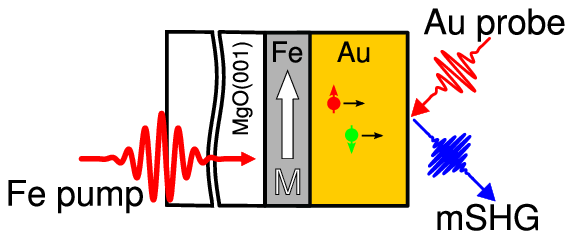}}
 \caption[Optically induced spin transfer]{A substrate of MgO(001) is the basis of the model setup that
   we investigate. With molecular beam epitaxy techniques, layers of iron and gold are deposited
   onto the substrate. Electrons are optically excited in the first ferromagnet and transfer their spin
   to a second ferromagnet which is separated by a paramagnetic spacer (a). To investigate spin transport in gold, we waive
   for the second ferromagnet and probe the magnetization at the gold surface (b).}
 \label{fig:principalscheme}
\end{figure}

However, such setups still have the disadvantage that the switching of
magnetization occurs electrically and thus we are still confined regarding
the speed of switching. This is the point where lasers come to the rescue.
If we are able to use pulsed lasers with pulse-lengths in the femtosecond
regime to excite spin-polarized electrons in a ferromagnetic layer,
let them pass through a non-magnetic layer and finally exert spin torque
in a second magnetic layer, we will achieve switching of magnetization at unprecedented speeds.
In fact, we have conceived such a setup which uses multi-layers
of iron as ferromagnets and gold as non-magnetic spacer layer. Figure
\ref{fig:principalscheme} (a) shows the principal setup of our concept.

Key tool in our experiments is a custom femtosecond laser system which provides ultrashort
and high-energy pulses necessary to excite spin-polarized electrons with sufficient
energy (\emph{hot electrons}). The setup allows time-resolved linear (MOKE) and non-linear
(SHG) optical spectroscopy. The technique behind the measurements is a pump-probe
scheme where the laser beam is split into a high-intensity ($80~$\% of beam power) pump
and low-intensity probe beam ($20~$\%). Time resolving is carried out with a delay
stage which varies the two beams relative path lengths on a picosecond time scale.
Detailed description of the experimental setup is found in section
\ref{sec:exprealisation}.

\section{Specific goals of this thesis}

Before the final goal of this project, the optical switching of magnetization can be
carried out, detailed investigations of the electron and spin dynamics involved
have to be performed. For this, the model setup is altered by waiving for the
second ferromagnetic layer (see Fig. \ref{fig:principalscheme}, (b)). This allows to measure
both the charge and the spin components of the carrier transport with
surface sensitive methods. Furthermore, the proper technique for producing samples has to be chosen by means
of measurements of film quality (see chapter \ref{chapter:samples}). The goals of
this thesis within the projects frame were both the production and
analysis of the samples and further investigations of the spin and electron dynamics.
This includes the following tasks:

\begin{itemize}
  \item construction of a shutter for controlled sample preparation
  \item construction of a magnet allowing both MOKE and SHG measurements
  \item development of techniques for sample preparation and of
        characterization procedures
  \item further investigation of electron and spin dynamics
\end{itemize}

\section{Experimental techniques}

Our all optical approach for exciting and probing the spin dynamics
in the specimen implements several linear and non-linear optical effects.
The latter ones are more important for us. 
In particular, we are focussing on the following effects:

\begin{itemize}
  \item Linear reflectivity to measure bulk electron and lattice dynamics
  \item Magneto optical Kerr effect (MOKE) to measure bulk spin dynamics
  \item Second Harmonic Generation (SHG) to measure electron, lattice and
        spin dynamics at surfaces and interfaces
\end{itemize}

\subsection{Linear Reflectivity}
\label{sec:linearref}
As already mentioned, we are using a pump-probe scheme to excite and
investigate the spin dynamics in our samples. The main concept behind
this technique is to excite the medium with a high power pump pulse
and probe the dynamics after a certain delay after excitation
with a much weaker probing pulse. Usually, the two beams are
obtained by separating the laser beam with a beam splitter.
When the delay between pump and probe is zero, however,
the situation changes and new effects arise. With zero
delay, pump and probe are superimposed and the excitation
depends of the coherent superposition of the beams.

\begin{figure}
\begin{center}
  \includegraphics[width=0.7\textwidth]{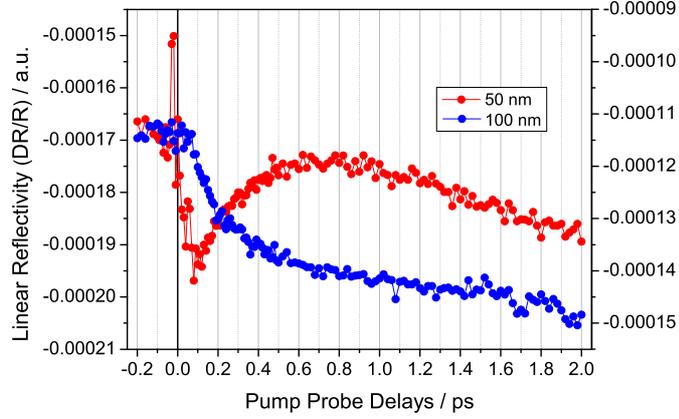}
\end{center}
 \caption[Sample plots for linear reflectivity]{Sample plots for linear reflectivity
  showing the linear reflectivity for 50 and 100 nm gold layers. The peak at zero
  delay arises from the spatial and temporal overlap of pump and probe
  and can hence be used as a calibration for zero delay.}
 \label{fig:linearrefplot}
\end{figure}

As we are primarily concerned with the pump-induced
changes in the second harmonic and not the fundamental signal,
we are not further discussing this topic here, more details
can be found in \cite{misochko_coherent}. However, the coherent
artifact becomes handy in our experimental setup. Since the
generation of the coherent artifact arises from the spatial
and temporal overlap of pump and probe beams, it can be used
to determine the \emph{zero} for the delay stage. Since the linear
reflectivity is measured with a separate photo diode instead
of the photo multiplier we have an independent mean to calibrate the
zero delay of the pump-probe setup since the delay $t = 0$ can be easily
distinguished by peaks arising from the coherent artifact
(see Fig. \ref{fig:linearrefplot}).

\subsection{MOKE}
\label{sec:moke}
Magneto-optical Kerr effect (MOKE) describes a phenomenon which allows optical detection
of magnetization. The basic effect is the interaction of polarized light with 
medium due to magnetization. The polarization of an incident light
beam is altered upon reflection. For transmission the principally same effect is referred
to as Faraday effect.

\medskip

To understand these effects on the macroscopic scale, we consider the dielectric constant
of a material as a tensor rather than a scalar:

\begin{equation}
  \epsilon (\omega) = \left(
    \begin{array}{ccc}
      \epsilon_{xx} & 0 & 0 \\
      0            & \epsilon_{yy} & 0 \\
      0            & 0            & \epsilon_{zz}
    \end{array}
  \right)
\end{equation}

For an optically isotropic medium, $\epsilon_{xx} = \epsilon_{yy} = \epsilon_{zz}$ (thus
$\epsilon$ is a scalar). If the magnetization is non-zero, the off-diagonal elements
will become non-vanishing. From \cite{zvezdin_magnetooptics}, the tensor for
a magnetized medium becomes:

\begin{equation}
  \epsilon (\omega) = \left(
    \begin{array}{ccc}
      \epsilon_{xx} & \epsilon_{xy} & -\epsilon_{xz} \\
     -\epsilon_{xy} & \epsilon_{xx} & \epsilon_{yz}  \\
      \epsilon_{xz} & -\epsilon_{yz} & \epsilon_{xx}
    \end{array}
  \right)
\end{equation}

and thus the medium is no longer isotropic.
\medskip
Macroscopically, MOKE can be explained as follows: We consider the incident light beam
(linear polarized wave) as a superposition of two circular polarized waves with opposite
rotation direction but equal amplitude. When the electromagnetic wave is reflected
from a medium with an anisotropic dielectric constant, the two superposing waves
experience different refractive indices and thus different velocities within
the medium. This results in a phase shift between the two constituents
and finally in rotation of the linear polarization plane. If the magnetization
is perpendicular to the the surface (polar MOKE) for example, the rotation due to the
Kerr effect is proportional to the magnetization $\mathbf{M}$ which is
proportional to the ratio of the components of the dielectric tensor:

\begin{equation}
  \Theta_K \propto \frac{\epsilon_{xy}}{\epsilon_{xx}} \propto \mathbf{M}
\end{equation}

\begin{figure}
\begin{center}
  \includegraphics[width=1.\textwidth]{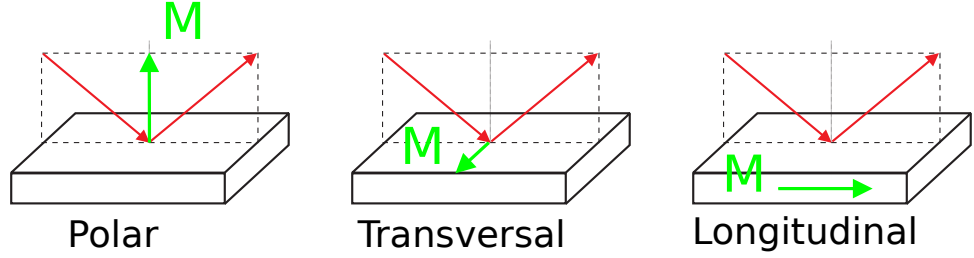}
\end{center}
 \caption[Possible configurations of MOKE]{Depending on the orientation
   of the magnetization in the specimen with respect to the incident
   light beam, one talks about \emph{polar}, \emph{transversal} or
   \emph{longitudinal mode}.}
 \label{fig:moke}
\end{figure}

Depending on the geometry, there are three configurations for MOKE (see
Figure \ref{fig:moke}): Either with in-plane magnetization (transversal
and longitudinal MOKE) or out-of-plane magnetization (lying within the plane of incidence, that is
along the surface normal: polar MOKE). To fully understand MOKE on the microscopic
scale beyond the given phenomenological explanation, quantum mechanics will play
the primary part. It is the combined effect of spin-orbit coupling and exchange
interaction with the selection rules for optical transitions which yield a proper
in-depth explanation of MOKE \cite{bennemann_nonlinear}.

Linear MOKE is generally bulk sensitive. For surface sensitive magnetization
measurements, the non-linear optical variant of MOKE has to be applied, also known as
magneto-induced second harmonic generation (mSHG) which will be discussed in Section
\ref{sec:shgvsmshg}. Moreover, linear MOKE is less sensitive since the off-diagonal
tensor elements which define the change of polarization and thus magnetization
of the sample, are usually very small. For mSHG, the even and odd tensor components
(with respect to reversal of magnetization, see Sec. \ref{sec:shg}) are of
comparable magnitude and therefore yield much high effects than linear MOKE.
As an example, in \cite{bennemann_nonlinear} it is shown for Ni that with
linear MOKE, the ratio of even and odd tensor components is around
0.03 whereas it is 0.27 for non-linear magneto-optics.

\subsection{Linear vs. non-linear optics}
\label{sec:linearvsnonlinear}
Non-linear optics hasn't evolved before the invention of the first lasers
in the 1960s since it requires high intensity fields to produce a detectable
non-linear optical response. Non-linear effects occur, for example, when focussing a
high-intensity ultrashort laser pulse onto a specimen which will
react to the irradiation with an \emph{optical response}. We consider
an incoming electric field $\mathbf{E}(\mathbf{r},t)$ and the specimen as a medium in
which a polarization $\mathbf{P}(\mathbf{r},t)$ is induced. The corresponding
non-homogeneous wave equation for the propagation of $\mathbf{E}$ through
a medium is derived from Maxwell's equations \cite{shen_nonlinear}:

\begin{equation}
\bigtriangleup \mathbf{E}(\mathbf{r},t) + \frac{1}{c^2}\frac{\partial^2}{\partial t^2}\mathbf{E}(\mathbf{r},t) = - \frac{4 \pi}{c^2}\frac{\partial^2}{\partial t^2}\mathbf{P}(\mathbf{r},t)
\end{equation}

This wave equation describes the optical response of the medium. This response
perturbs the further propagation so that a mutual dependence exists,
which results in the various linear and non-linear optical effects. For
linear effects, the polarization is usually expressed as follows:

\begin{equation}
  \mathbf{P}(\mathbf{r},t) = \epsilon_0 \int^{+t}_{-\infty} d t' \int^{+\infty}_{-\infty} d^3r' \chi^{(1)} (\mathbf{r} - \mathbf{r}', t - t') \mathbf{E} (\mathbf{r}',t')
\end{equation}

with $\epsilon_0$ being vacuum permittivity and $\chi^{(1)}$ the first-order susceptibility tensor which
accounts for the fact that the medium might be anisotropic, i.e. the response field might not be parallel
to the electric field. This linear expression is valid as long as the intensity of the irradiating
electric field $\mathbf{E} (\mathbf{r},t)$ is not too high. For high intensities, non-linear effects
will arise and the above expression has to be expanded in series of the electric field to account
for the non-linear terms, thus for the n-th order:

\begin{eqnarray}
  \mathbf{P}^{(n)}(\mathbf{r},t) & = & \epsilon_0 \int^{+t}_{- \infty} d t_1..d t_n \int^{+\infty}_{-\infty} d^3r_1..d^3r_n \times \\ \nonumber
                                & &\chi^{(n)} (\mathbf{r} - \mathbf{r}_1, t - t_1,..,\mathbf{r} - \mathbf{r}_n, t - t_n) \mathbf{E} (\mathbf{r}_1,t_1).. \mathbf{E}(\mathbf{r}_n,t_n)
\end{eqnarray}

and the total polarization has to be summed over all $n$:

 \begin{equation}
   \mathbf{P}(\mathbf{r},t) = \sum_n \mathbf{P}^{(n)} (\mathbf{r}, t)
\end{equation}




First order terms correspond to linear, all higher terms to non-linear effects. Since we
are dealing with linear and second harmonic effects only, we neglect terms of order
3 and higher. We calculate the Fourier transformation of to yield polarization in
frequency domain, restricting us to \emph{Sum Frequency Generation} (i.e. only
considering sums of the composing frequencies):

\begin{eqnarray}
  \mathbf{P}(\omega) & = & \mathbf{P}^{(1)}(\omega) + \mathbf{P}^{(2)}(\omega_1 + \omega_2) + ... \label{eqn:ptaylor2_1} \\
                  & = & \epsilon_0 [\chi^{(1)} (\omega) \mathbf{E}(\omega) + \chi^{(2)} (\omega_1 + \omega_2) \mathbf{E}(\omega_1) \mathbf{E}(\omega_2) + ... ]  \label{eqn:ptaylor2_2} 
\end{eqnarray}

For the case of $\omega_1 = \omega_2$ we are dealing with \emph{Second Harmonic Generation} which is
underlying principal physics of the main tool for our measurements, see section \ref{sec:shg}.

\subsection{Second Harmonic Generation}
\label{sec:shg}
Second Harmonic Generation (SHG) is a special case of sum frequency generation (SFG) where two photons unite
to generate a new photon with the energy being the sum of the energies of the single
photons. SHG is the degenerate case of SFG in the sense that the frequencies of the two
photons are identical, thus $\omega_1 = \omega_2$. From \ref{eqn:ptaylor2_1} and \ref{eqn:ptaylor2_2},
the polarization for SHG becomes in electric-dipole approximation:

\begin{equation}
  \mathbf{P}^{(2)}(\omega_1 + \omega_2 = 2\omega) = \epsilon_0 \chi^{(2)} (2\omega) \mathbf{E}(\omega)\mathbf{E}(\omega)
\end{equation}

or for just one component:

\begin{equation}
  P^{(2)}_i(2\omega) = \epsilon_0 \chi_{ijk}^{(2)} (2\omega) E_j(\omega)E_k(\omega)
\end{equation}

The interesting quantity in this context is the susceptibility
$\chi$ which is a third rank tensor with 27 components. However, since the frequencies of the two
contributing waves are identical, we have a symmetry regarding the permutations of $j$ and $k$
and thus the tensor $\chi$ reduces the amount of its non-vanishing components to
18 since $\chi_{ijk}^{(2)} = \chi_{ikj}^{(2)}$. The amount of components is reduced further
when choosing a particular crystal symmetry, polarization geometry or by choosing a
particular crystallographic orientation in the optical setup.

\medskip

It remains to clarify the importance of SHG for surface investigations. SHG has a very high sensitivity
to the properties of surfaces and interfaces of centrosymmetric media such as their symmetry structure,
charge distribution and magnetization and so on. This specific sensitivity of SHG can be understood
easily, let's consider equation \ref{eqn:ptaylor2_2}. In case of a centrosymmetric crystal, \ref{eqn:ptaylor2_2}
has to be invariant against inversion symmetry. Thus exchanging $\mathbf{E}$ and $\mathbf{P}$ for
$-\mathbf{E}$ and $-\mathbf{P}$ should yield the same mathematical result, thus \ref{eqn:ptaylor2_2}:

\begin{equation}
  \mathbf{P} =  \epsilon_0 [\chi^{(1)} (\omega) \mathbf{E}(\omega) + \chi^{(2)} (2\omega) \mathbf{E}(\omega) \mathbf{E}(\omega) + ... ] 
\end{equation}

becomes:

\begin{equation}
  -\mathbf{P} =  \epsilon_0 [- \chi^{(1)} (\omega) \mathbf{E}(\omega) + \chi^{(2)} (2\omega) \mathbf{E}(\omega) \mathbf{E}(\omega) + ... ] 
\end{equation}

However, both equations can be valid simultaneously only if all even orders of $\chi$ are zero. Therefore
in a centrosymmetric medium, the contributions to SHG are \emph{zero}. At surfaces, however, the
inversion symmetry of centrosymmetric crystals is broken and the odd components of contributions
to polarization become non-zero.

\medskip

\emph{Within electric-dipole approximation, for centrosymmetric media, SHG processes are
sensitive to surfaces and other interfaces where inversion symmetry is broken.}

\subsection{Non-magnetic vs. magnetic SHG}
\label{sec:shgvsmshg}
Upon irradiating the medium with a high-intensity laser pulse, a SHG signal
is generated which is used to probe electronic and magnetic properties. This
is possible since the electric field at second harmonic is compound of
two signals:

\begin{itemize}
  \item a non-magnetic or ``crystallographic'' component which reflects changes
        of the band structure and the electronic distribution function; both are
        monitors for electron and lattice dynamics since vibrations and lattice
        deformations are reflected there
  \item a magneto-induced component which is, accordingly to all known experimental
        results, proportional to the local magnetization $\mathbf{M}$
\end{itemize}

We refer to the two components as \emph{even} and \emph{odd} component, forming
the total electric field at $2 \omega$. Even and odd here refer
to the symmetry of the components regarding inversion of the external magnetic
field: the even component remains unchanged while the odd component
flips its sign.

\medskip

We write down the components of the SHG electric field:

\begin{equation}
  \mathbf{E} (2 \omega) = \mathbf{E}_{even}(2 \omega) + \mathbf{E}_{odd}(2 \omega)
\end{equation}

In our optical experiment, we measure intensities rather than electric
fields, which can be expressed as:

\begin{eqnarray}
  I^{\uparrow \downarrow}(2 \omega) & = & \left|\mathbf{E}_{even} (2 \omega) + \mathbf{E}_{odd} (2 \omega)\right|^2 \nonumber \\
              & = & E^2_{even} (2 \omega) + E^2_{odd} (2 \omega) \pm 2 E_{even} (2 \omega) E_{odd} (2 \omega) \cos \varphi
\label{eqn:shgintensity1}
\end{eqnarray}

Here $I^{\uparrow \downarrow}(2 \omega)$ refers to the direction of the saturating external magnetic field which
determines the sign of the cross term (thus $\pm 2 E_{even} (2 \omega) E_{odd} (2 \omega) \cos \varphi$).
With the experimentally proved approximation $\left| \mathbf{E}_{even} \right| \gg \left| \mathbf{E}_{odd} \right|$,
we find an expression which involves the magnetization:

\begin{eqnarray}
  I^{\uparrow \downarrow}(2 \omega) & \propto & \left|\mathbf{E}_{even} (2 \omega) + \mathbf{E}_{odd} (2 \omega)\right|^2 \nonumber \\
                                  & \approx & \left|\vec{\beta}\right|^2 + 2 \vec{\alpha}\vec{\beta} M = k M + c
\label{eqn:shgintensity2}
\end{eqnarray}

\begin{figure}
\begin{center}
  \includegraphics[width=0.7\textwidth]{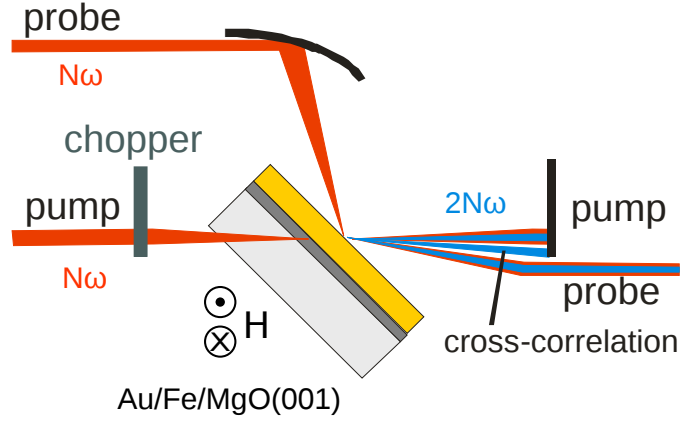}
\end{center}
 \caption[Principal experimental setup]{Principal experimental configuration of the
 pump-probe scheme. The pump beam hits the sample from the backside after passing through
 a chopper. The chopper is introduced to measure the pump-induced effects in the probe
 signal with low noise. The probe signal is measured from the front-side at the gold
 surface. The sample sits in a magnetic field which is switched in its direction
 back and forth for every delay step. The resulting SHG signals are pump,
 cross correlation and probe. Only the latter is detected,  the other two
being blocked by a knife. The probe beam passes through a monochromator
 and is detected with a photo-multiplier (not depicted).}
 \label{fig:experimentalsetup}
\end{figure}

Now, when we want to use SHG as a tool to measure \emph{pump-induced variations} (both non-magnetic and
magnetic) of the non-linear response under laser irradiation, we have to define the quantities for these components and then
derive them from the equations for the intensities (equations \ref{eqn:shgintensity1} and \ref{eqn:shgintensity2}).
Since we want to measure the variations \emph{time-resolved}, we have to measure intensities within
defined time intervals. The standard scheme for time-resolved measurements of pump-induced variations
of the SHG usually involves a modulation technique which uses a chopper in the pump beam working
at a frequency lower than the repetition rate of the laser but faster than any other experimental
variations. With the chopper, we can measure the probe signal both with and without excitation
by the pump, allowing to cancel noise with the help of a lock-in amplifier which is triggered by
the chopper (see Fig. \ref{fig:experimentalsetup} and Sec. \ref{sec:exprealisation}). The evolution
of the variations in time are measured by delaying pump and probe beam with respect to each other.
This is realized through a delay stage which varies the path length of the pump beam with high precision
allowing to set the delay between the two beams as they hit the sample.

For each data point on the time axis and thus position of the delay stage defining the time
delay between pump and probe, we measure the SHG intensity for opposite directions
of the external magnetic field $\mathbf{H}$ with a detector synchronized with the chopper
yielding the intensities in the presence $I^{\uparrow \downarrow} (t)$ and in the
absence $I^{\uparrow \downarrow}_0 (t)$ of excitation (the last quantity does not
depend on the pump-probe delay but may vary in time due to external fluctuations
of laser intensity, alignment, etc.). We combine the measured intensities into:

\begin{equation}
  D^{\pm} (t) = \frac{I^{\uparrow} (t) \pm I^{\downarrow} (t)}{I^{\uparrow}_0 (t) \pm I^{\downarrow}_0 (t)}
\end{equation}

It has been shown in \cite{radu_laser} that the phase $\varphi$ is smaller than 15$^\circ$ when measured in single-beam
interferometry and therefore we can use $\cos \varphi \approx 1$ in Equation \ref{eqn:shgintensity1}. Together
with $E_{odd}^2 \ll E_{even}^2$, the even and odd components, which describe
electron/lattice and spin dynamics respectively, become:

\begin{equation}
  \Delta_{even} (t) = \sqrt{D^+(t)} - 1 \approx \frac{E_{even} (t) - E_{even}^0}{E_{even}^0}
  \label{eqn:deltaeven}
\end{equation}

for the electronic variations (electron/lattice dynamics) and

\begin{equation}
  \Delta_{odd} (t) = \frac{D^- (t)}{\sqrt{D^+ (t)}} - 1 \approx \frac{E_{odd} (t) - E_{odd}^0}{E_{odd}^0} \approx \frac{M(t) - M_0}{M_0}
  \label{eqn:deltaodd}
\end{equation}

for the magnetic variations.

\medskip

The expressions \ref{eqn:deltaeven} and \ref{eqn:deltaodd} yield the non-magnetic and magnetic variations and are
thus our main tools for measuring pump-induced variations. However, when dealing with non-ferromagnetic media,
equation \ref{eqn:deltaodd} cannot be used to account for the magnetic variations since the material does
not exhibit any magnetization. In order to be still able to measure changes induced by magnetization
dynamics, we measure the \emph{magnetic contrast} or magnetic asymmetry. It is defined as the relative variations
of the SHG intensity for opposite directions of the external magnetic field. Thus:

\begin{equation}
  \rho = \frac{I^{\uparrow} (2 \omega) - I^{\downarrow} (2 \omega)}{I^{\uparrow} (2 \omega) + I^{\downarrow} (2 \omega)}
  \label{eqn:magncontrast}
\end{equation}

using Equation \ref{eqn:shgintensity1}, we can also write:

\begin{equation}
  \rho \approx 2 \frac{\left|E_{odd}\right|}{\left|E_{even}\right|} \cos \varphi
\end{equation}.

\begin{figure}
  \subfigure[Pump-induced effects]{\includegraphics[width=0.5\textwidth]{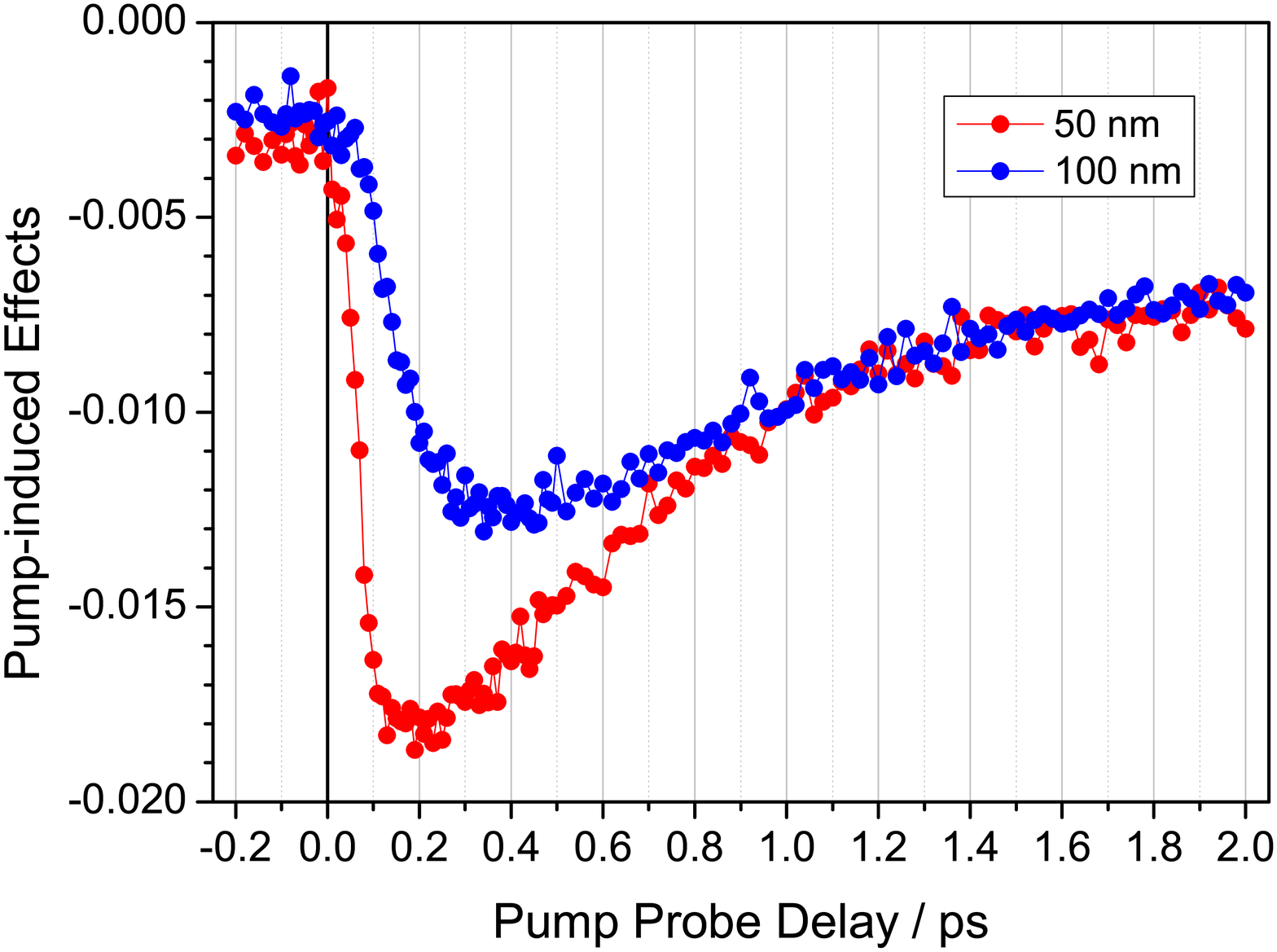}}
  \hfill
  \subfigure[Transient magnetic contrast]{\includegraphics[width=0.5\textwidth]{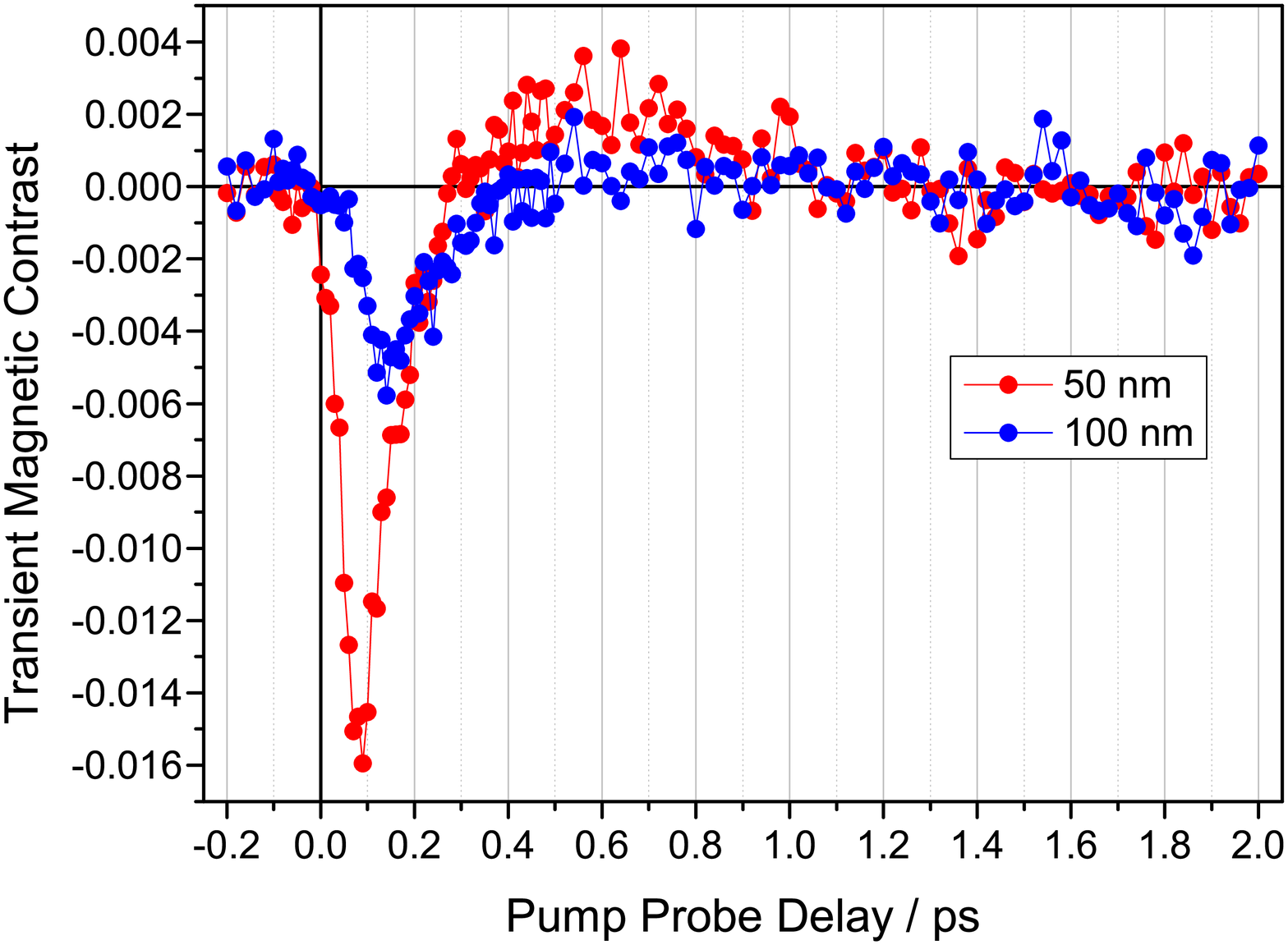}}
 \caption[Sample SHG measurements graphs]{Excerpt from SHG measurements on MgO(001)/Fe/Au multi-layers. Plot
 (a) shows the pump-induced effect of the SHG signal in dependence on the pump-probe delay for 50 and 100
 nanometer thick Au layers. The maximum amplitude indicates the arrival of the ballistic electrons at
 the gold surface. (b) shows the transient contrast over the delay. The first, sharp peak indicates
 the arrival of the ballistic electrons, the overshoot is caused by electrons in the diffusive transport
 regime.}
 \label{fig:sampleplots}
\end{figure}

Figure \ref{fig:sampleplots} shows typical result plots for the \emph{even component} (pump-induced effects)
and \emph{transient magnetic contrast} from our measurements. The pump-probe delay usually ranges between
$-0.5$ and $5.0~\text{ps}$ which covers the most interesting regions for the evolution of the
SHG signals\footnote{As we will see later, it turned out, that despite that there are other interesting
effects detectable when scanning with longer pump-probe delays.}. The relative signal changes
are usually within a percent of the electric field. This is enough to detect the ballistic electrons
and distinguish them from the diffusive electrons, the former are manifested in the first peak while
the latter induce a change of sign and an overshoot in the signal (Figure \ref{fig:sampleplots}).
The position of the peaks is mainly determined by the thickness of the films whose correlation
is one of the major points of interest to be investigated in this work. Accurate measurements
for different thicknesses will allow to determine the velocities of both ballistic and
diffusive transport in the metal films.

\section{Experimental Realization}
\label{sec:exprealisation}
The experimental setup consists of the following main components:

\begin{itemize}
  \item the femtosecond laser as excitation source and signal probe
  \item the optics on the laser table to split the laser beam into
        pump and probe, compensate dispersion, control pump-probe
        delay and a non-linear optical crystal (BBO) for SHG reference
  \item a home-made magnet in Helmholtz configuration to produce
        homogeneous fields in the sample plane (both vertically
        and horizontally)
  \item a photodiode to detect linear response and a photo multiplier
        with a monochromator for the detection of non-linear response
  \item a lock-in amplifier and a photon counter connected to the photo
        diode and photo multiplier respectively, both triggered
        by the chopper in the pump-beam
  \item a computer running a \emph{LabView} program code
        to control the delay stage and read and evaluate the
        outputs from the lock-in and photon counter
\end{itemize}

\begin{figure}
\begin{center}
  \includegraphics[width=\textwidth]{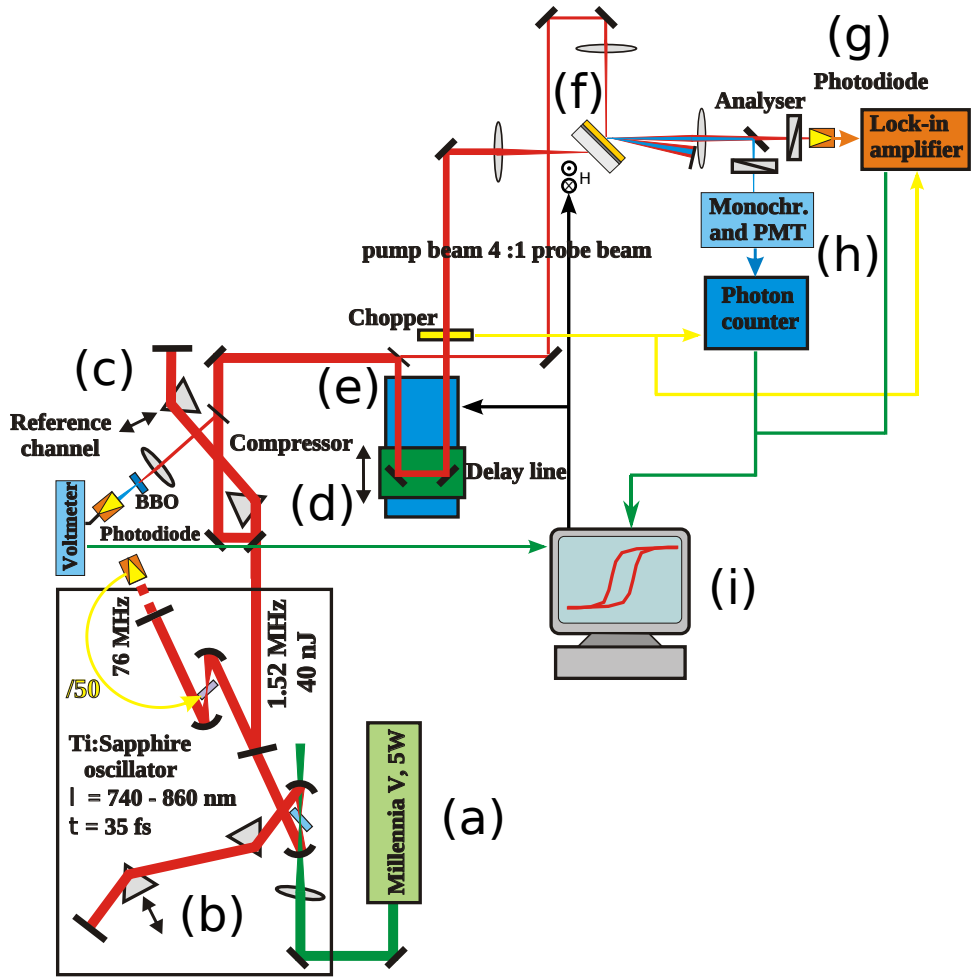}
\end{center}
 \caption[Schematic view of complete setup]{Overview of the complete setup for
   measurements. A 5W cw-laser (a) is used to pump a Ti:Sapphire crystal in the
   home-made femtosecond laser (b). The cavity is dumped at a rate of $1.52~\text{MHz}$.
   A reference channel (c) is established with a BBO crystal to check the SHG signal.
   To pre-compensate the positive group dispersion of the laser pulse due to the optical
   elements, two prism compressors are employed, one within the cavity box (b)
   and a second one outside the box (d). The pump-probe delay can be tuned with
   a delay line (e) which varies the length of the beam path of the pump beam,
   pump and probe are split right before the delay line, with a ratio of 4:1. The pump beam runs through a chopper
   triggering both the photon counter (h) and the lock-in amplifier (g).
   The sample is located within the magnet at position (f), where
   both pump and probe adjoin again and the generated SHG signal is filtered
   with a knife and a dichroic mirror reflecting the SHG signal towards the
   slit of the monochromator and transmitting the fundamental to the photo-diode.
   The measured signals are fed into a computer which runs the software \emph{LabView} (i).
   The computer controls both the delay line (e) and the magnet (f).}
 \label{fig:completesetup}
\end{figure}

A schematic view of the complete setup can be seen in Figure
\ref{fig:completesetup}. The caption of this figure contains a
detailed description of the interaction of the single components,
the most important of these are described in the following
sections.

\subsection{Femtosecond Laser}
\label{sec:femtosecondlaser}
The laser employed in our setup is a \emph{mode-locked} femtosecond laser system. This
means, that unlike a normal continuous wave laser which produces a time-steady signal,
the femtosecond laser generates single pulses. The pulses are necessary to achieve
the minimum time resolution to measure ultra-fast electronic and magnetic
phenomena whose durations lie within the pico- to femtosecond regime.

\medskip

Our laser is a home-made Titanium:sapphire (Ti:Al$_2$O$_3$) oscillator which provides
pulses as short as 35 fs\footnote{Since there are additional optical components involved,
without the cavity dumping, pulses could be as short as 25 fs \cite{conrad_phdthesis}).}.
The laser was built by a former PhD student of this group and has been employed in a
number of studies in surface science. It has two principal operating modes:

\begin{itemize}
  \item normal oscillator mode (76 MHz repetition rate)
  \item cavity-dumped mode (1.52 MHz repetition rate)
\end{itemize}

In normal oscillator mode (@76 MHz), the pulse energy is around 6 nJ. By enabling
cavity dumping, the pulse energy can be increased up to 42 nJ. The pulse duration
always remains 35 fs. Since cavity dumping is performed by electronic
switching (acousto-optical modulator, AOM), the dumping rate can be tuned
externally and thus one has variable repetition rates.

\begin{figure}
\begin{center}
  \includegraphics[width=1.\textwidth]{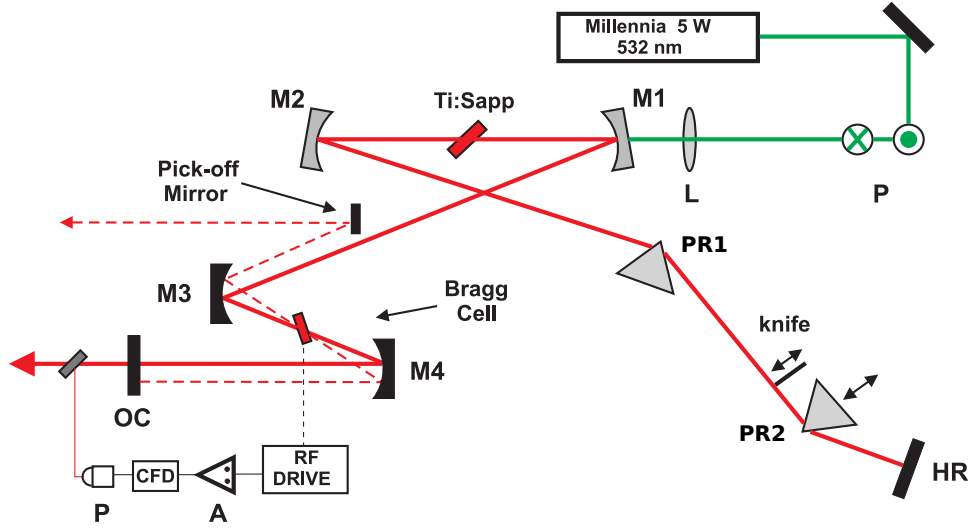}
\end{center}
 \caption[Configuration of the custom FS laser]{Configuration of the home-made laser
   employed in our setup. The oscillator system is pumped by a Nd:YVO$_3$ cw laser
   (Millennia). A periscope (\textbf{P}) rotates the plane of polarization of the beam parallel
   to the table, the lens (\textbf{L}) focusses it into the Ti:Sapp crystal which defines an oscillator
   together with the two mirrors \textbf{M1} and \textbf{M2}. The cavity dumping is performed
   by the Bragg-cell and mirrors \textbf{M3} and \textbf{M4}. The prisms \textbf{PR1} and
   \textbf{PR2} compensate negative group dispersion, the knife is used to tune the wavelength since it cuts
   parts of the spectrum. The main cavity is defined by the mirrors \textbf{OC} and \textbf{HR}.}
 \label{fig:laserscheme}
\end{figure}

\subsubsection{Mode locking}

Mode locking is the basic concept behind a pulsed laser system in the femtosecond
regime. Within the laser oscillator cavity of a \emph{continuous wave} laser (steady
output as opposed to pulsed lasers), there are only a few modes and they oscillate
independently from each other. In a mode-locking system, however, modes become
coupled with each other so they form a wave package with
high energy and large bandwidth instead of single continuous waves
with discrete, nearly vanishing bandwidths\footnote{In a cw laser the bandwidth
is only determined by the natural bandwidth and other bandwidth broadening effects
like the Doppler effect.}and small energies. To give some numbers: the mode-locked
Ti:Sa laser has up to 250000 modes and 10 THz bandwidth while the conventional He:Ne
laser yields only a couple of modes and a few GHz for the bandwidth.

\medskip

Mode locking occurs when a mode of frequency $\nu$ is modulated with the
frequency $f$ which results in additional modes in the sidebands
with frequencies $\nu + f$ and $\nu -f$. There are generally two
types of mode locking: \emph{active} and \emph{passive}. In active
mode locking, the laser oscillator is modulated externally
with an electronic device (e.g. piezo, loudspeaker, servo, etc).
Passive mode locking, on the other hand, takes place directly
in the cavity of the laser. An optically active crystal
works as an optical Kerr lens (\emph{optical Kerr effect})
and amplifies locked modes.

\medskip

The optical Kerr effect (not to be confused with the magneto-optical
Kerr effect) is a phenomenon of non-linear optics. What happens
is that the crystal changes it refractive index upon irradiation
with laser light of high intensity. The refractive index can thus
be written as:

\begin{equation}
  n(E) = n_0 + c E^2
\label{eqn:kerr}
\end{equation}

To initiate mode locking, one has to introduce a perturbation into the
system which can be easily achieved by knocking against one of the
end mirrors of the cavity\footnote{In fact, mode locking was discovered
by accident as one day one laser experimenter hit his laser setup
which initiated mode locking.}. This perturbation will start to
modulate the main mode of the cavity so that side bands arise. The more
sidebands become locked to the main mode, the higher will be the intensity
of this wave package generated. Provided that the intensity of the package is
high enough, it will change the refractive index of the crystal (see Eq. \ref{eqn:kerr})
so that the package is focussed and amplified even further. For wave
packages which do not have a large number of modes, the Kerr-focussing
is minimal and thus those packages aren't amplified further and therefore
discarded.
\subsubsection{Cavity dumping}
As already mentioned, cavity dumping is a technique used to increase the pulse
energy of the oscillator by almost a magnitude. The idea behind the concept of
cavity dumping is to let a wave package run within the oscillator until
it has accumulated enough energy. The dumping is performed by a Bragg
crystal where the laser beam is diffracted. This is realized
through acoustically induced optical grating which means that the refractive
index of the crystal is spatially modulated. The crystal is mounted on a
piezo which is fed by an external RF source (microprocessor-controlled
driving unit from \emph{APE Berlin} which controls the whole cavity
dumping). Thus the laser pulse is coupled out of the oscillator every
time the acoustic wave is turned on. The cavity dumping is feedback
controlled, this means that a fast photodiode is employed to synchronize
the driving unit with the repetition rate of the laser. For (fine) tuning,
the driving unit allows to set the \emph{phase}, \emph{time delay},
\emph{dumping rate} and \emph{power output for the RF signal}. Since only
every 50th pulse is dumped out of the cavity, the repetition rate
is reduced compared to normal oscillator mode (1.52 MHz vs. 76 MHz).
Other methods to increase the pulse energies are \emph{single pulse
amplifying} and \emph{regenerative amplifying}. The former allow
to keep the original repetition rate of the normal oscillator mode
but since each single pulse has only a limited time\footnote{In single
pulse amplifying systems, each single pulse runs through a second cavity
and has only the time span between two pulses for amplification.} to accumulate
energy, the pulse energies are lower compared to the cavity dumping
technique we employ. On the other hand, regenerative amplifying allows
high amplification of pulse energies with the caveat of a heavily
reduced repetition rates since the pulses remain in the amplifying cavity
for a long time before being dumped\footnote{In fact, regenerative amplifiers
stretch the pulses, couple them into a cavity, dump the pulse once the
pulse energies are high enough, then compress the pulses again. Thus it
is obvious why the repetition rate is reduced so much.}. We chose
\emph{cavity dumping} since we wanted to have increased pulse energies
to maximize the pump-induced effects (which are proportional to the
pulse energy of the pump beam) but not as high as we could
achieve with regenerative amplifiers since there would be risk
of ablation (thermal destruction of the films). Single pulse amplifying
would not yield high enough pulse energies and additionally
the high repetition rates would risk the quick destruction of the films
due to heat accumulation.

\subsection{Chopper and lock-in amplifier}

The chopper, photon counter and lock-in amplifier are essential
parts of our pump-probe setup. The chopper operates at a frequency
of approximately 450 Hz and modulates the pump beam and is necessary
for the measurement of the pump-induced signal.
\begin{figure}
\begin{center}
  \includegraphics[width=0.5\textwidth]{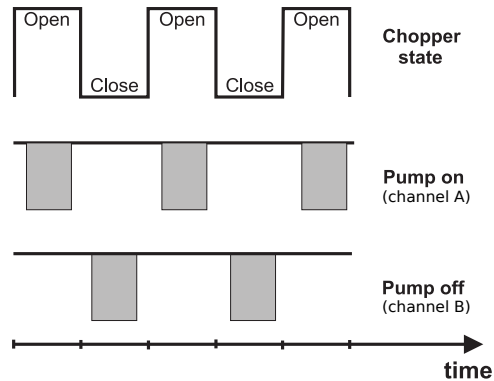}
\end{center}
 \caption[Explanation of the chopper signal]{Schematic view of the chopper states
   on the time axis (upper part). The two lower parts show the time frames when
   the signal is measured for pump on and off. These time frames are shorter than
   the states of the chopper when it's open or closed. This makes sure that the
   chopper is always fully open or closed respectively when measuring the probe
   signal. The divisions on the time scale correspond to $1/f_{chopper}$.}
 \label{fig:chopperscheme}
\end{figure}
The setup works as follows (see Fig. \ref{fig:chopperscheme}):
The chopper generates a rectangular output signal which is
synchronized to its operation frequency. This signal
is fed both to the photon counter (to measure the SHG signal) and
to the lock-in amplifier (to measure the fundamental signal).
The photon counter has two channels, one which counts the
photons for the chopper state \emph{open} and one for \emph{closed}. The
phase of the input signal from the chopper to the photon counter is adjusted
such that the photon counter channel for \emph{pump closed} (channel B) matches
the rectangular output signal \emph{low} and the channel for
\emph{pump open} (channel A) matches the \emph{high} signal. Thus the photon
counter is synchronized to the chopper and counts the states
\emph{chopper open} and \emph{chopper closed} in the proper channels.
\medskip
The pump-induced effects are then proportional to:
\begin{equation}
  \Delta_{even} \propto \sqrt{\frac{A}{B}} - 1
\end{equation}
To measure the fundamental, a photodiode is used which connects to a separate
lock-in amplifier over a home-made amplification stage which consists of
a fast (channel C) and slow (channel D) channel. The slow channel measures
the average signal of the photodiode while the fast channel measures the fluctuating
signal for chopper states \emph{open} and \emph{closed}. The former
is fed into an analog-digital converter and the latter into the lock-in
which suppresses the DC-part (average signal). The relative linear
reflectivity is then obtained with:
\medskip
\begin{equation}
  \frac{\Delta R}{R} \propto \frac{C}{D}
\end{equation}

\subsection{Sample holder and magnet}

\begin{figure}
\begin{center}
  \includegraphics[width=1.0\textwidth]{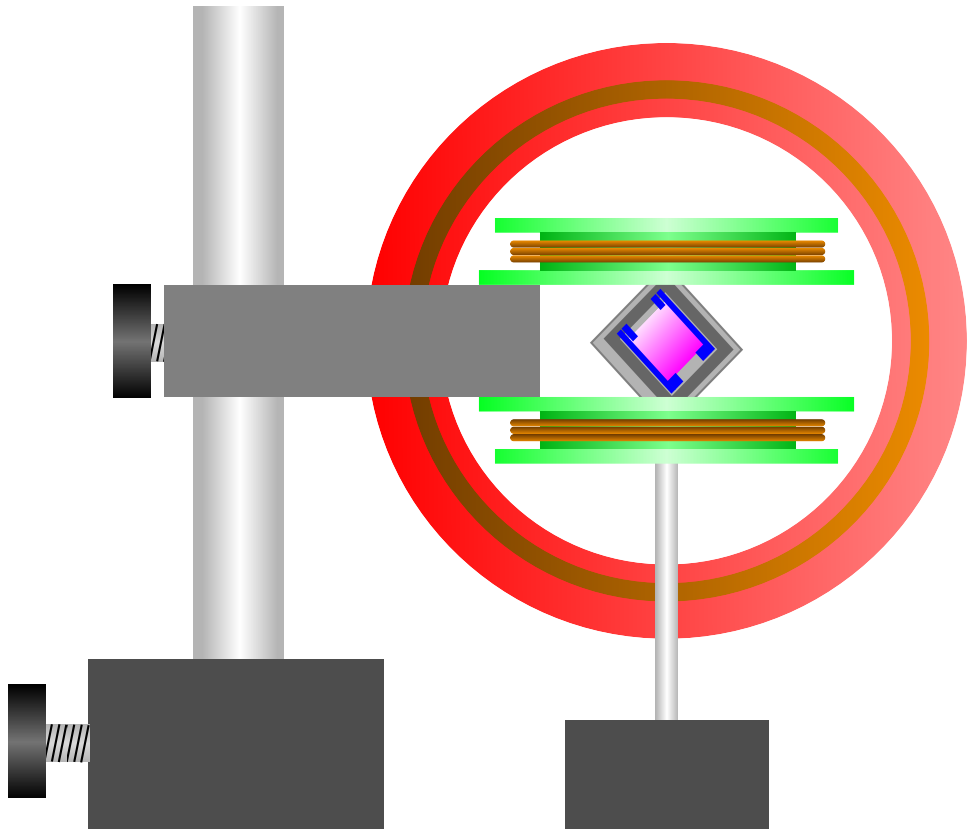}
\end{center}
 \caption[Setup of dual nested Helmholtz-coils]{Schematic view of the dual nested
 Helmholtz setup used to produce two perpendicular magnetic fields in the sample
 plane. The coil-reels are given in red and green color (outer and inner pair). The
 sample is mounted onto a plate which has an omicron-clamp mounted to it (see Fig.
 \ref{fig:omicron-plate}). The coils (orange) are held in fixed position, while the
 sample-holder can be translated along all three axes. Note that the sample-plane is
 actually rotated by 90 degrees with respect to the outer coils' plane. The sample itself
 is colored purple.}
 \label{fig:helmholtzmagnet}
\end{figure}

To position the sample in the laser beam-path, a custom sample holder was designed
(see fig. \ref{fig:samplemount}, appendix \ref{chapter:drawings}). It consists of
an aluminum holder which has a molybdenum clamp mounted to it. The design was
derived from the Omicron-brand sample holders which were used in the UHV chamber
for sample preparation and analysis (see fig. \ref{fig:omicron-plate},
chapter \ref{chapter:samples}).
\medskip
The sample itself is located inside two pairs of Helmholtz coils which produce a magnetic
field in the plane of the sample and thus the thin films (see fig. \ref{fig:helmholtzmagnet}).
This is the external field necessary to produce the magnetization in the ferromagnetic
films to yield spin-polarized electrons upon excitation. The Helmholtz configuration
yields homogeneous fields and was therefore chosen for these experiments. The idea
of using two nested Helmholtz pairs was to be able to produce magnetic fields both in transversal
and longitudinal geometry regarding the sample surface. As we will learn later, measuring
magnetic SHG requires the field to be transversal, measuring MOKE requires longitudinal
configuration.
\medskip
The coils are custom-made, they were designed using a CAD-software and then handcrafted by
the precision-engineering group of the department. Construction schematics can be
found in the appendix.

\medskip

\emph{All measurements were performed with fields high enough to saturate the magnetization
of the ferromagnetic films.}

\section{Calibration of the magnet}

Since the magnet is a home-made construction, before actually using it, we wanted to test whether
it would satisfy our requirements to produce the external magnetic field to induce magnetization
in the ferromagnetic layers of the samples. Furthermore, since we are driving the magnet
with remote-controlled power supply, we only know the current through the coils but not the
actual field. Thus, we have to calibrate the magnet to be able to match the field with
the current set by the power supply. The specifications of the magnet are listed in
Tab. \ref{tab:magnet}.

\begin{table}[htbp]
        \centering
                \begin{tabular}{|c|c|c|c|} \hline
                \textbf{Parameter}          & \textbf{Inner Coil Pair}   & \textbf{Outer Coil Pair}   \\ \hline
                Number of Turns             &        75                  &          120               \\ \hline
                Electrical Resistance       & $1.641 \pm 0.278~\Omega$   & $4.469 \pm 0.089~\Omega$   \\ \hline
                Radius                      & $0.038 \pm 0.002~\text{m}$ & $0.068 \pm 0.002~\text{m}$ \\ \hline
                \end{tabular}
        \caption{Specifications of the two home-made Helmholtz magnets}
        \label{tab:magnet}
\end{table}

We used a hall detector to measure the field of the operating magnet at various currents.
Then we calculated the theoretical fields for both coils in the Helmholtz configuration.
It is known from standard literature, that for a Helmholtz configuration the field along the
axis adjoining the coils of a Helmholtz pair, in the center is:

\begin{equation}
  B = \mu_0 \times \frac{8 \times I \times N}{\sqrt{125} \times R}
\label{eqn:biotsavart}
\end{equation}

with $\mu_0$ being the permeability of the vacuum, $N$ the number of turns per coil, $I$ the
current through the coils and $R$ the radius of the coils. In perfect Helmholtz configuration,
the spacing of the coils matches the radius of the coils, thus $R$.

\begin{figure}
  \subfigure[Inner coil]{\includegraphics[width=0.5\textwidth]{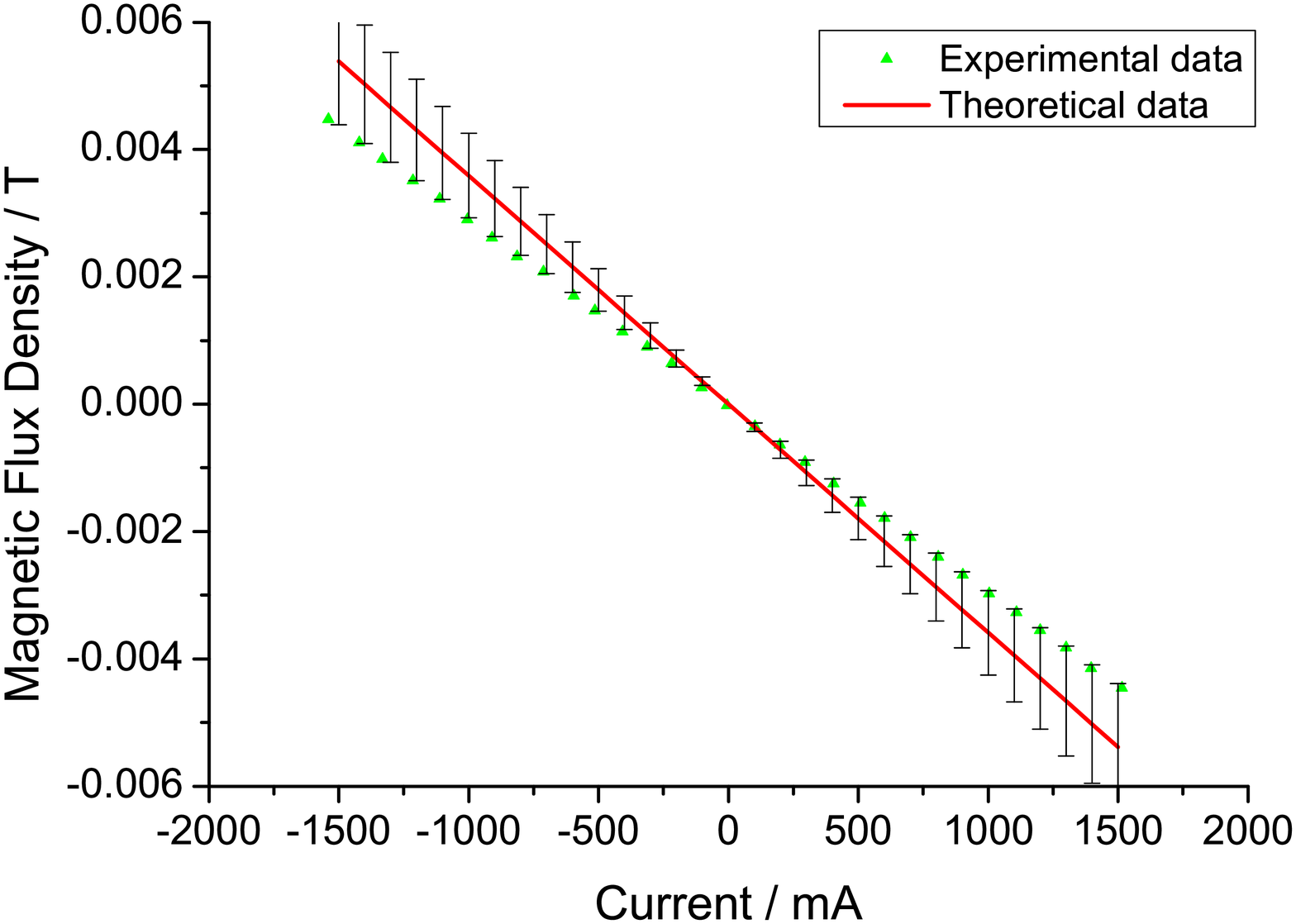}}
  \hfill
  \subfigure[Outer coil]{\includegraphics[width=0.5\textwidth]{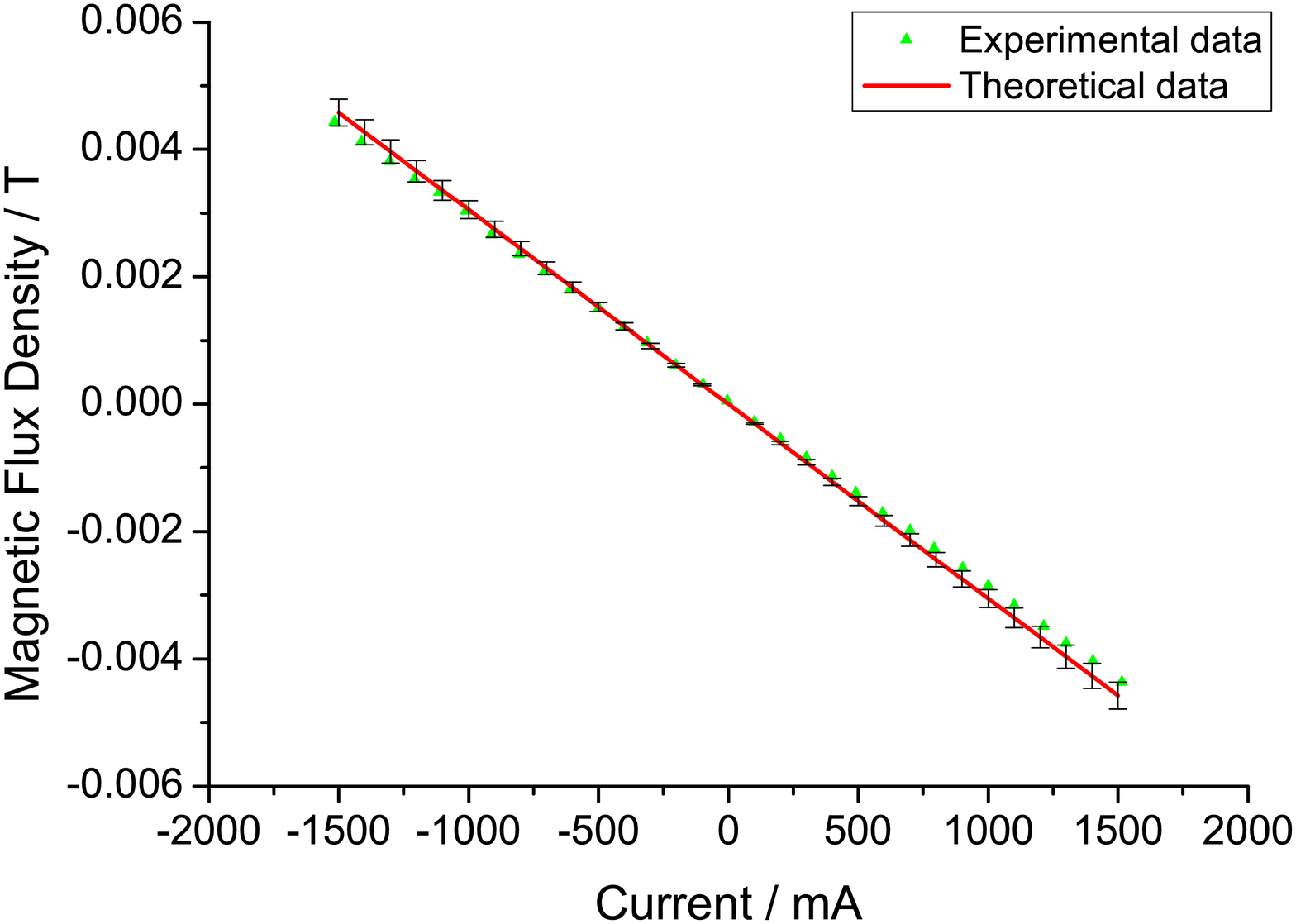}}
 \caption[Calibration of the magnet coils]{Comparison of the experimental data measured with a hall detector
   and theoretical data determined from the coils parameters (diameter, coil spacing, number of turns).
   Current varies from -1500 to 1500 mA.}
 \label{fig:magnetcalibration}
\end{figure}

The calibration data is shown in Figure \ref{fig:magnetcalibration}. The plots show that
the magnetic flux densities measured with the hall detector match nicely with the
theoretical values calculated from the coil parameters in Table \ref{tab:magnet}. Thus
the magnet is ready for the actual measurements and we are always able to determine
the exact flux density from the current displayed on the power supply.
\medskip
From the plots in Figure \ref{fig:magnetcalibration}, we read the following calibration:

\begin{itemize}
  \item Inner coils: 1 Ampere $\equiv$ 35 Gauss
  \item Outer coils: 1 Ampere $\equiv$ 30 Gauss
\end{itemize}

\subsection{LabView evaluation software}

\begin{figure}
\begin{center}
  \includegraphics[width=1.\textwidth]{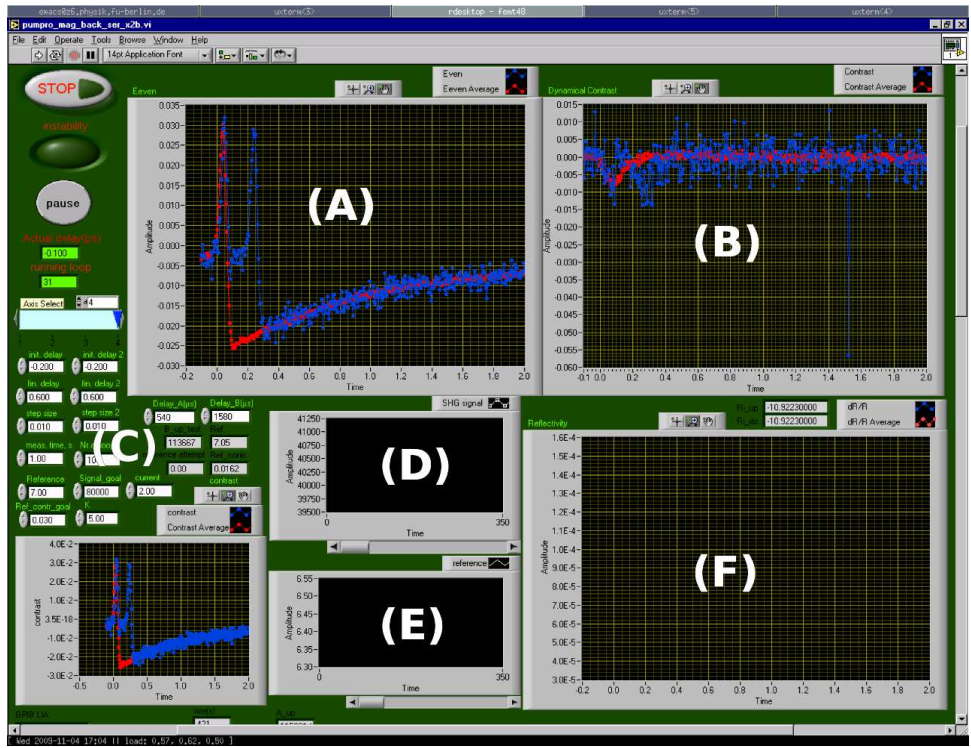}
\end{center}
 \caption[Screenshot of the LabView program]{Screenshot of the main program
   used for our SHG measurements. The pump-induced effects are measured
   with the graph instrument (\textbf{A}), the transient contrast with (\textbf{B}).
   All parameters (scan range and resolution for pump-probe delay, minimum signal,
   minimum contrast in the reference channel, current for the coils, measurement times per step and
   chopper parameters) can be set with the edit boxes in (\textbf{C}). The
   graphs in (\textbf{D}) and (\textbf{E}) measure the plain SHG intensity
   and the SHG signal in the reference channel (BBO crystal). Graph (\textbf{F})
   measures the linear reflectivity (see section \ref{sec:linearref}).}
 \label{fig:labviewshot}
\end{figure}

Since the measurement results are always subject to fluctuations due to
variations in the optical alignment, external influences (temperature,
humidity, etc), we are performing several iterations of the same scans
to reduce the statistical noise. As one single measurement can already take quite long to
finish\footnote{Scanning from $-0.2~$ps to $5.0~$ps delay with $0.01~$steps,
for example, will take around 1200 seconds or 20 minutes (1 second each for measuring at
each direction of the magnetic field and some time is required to move the
delay stage further), running 10 loops for the statistics will take about 3 hours.},
the measurement progress is completely automated and controlled with a program written in
\emph{LabView}. Thus long measurements can run completely unattended,
only occasional checks of the alignment are naturally necessary.
The program controls the delay stage, the magnet and keeps track
of all measured parameters in several virtual instruments on its
interface as seen in Figure \ref{fig:labviewshot}.

Furthermore, the software counter-checks the measured SHG signal with a reference
to make sure the optical alignment of the laser still yields output with
high enough intensities\footnote{Usually the phase of the cavity dumping
shifts away during longer measurements which results in reduced power.}. If the
SHG signal in the reference channel or in the actual measured SHG channel
is too low, the software discards that data point and tries to measure it
again. This makes sure that only data with proper alignment is recorded
and all changes in the signal are only due to real effects in the sample and
not due to some external fluctuations. Of course, there is some small tolerance
for variations of the output signal but those are covered by the statistics
and will not distort our data.

\medskip

Internally, the program also performs the necessary computations of the
input data so that the data stored on disk is ready for further analysis
with additional software like \emph{Origin} (i.e. pump-probe delay is calculated in picoseconds,
intensities are for the pump-induced effects and transient contrast are
calculated as relative changes, data is averaged, etc.).
Other custom Labview programs include some to determine the cross correlation
to setup the proper overlap and measure hysteresis loops of ferromagnetic
films.

\chapter{Sample Preparation}
\label{chapter:samples}

\begin{figure}
\begin{center}
  \includegraphics[width=1\textwidth]{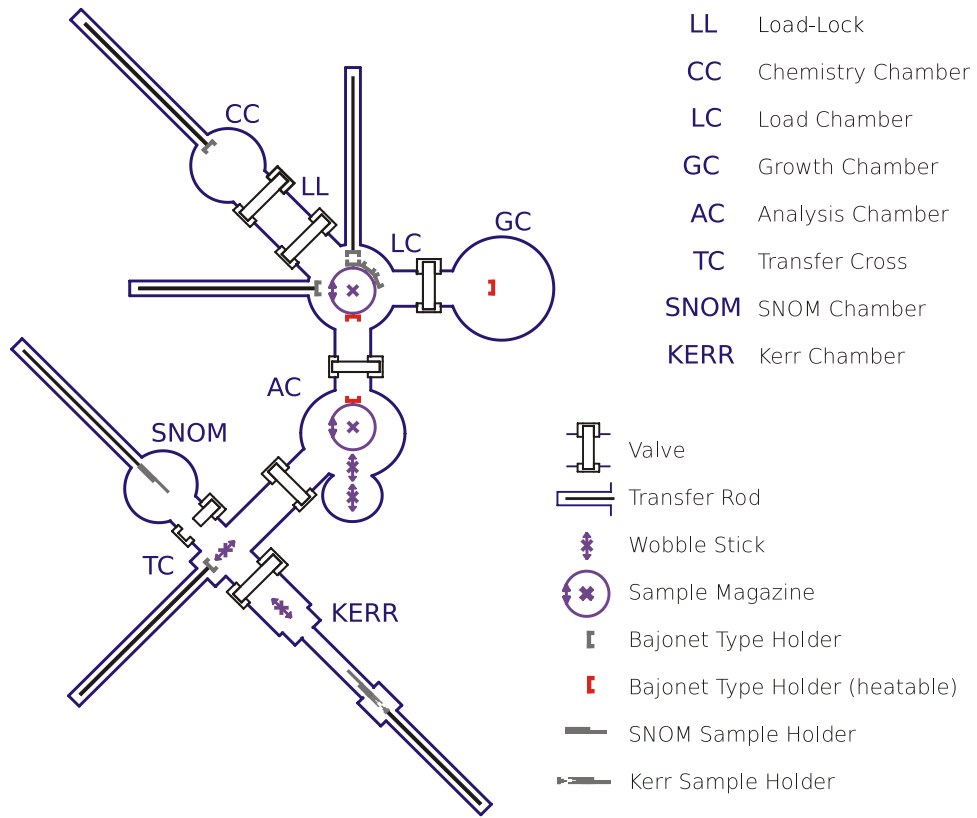}
\end{center}
 \caption[MBE Chamber]{Schematic view of the MBE chamber used for sample preparation
 and quality analysis. The fresh MgO substrates are transferred into through the load lock (LL),
 then heated and annealed in the load chamber (LC). For film preparation they are
 transferred into the growth chamber (GC). All quality measurements of the films (STM, AES)
 are made in the analysis chamber (AC). RHEED pictures are taken \emph{in-situ} in the
 growth chamber though. Figure courtesy diploma thesis Bj\"orn Lewitz \cite{lewitz_thesis}.}
 \label{fig:mbechamber}
\end{figure}

The quality of the samples has key importance for our experiments. Previous experiments in \cite{shen_shgintroduction}
have shown that the intensity of the SHG signal depends on the topography of the surface, rough
surfaces usually yield higher signals. On the contrary, impurities as well as high corrugations
may cause unwanted scattering of the laser beams which can deteriorate the shape of the signal
or impede finding the overlap in the pump-probe setup. Therefore we wanted to be able to control the
topology of the surface from the evaporation and with possible additional treatments like sputtering
to achieve surfaces which yield high intensity signals in the optical measurements with highest possible
reproducibility.
In order to achieve high purity, the samples were prepared in a ultra high vacuum (UHV, less than $10^{-9}$ Torr),
after the substrates had been cleaned in an ultra sonic bath with various organic solvents and water.
The whole process of preparing samples in UHV and depositing thin films using evaporators is usually
referred to as \emph{Molecular Beam Epitaxy} (MBE). Since our group does not possess such an MBE
chamber, the group of Professor Fumagalli kindly granted us access to their setup which is illustrated
in figure \ref{fig:mbechamber}. The setup is very typical and reflects the common setup described
in \cite{lueth_surfaces}.

Topography analysis is performed utilizing well developed and proven methods of surface
examination like reflection high-energy electron diffraction (RHEED) and scanning tunneling
microscopy (STM). While the RHEED analysis can be carried out \emph{in-situ}
(i.e. the measurements can be made during the evaporation process onto the substrate), the
examination of the samples with STM require the transfer of the prepared samples into the
analysis chamber (AC) where the STM is located together with other setups (AFM, LEED) for
\emph{ex-situ} surface analysis.

\section{Choice of substrates and film}

The choice of materials of the model system for our experiments was made for several reasons.
First, we needed a substrate which is transparent for the spectrum of our laser. Secondly
we chose a substrate on which iron grows epitaxially. Studying previous works in
\cite{muehge_structural}, \cite{fahsold_epitaxial} and \cite{demokritov_morphology}
we decided to use MgO$(001)$ substrates. Fe grows in the $[001]$-direction with the Fe $[100]$-axis
parallel to the MgO $[110]$-axis. As we will see later from our STM measurements, Fe shows
a very small surface and interface roughness as also noted by M\"uhge et al. in
\cite{muehge_structural}.

\medskip

The reason for the high quality of the Fe-films on MgO$(001)$ is that the lattice constants
of Fe along $[100]$ and $[001]$ match very nicely to that of MgO along the $[110]$-direction:

\begin{eqnarray}
  d_{\text{Fe}[100]} & = & 0.2866~\text{nm,} \nonumber \\
  d_{\text{MgO}[110]} & = & 0.29817~\text{nm.} \nonumber
\end{eqnarray}

This means that there is only low distortion of the Fe lattice, which was determined by
M\"uhge at al. to be the following:

\begin{eqnarray}
  d_{\text{Fe}[100]} & = & (0.2859 \pm 0.0005)~\text{nm,} \nonumber \\
  d_{\text{Fe}[001]} & = & (0.2874 \pm 0.0003)~\text{nm.} \nonumber \\
\end{eqnarray}

Regarding the Au film, investigations were made by Rieckart et al. in \cite{demokritov_morphology}.
They have shown that Au grows on MgO with the $[100]$-direction parallel to the
$[100]$-direction of the substrate. However, LEED measurements showed that Au
does not grow layer-by-layer but in forms of islands. Depositing a buffer
layer of $1~\text{nm}$ Fe was the solution which allowed smooth and flat surfaces
similar to Fe on MgO. Thus, the Fe on MgO does not only serve as a ferromagnetic
layer in our experiments but also as a buffer layer to improve the epitaxy
of the Au layer. Regarding Fe on Au (as we will need for the later spin transfer
experiments), it is known from \cite{gruenberg_interlayer} and \cite{krebs_properties}
for example, that Fe$(001)$ grows in $45^\circ~$rotation regarding Au$(001)$, similar
to Fe on MgO$(001)$. Thus, in the sandwich configuration of MgO/Fe/Au/Fe, MgO$(001)$
and Au$(001)$ have parallel lattice alignments while the itinerant ferromagnetic
Fe$(100)$ are rotated by $45^\circ$ with respect to MgO and Au.

\section{Equipment for Evaporation}

\subsection{Electron Beam Evaporator}
\label{sec:esv}
\begin{figure}
\begin{center}
  \includegraphics[width=0.7\textwidth]{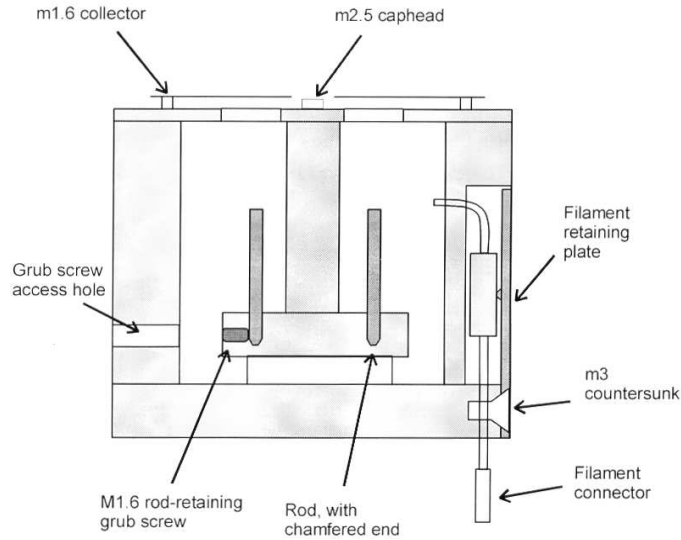}
\end{center}
 \caption[Electron Beam Evaporator]{Electron Beam Evaporator, head assembly. The evaporation
  material is mounted as rods inside a copper assembly. Material may also be evaporated from
  $85 ~\text{mm}^3$ crucibles which provide higher efficiency. The filaments are driven
  with currents up to $5~\text{A}$ allowing electron emission currents up to $100~\text{mA}$. The
  electrons are accelerated through a high voltage of $2~\text{kV}$. From the manual
  of Oxford mini e-beam evaporator \cite{oxford_evaporator}.}
 \label{fig:esvschema}
\end{figure}

Before we commenced with the actual evaporation we had to decide what evaporators to use. For iron we
chose an electron beam evaporator (EBE, Fig. \ref{fig:esvschema}) from Oxford Research, UK \cite{oxford_evaporator}.
It evaporates rod material by targeting high energy electrons at it, allowing to evaporate
high melting point materials at rates between $<1~$monolayer per minute to over $5~$nm per minute.
For first attempts we also evaporated gold with the same evaporator. However, since the amount
of material which can be evaporated with the beam evaporator is limited for each load\footnote{Load
here refers to the evaporator being filled with new evaporation material and being ready
for evaporation. Loading the evaporator requires venting the UHV and detaching
the evaporator from the chamber, a tedious process.}, we evaporated gold from a Knudsen-cell
(see sec. \ref{sec:htc}) which can evaporate large amounts of (mid to high vapor pressure)
materials.

A high-voltage potential of $2~$kV is applied to the target rod in order
to accelerate the electrons emitted from a dedicated filament which is held at ground
potential. The control unit of the evaporator allows setting of both filament current
(up to $6~$A) and electron energy. Emission current must be maintained through these
two parameters (it cannot be controlled directly). However, the control unit allows
online check of \emph{high voltage, filament current, emission current and ion flux}
(a certain percentage of the vapor ionizes and can thus be measured as an anode current)
and thus we could control emission all the time during evaporation.

Material can be loaded into four individual pockets, either as a rod (max. $2~$mm diameter,
$13.5~$mm length) or in special crucibles. We tested rods of different diameters and different
types of crucibles. Both crucibles and rods have benefits and drawbacks. Crucibles offer
higher efficiency of the evaporation process than rods. While evaporation from a rod of
$1.6~$mm diameter yields films of up to $15~$nm per pocket, the crucible allowed us to
evaporate films thicker than $50~$nm. The higher efficiency is a result of the smaller
beam aperture as compared to the rod. As a result, there is less flaked evaporated
material\footnote{Flaked material is weakly adhesive to the pocket walls and thus
it may fall off easily. If this excess material is not removed periodically
there is the risk of shorting the rods to ground, rendering the high-voltage and thus
the whole evaporator useless.} within the pockets. On the other hand, however,
we had problems with the thermal stability of the crucibles. At high emission currents
we observed melting of the crucibles with the evaporation material which impedes the purity
of the films deposited. The risk of melting can be avoided by using crucibles from a
material with very low vapor pressure, i.e. tungsten. However, we did not have any
such crucibles and producing them at the precision engineering department requires a
setup for electrical discharge machining which was not available at that time. The crucibles
that we tested were made from tantalum and molybdenum and were milled and drilled. To maintain
the purity of the films and still be able to use high evaporation rates (i.e. high
emission current) we chose rods over crucibles.

Prior to any evaporation, the unit and each of its individual pockets have to be
degassed. First, we degassed the filaments by running them at up to $4-4.5~$A, the filament
current which is used during evaporation (see tab. \ref{tab:esvparams}). While the filaments degas,
background pressure will rise up to a lower $10^{-8}~$mbar range. Once the pressure has returned
to the $10^{-10}~$regime, degassing of that pocket is complete.

After all four filaments have been degassed, the high voltage is set to maximum ($2~$kV) while
the filaments are still off. Increasing the filament current triggers electron emission
and thus electron bombardment of the target. When the emission current reading registers $\approx$2-5 mA,
the target material and its turret or crucible will start to degas. This is registered
by an increase of the measured ion flux and increase of the chamber pressure up to
$10^{-7}$. Once again, after a certain time-span (usually 10-20 minutes), degassing ceases
when pressure is back to $10^{-10}$.

\medskip

The observed ion flux is determined by two effects:

\begin{itemize}
  \item \textbf{Background ionization:} The background ionization results from the residual gas in the
                                        chamber and therefore depends on the chamber pressure. It causes
                                        the ion flux to rise as the emission current is increased,
                                        \textbf{instantly and linear with emission current}. After a longer
                                        period of evaporation, the contributed flux decreases as both the evaporator
                                        and the charge material will outgas even more.
  \item \textbf{Evaporation:}           Once the emission current is high enough, evaporation starts which
                                        causes a drastic increase in ion flux. Since evaporation requires
                                        heating of the charge material, \textbf{the evaporation comes with
                                        an associated time lag}. When using crucibles, the observed delay
                                        is even longer and it may take $1$-$2~$minutes until the ion flux
                                        due to evaporation can be measured after reaching the necessary
                                        emission current.
\end{itemize}

In case the background ionization rises too much which can be registered as a high flux of several
micro amperes, the high voltage flashes over to ground potential. Since the delicate electronics
of this particular electron beam evaporator model is poorly protected against such flash-overs, the
glitch usually puts the micro controller of the control unit into a undefined state requiring a restart
of the whole unit. Therefore a proper degassing and cooling of the evaporator was critical for
reliable operation.

Since we needed to determine the exact evaporation rates in sensible physical units (thickness per time), a
quartz micro balance (QMB, see section \ref{sec:qmb}) was used for feedback during degassing and evaporation.
The flux can be used as a coarse indicator to determine if there is any evaporation taking place and
the flux approximately reflects to a certain rate on the QMB for identical materials and target configurations.

\medskip

For the evaporation of iron, the following parameters had been found to deliver best results at highest
stable rates:

\begin{table}[htbp]
        \centering
                \begin{tabular}{|c|c|c|c|} \hline
                \textbf{Source}             & \textbf{Filament Curr.}    & \textbf{Electron Energy} & \textbf{Emission Curr.} \\
                                            &            A               &           kV             &           mA          \\ \hline
                Rod ($1.6~$mm)               &        $4.2-5.0$           &          $2$             &           $12$        \\ \hline
                Rod ($2.0~$mm)              &        $4.4-5.0$           &          $2$             &           $15$        \\ \hline
                Ta-Crucible                 &        $4.2-4.5$           &          $2$             &         $18-20$       \\ \hline
                Mo-Crucible                 &        $4.3-4.5$           &          $2$             &         $30-35$       \\ \hline
                \end{tabular}
        \caption{EBE parameters for Fe/Au evaporation from rod/crucible.}
        \label{tab:esvparams}
\end{table}

\subsection{High Temperature Cell}
\label{sec:htc}
Although the electron beam evaporator is suitable for evaporation of even high melting point
materials it has two significant disadvantages which are its limited loading capacity and
an upper limit for evaporation rates. Using crucibles to maximize efficiency, a charge
of evaporation lasts up to $50~$nm per pocket. However, since we wanted to grow thick
gold films of up to $150~$nm, we would have been subject to frequent refills of the
pockets which requires the tedious process of opening the UHV chamber. Also, the limited rates make
evaporation of thicker films uncomfortable and quite time-consuming.

For this reason, we were using a Knudsen-type, high-temperature effusion cell for the evaporation
of gold. The name ``high-temperature'' indicates that even materials with relatively
low partial pressures can be evaporated, but not tungsten contrary to the electron beam
evaporator (\ref{sec:esv}). The effusion cell consists of a self-supporting tungsten
wire as heating system and an exchangeable $10~\text{cm}^3$ crucible. A built-in thermocouple
allows temperature feedback to its control unit.

Compared to the electron beam evaporator, operation of the high-temperature cell
(HTC) is relatively simple: once the material is loaded, the chamber pumped to UHV conditions
and the unit is degassed\footnote{Degassing is performed analogous as for the electron
beam evaporator. The unit is degassed at temperatures around the target temperatures
until pressure has stabilized.}, we just need to set a temperature high enough for the
evaporation material to reach a sufficient vapor pressure which is indicated by the
rate registered on the QMB (Section \ref{sec:qmb}).

The only concerning issue with the HTC is the choice of a proper crucible. There
are several types available, made from different materials:

\begin{itemize}
  \item Boron Nitride ($PBN$):
    \subitem pro: high thermal stability, low adhesivity of evaporation material
    \subitem contra: outgassing of nitrogen possible at high temperatures (above $1300~^{\circ}\text{C}$)
  \item Aluminium Oxide ($Al_2O_3$):
    \subitem pro: high thermal stability
    \subitem contra: outgassing of Al and O possible
  \item Tantalum ($Ta$):
    \subitem pro: no outgassing at lower and mid-range temperatures
    \subitem contra: evaporation material can melt into crucible at high temperatures, low thermal stability (risk of crucible-burst)
  \item Pyrolytic Graphite, Vitrious Carbon ($C$):
    \subitem pro: high thermal stability, allows very high temperatures
    \subitem contra: outgassing of carbon possible
\end{itemize}

We have chosen a PBN crucible since an increased partial pressure of $N_2$ is not critical
for the sample quality. It has a low sticking coefficient on metal surfaces, thus our metal
films and also on the chamber walls. The latter meaning that $N_2$ can be pumped out very
quickly.


\subsection{Quartz Micro Balance}
\label{sec:qmb}
The quartz micro balance (QMB) is an essential measurement tool used in the evaporation processes.
It allows accurate and independent\footnote{The electron beam evaporator has some
limited means of controlling the evaporation rate by means of measuring its ion flux.
However the flux determined there is rather a coarse indicator since the background ionization
contributes as well here. Thus the actual flux of metal ions is smaller than the
value displayed.} measurements of the film thicknesses and calculates the current
evaporation rate by averaging the change of thickness over time. It is based on the
works of G\"unther Sauerbrey in 1957 \cite{sauerbrey_qmb}.

\begin{figure}[h]
\begin{center}
  \includegraphics[width=0.5\textwidth]{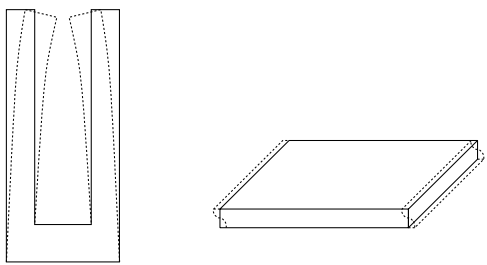}
\end{center}
 \caption[Various forms of quartz-crystals]{Various forms of quartz crystals. Left
  shows the fork shape used in most clock quartz crystals, right the plate shape
  employed for the quartz micro balances. The fundamental modes are indicated with
  dashed lines.}
 \label{fig:quartzcrystals}
\end{figure}

The balance consists of a small, oscillating ($f \approx 6~\text{MHz}$) quartz, shaped
as a plate unlike most clock quartz crystals which have fork shapes and an
oscillation frequency of $32.768~\text{kHz}$ (see Fig. \ref{fig:quartzcrystals}).
It is mounted inside the UHV chamber next to the sample holder so it is
exposed to the evaporation beam as well. The films increase the mass of the oscillating
plate which is detected as a drop of the oscillation frequency of the plate
($\omega = \sqrt{\frac{D}{m}}$). Frequency shifting of the quartz crystal due
to deposited films which are thin compared to the thickness of the plate can be determined
by considering the films as an increase of the thickness of the crystal plate.

The fundamental frequency $f_0$ of such a crystal plate with thickness $d_0$ is:

\begin{equation}
f_0 = \frac{N}{d_0}
\end{equation}

where $N = 1670 ~\text{kHz mm}$, a material-dependant parameter which derives
from the acoustic impedance $Z$ of the material. $Z$ incorporates the speed of
transversal waves through the medium.

Evaporating a thin film will increase the thickness by $\Delta d \ll d_0 $ and
we can use the approximation $\frac{1}{1 + x} \approx 1 - x$:

\begin{equation}
f = \frac{N}{d} = \frac{N}{d_0 + \Delta d} = \frac{N}{d_0}\frac{1}{1 + \frac{\Delta d}{d_0}} \approx f_0 (1 - \frac{\Delta d}{d_0})
\end{equation}

However, internally the control unit of the balance determines the mass of the evaporated film,
therefore the density of the evaporation material has to be programmed into the unit
prior any evaporation process to eventually yield the actual thickness of the film
from $d = \frac{Mass ~ m}{Area ~ A \times Density ~ \rho}$.
\medskip

As the balance is positioned next to the sample, the thicknesses determined vary from
the actual thicknesses on the target substrate by a constant factor called the
\emph{Tooling Factor} \cite{maxtek_qmb}. This factor is programmed into the
control unit of the balance during initial setup of the chamber. In order
to increase the displayed accuracy of the balance, we have set the tooling factor
to 1000, the maximum value\footnote{By setting the tooling factor to 1000, the QMB displays an
additional digit for both rate and thickness.}. The balance has been calibrated
for both the electron beam evaporator (Fe evaporation) and the high temperature
cell (Au evaporation) and thus the displayed values have to be multiplied by a
different scaling factor each. Table \ref{tab:qmbparams} lists all parameters
for the quartz micro balance for evaporation of Fe and Au.

\begin{table}[htbp]
        \centering
                \begin{tabular}{|c|c|c|c|} \hline
                \textbf{Material} & \textbf{Density}  & \textbf{Impedance} & \textbf{Scaling Factor} \\
                   & $\text{g} ~ \text{cm}^{-3}$ & $10^5 / \text{cm}^2 ~ \text{s}$ & \\ \hline
                Fe & 7.86  & 25.30 & 0.1122  \\ \hline
                Au & 19.30 & 23.18 & 0.2375  \\ \hline
                \end{tabular}
        \caption{QMB parameters for evaporation of Fe/Au films.}
        \label{tab:qmbparams}
\end{table}

During evaporation of the wedges, the deposited thicknesses were logged regularly to
keep track of the evaporation processes. When growing a wedge, the translation of the
shutter is performed once a certain intermediate thickness has been reached
(vertical step size of the wedge terraces).

\section{Cleaning the Substrates}

\begin{figure}
\begin{center}
  \includegraphics[width=0.4\textwidth]{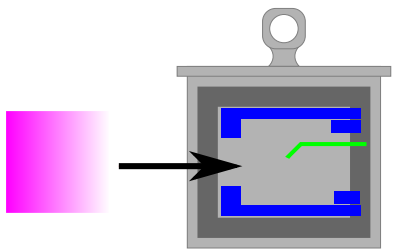}
\end{center}
 \caption[Omicron Sample Plate]{Omicron sample plate. The MgO substrates (purple) are put into
 the sample plate which fit into the 2-inch sample holders (Fig. \ref{fig:2inch-holder}) and
 are fixed with molybdenum clamps (blue).  They can be grasped at the tag on the top with
 pincer grip wobblestick in order to be pulled out of the 2-inch holder in the UHV chamber
 and then transferred into the STM for further analysis for example. A sample
 carousel allows temporary storage of up to 8 sample plates. An additional tantalum needle (green) which
 point-contacted the substrate  was welded to the sample plate to make sure there was an
 electrical contact between the evaporated film and the sample holder.}
 \label{fig:omicron-plate}
\end{figure}

Our samples are based on single-crystalline substrates of Magnesium Oxide
(MgO) in $(001)$ orientation. They fit into the Omicron-brand sample holders
(see figure \ref{fig:omicron-plate}) and have the dimensions $10\times10\times0.5 ~ \text{mm}^3$.
Furthermore they are polished on both sides by the manufacturer. As one aspect, only a
polished surface will allow the thin films to grow with high symmetry and coherence lengths.
On the other hand, a polished surface on the backside of the sample is necessary to
reduce optical scattering in the laser experiments.

Before we could transfer the samples into the UHV-chamber for film evaporation they needed be
cleaned thoroughly, both chemically and by degassing through heating. As noted by Blomqvist
et al. in \cite{blomqvist_structural} for example, the best possible results are achieved by cleaning the
MgO substrates in an ultrasonic bath of acetone, isopropyl alcohol, ethanol for
10 minutes each. To remove any chemical residuals (\emph{Carbon} !), in this case we also rinsed the substrates
for another 5 minutes in deionized and highly cleaned water with ultrasonic cleansing. Since
Magnesium Oxide is highly hygroscopic, the last bath must not last too long and the substrates
are to be dried immediately with gaseous nitrogen. In order to prevent any contamination of the
now clean substrates they were handled with ceramic tweezers instead of metallic ones
and latex gloves\footnote{In fact, during the sample preparation metallic tweezers
were accidentally used once. AES spectra taken of these samples clearly showed contamination
with nickel.}.

\begin{figure}
\begin{center}
  \includegraphics[width=0.4\textwidth]{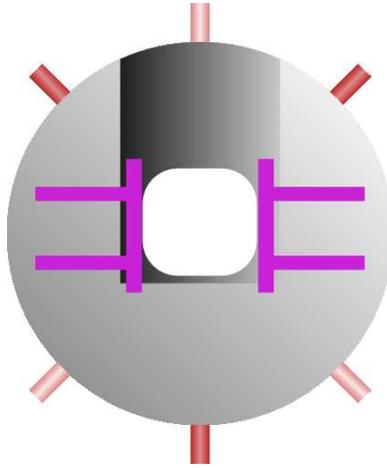}
\end{center}
 \caption[2-inch Sample Holder]{Illustration of the 2-inch sample holders used for transfer
 within the UHV chambers. The round plate has 6 mounting bolts, 3 of each along one perimeter
 circle (at $0^\circ, \pm 135^\circ$ and $180^\circ, \pm 45^\circ$), allowing the plate to be mounted
 onto the transfer rods and stationary holders in each chamber with a bayonet-type connector.
 The holder is put onto or picked up from a stationary holder by sliding the rod into it and twisting
 the rod grip to lock or unlock the 2-inch holder from the rod or the stationary holder and vice versa. }
 \label{fig:2inch-holder}
\end{figure}

After cleaning the substrates chemically they were ready to be transferred into the UHV chamber
for MBE and analysis. Since fully opening the chamber requires a venting of
the vacuum to ambient pressure and, after installing the samples, a time-consuming
and tedious (multi-staged pumping and bake-out) restoring of the vacuum, the samples
are transferred into the UHV through a load lock. The lock is usually a very small chamber
whose pressure can be restored to a good vacuum within a few hours after opening for loading.
From the load lock the samples can be transferred throughout the whole UHV chamber
using transfer rods (figure \ref{fig:mbechamber}) which grasp the samples with 2-inch
sample holders of molybdenum (see figure \ref{fig:2inch-holder}).

\medskip

\begin{figure}
\begin{sideways}
  \includegraphics[height=1.\textwidth]{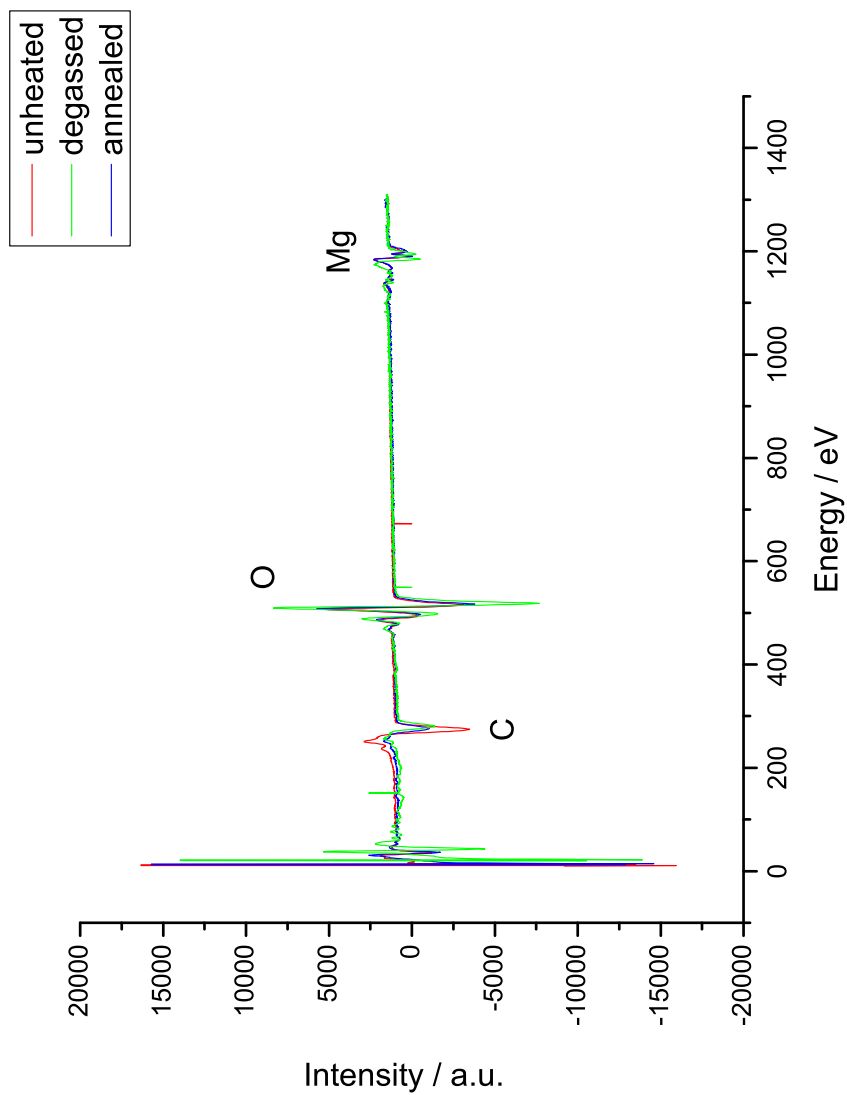}
\end{sideways}
\caption[AES spectra of MgO during preparation]{AES spectra of MgO$(001)$-substrates showing the
 effectiveness of degassing and annealing. The spectra were taken directly after transferring
 the substrates into the UHV (\emph{unheated}), after being degassed for $12$ hours at $100~^\circ$C
(\emph{degassed}) and after annealing (\emph{annealed}) at $700~^\circ$C for $180$ minutes. A remarkable drop of the
 intensity of the carbon peak at around $300~$eV is detected after degassing and annealing respectively.}
 \label{fig:aesspectra}
\end{figure}

The purity of the substrates was checked with AES analysis before evaporating the thin films.
Figure \ref{fig:aesspectra} shows the effectiveness of degassing and annealing the substrates.
The carbon peak shows a remarkable drop in intensity proving that we were able to get rid
of at least half of the carbon on the substrate. However, since there is also some
carbon in the bulk of the crystal which ascends to the crystal surface while heating
the sample, there will be still carbon visible in the spectrum.

Therefore we concluded that with the methods employed in this work (chemical cleansing, outgassing and
annealing), a clean and perfect surface could not be achieved. For perfect results, the MgO substrate
needs to be sputtered with appropriate methods, more precisely one needs
\emph{low-energy sputtering} with a plasma source \footnote{Conventional
sputter guns do not allow low-energy sputtering with insulators like
MgO since the sputtering ions charge the target which will deflect most
of the ions effectively. This plasma source compensates the total charge
by shooting additional electrons to the target.}. In \cite{demokritov_morphology}
Rieckart et al. have presented the use of such a source with which they were
able to get rid of the carbon contamination below the sensitivity of AES analysis.
We have taken ownership of the plasma source, its sophisticated setup and calibration,
however, would go beyond the scope of this diploma thesis and therefore its
use is subject to future works in this project.

\section{Evaporation Process}

The evaporation processes of iron and gold can be concluded as follows.

\subsection{Evaporation of Iron Films}

After cleaning, outgassing and annealing the samples, they were transfered
into the growth chamber. The sample holder in the chamber was faced away
from the evaporators while the electron beam evaporator with its shutter
closed was running up. The parameters for the evaporator were set according
to Table \ref{tab:esvparams}, then after the flux had stabilized the shutter
of the electron beam evaporator (EBE) was opened and the rate was checked with the QMB which was
set to the parameters for iron (\ref{tab:qmbparams}). When the designated
evaporation rate was gained, the shutter of the EBE was closed again
and the sample was faced downwards to the evaporators. The shutter
to cover the sample (Fig. \ref{fig:shutter}) was moved in and positioned
over sample as necessary. The EBE shutter was opened again and evaporation
starts and is registered with the QMB. Once the target thickness has been
reached, the EBE shutter is closed again and the EBE is shutdown afterwards.

\subsection{Evaporation of Gold Films}

The process of gold evaporation was similar to that for iron. First turn
the sample holder away from the evaporators, then close the shutter of
the high temperature cell (HTC) and heat up the cell to 1350-1400$^\circ$C.
Then the shutter of the HTC was opened and the rate was checked with the
QMB. If necessary, the temperature of the HTC was tuned until the designated
rate on the QMB was read. Then the HTC shutter was closed again and
the sample was brought into position as well as the sample shutter. Upon
opening the HTC shutter, evaporation starts immediately but the QMB
takes a few seconds until it reaches thermal equilibrium (the heat
from the HTC produces false readings on the QMB since the crystal
is out of thermal equilibrium).



\section{Thin Films with Gradient Thickness}
\label{sec:wedge-production}
One goal of our measurements was to determine the exact mean free path of the
ballistic electrons travelling through the gold films and to determine the optimum
thickness for the iron film in order to achieve the highest possible efficiency
in electron injection.

One could now prepare films with various thicknesses, measure the values of the
aforementioned parameters for these and determine their optimal values by means of
data evaluation methods like interpolation. Since the preparation of several
samples is very time-consuming\footnote{Remember that the amount of evaporation
material like iron is limited in the evaporators. Only a few films can be prepared
until the the evaporators have to be refilled which requires the UHV chamber
to be opened.} and the conditions for each sample and measurement may vary\footnote{Varying
conditions are slightly different substrates, ambient temperatures, contaminations, alignments
of the optical setup etc.}, we are taking a different approach for these measurements:
films with gradient thicknesses (wedges).

The process of evaporation can be seen analogous to the process of exposition
of photographic paper towards a light source. The longer the exposition time
of the photographic paper, the darker the exposed areas will be. In the case
of Molecular Beam Epitaxy (MBE), the film thickness grows with exposition time.
Thus in order to achieve a gradient of thickness along one lateral axis
of the film, we would have to start exposing only one part of the sample,
covering the rest of it with a shutter, then translate the shutter along
the desired axis to expose more area to the evaporation beam over the time.
The area which was exposed from the beginning of the evaporation process
will naturally gain the highest thickness whereas the one exposed at
last will have thinnest. One can also start with a completely exposed
sample and then gradually close the shutter over the sample. This
technique has the advantage that we can directly read the current
thickness off the quartz micro balance during evaporation. Performing
it vice versa, i.e. opening the sample gradually, means that we have
to subtract the current thickness measured off the QMB from the total
thickness in the end to determine the thickness of the wedge at
a specific point. We therefore favored closing the shutter gradually.
Moreover, since the sample holder in the UHV chamber has more degrees
of freedom than the shutter and since the latter extends into the
chamber in an uncomfortable angle, we decided to keep the shutter
fixed and move the sample holder instead.

For this specific task a custom shutter was designed and handcrafted from
high-grade steel\footnote{Materials for constructions of UHV parts
must be carefully chosen. Metals like aluminum have the property to
degas atoms from their bulk and may thus deteriorate the vacuum in the
chamber. High-grade steel is in most cases the best choice, in fact, the chamber
itself is made from it.}. The shutter has to serve two tasks. First
it must be able to cover the whole substrate so that a gradient thickness
of the film can be grown along the full size of the substrate, and secondly,
it must be able to cover an area of the substrate during evaporation
permanently. This was necessary to achieve a region with small thicknesses
at the verge of the substrate while we can still grow the gradient
film on the remaining area of the substrate. The thin verge is required
for optical adjustments in the laser setup later on: the pump and
the probe-beam have to be aligned to meet in one spot which is only
possible as long the transmission of the beam through the film is high
enough (i.e. finding the overlap and thus cross correlation signal).


\begin{figure}
\begin{center}
  \includegraphics[width=1\textwidth]{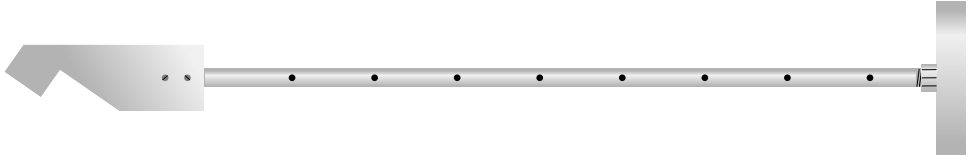}
\end{center}
 \caption[Construction of the shutter used for evaporation.]{Schematical sketch of
   the shutter constructed for sample preparation. The arm is made from a
   high-grade steel tube (length: 350 mm, diam.: 8 mm) with vent holes. The sail is made
   from an iron sheet which was cut with a mill and mounted with two screws to
   the arm. A blind flange was chosen to mount the shutter to a translation
   stage to the MBE chamber.}
 \label{fig:shutter}
\end{figure}

\begin{figure}
\begin{center}
  \includegraphics[width=0.7\textwidth]{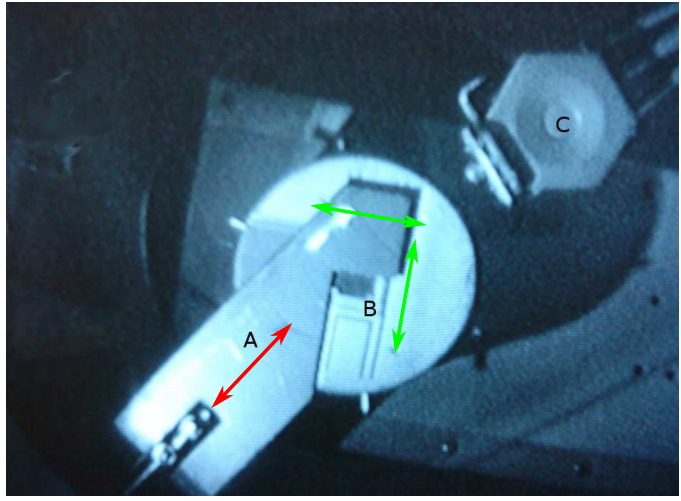}
\end{center}
 \caption[Shutter and sample holder in the chamber during evaporation.]
  {Live shot from the CCD camera through the bottom window in the growth chamber.
  It shows the shutter (A), the sample-holder (B) and the quartz-micro-balance (C). Green and red
  arrows indicate degrees of freedom of sample holder and shutter respectively. }
 \label{fig:shutter-ccdshot}
\end{figure}

The shutter consists of an iron tube (length: 350 mm, diameter: 8 mm) with vent holes
which is mounted onto a blind flange (see Fig. \ref{fig:shutter}). A
tube was chosen in favor over a solid arm to reduce the risk of dead
volumes which cannot be pumped properly and may deteriorate the vacuum
continuously. The flange itself is mounted to a translation stage which
is mounted to a $63~$mm flange onto the chamber, coming in a angle
of $35^\circ$ regarding to the front flange of the chamber.
Since the manipulator holding the substrate is located at an angle of $180^\circ$ towards
the front flange, the shutter sail had to be cut respectively. Figure
\ref{fig:shutter-ccdshot} shows a photograph with the shutter sail
covering the sample holder, taken with a CCD camera outside the chamber.

In order to achieve a linear gradient in thickness for a given start and
end value, sample and shutter either have to be translated with respect to each
other at a constant speed which depends on the evaporation rate and
the total length of the gradient on the sample. Or, what we eventually favored,
one keeps track of the current thickness with the help of the QMB (Sec. \ref{sec:qmb})
and translates the shutter\footnote{In fact, we kept the shutter at a fixed position
and moved the sample holder with respect to the shutter. Moving the sample holder
allows to use more than one degree of freedom for sample movement
(see Fig. \ref{fig:shutter-ccdshot}).} each time a predefined thickness step has
been evaporated. This choice was made since during the several cycles of evaporations
we have performed, it showed that the rate of the evaporators regularly varied by up to half a
magnitude from time to time, so that the choice of fixed time intervals would
mean that we get non-uniform step sizes and therefore a bad linearity
in the wedge.

\medskip

Since the dimensions of the MgO substrates were $10 \times 10~$millimeters and
we evaporated the wedge along the diagonal axis, the length of the projection
of the wedge on the substrate is $10 \times \sqrt{2} \approx 14~$millimeters.
We wanted to evaporate with the smallest sensible step size, this yields the
smoothest possible gradient: one mono-atomic layer, i.e. $3~$\AA ngstr\o ms
or $0.3~$nanometers.

\medskip

Let's assume an end thickness of $50~$nm for example, then we get for the number of steps,

\begin{equation}
  n_{steps} = \frac{\text{d}_{total}}{\text{step size}} = \frac{50~\text{nm}}{0.3~\text{nm}} \approx 167
\end{equation}

since the lateral length of the wedge is $14~\text{mm}$, we have to translate
the shutter by $\approx 0.08~\text{mm}$ each time

\begin{equation}
 \frac{14~\text{mm}}{167~\text{steps}} \approx 0.08~\frac{\text{mm}}{\text{step}}
\end{equation}

For a convenient operation of the shutter, we chose $0.10~\text{mm}$ per step.

\medskip

The evaporation rate was chosen such that the pressure in the chamber would not deteriorate too
much but still a rate is achieved which allows to grow the wedge within
a reasonable amount of time with a comfortable step rate for the
shutter translation\footnote{The step rate here refers to the time interval
after which the shutter or sample are manually translated with respect
to each other.}.

The evaporation rate for gold was empirically set to $\approx0.1~\text{\AA}~\text{s}^{-1}$.
At this rate, the pressure will not rise above $\approx$ 1-2 $\times 10^{-8}$ mbars
during evaporation.

\medskip

So, assuming a constant rate during evaporation, the total duration of the process
will be:

\begin{equation}
  t_{total} = \frac{\text{d}_{total}}{\text{rate}} = \frac{50~\text{nm}}{0.01~\text{nm}} = 5000~\text{s}
\end{equation}

and the shutter will have to be moved each

\begin{equation}
  t_{interval} = \frac{\text{step size}}{\text{rate}} = \frac{0.3~\text{nm}}{0.01~\text{nm s}^{-1}} = 30~\text{s}
\end{equation}

\section{Examination of the Sample Quality}

The samples are analyzed both in real space (STM) as well as in reciprocal space (RHEED).
Real space analysis allows to check the surface for local defects as well
investigate corrugations and determine the actual dimensions. Reciprocal space
tells us about the periodicity over large scales and gives a principal idea
about the surface geometry.

\subsection{RHEED Analysis}

RHEED stands for \emph{Reflection High Energy Electron Diffraction}. Thus electrons with
high kinetic energy (5 to 50 keV) are diffracted in reflection at the sample surface.

\subsubsection{Basics of RHEED}

\begin{figure}
\begin{center}
  \includegraphics[width=1\textwidth]{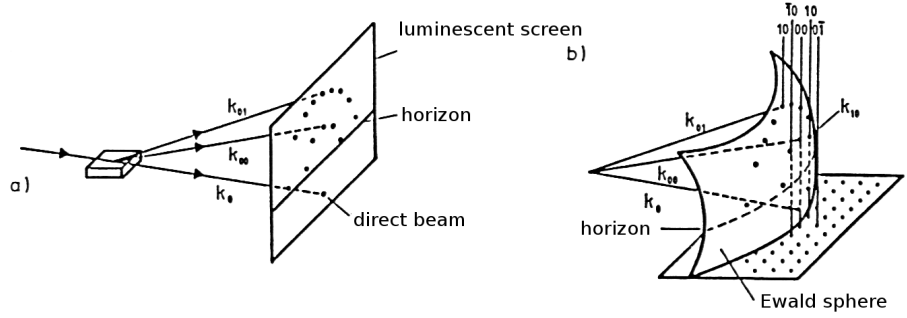}
\end{center}
 \caption[Schema of RHEED]{A high-energy electron beam hits the sample in a small gracing
 angle. Electrons are diffracted elastically and hit a luminescent screen. The direct
 beam is not visible on the screen as it lies below the horizon defined by the
 sample's shadow (a). The diffraction pattern can be constructed with the help of the
 Ewald sphere (b). Constructive  interference occurs only when the rods indicating the
 allowed momentum vectors of the incident beam intersect with the Ewald sphere.
 Figure taken from M. Henzler, \emph{Oberfl\"achenphysik des Festk\"orpers}
 \cite{henzler_oberflaechenphysik}.}
 \label{fig:rheed_schema}
\end{figure}

In contrast to LEED (\emph{Low Energy Electron Diffraction}), electrons are accelerated to
high kinetic energies of up $50~\text{keV}$ and hit the target in a small gracing angle
($0.1^{\circ}$-$4^{\circ}$). The diffracted beam is reflected to an adjacent fluorescent
screen where the diffraction pattern can be seen. A CCD camera located behind the screen
connected to a computer allows to take shots of the pattern to analyze intensities and
lattice constants. The principal setup can be seen in figure \ref{fig:rheed_schema}, (a).

To understand why diffraction occurs and to be able to interpret the observed pattern, one uses the
concept of the reciprocal lattice, reciprocal lattice vectors and the Ewald sphere (see Fig.
\ref{fig:rheed_schema} (b)). The reciprocal wavelength $\frac{1}{\lambda}$ of the wave vector
of the incident beam defines the radius $k$ of the Ewald sphere in the reciprocal lattice:

\begin{equation}
  k = \frac{2 \pi}{\lambda}
\end{equation}

\begin{figure}
  \begin{center}
    \includegraphics[width=0.5\textwidth]{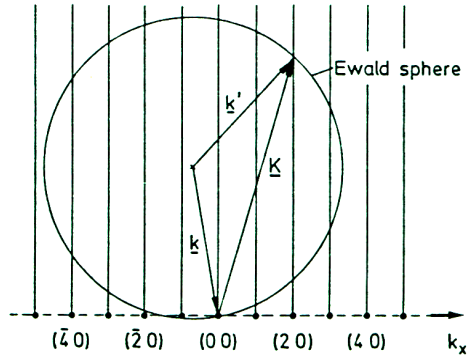}
  \end{center}
  \caption[Ewald sphere in 2D]{Ewald sphere in a two-dimensional reciprocal lattice.
    The sphere is a helpful construction to understand why diffraction occurs in a
    crystal lattice. Its radius is defined by the incident wave vector $\mathbf{k}$. Constructive
    interference takes place if the both the diffracted beam end on a lattice point or
    if two lattice point lie on the perimeter of the Ewald sphere. Both equals that
    both vectors satisfy Equation \ref{eqn:ewald}. Figure taken from H. L\"uth,
    \emph{Surfaces  and Interfaces of Solid Materials} \cite{lueth_surfaces}.
  }
  \label{fig:ewald_sphere}
\end{figure}

Thus, if one varies the energy (and therefore the wavelength $\lambda$) of the electron beam,
the radius of the Ewald sphere will change anti-proportionally to it. This
is helpful to tune the amount of spots visible on the screen, the larger the radius
the larger will the solid angle of the sphere be and thus the number of spots
on the projected area. Now, if two points of the reciprocal lattice lie on the
surface of the Ewald sphere,

\begin{equation}
  \mathbf{K} = \mathbf{k'} - \mathbf{k}
  \label{eqn:ewald}
\end{equation}

the requirements for diffraction are met and reflected spots can be observed on the screen.
Figure \ref{fig:ewald_sphere} shows the picture of the Ewald sphere in reciprocal lattice,
indicating the two vectors (incident and diffracted beam and the the reciprocal
lattice vector $\mathbf{K}$). Technically speaking, this means that the incident has the proper momentum
and angle of incidence so that constructive interference between several elastically
scattered beams occurs.

The striking advantage of RHEED over LEED is the geometrical setup where the electrons
streak the sample parallel to its surface. Access to the sample in perpendicular direction
is therefore still possible and one can evaporate films onto the sample while performing
RHEED analysis. Since the perpendicular component of the electron beam energy is very small
(a few hundreds eV: for $50~$keV and $0.5^{\circ}$ one gets $E = \sin(0.5^{\circ}*50~\text{keV}
\approx 436~eV$), the overall electron energy has to be very high compared
to LEED to achieve the effectively same energy of incidence.

\medskip


We used a \emph{Createc HP4} RHEED system. The HP4 is an advanced
apparatus which is specially designed as a \emph{in-situ} characterization tool in MBE
analysis for studying growth rates and growth dynamics by measuring RHEED intensity
oscillations. The system consists of the RHEED gun which is mounted to the MBE chamber
(see Fig. \ref{fig:mbechamber}), a power supply and a control unit for the magnetic lens and
electrostatic deflection system. The power supply allows energies up to $50~\text{keV}$ and beam currents of up to
$100~\mu\text{A}$. The lens and deflection system is directly integrated into the RHEED gun,
the magnetic lens focuses the beam on the luminescent screen while the electrostatic deflection system
positions the beam. Table \ref{tab:rheedparams} lists the beam parameters used for
our RHEED measurements. Settings for the lens and deflection system change for every
single measurement and have to be adjusted anew every time by tuning the pattern
to best possible contrast and visibility.

\begin{table}[htbp]
        \centering
                \begin{tabular}{|c|c|} \hline
                  Parameter                  & Setting      \\ \hline
                  High voltage / kV          & $50$         \\ \hline
                  Filament Current / A       & $2$          \\ \hline
                  Emission Current / $\mu$A  & $35-40$      \\ \hline
                \end{tabular}
        \caption{RHEED parameters used for our measurements.}
        \label{tab:rheedparams}
\end{table}

\subsubsection{Interpretation of RHEED images}

\begin{figure}
\begin{center}
  \includegraphics[width=0.7\textwidth]{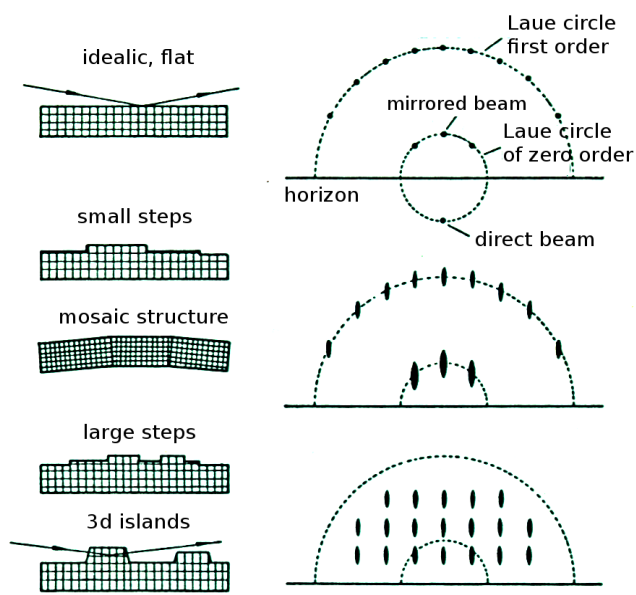}
\end{center}
 \caption[Interpretation of RHEED patterns]{The morphology of the surface determines the
   pattern which can be observed on the luminescent screen during a RHEED measurement. An
   ideal, flat surface with high symmetry shows a pattern with sharp spots which lie
   on Laue circles. Small steps or a mosaic structure wash the spots out. If the surface
   consists of large, more or less regular steps (i.e. islands), the spots do not lie
   on the Laue circles anymore but instead form a regular, rectangular pattern.
   Figure taken from M. Henzler, \emph{Oberfl\"achenphysik des Festk\"orpers}
   \cite{henzler_oberflaechenphysik}.}
 \label{fig:rheed_interpretation}
\end{figure}

In the case of an atomically flat and highly symmetric surface, the diffraction pattern
shows bright, highly focussed spots which lie on different circles representing the different
order of diffraction (\emph{Laue circles}, see Fig. \ref{fig:rheed_interpretation}).
However, in most cases the symmetry of the surface is violated so that the observed
pattern differs from that of an ideal surface. If there are small steps, or areas of the
surface are slightly tilted (\emph{mosaic structure}), the spots have elliptical shapes
and less intensity. In the sense of the Ewald sphere, this means that the rods which
intersect the sphere have a non-zero thickness.

\subsubsection{RHEED Images}


We measured Fe on MgO$(001)$ films at electron energies of $50~\text{keV}$, captured the luminescent screen
with a CCD camera and analyzed the images with the software \emph{Specs Safire 4} \cite{specs_safire}.

\medskip

\begin{figure}
\subfigure[Initial angle]{\includegraphics[width=0.5\textwidth]{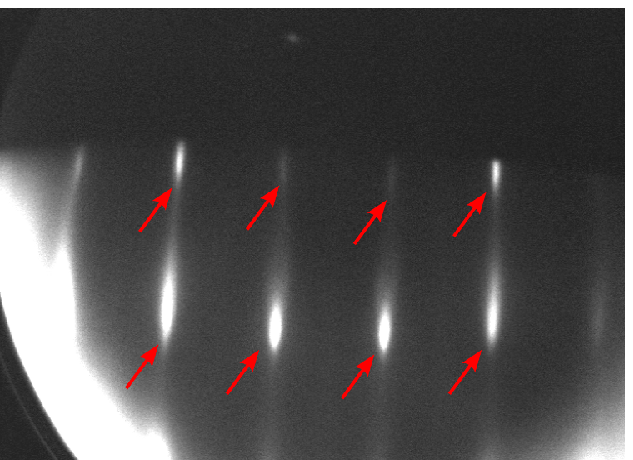}}
\subfigure[Relative angle of $45^\circ$]{\includegraphics[width=0.5\textwidth]{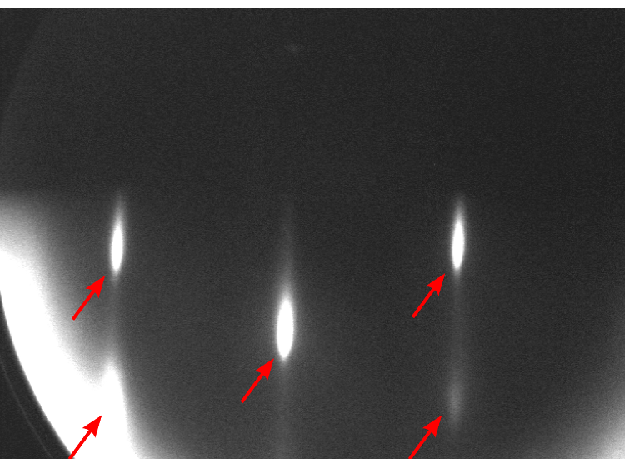}}
\caption[RHEED image showing islands]{RHEED image of Fe(001) on MgO taken at $E_{kin} = 50~\text{keV}$. The
   displayed pattern suggest three-dimensional islands after \cite{henzler_oberflaechenphysik},
   at first sight (red arrows) (see Fig. \ref{fig:rheed_interpretation}). To support this assumption, an image
   along the same direction as (a) is analyzed in figure \ref{fig:rheed_analysis}. The image
   in (a) indicates a distorted 4-fold symmetry and the one in (b) a 6-fold symmetry meaning that we were
   looking along the [10]- and [11]-direction of the island superstructure.}
\label{fig:rheed_islands}
\end{figure}

\begin{figure}
\begin{center}
\includegraphics[width=0.5\textwidth]{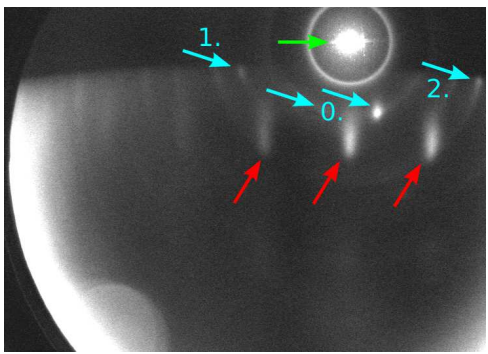}
\end{center}
\caption[RHEED image showing several orders of diffraction]{Another RHEED image of Fe(001) on MgO
  taken at $E_{kin} = 50~\text{keV}$ at the initial angle position (see Fig. \ref{fig:rheed_islands}, (a)).
  In this image, the direct beam is clearly visible (green arrow) and can be distinguished from the
  other spots due to its central position and bright intensity. Blue arrows
  indicate the diffraction spots of zero, first and second order (see \ref{fig:rheed_interpretation}, (a))
  of a flat and periodic structure with sharp spots. Red arrows indicate the superstructure again
  (Fig. \ref{fig:rheed_islands}).}
\label{fig:rheed_analysis}
\end{figure}

Figure \ref{fig:rheed_islands} shows two RHEED images, taken at a relative azimuthal angle
of $45^\circ$. The images are distorted in their geometry but the patterns suggest
that the surface has islands according to \cite{henzler_oberflaechenphysik}.
Looking at figure \ref{fig:rheed_analysis} which is another RHEED image of Fe on MgO$(001)$
at an initial azimuthal angle, one can see that there are actually two diffraction
patterns, from Fig. \ref{fig:rheed_interpretation} we concluded that this compound
RHEED image can only be explained when considering the surface to be a crystalline
structure (which creates the periodic pattern) which has small islands all over the surface
(which creates the 4-fold symmetry pattern). The overlaid 4-fold pattern can only be a result of small islands
of the iron film unless there are other gaseous materials in the chamber which were adsorbed
during evaporation. Since the principal residual gasses in the chamber are nitrogen, hydrogen and
water, which do not adsorb crystalline at room temperature, adsorbed residual gases
cannot induce the observed diffraction pattern. Further qualitative analysis of the islands is
done in section \ref{sec:stm}.


\subsubsection{RHEED oscillations}

When measuring the intensity of a diffraction spot while evaporating a thin film onto
the substrate, a periodic change can be observed, the so-called \emph{RHEED oscillations}.
The cause of this phenomenon is that each time a new monolayer gets assembled from
single atoms, the symmetry of the surface is lowered compared to that of a complete
monolayer. This means, that when the intensity has a local maximum, we have just
evaporated a full monolayer and when the intensity has a local minimum, we are
just right in the process of building the next monolayer. RHEED oscillations can
be used to cross-check the rate measured with the quartz micro balance
(sec. \ref{sec:qmb}). They also depend on the temperature of the substrate
during evaporation \cite{schatz_rheed}.

\begin{figure}
\subfigure[First run]{\includegraphics[width=0.5\textwidth]{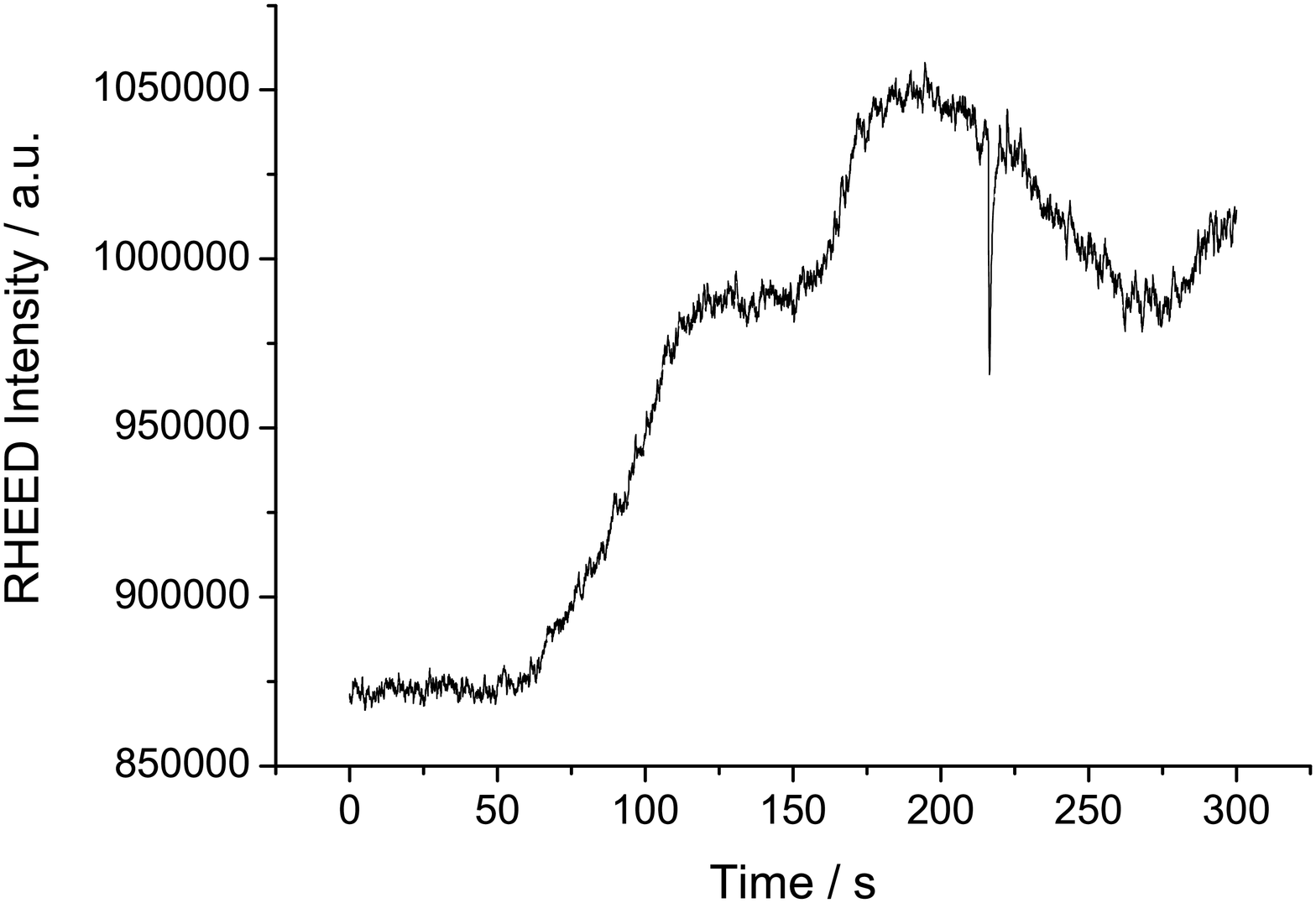}}
\hfill
\subfigure[Second run]{\includegraphics[width=0.5\textwidth]{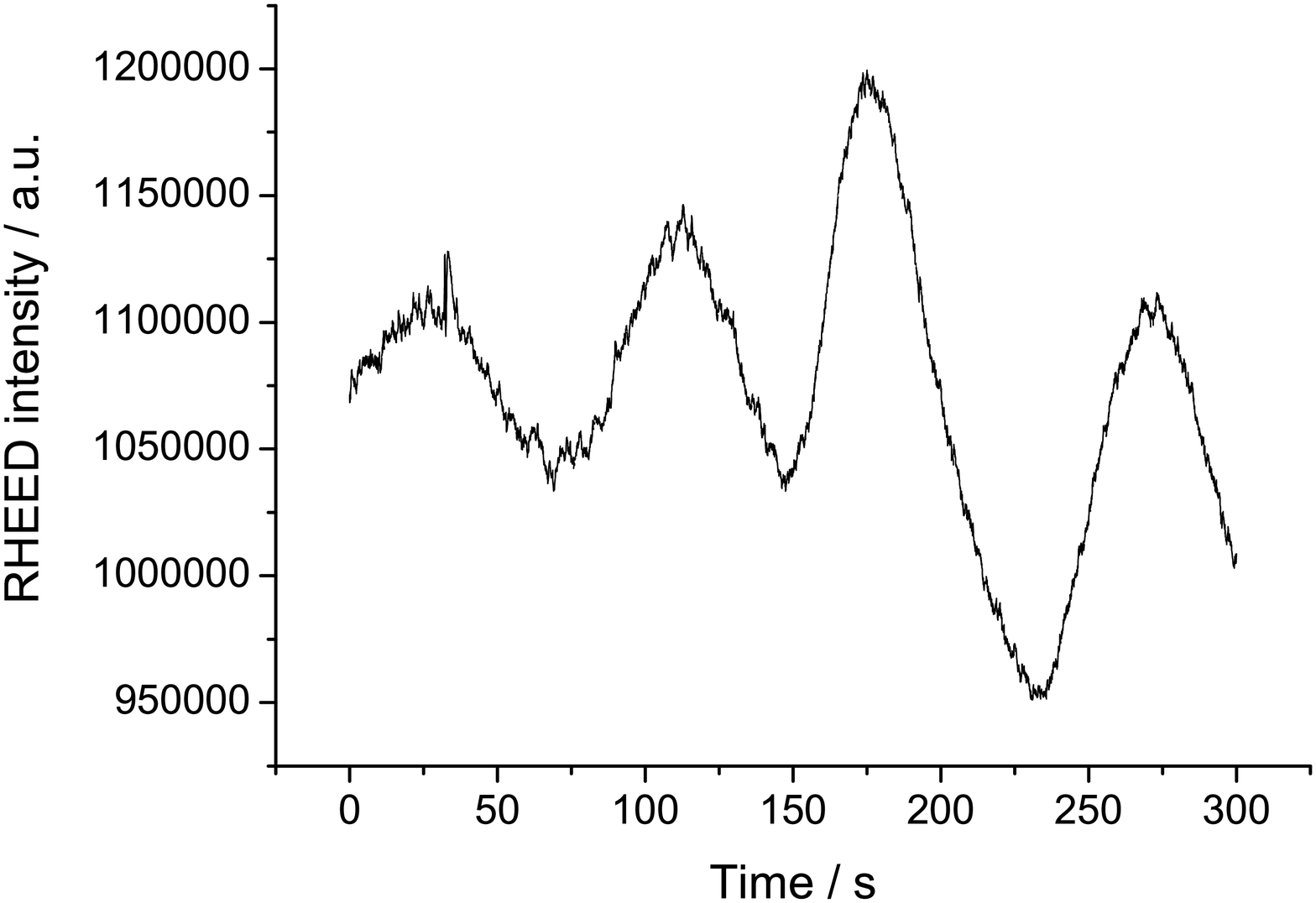}}
\subfigure[Third run]{\includegraphics[width=0.5\textwidth]{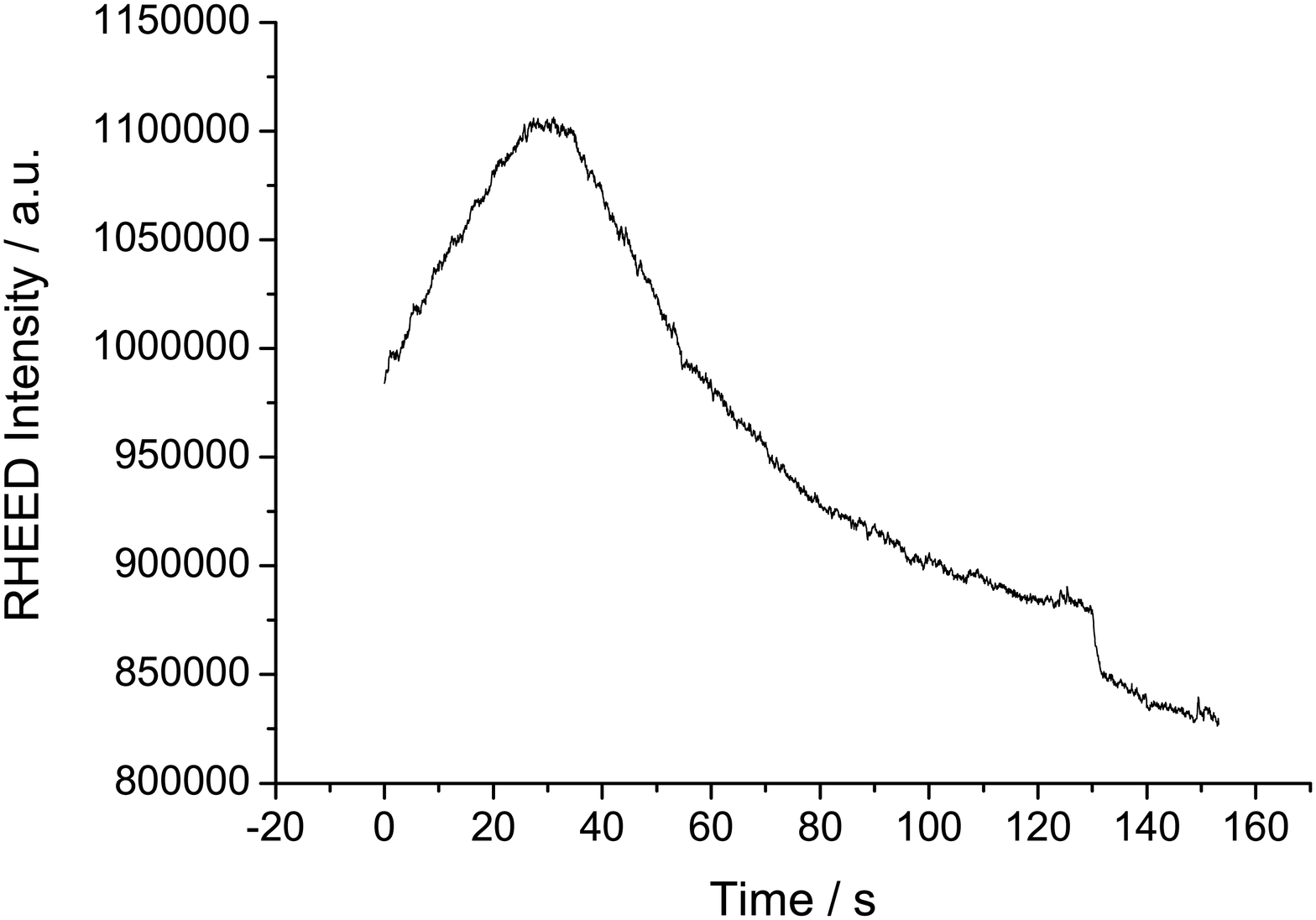}}
\hfill
\subfigure[Oscillations from \cite{lueth_surfaces}]{\includegraphics[width=0.5\textwidth]{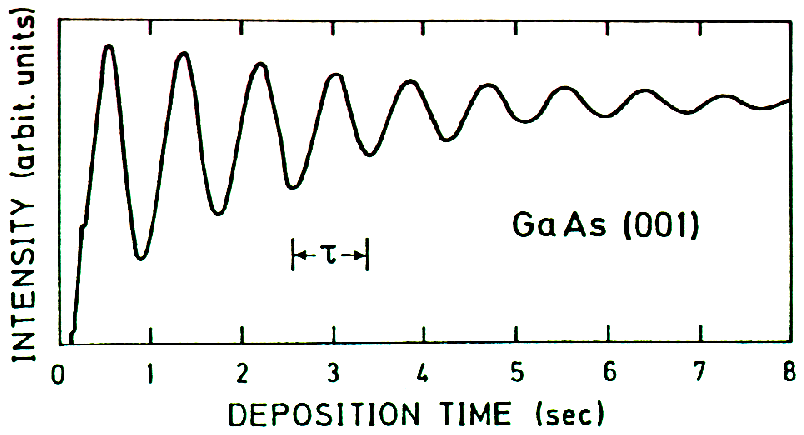}}
\label{fig:rheedoscillations}
\caption[Measured RHEED oscillations]{Several runs of measured RHEED oscillations illustrating
  the layer-by-layer growth of Fe on MgO$(001)$. During the first and the third run (plots (a) and (c)),
  we had problems maintaining a stable rate from the electron beam evaporator. The variations in amplitude do
  not result from the growth but from the changing background luminescence of the evaporators
  filament. During the second run (b), the evaporation rate was stable most of the time and so
  we think the oscillations observed really result from the growth while there is still
  a background from the filament which changes the amplitude. Figure (d) shows a measurement
  from literature \cite{lueth_surfaces} which exhibits perfect periodic behavior with exponentially
  decaying amplitude meaning the film grows thicker and out of focus from the electron beam.}
\end{figure}

Since the measurement of RHEED oscillations was not essential for our experiments it was not subject
to further investigations. However, we measured 3 curves for a diffraction spot while
evaporating Fe on MgO during the first evaporation tests. The rate was about $0.05-0.1\text{\AA s}^{-1}$
as measured with the QMB. The RHEED oscillations measured in the first and third run did not yield
any periodic change in intensity due to fluctuations of the electron beam evaporator\footnote{Fluctuations
are usually compensated by adjusting the filaments whose light is seen as background illumination
on the RHEED screen.}, however the second run showed an oscillation that might come from a
layer-by-layer growth.

\begin{figure}
  \begin{center}
  \includegraphics[width=0.7\textwidth]{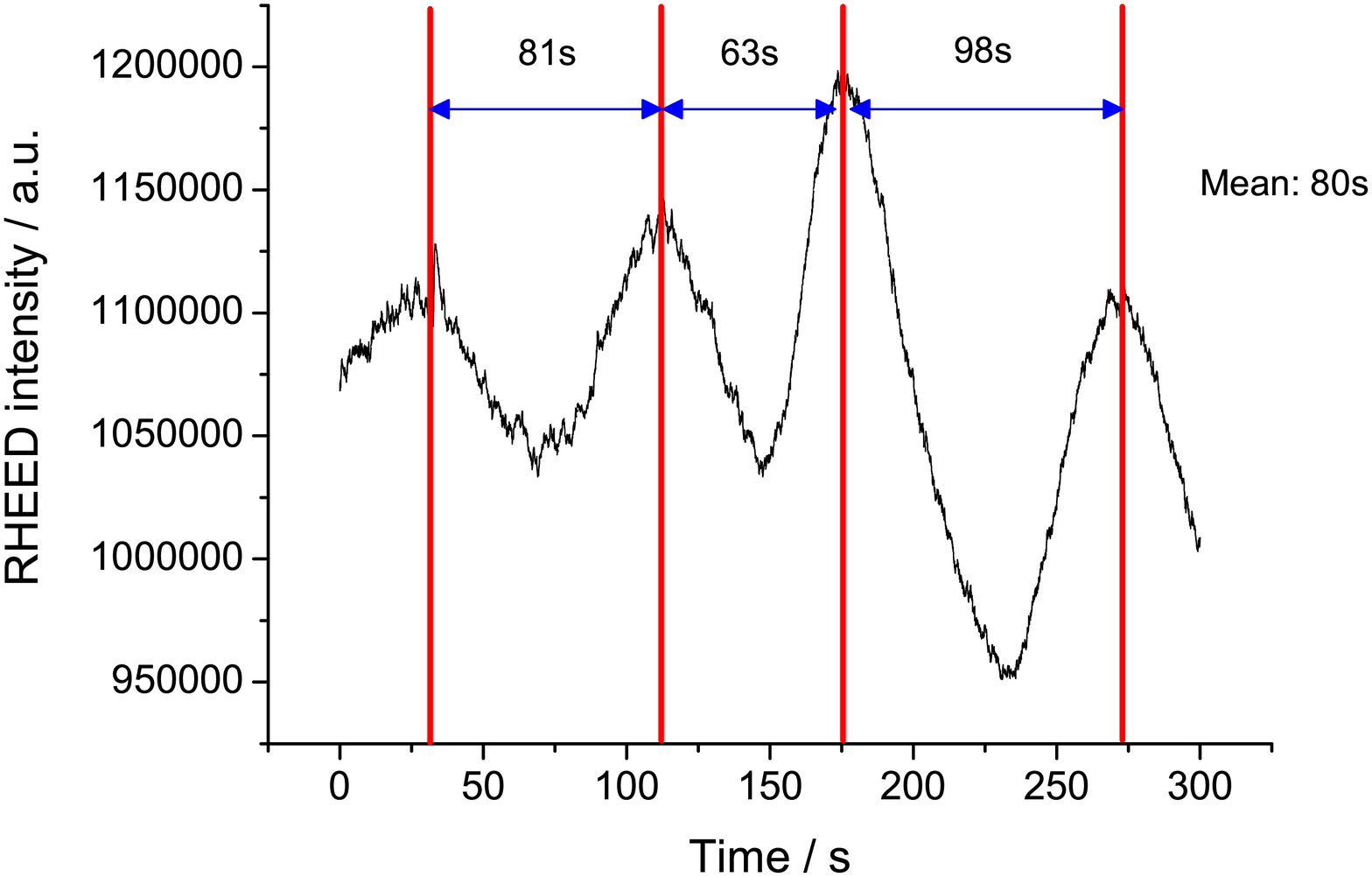}
\end{center}
\caption[Time analysis of RHEED oscillations]{Each intensity peak has been marked on
  the time axis allowing to determine the distance between two peaks which reflects
  the amount of time that was necessary to grow one monolayer, we took the mean
  value of these intervals. The increase in amplitude in the last peak was due to
  a mandatory increase of the EBE filament current (the rate was decreasing)
  which generates an offset (filaments glow brighter).}
\label{fig:rheedoscillations_analysis}
\end{figure}

\medskip

Figure \ref{fig:rheedoscillations_analysis} shows the evaluated RHEED oscillation from the
second run. The mean value of the duration to finish one monolayer (i.e. $10~\text{\AA}$) is
is determined to be $80~\text{s}$ which corresponds to a rate of $0.1~\text{\AA s}^{-1}$
which is in agreement with the rate measured with the QMB.

\subsection{STM Analysis}
\label{sec:stm}
\subsubsection{Basics of STM}

STM stands for Scanning Tunneling Microscopy and is a technique for real-space analysis of samples
at an atomic scale. Unlike methods which use reciprocal space, it yields an image which reflects
almost exactly the true morphology of a surface\footnote{Since STM measures the electron densities
instead of the atoms, the picture might still look a bit different from the real surface.}. The
basic physics behind an STM is the \emph{Quantum Tunneling Effect}, the phenomenon which describes
the ability of particles on the atomic scale to tunnel through energy barriers which they
could never overcome when applying classical physics. Since the morphology is analyzed by
measuring the tunneling current between STM tip and sample while scanning over the sample,
its surface has to be electrically conducting. In order to be able examine non-conducting surfaces, one
either has to cover the surface with a conducting layer (in most cases a thin evaporated
gold film will suffice) or use the related \emph{Atomic Force Microscopy} instead. As our
substrates are originally insulators (MgO), STM cannot be performed before the metallic films are evaporated
onto the surface. Electrical contact with the STM electronics is made with the Omicron sample
plate, the clamps of the holder must therefore have electrical contact with the films. To
improve the contact further, we welded a contact needle to the sample holder which would
poke into the substrate (see Fig. \ref{fig:omicron-plate}).

\medskip

\begin{figure}
  \begin{center}
    \includegraphics[width=0.7\textwidth]{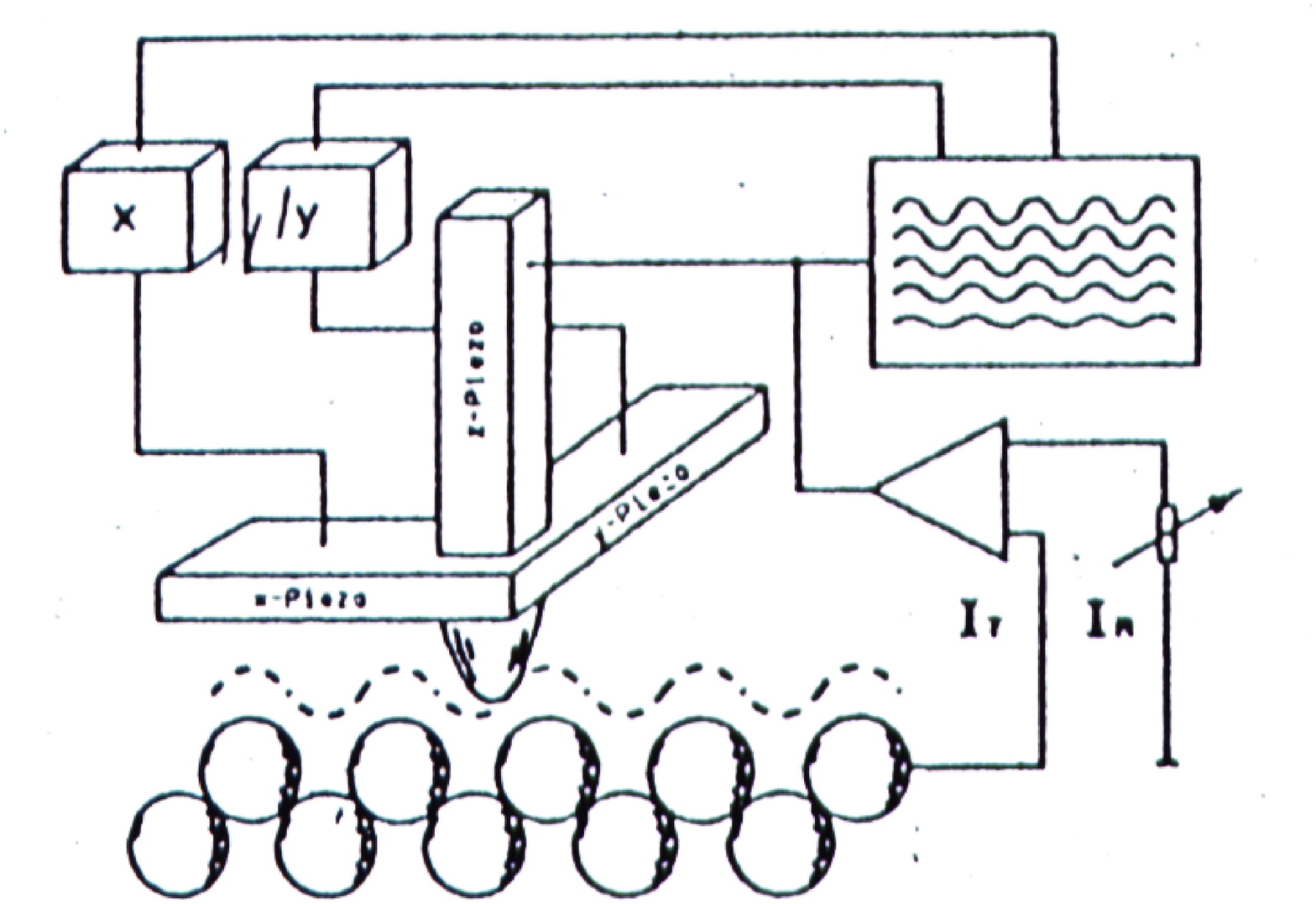}
  \end{center}
  \caption[Generic scheme of an STM]{Generic scheme of an STM. A very fine tip (usually tungsten or
  platinum) is scanned along the topography of the surface. As the distance between tip and sample
  changes while the tip is held at \emph{constant height}, the tunneling current will change. In
  \emph{constant current mode}, the tunneling current is kept constant over a feedback circuit which
  adjusts the height of the tip over the sample by controlling the piezo manipulators.}
  \label{fig:stmprincip}
\end{figure}

For a generic wave function which represents a particle, for the transmission probability it can be
shown \cite{schwabl_quanten} that:

\begin{eqnarray}
T = |S(E)|^2 ~ = ~ \frac{16E(V_0 - E)}{V_0^2} ~ \exp \left(-4 \sqrt{2m(V_0 - E)} ~ \frac{d}{\hbar}\right)
\end{eqnarray}

where $S$ is the ratio of the amplitudes of the incoming and transmitted wave functions, $E$ the
kinetic energy of the particle and $V_0$ the energy of the potential barrier. The equation is valid
in the approximation of wide and high barriers. Thus the probability of a particle to tunnel through
a barrier decays exponentially with the width $d$ of the barrier. Using \emph{Fermi's Golden Rule},
we can derive an expression for the tunneling current:

\begin{equation}
I_{\text{tunnel}} \propto \exp (-2 \kappa d)
\label{eqn:tunnelingcurrent}
\end{equation}

with $\kappa = \frac{\sqrt{2m_e\Phi}}{\hbar}$, $\phi$ being the work function of the conducting material.
From Equation \ref{eqn:tunnelingcurrent} it is clear, that $I_{\text{tunnel}}$ depends exponentially on the distance
$d$ between tip and surface: if that distance changes by just one \AA ngstr\o m, the current will change
by one magnitude already. Thus even smallest corrugations around $0.01~\text{\AA}$ can be detected
at best. Figure \ref{fig:stmprincip} shows the generic scheme of an STM.

\medskip

We used a combined STM/AFM setup by \emph{Omicron Vakuumphysik GmbH} \cite{omicron_userguide}. It's located
within the analysis chamber of the MBE chamber (see Fig. \ref{fig:mbechamber}) and can be operated both
as an STM as well as an AFM. The images were stored with the software provided by Omicron (\emph{Scala Pro 4.1}),
for analysis of the images we used the freeware \emph{wsxm} \cite{horcas_wsxm}. It allows plane correction
(for taking into account when the sample is not located perpendicular to the STM tip), Fourier analysis,
smoothing, roughness analysis, distance calculations and a lot of methods to spruce up the images for better
presentation.

\medskip

\begin{figure}
  \subfigure[First STM image of Fe]{\includegraphics[width=0.4\textwidth]{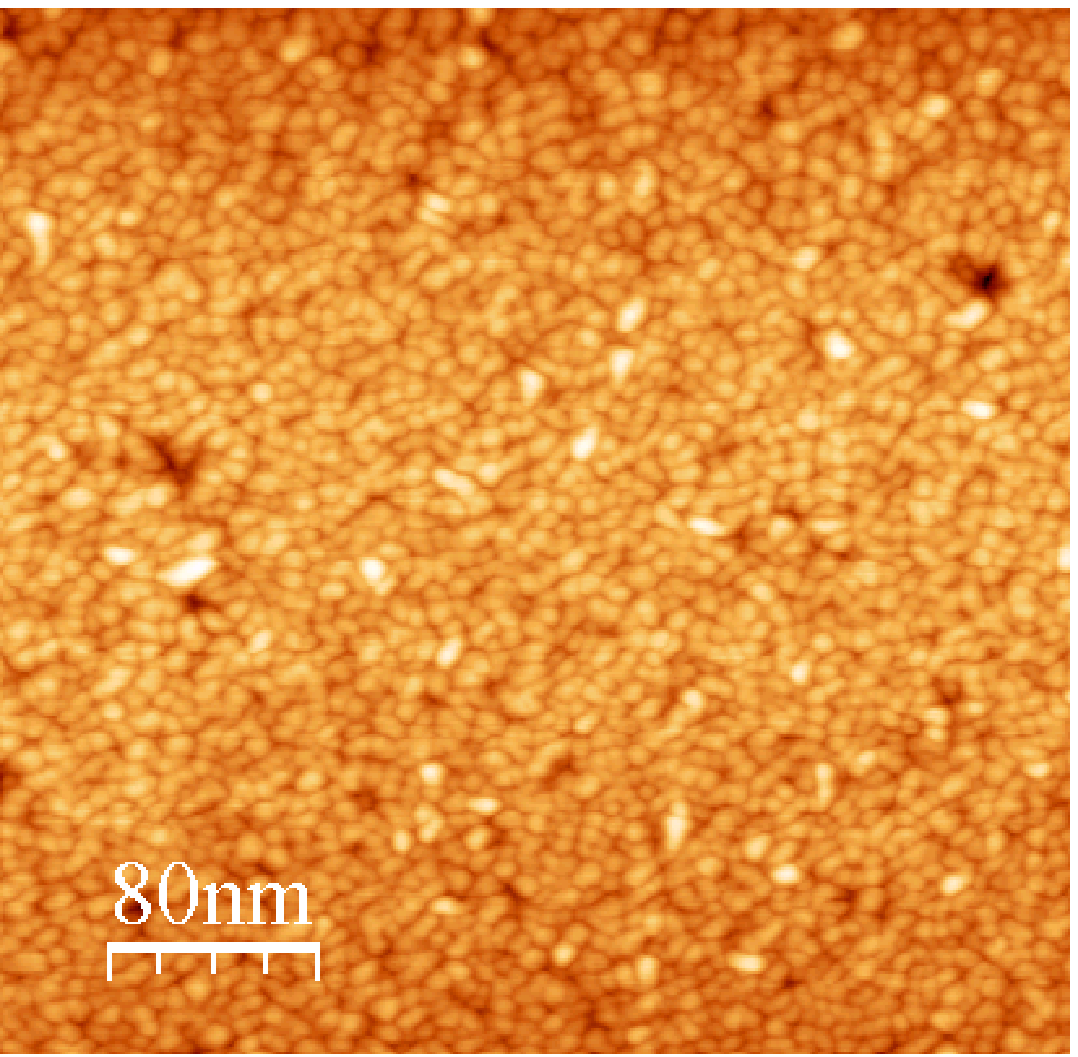}}
  \hfill
  \subfigure[First STM image of Au]{\includegraphics[width=0.4\textwidth]{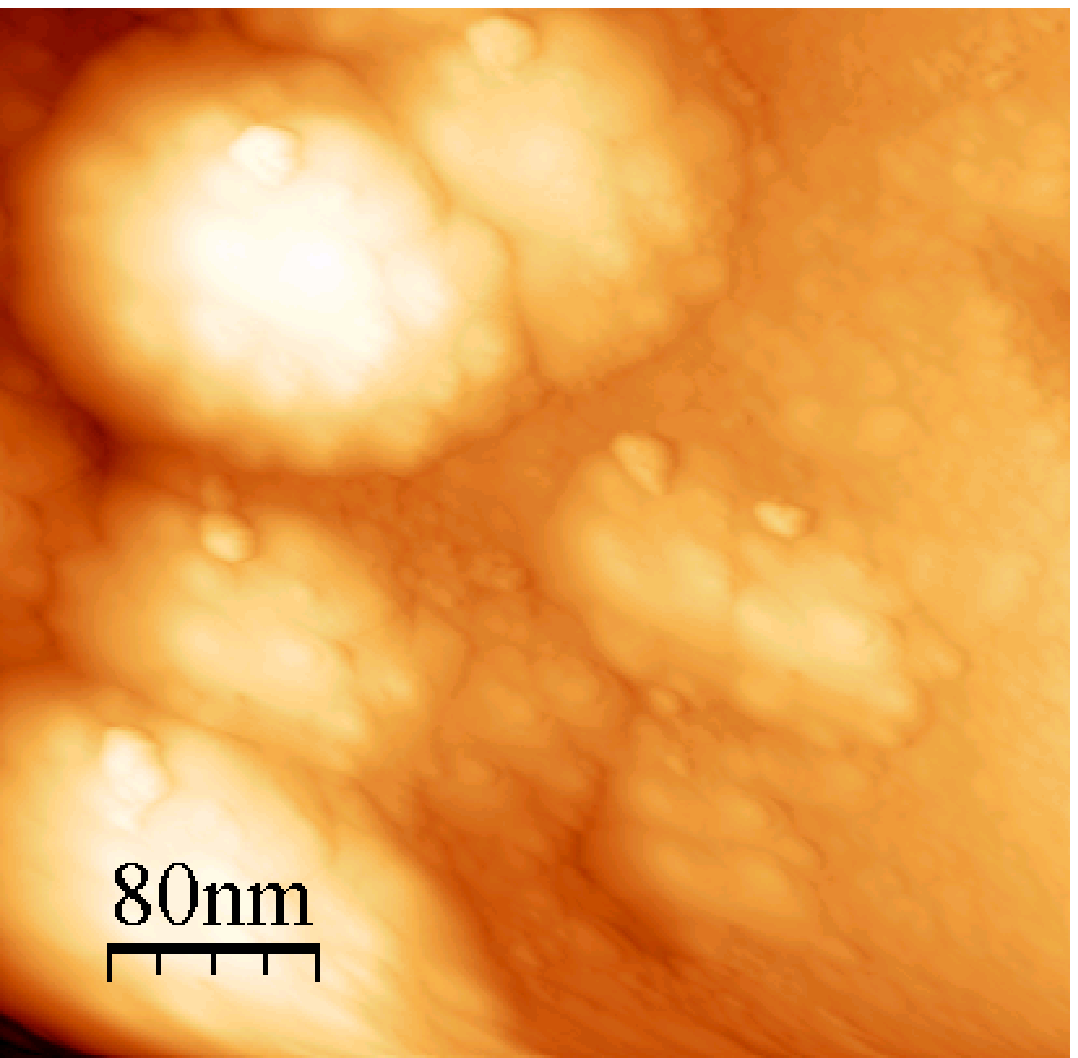}}
  \caption[First STM images of Fe and Au]{First STM images taken of iron and gold surfaces evaporated with the electron
  beam evaporator. While iron (a) has a rather smooth surface with a small grain size, the gold surface is marked by
  several large blobs (b). For detailed analysis, see Sec. \ref{sec:corrugationanalysis}, \ref{sec:feaugrainsize},
  \ref{sec:feaugrainsizecomp}.}
  \label{fig:stmfeau}
\end{figure}

Before we started with the real surface analysis with the STM, we ran a few tests to get familiar with the operation
and usage of the STM. It turned out that both positioning the samples with the wobblestick into the STM
as well as the operation of the STM require quite some practice and experience. For example, every tip will
always produce a different quality of images. So does an old tip yield different results. In principle, best
results are achieved with a single-atomic tip. In order to get a single atom at the tip, one needs to carefully
crush the tip into surface in a controlled manner by increasing the \emph{Loop Gain} (a relative value which
controls how strong the feedback control for height correction is set in constant height mode) and the
\emph{Bias Voltage} (the applied voltage between tip and sample) both by a factor of 10 to pick
up an atom from the sample surface. After a minute, both settings can be restored to normal and
the tip is moved to a fresh spot. Usually this will yield usable results if there were problems
acquiring a proper image before. Before acquiring an image, both the forward and backward scan
display windows of the software should be set to \emph{line scan}, allow to control
that both scans of the same line deliver the same profile. If these profiles don't match, either the tip
has still a blunt shape or the current spot on the sample is inappropriate for measurements.

\medskip

For both iron and gold, Table \ref{tab:stmparams} shows the parameters for the Omicron STM which were
used as initial values when tuning the STM for image acquisition.

\begin{table}[htbp]
        \centering
                \begin{tabular}{|c|c|} \hline
                Parameter             & Initial Value  \\ \hline
                Bias Voltage / V      & $0.3$-$0.7$      \\ \hline
                Feedback Current / nA & $0.1$        \\ \hline
                Loop Gain / \%        & $0.5$-$1$        \\ \hline
                Scan Area / nm$^2$    & $400\times400$ \\ \hline
                Scan Speed / nm$\times$s$^{-1}$& $100$  \\ \hline
                \end{tabular}
        \caption{Initial parameters for STM image acquisition.}
        \label{tab:stmparams}
\end{table}


\subsubsection{Corrugation Analysis}
\label{sec:corrugationanalysis}
In the quest to find the optimal parameters for sample preparation we wanted to investigate how the
temperature of the substrate influences the roughness of the surface when evaporating. In
\cite{thuermer_dynamic} Rieder et al. showed that the temperature of the substrate has to be
high enough to overcome the \emph{Schwoebl barrier}. An energy which defines the lower limit
at which large, atomically flat surfaces can be achieved. The additional energy is required
for the atoms to have the necessary mobility to distribute evenly along the surface. Below
this energy, the film is expected to grow islands instead of flat terraces. On the other hand,
high temperatures mean that the atoms have an increased mobility so the chances that
disordering of the surfaces takes place is higher. Apparently both effects are competing
and thus we want to check the situation for the case of iron on MgO$(001)$.

\medskip

The flatness of the surface was important for our iron-gold interface. Large corrugations at the
interface can increase scattering and lower the efficiency of hot electron injection from the ferromagnetic
layer into the gold layer. To investigate the temperature dependence, we evaporated iron from the electron beam
evaporator (see Sec. \ref{sec:esv}) at different substrate temperatures, namely $100^\circ$C, $200^\circ$C,
$300^\circ$C, $400^\circ$C, $500^\circ$C and $600^\circ$C. Several STM images of these samples were
taken directly after evaporation and the data was again analyzed with \emph{wsxm} \cite{horcas_wsxm}.
However, due to problems with the electrical contact, we couldn't take any STM images of the sample
evaporated at $500^\circ$C substrate temperature.

\begin{figure}
\begin{center}
  \includegraphics[width=0.7\textwidth]{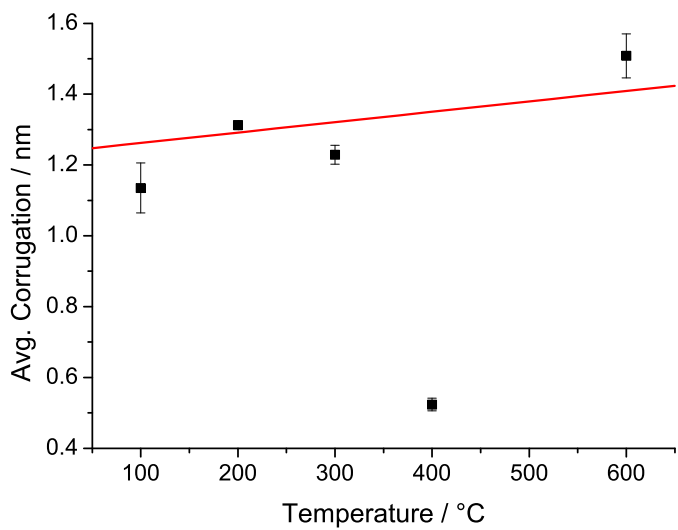}
\end{center}
 \caption[Temperature dependence of corrugation during evaporation.]{Temperature dependence of
   corrugation during evaporation. Data was taken from the FWHMs of the histograms in figures
   \ref{fig:stmtemp100}-\ref{fig:stmtemp600}. The data point at $400^\circ$C was omitted in
   the linear fit since the STM measured unrealistic corrugations (see text), the rest was
   fitted with a linear function.}
 \label{fig:stmtempcompplot}
\end{figure}

The software \emph{wsxm} allows for roughness analysis by plotting histograms of the surface corrugation. This
yields a column bar plot which plots the number of events over the height of each pixel in the image.
Importing these histograms into \emph{Microcal Origin} and fitting them with a Gaussian yields a
value for the roughness from full width at half maximum of (FWHM) the peaks. Four images of each temperature
were taken and analyzed. We took the mean value of the four FWHM values calculated and finally
plotted these over the temperature. The result is shown in figure \ref{fig:stmtempcompplot}.

\medskip

To conclude, the corrugation is nearly constant over the temperature range we investigated.
The corrugation shows a slight increase with growing temperature, however, this might be
owed to other external fluctuations during evaporation or STM measurements. We will see
in the next chapter though that the temperature still influences the film quality as
higher temperatures increase the grain sizes. As a conclusion, we evaporated the films
at room temperature to achieve a well-defined interface between iron and gold.

\begin{figure}
\begin{flushleft}
\subfigure{\includegraphics[width=0.5\textwidth]{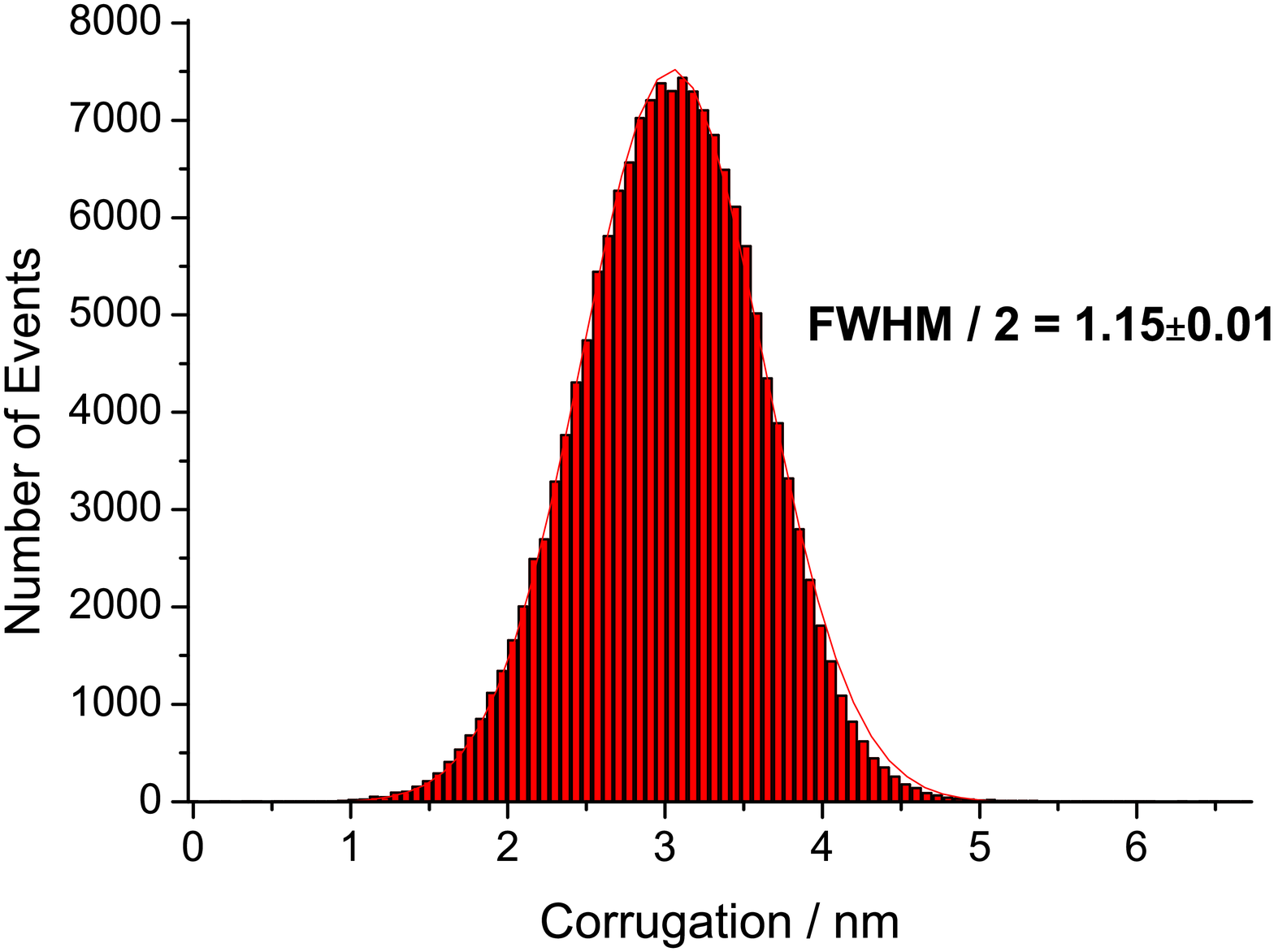}}
\hfill
\subfigure{\includegraphics[width=0.3\textwidth]{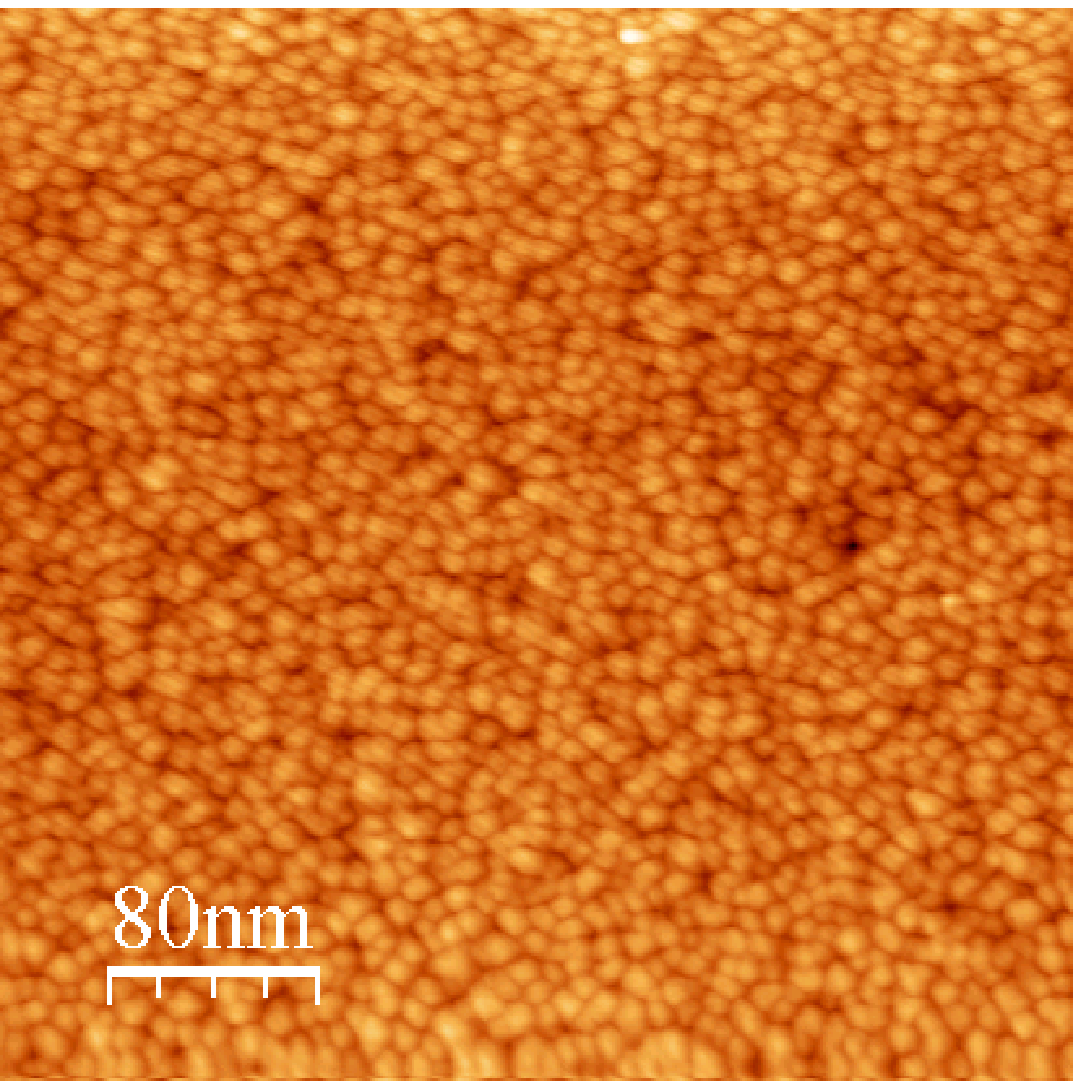}}
\subfigure{\includegraphics[width=0.5\textwidth]{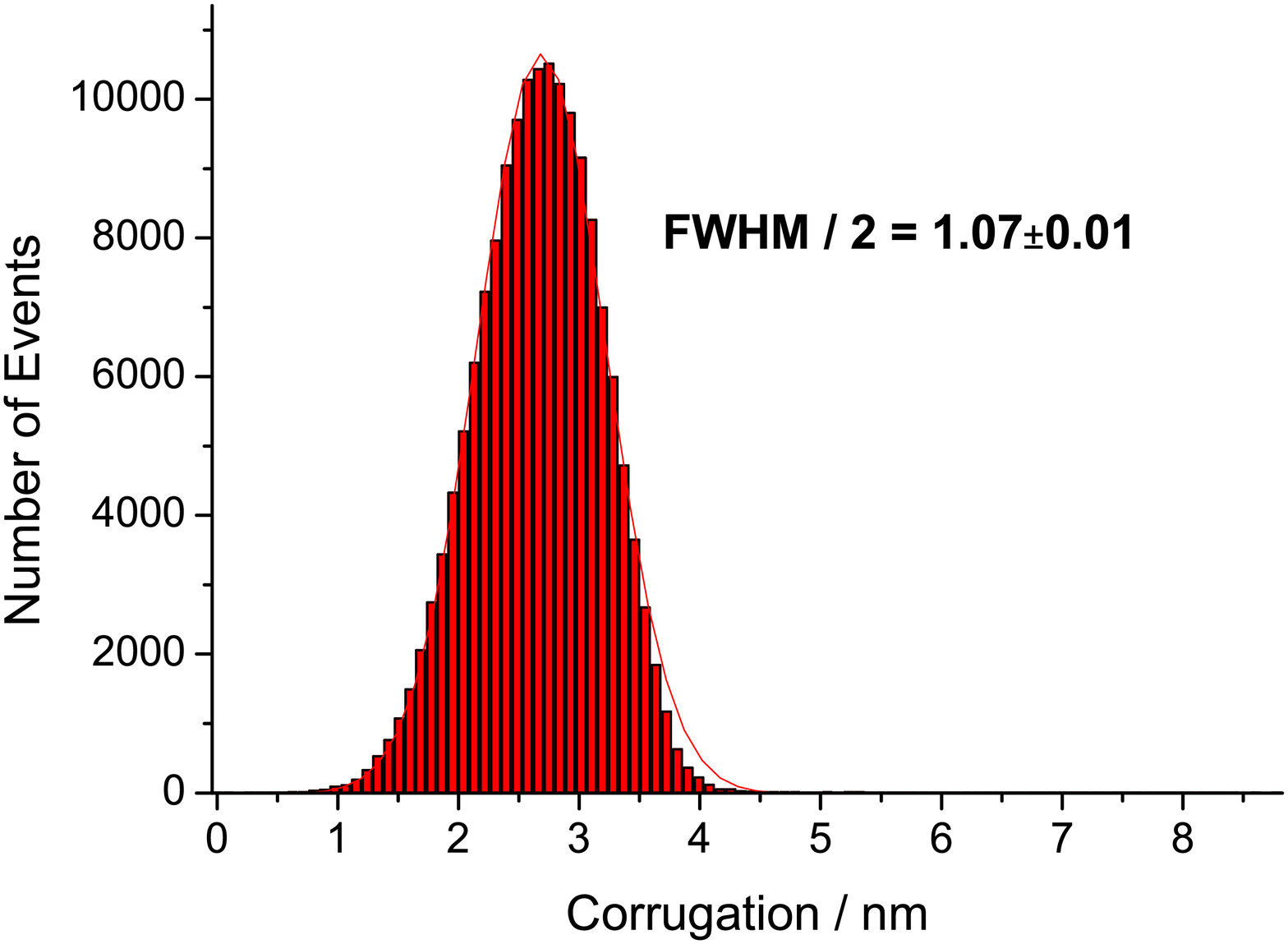}}
\hfill
\subfigure{\includegraphics[width=0.3\textwidth]{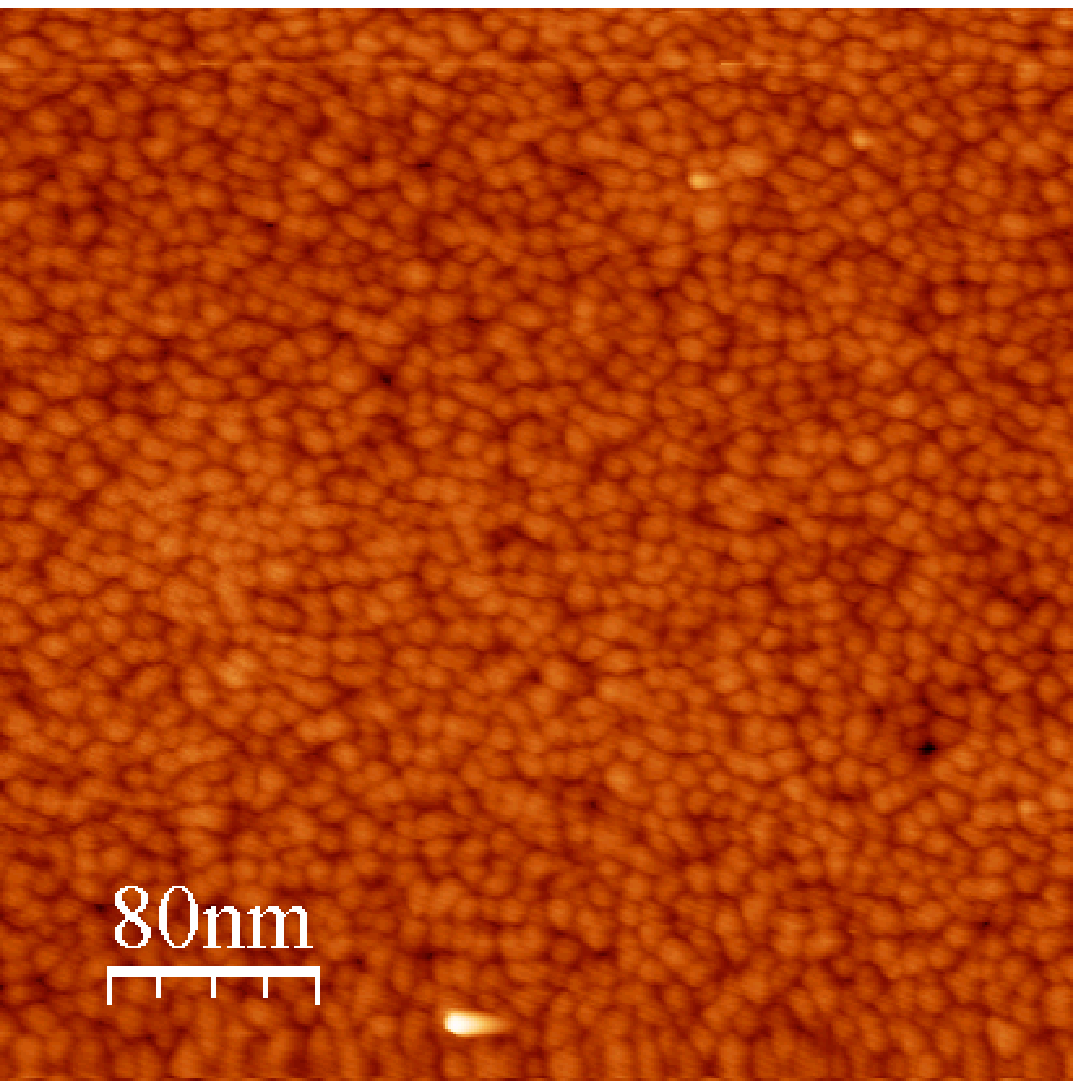}}
\subfigure{\includegraphics[width=0.5\textwidth]{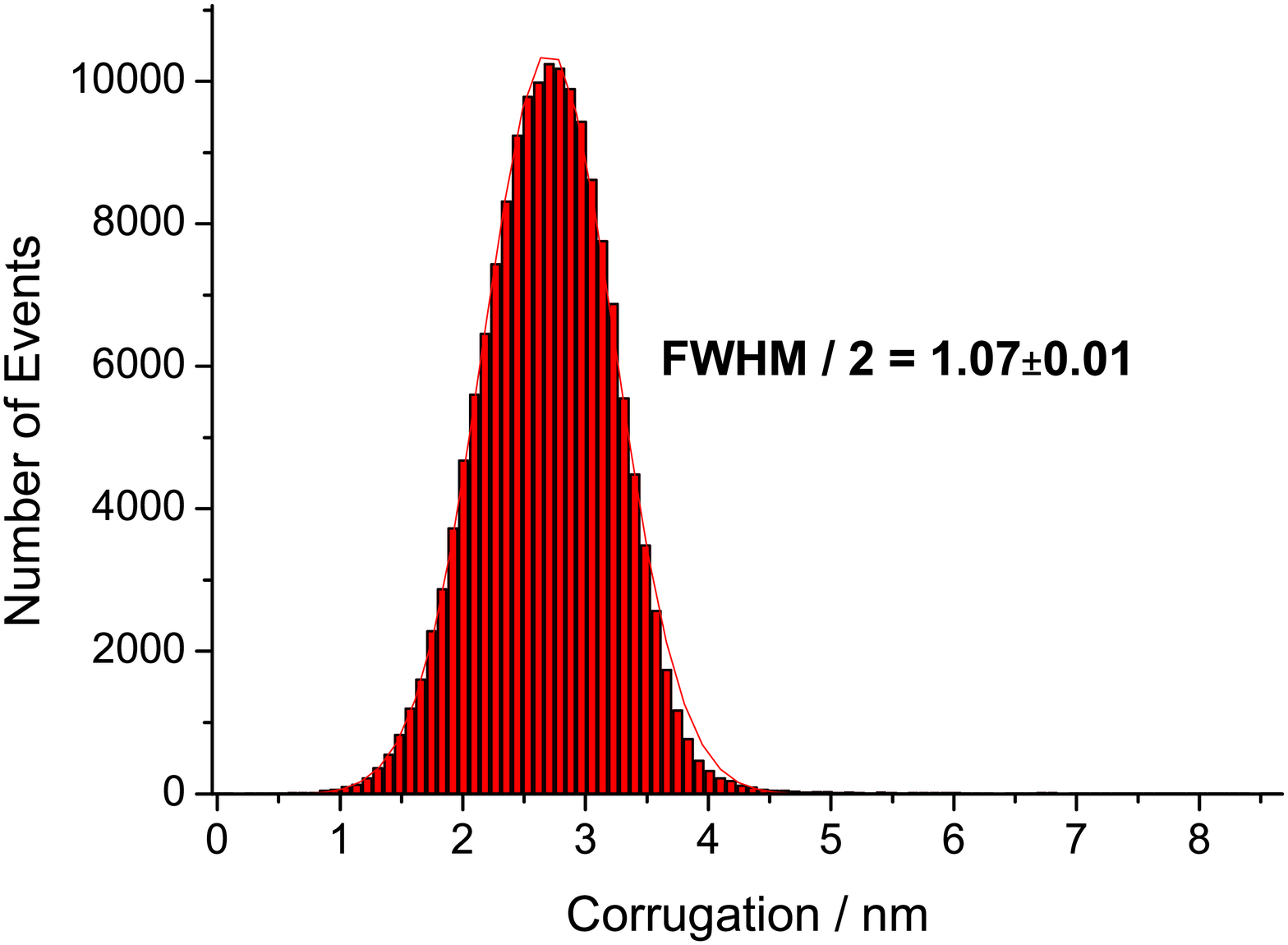}}
\hfill
\subfigure{\includegraphics[width=0.3\textwidth]{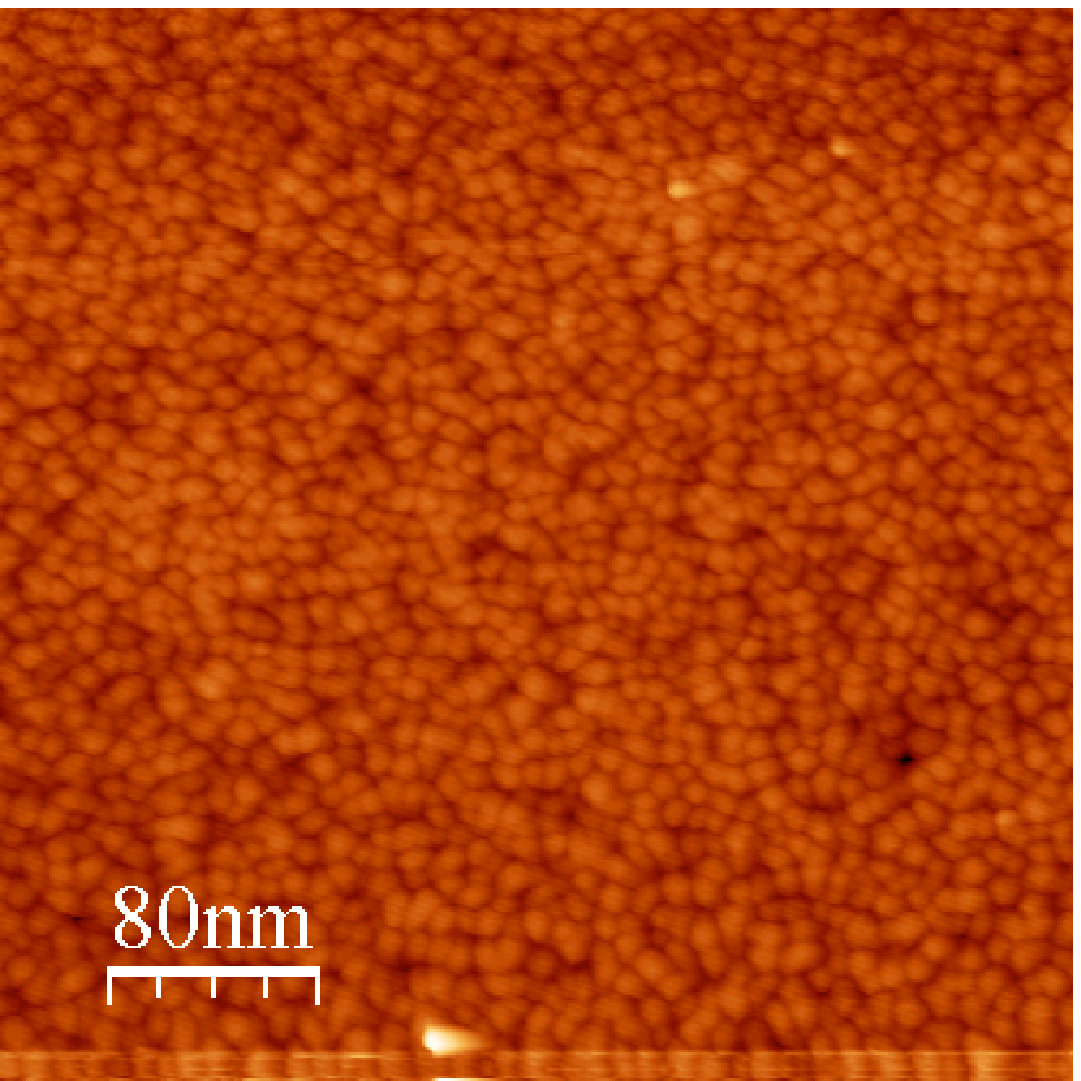}}
\subfigure{\includegraphics[width=0.5\textwidth]{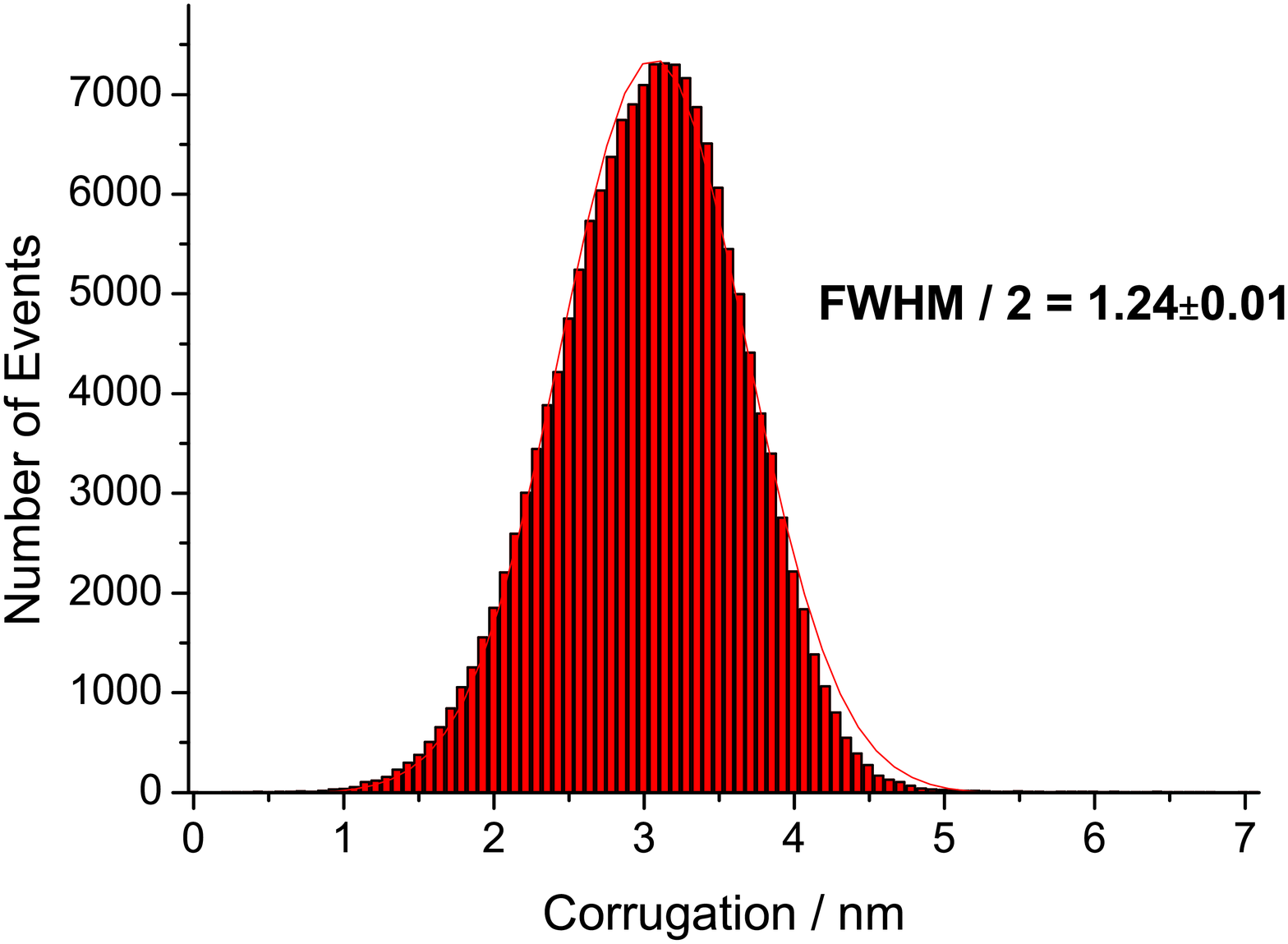}}
\hfill
\subfigure{\includegraphics[width=0.3\textwidth]{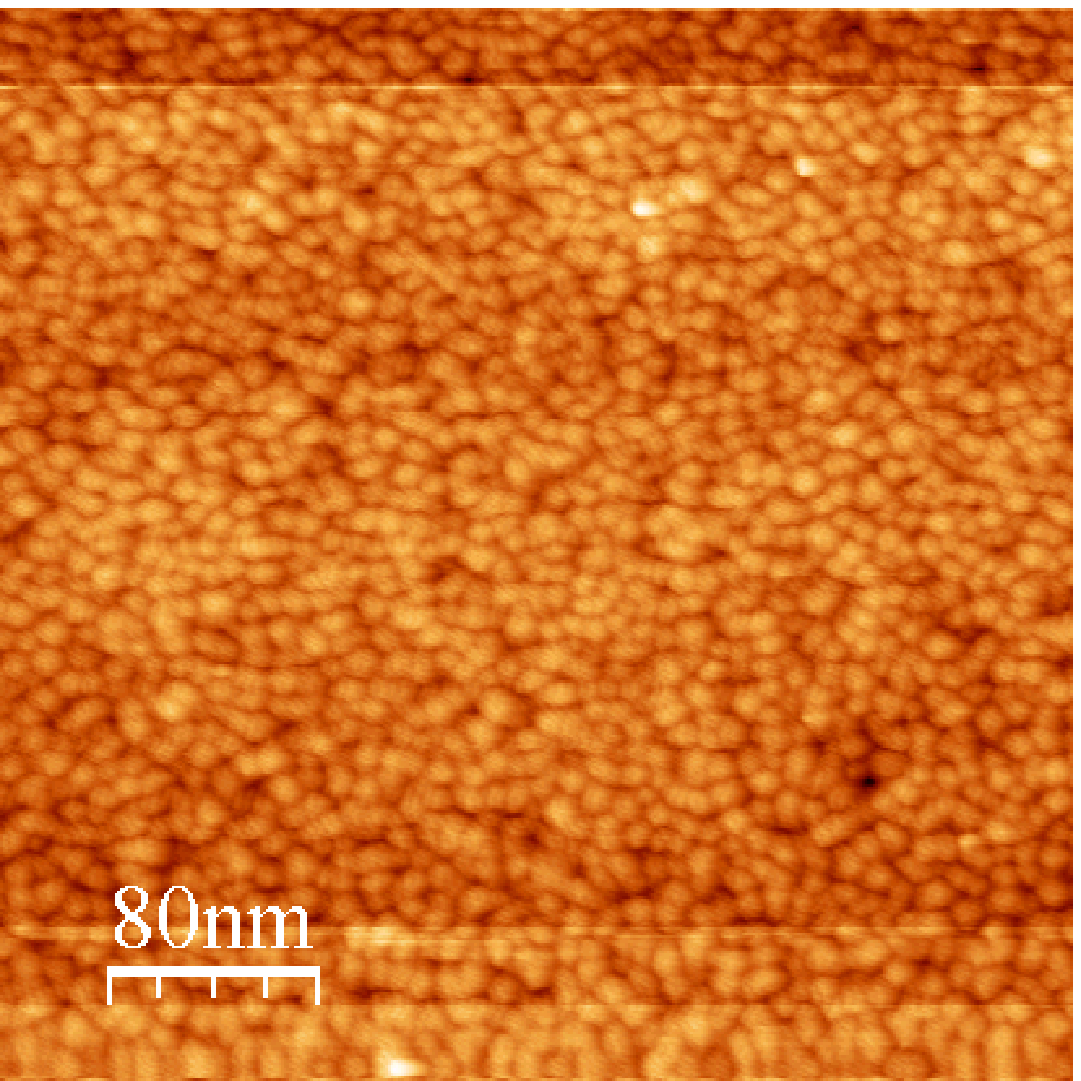}}
\caption[Corrugation histograms of Fe evaporated at $100~^\circ$C]{Corrugation histograms of
  Fe evaporated at $100~^\circ$C}
\label{fig:stmtemp100}
\end{flushleft}
\end{figure}

\begin{figure}
\begin{flushleft}
\subfigure{\includegraphics[width=0.5\textwidth]{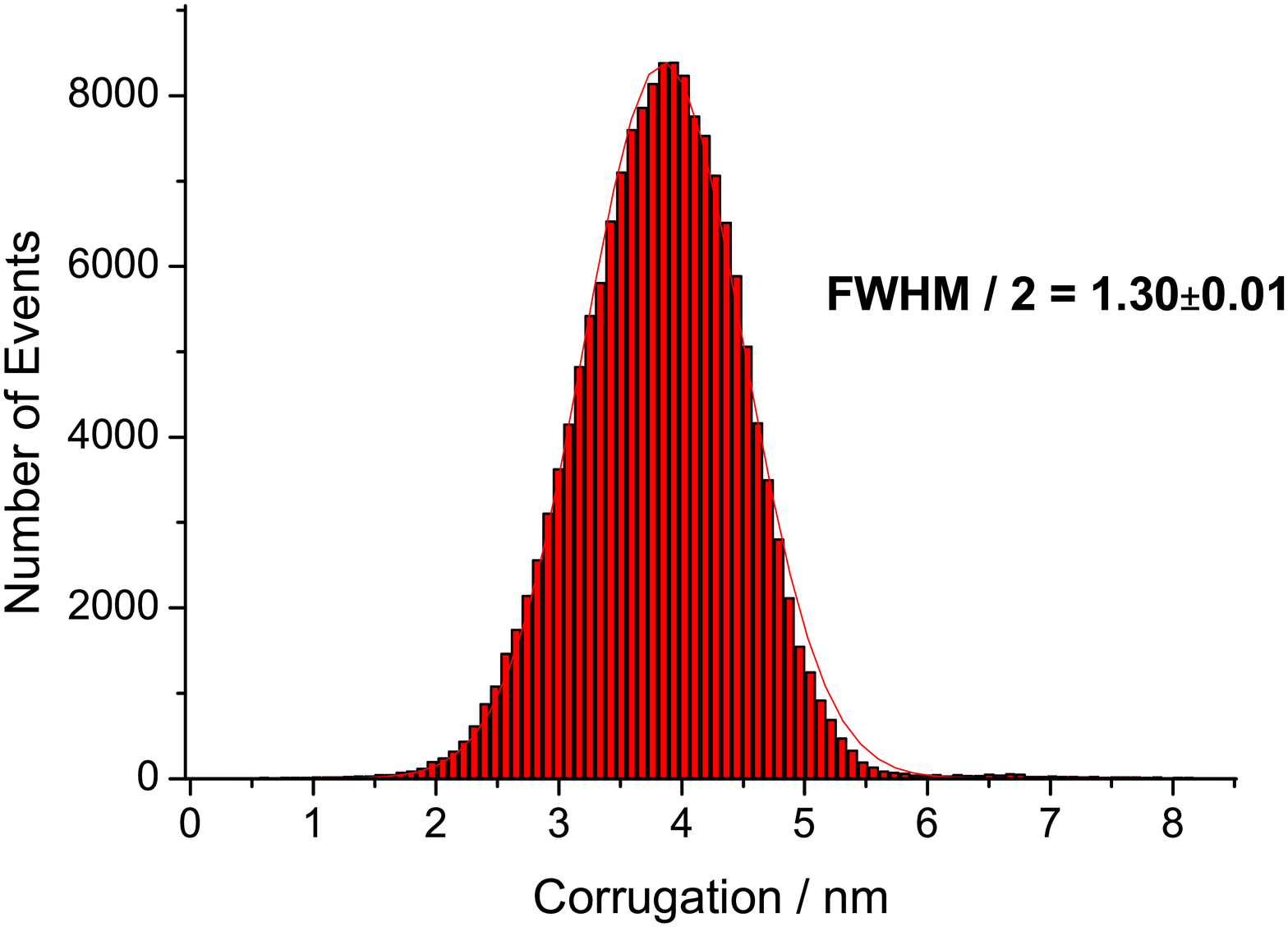}}
\hfill
\subfigure{\includegraphics[width=0.3\textwidth]{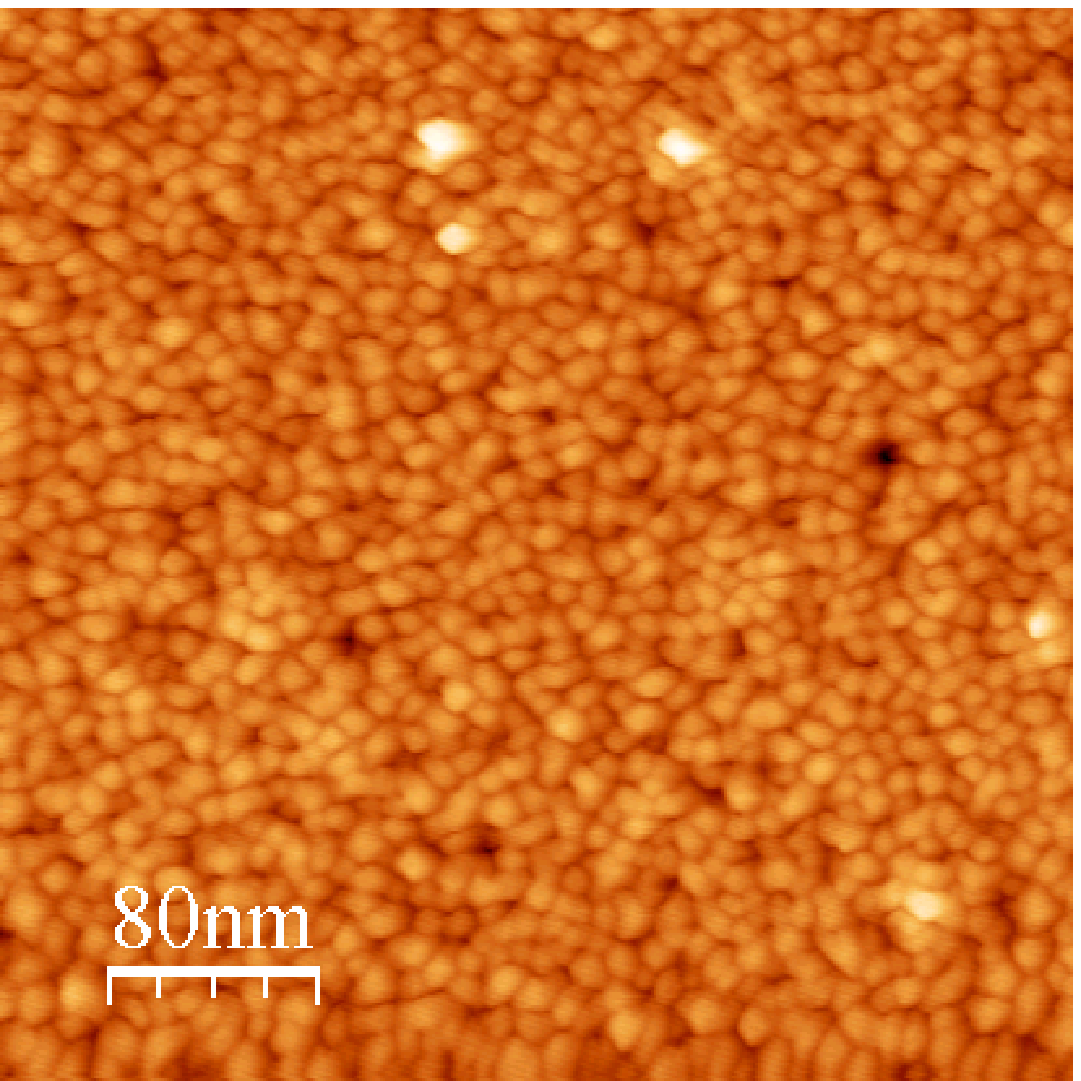}}
\subfigure{\includegraphics[width=0.5\textwidth]{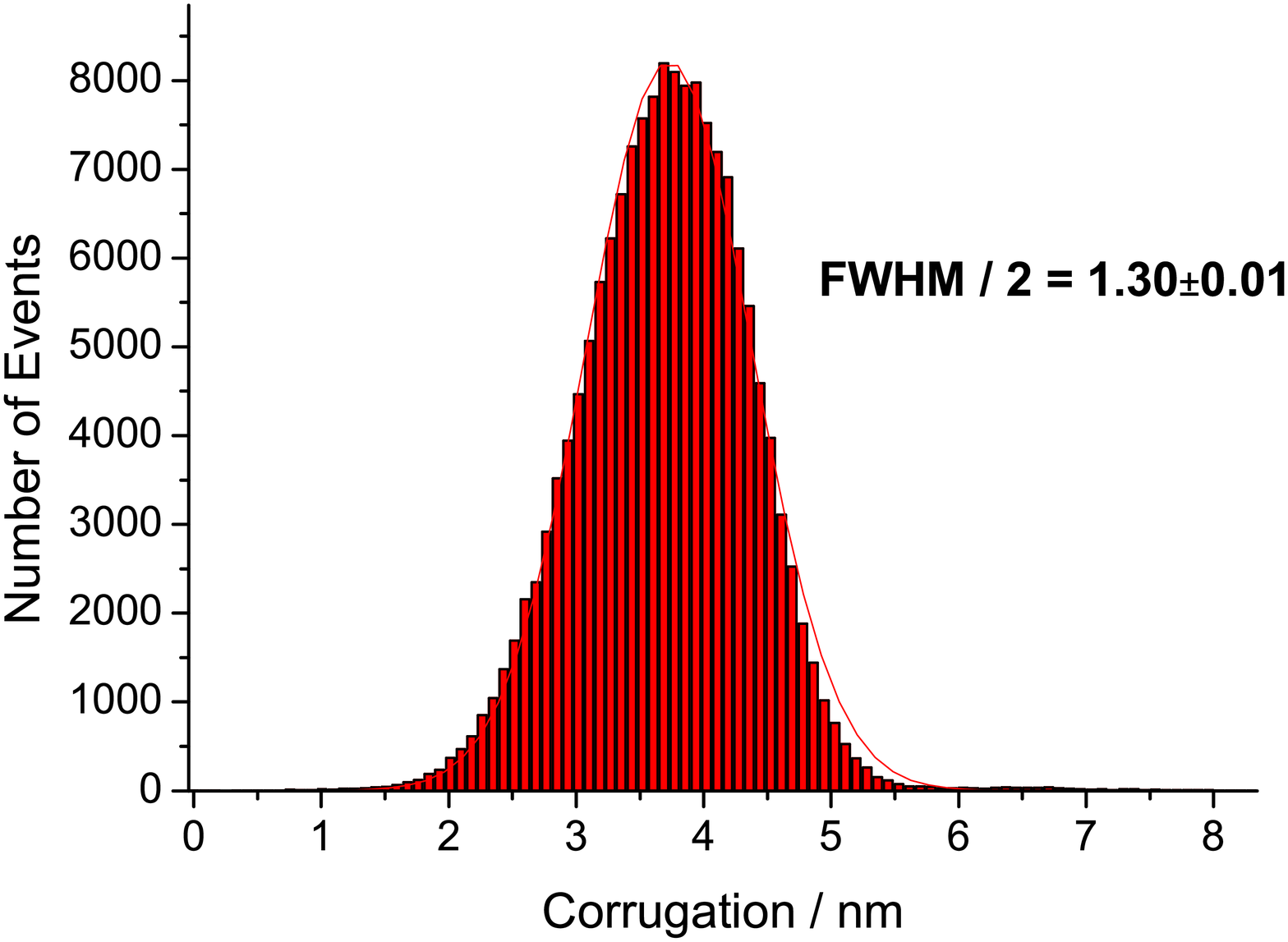}}
\hfill
\subfigure{\includegraphics[width=0.3\textwidth]{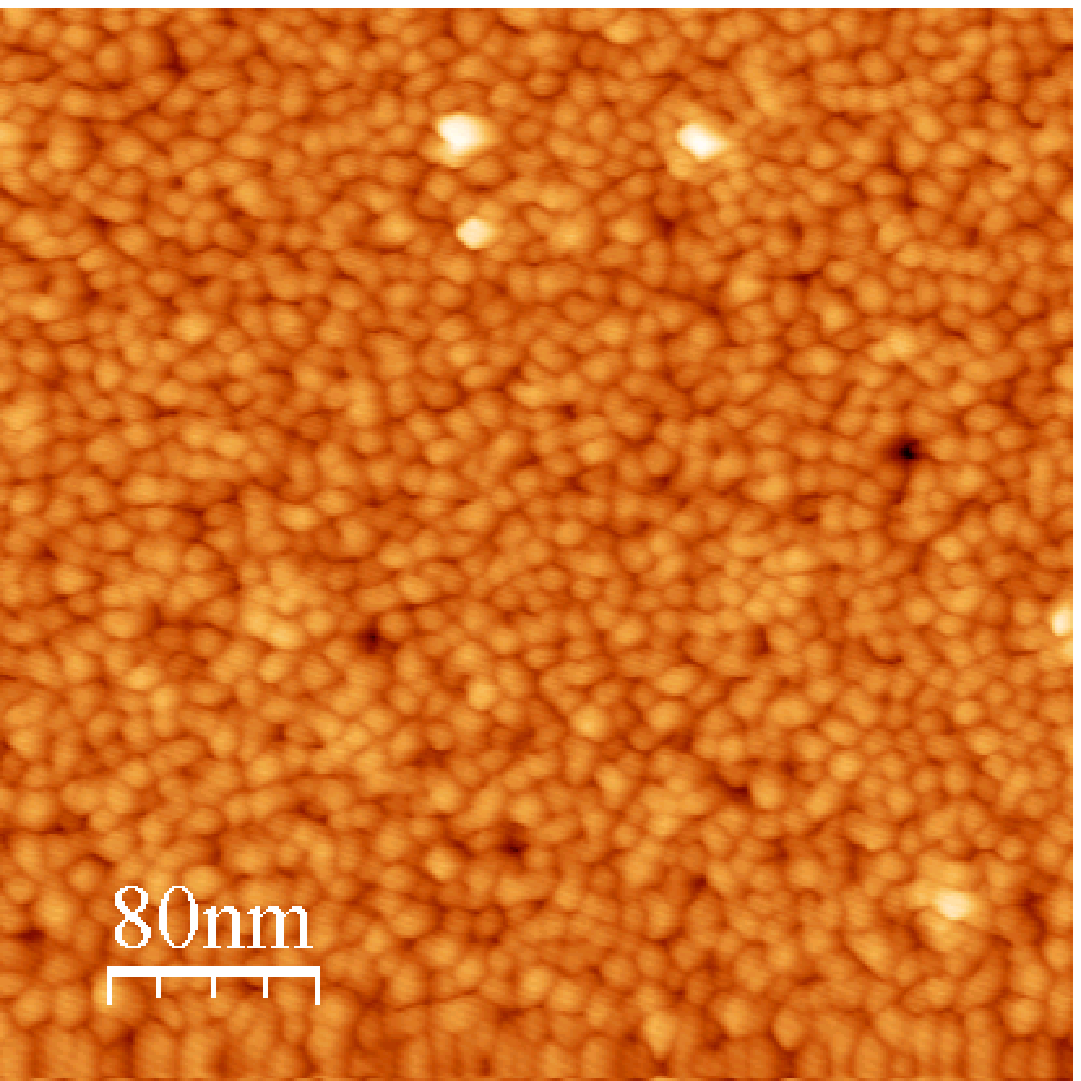}}
\subfigure{\includegraphics[width=0.5\textwidth]{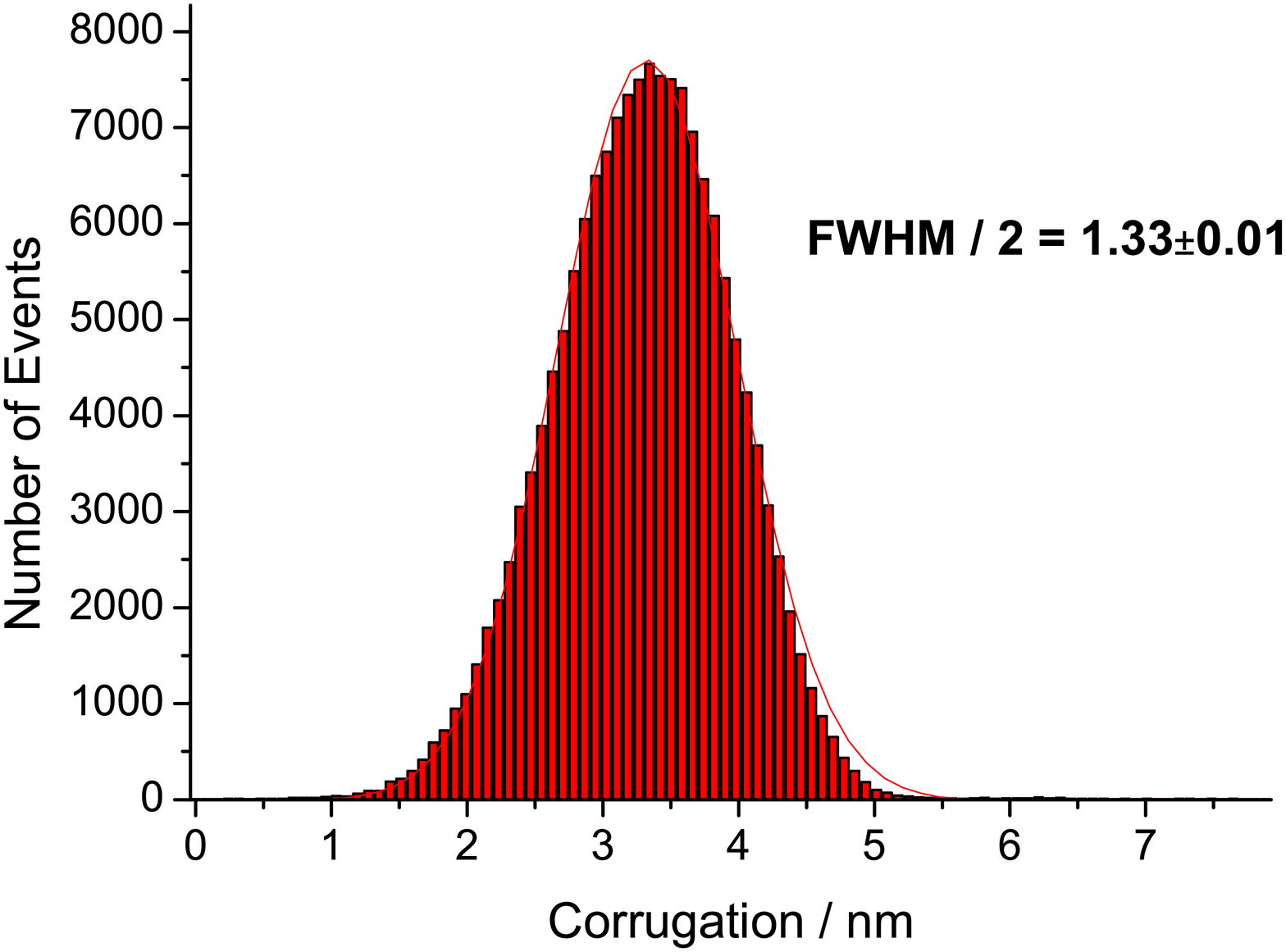}}
\hfill
\subfigure{\includegraphics[width=0.3\textwidth]{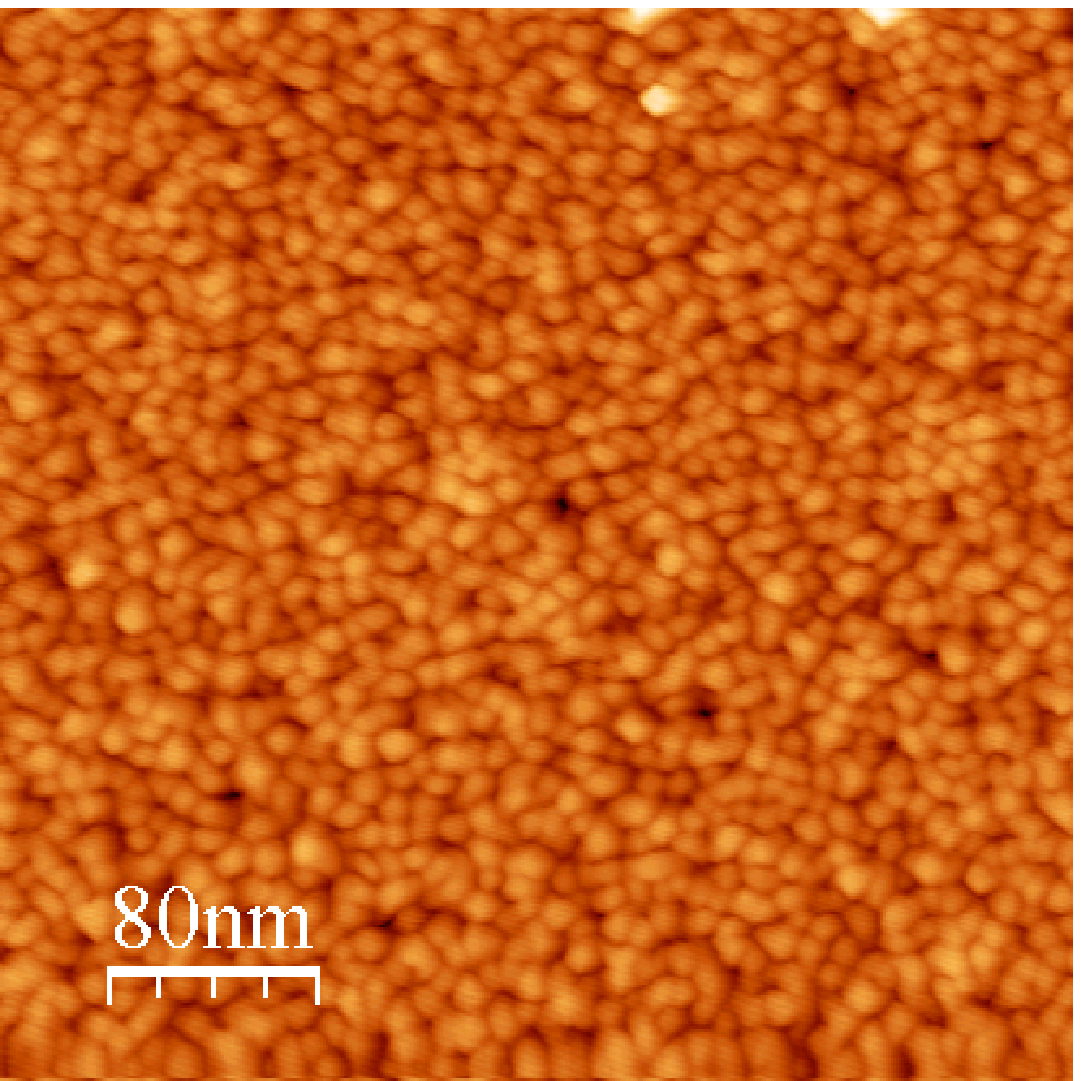}}
\subfigure{\includegraphics[width=0.5\textwidth]{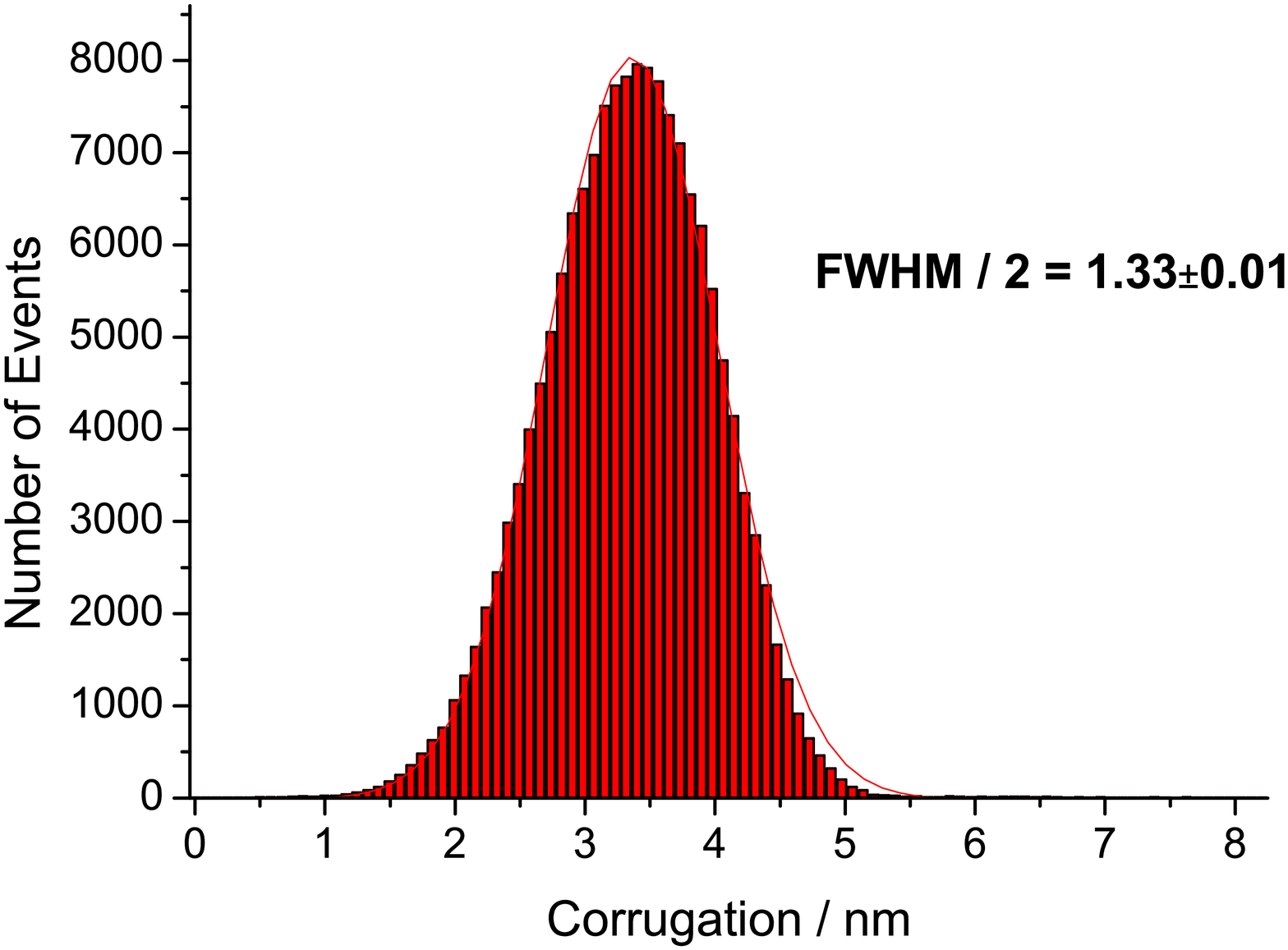}}
\hfill
\subfigure{\includegraphics[width=0.3\textwidth]{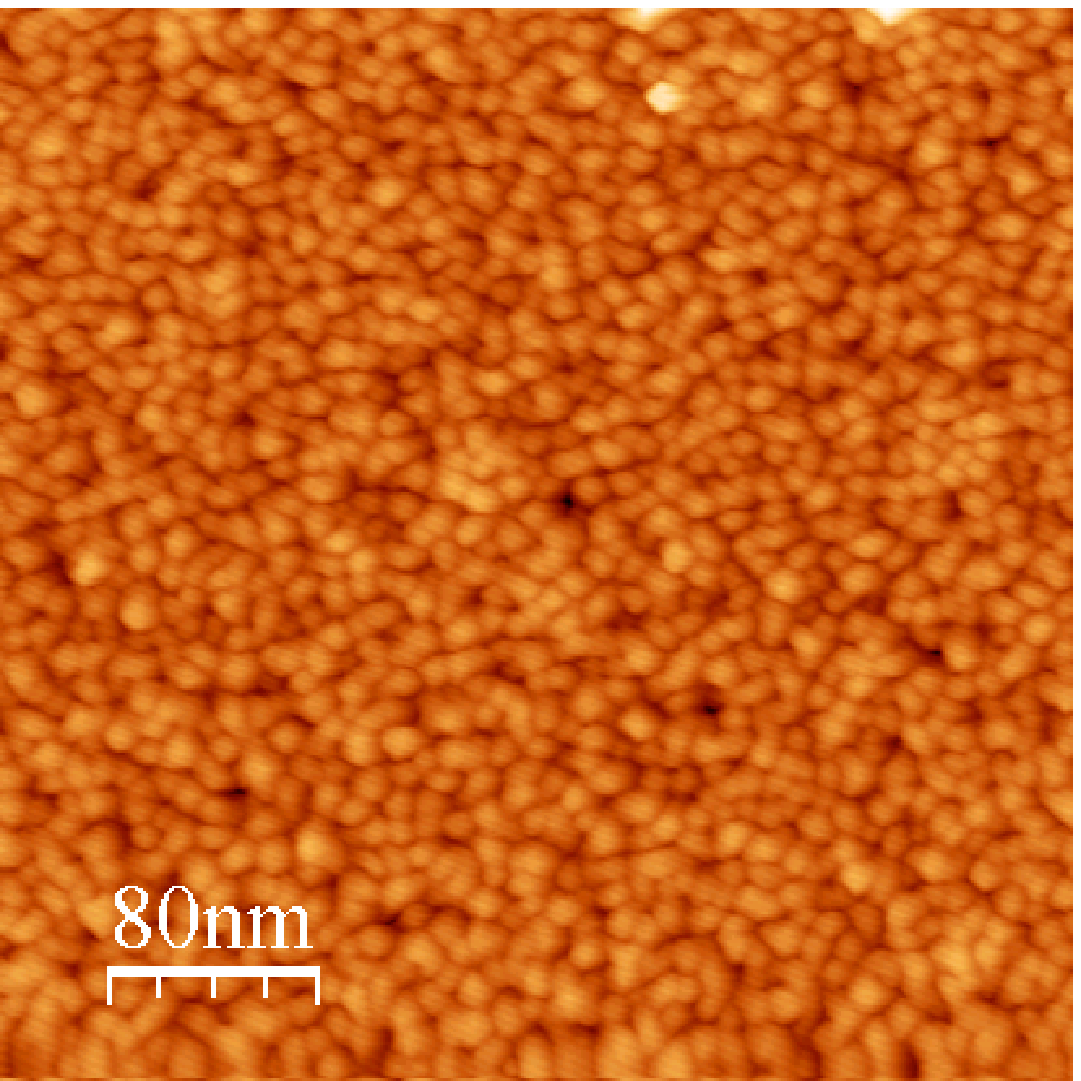}}
\caption[Corrugation histograms of Fe evaporated at $200~^\circ$C]{Corrugation histograms of
  Fe evaporated at $200~^\circ$C}
\label{fig:stmtemp200}
\end{flushleft}
\end{figure}

\begin{figure}
\begin{flushleft}
\subfigure{\includegraphics[width=0.5\textwidth]{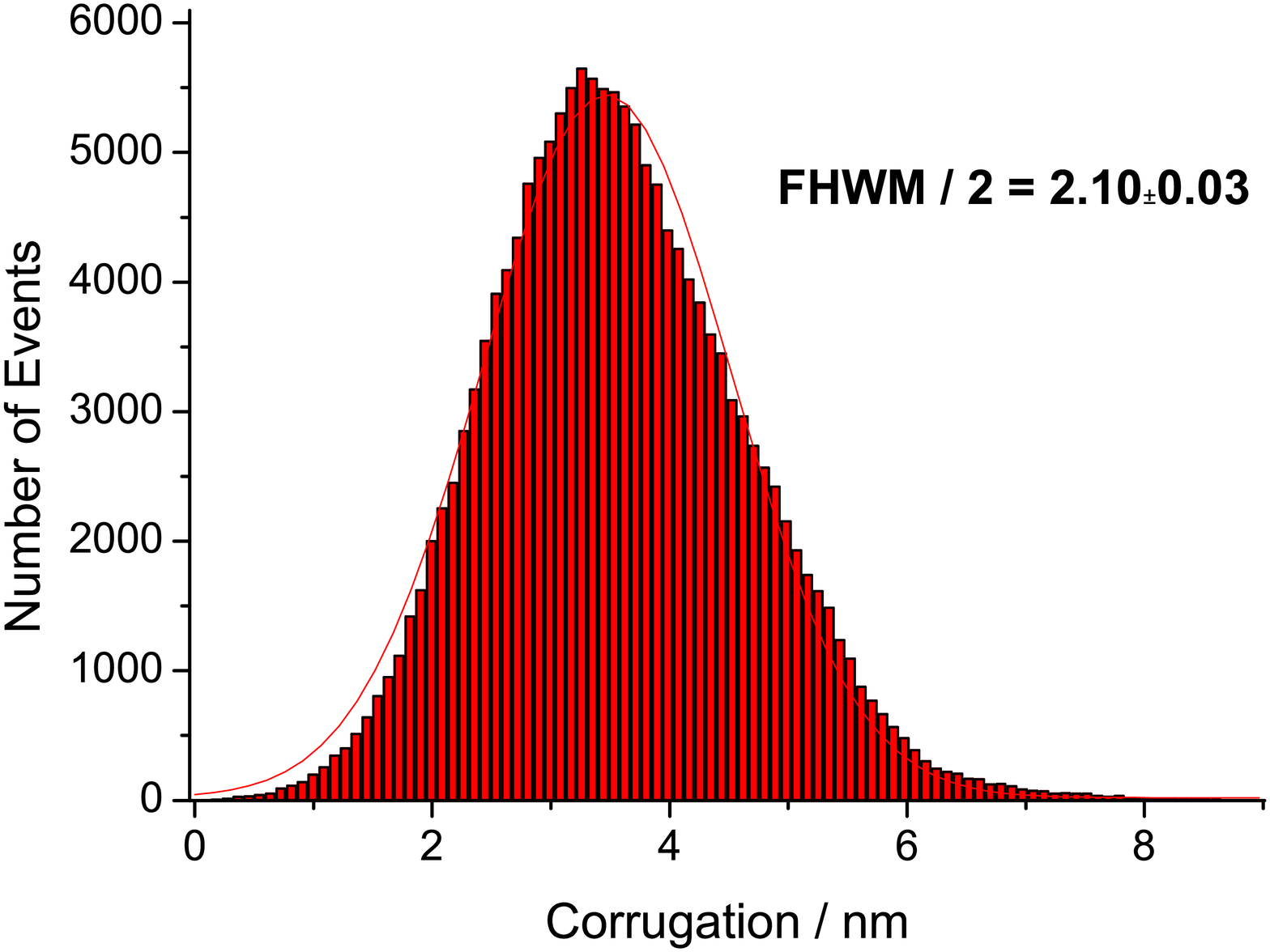}}
\hfill
\subfigure{\includegraphics[width=0.3\textwidth]{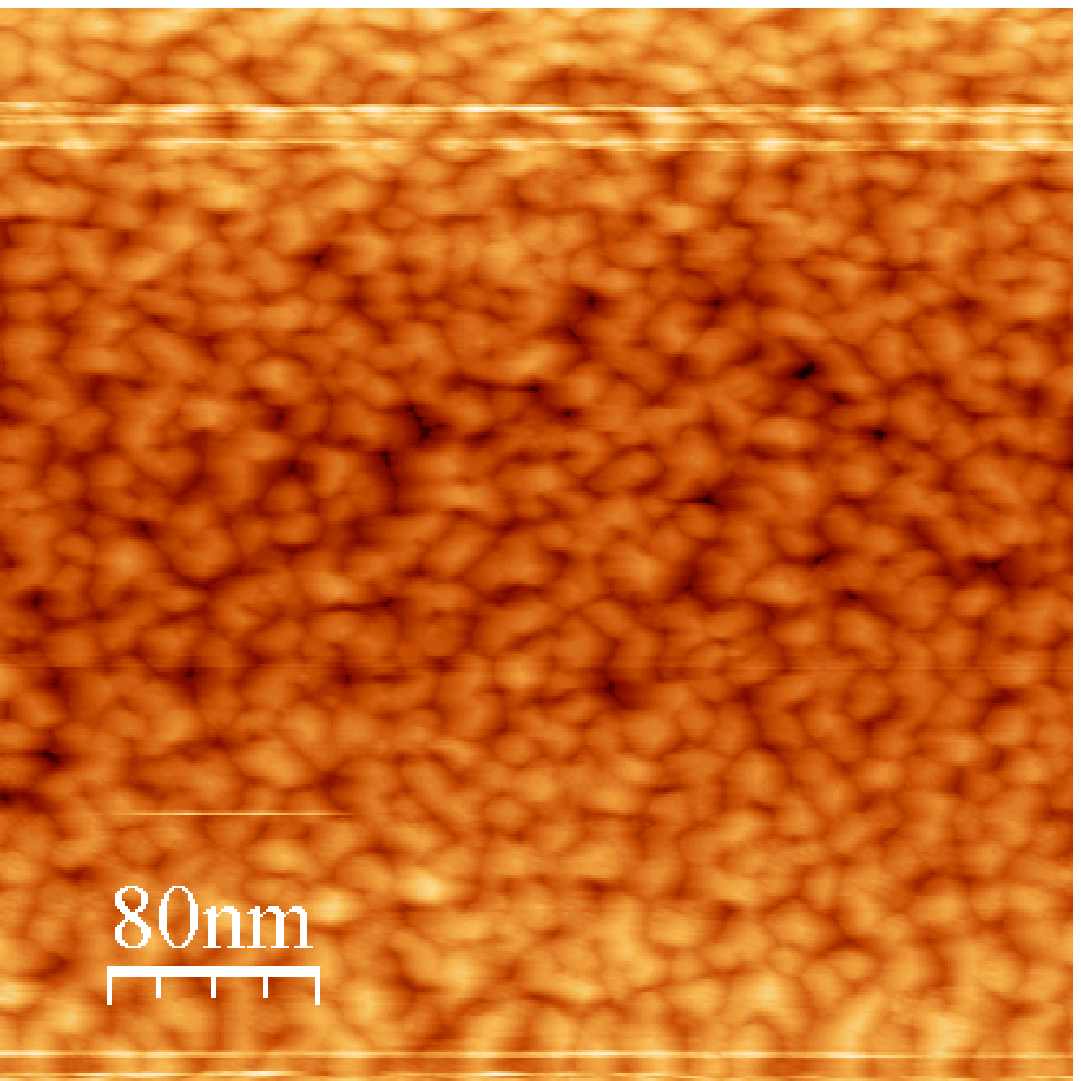}}
\subfigure{\includegraphics[width=0.5\textwidth]{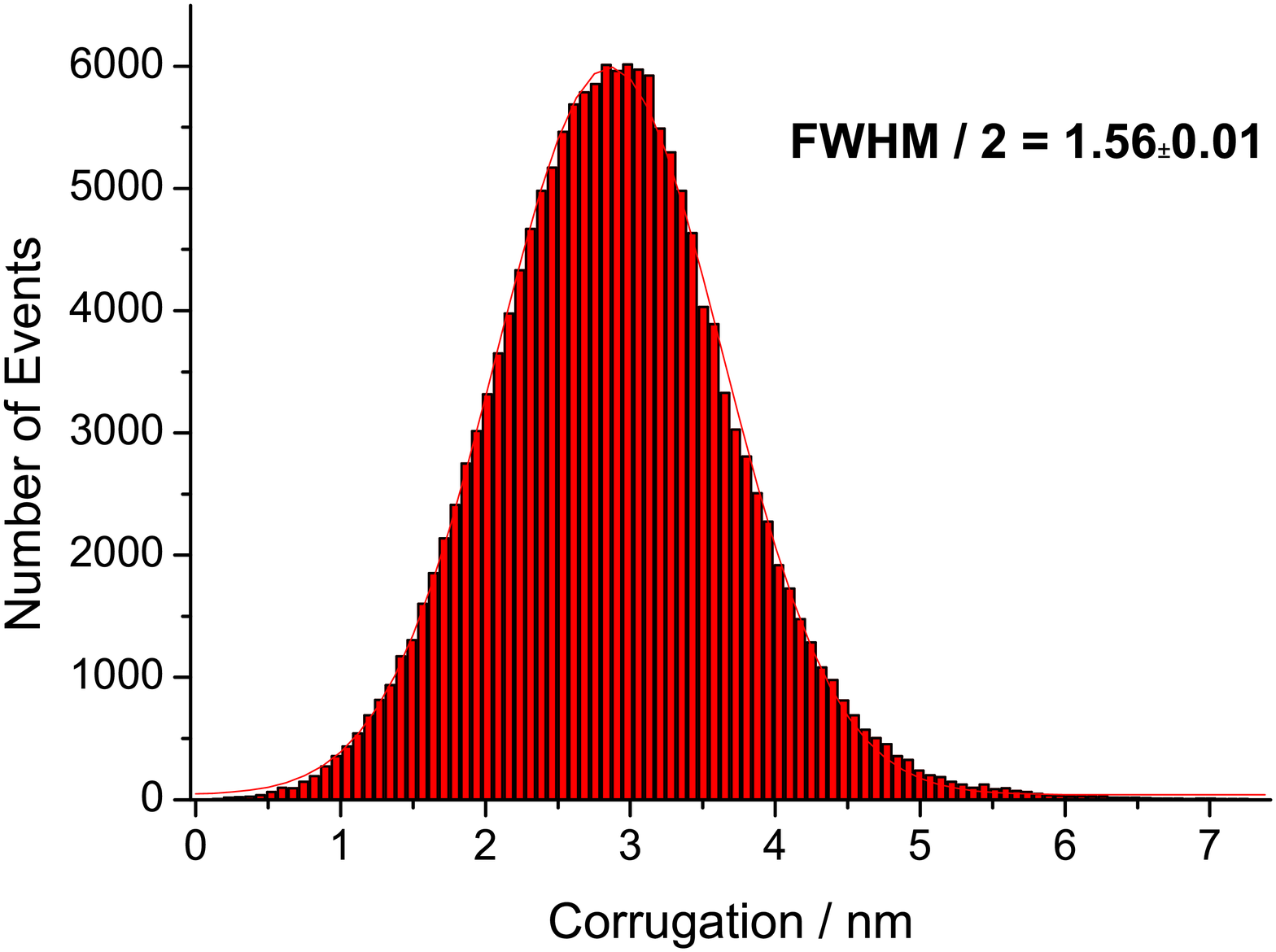}}
\hfill
\subfigure{\includegraphics[width=0.3\textwidth]{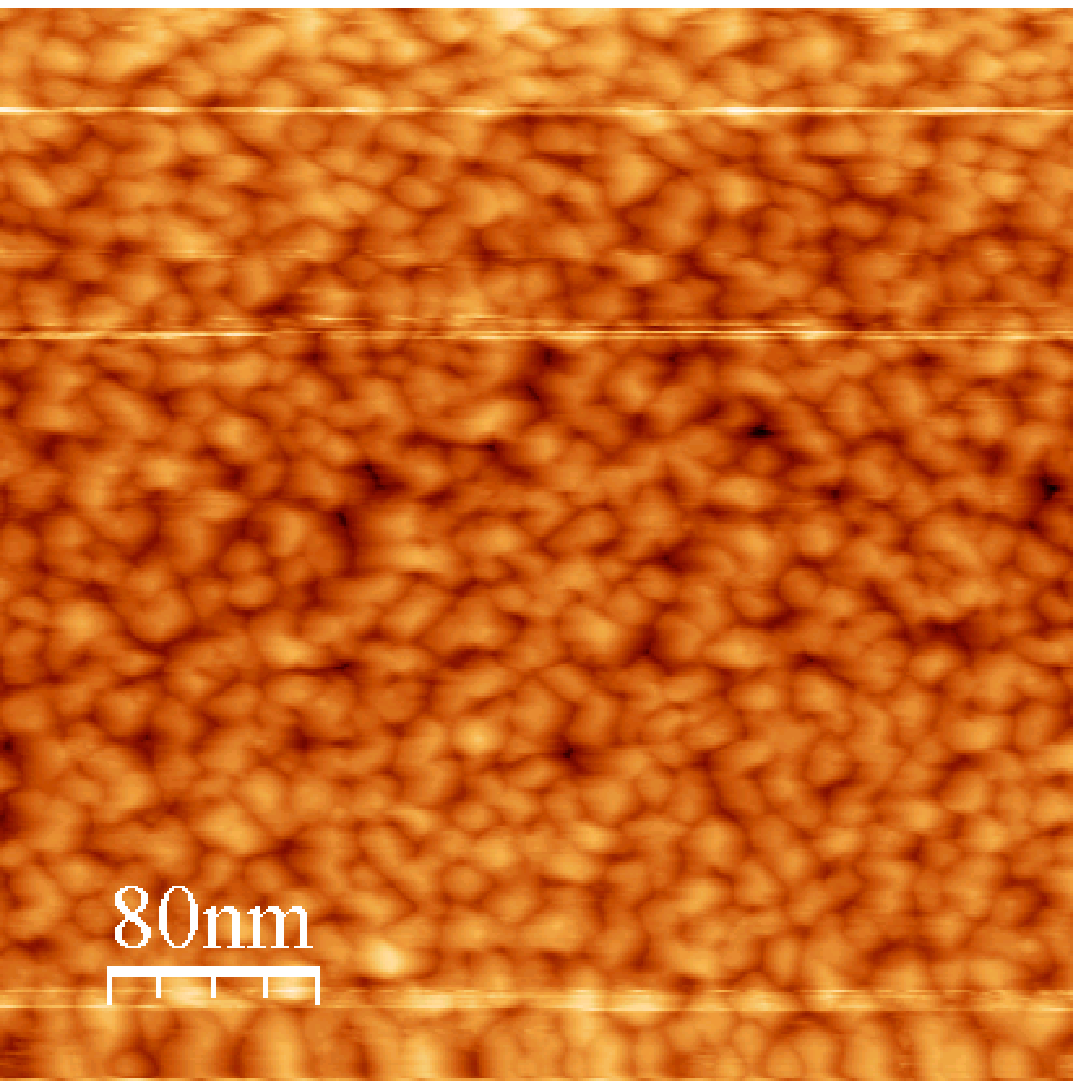}}
\subfigure{\includegraphics[width=0.5\textwidth]{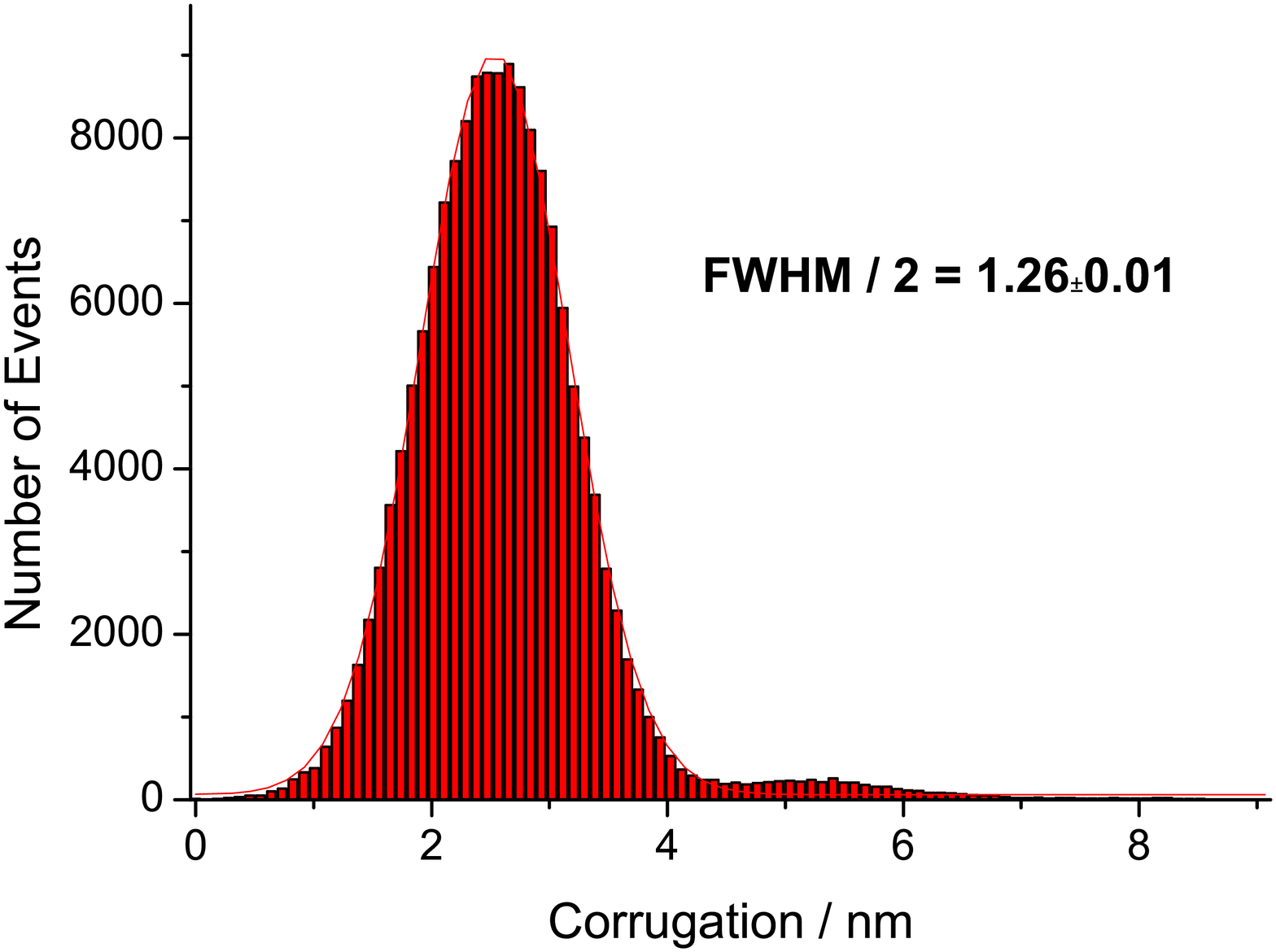}}
\hfill
\subfigure{\includegraphics[width=0.3\textwidth]{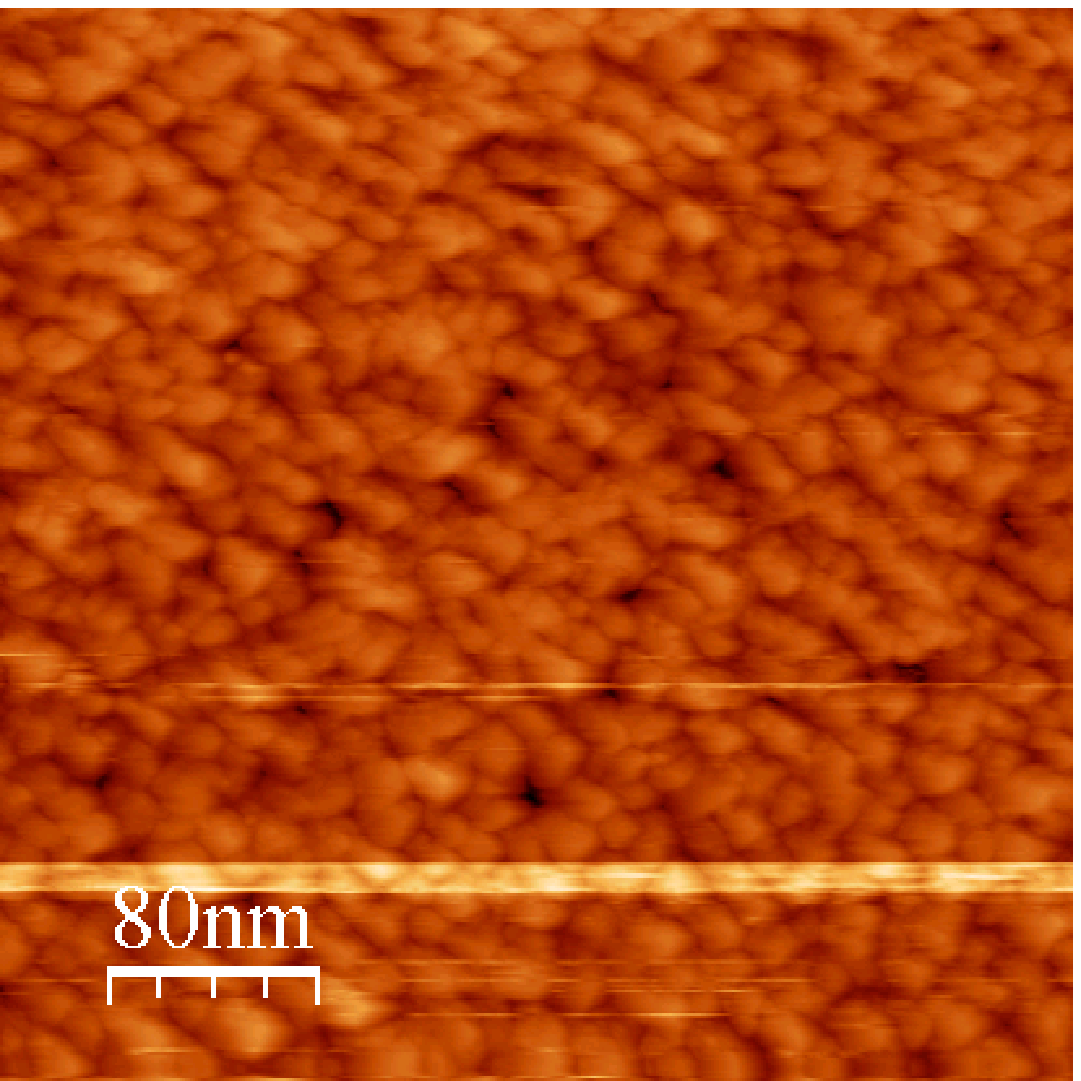}}
\subfigure{\includegraphics[width=0.5\textwidth]{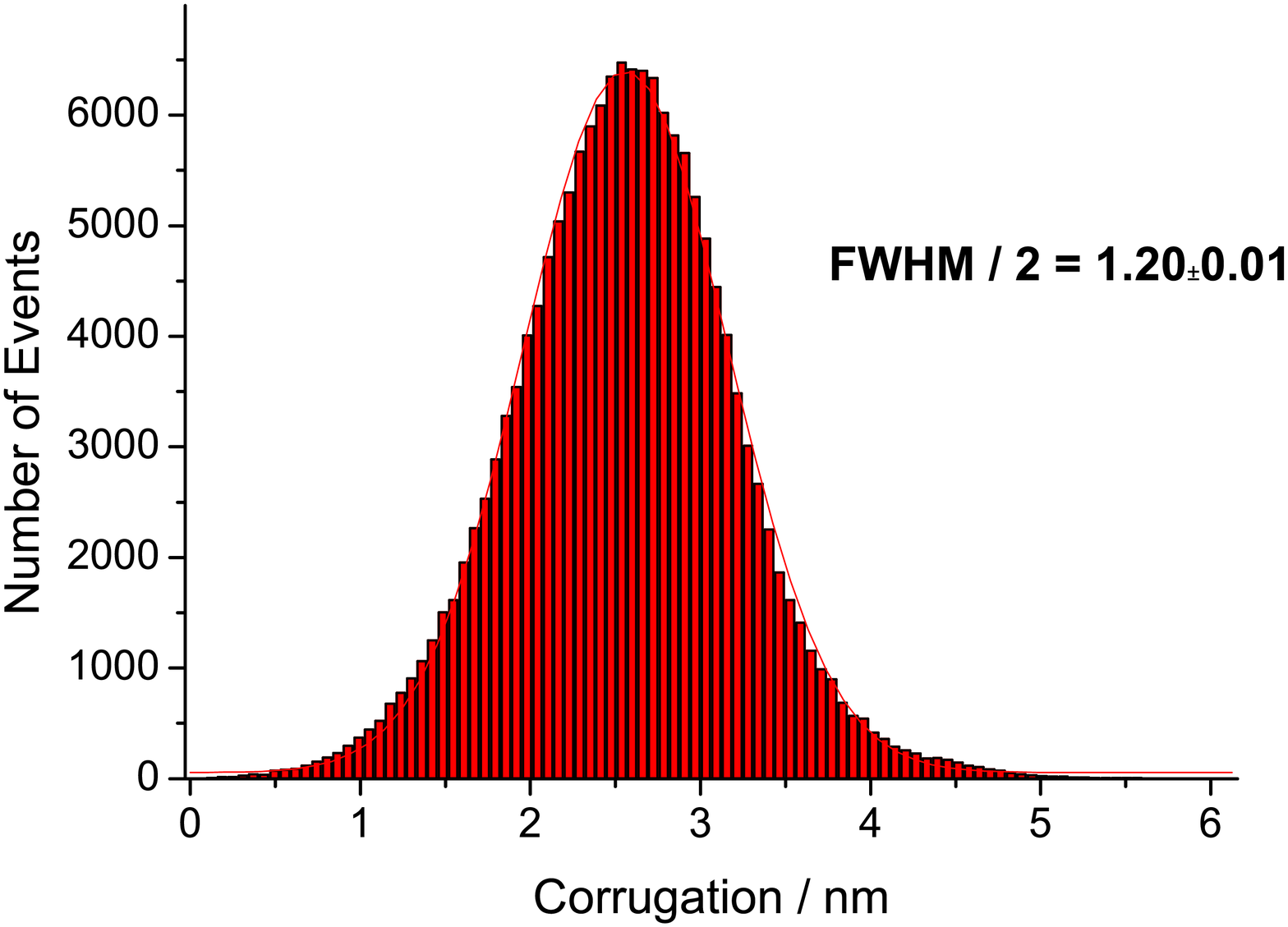}}
\hfill
\subfigure{\includegraphics[width=0.3\textwidth]{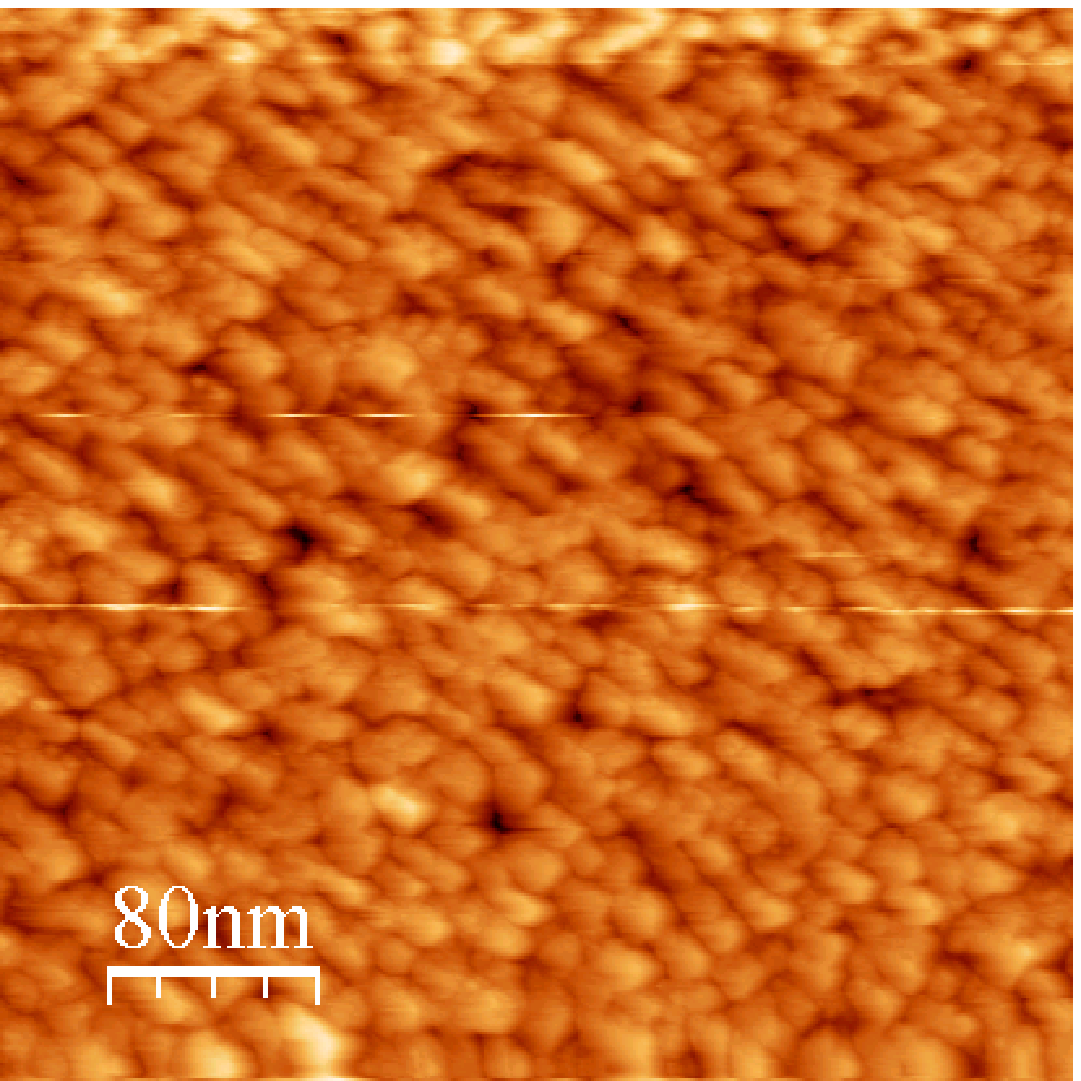}}
\caption[Corrugation histograms of Fe evaporated at $300~^\circ$C]{Corrugation histograms of
  Fe evaporated at $300~^\circ$C}
\label{fig:stmtemp300}
\end{flushleft}
\end{figure}

\begin{figure}
\begin{flushleft}
\subfigure{\includegraphics[width=0.5\textwidth]{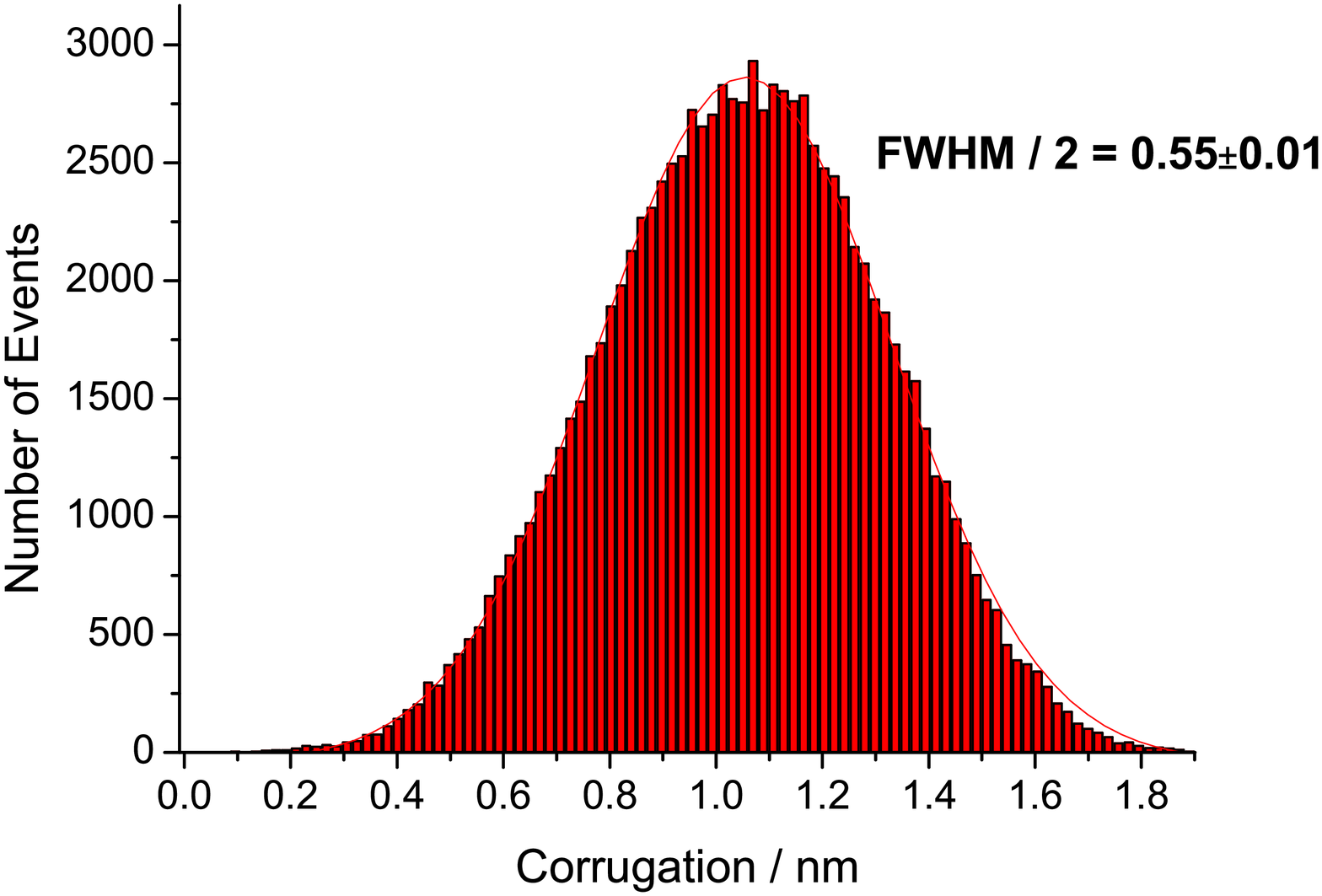}}
\hfill
\subfigure{\includegraphics[width=0.3\textwidth]{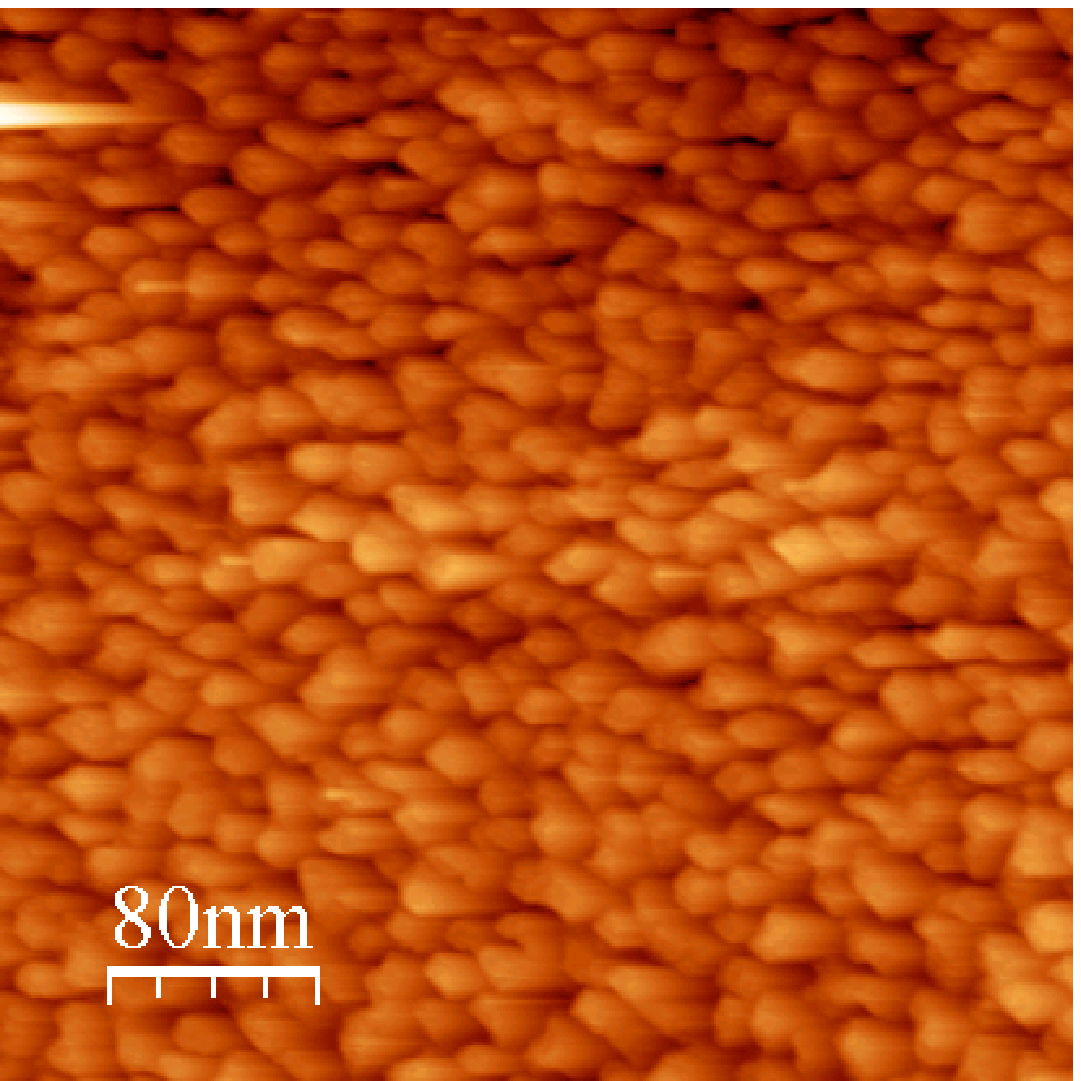}}
\subfigure{\includegraphics[width=0.5\textwidth]{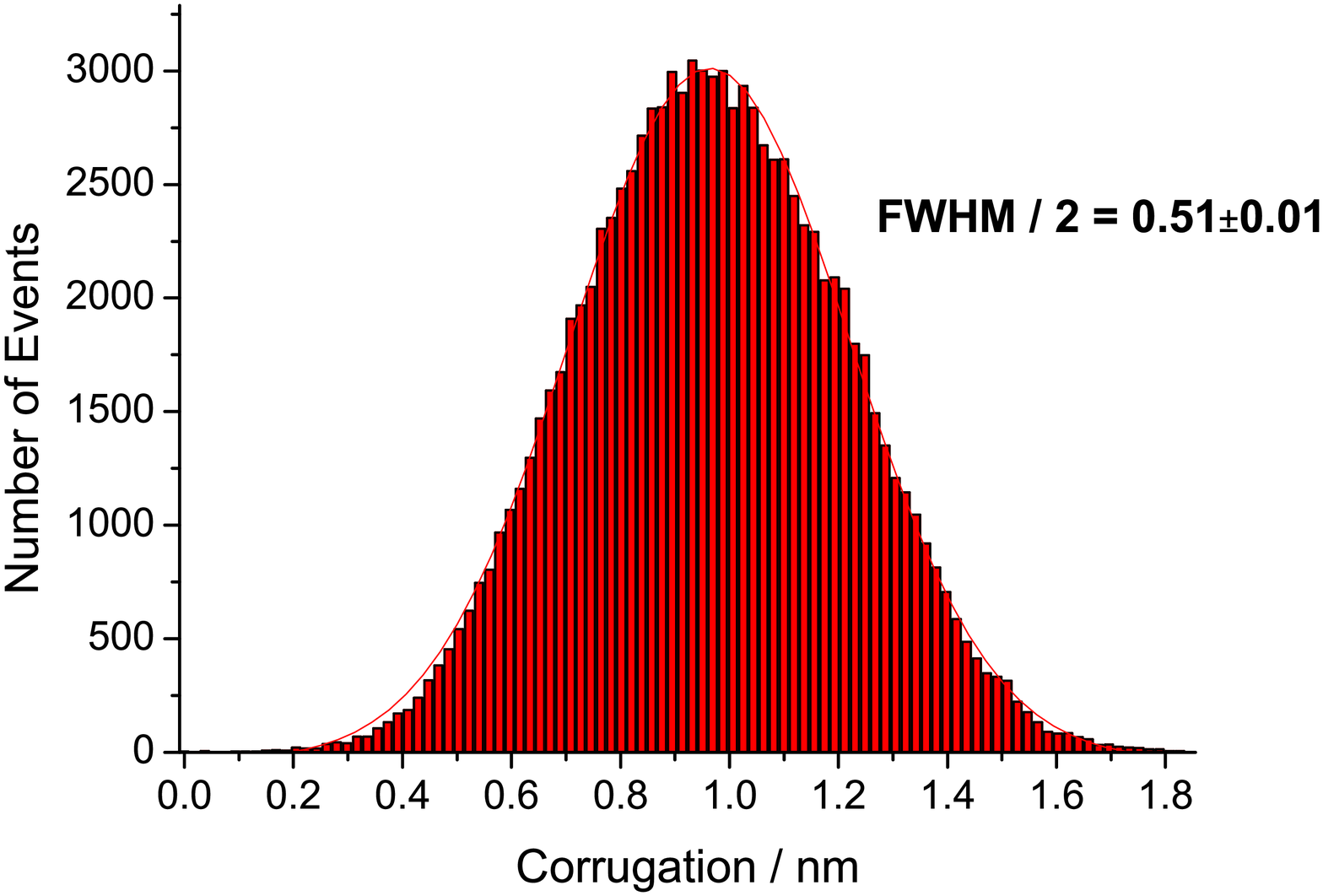}}
\hfill
\subfigure{\includegraphics[width=0.3\textwidth]{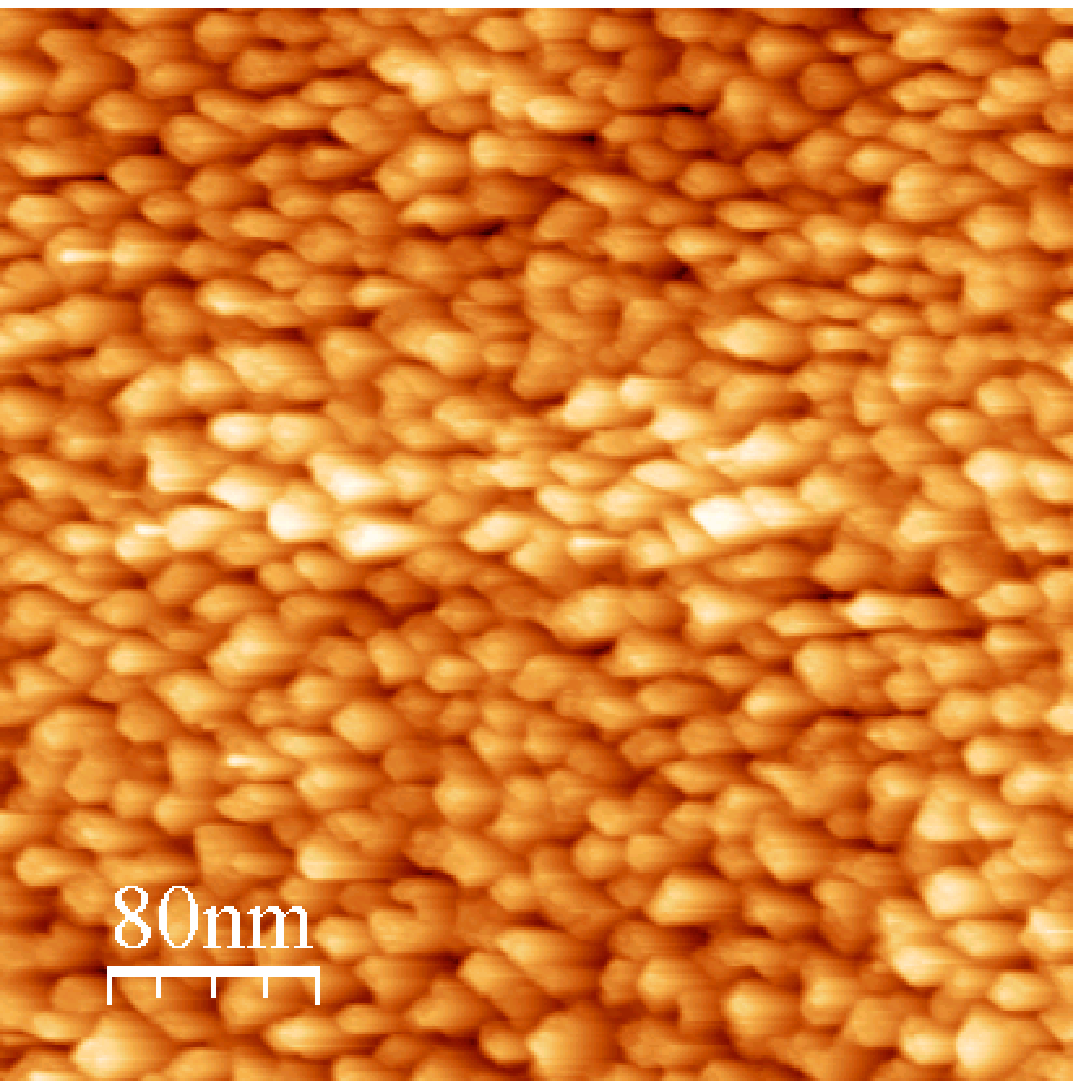}}
\subfigure{\includegraphics[width=0.5\textwidth]{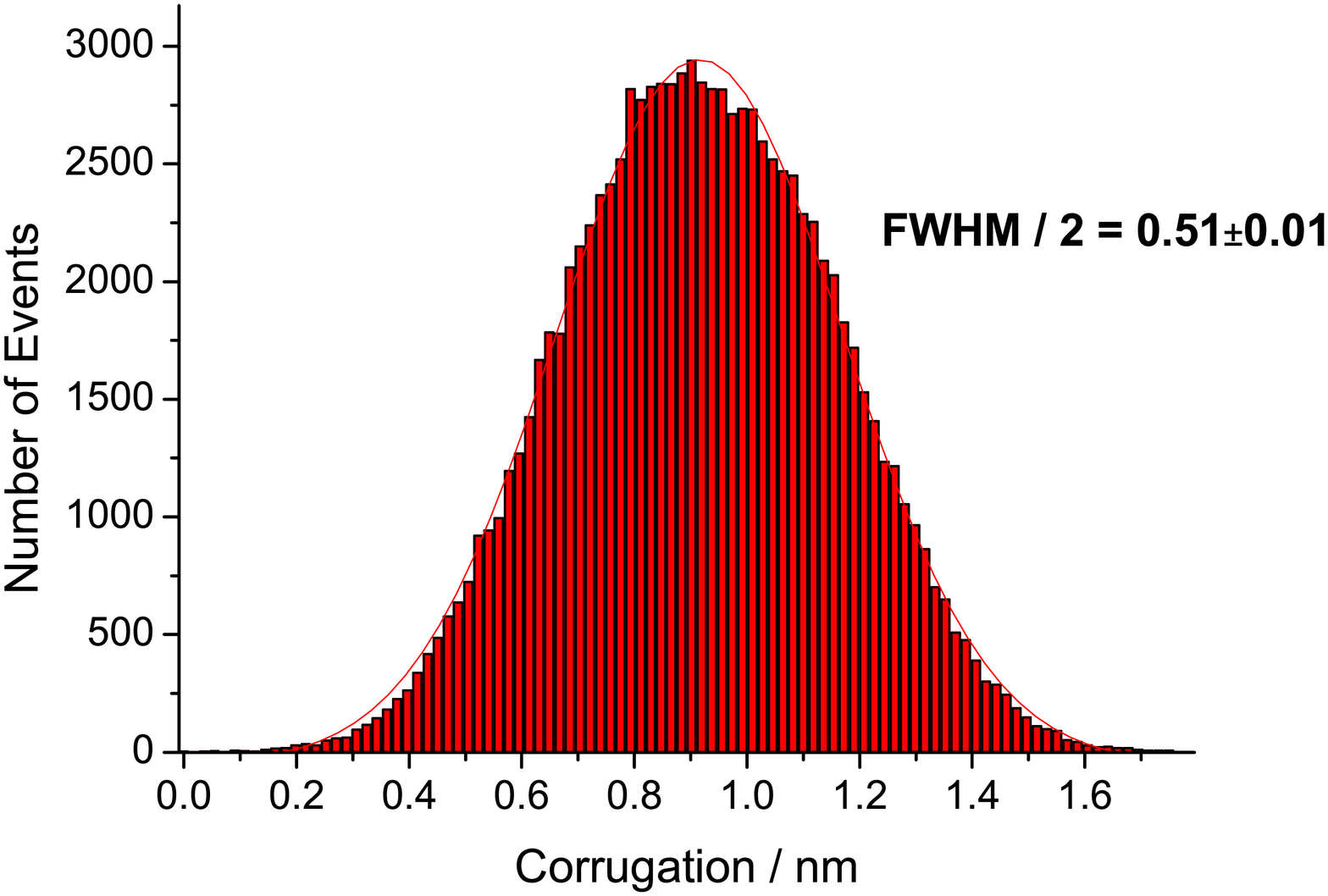}}
\hfill
\subfigure{\includegraphics[width=0.3\textwidth]{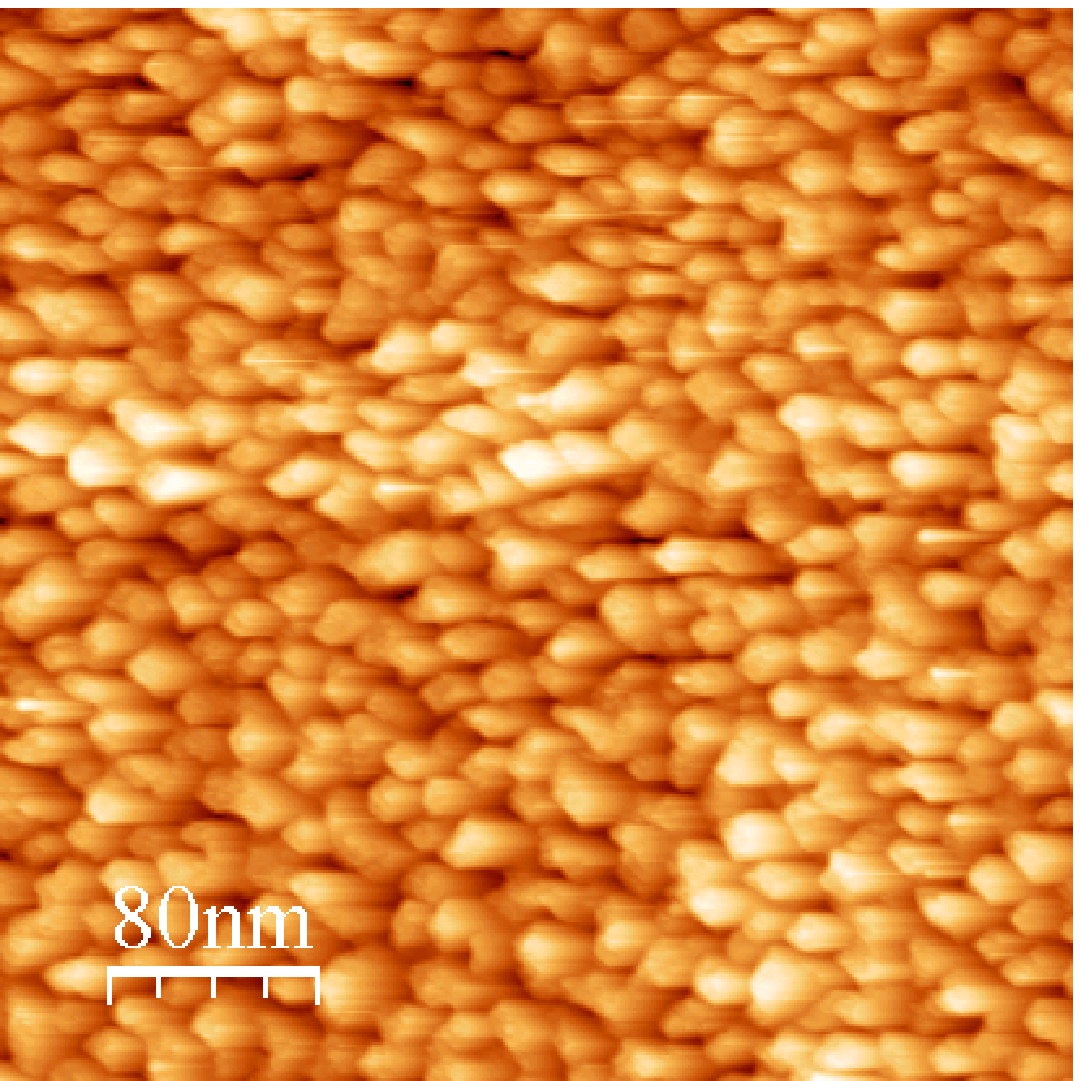}}
\subfigure{\includegraphics[width=0.5\textwidth]{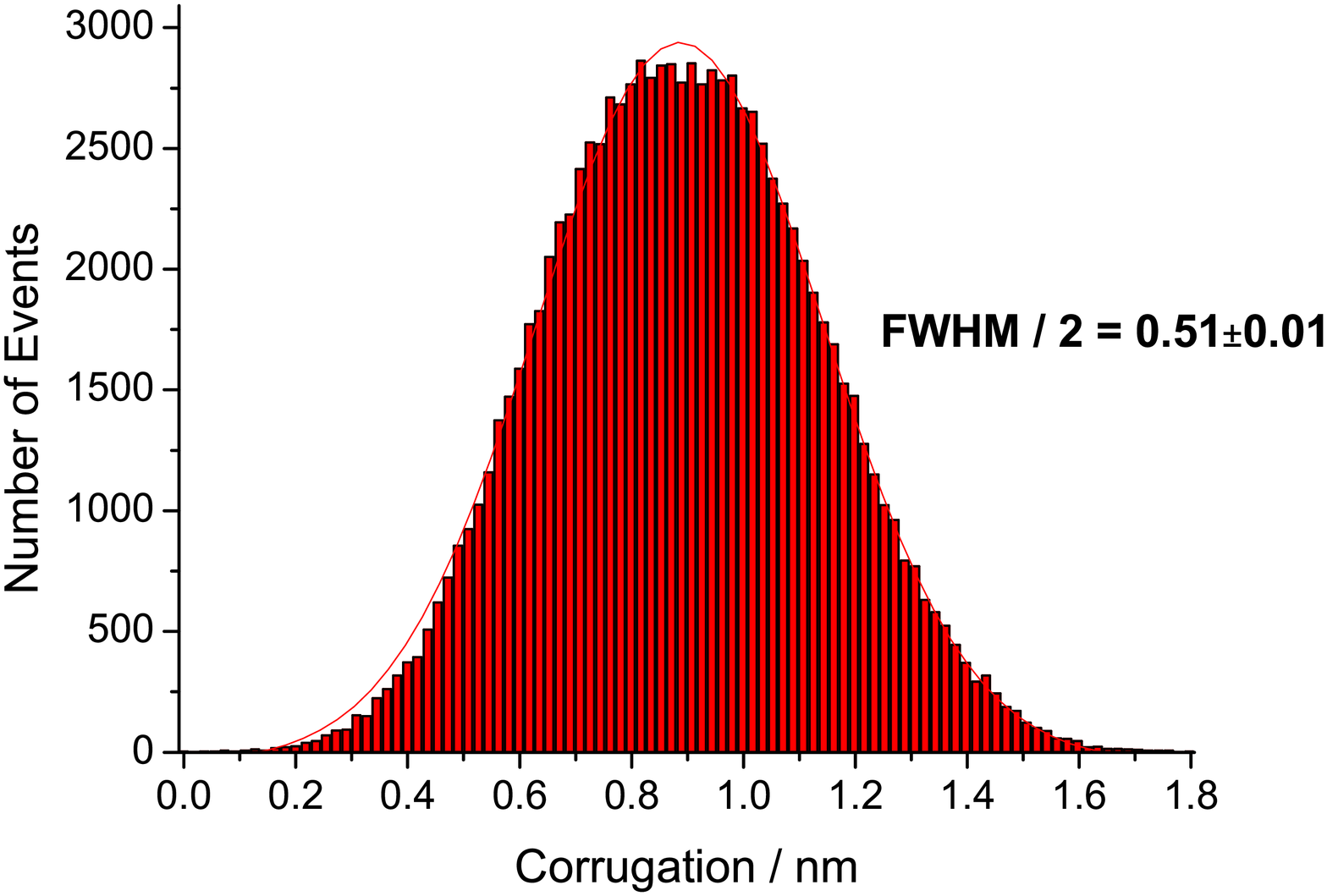}}
\hfill
\subfigure{\includegraphics[width=0.3\textwidth]{figures/sample-preparation/stm-temp-check/m23_400}}
\caption[Corrugation histograms of Fe evaporated at $400~^\circ$C]{Corrugation histograms of
  Fe evaporated at $400~^\circ$C}
\label{fig:stmtemp400}
\end{flushleft}
\end{figure}

\begin{figure}
\begin{flushleft}
\subfigure{\includegraphics[width=0.5\textwidth]{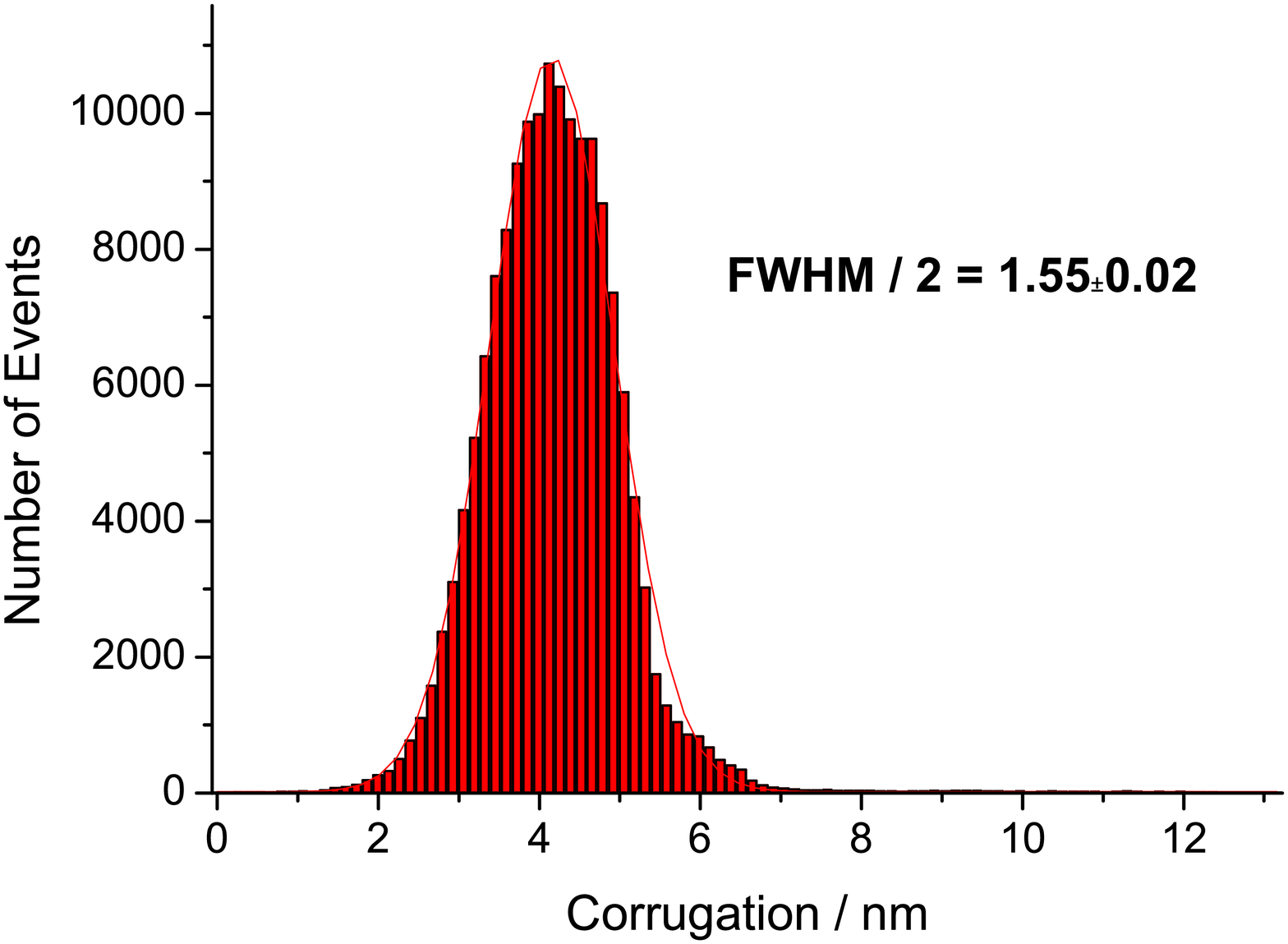}}
\hfill
\subfigure{\includegraphics[width=0.3\textwidth]{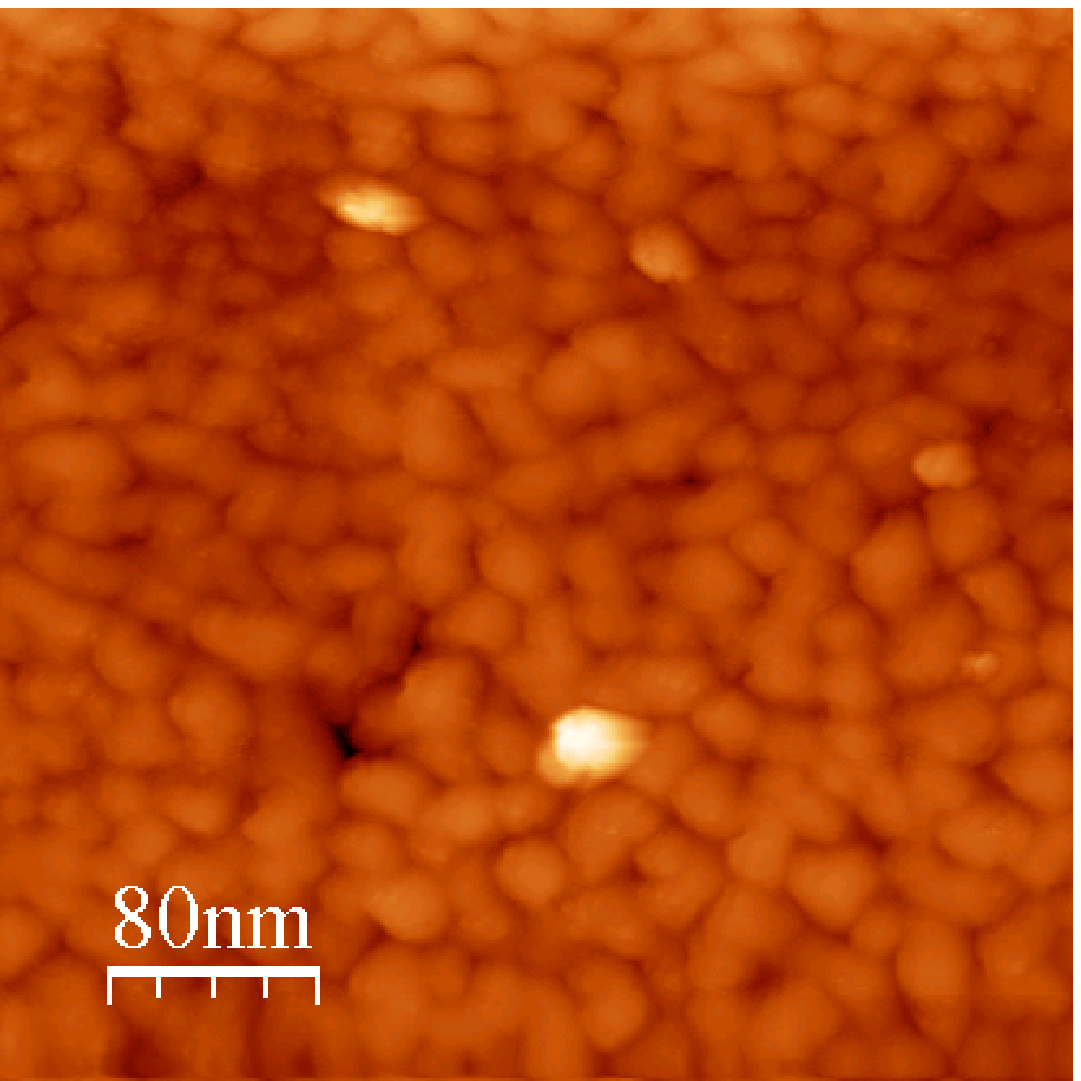}}
\subfigure{\includegraphics[width=0.5\textwidth]{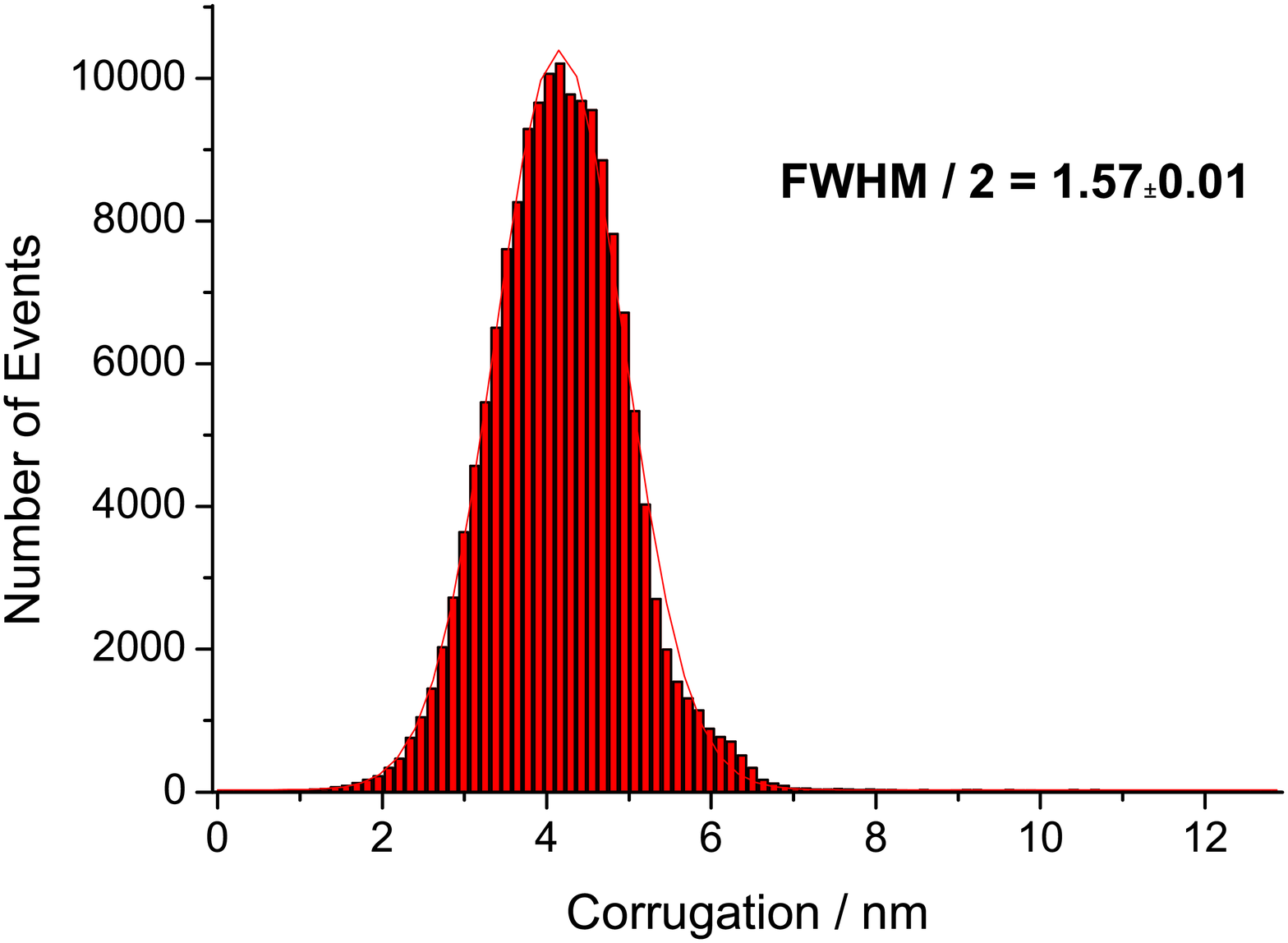}}
\hfill
\subfigure{\includegraphics[width=0.3\textwidth]{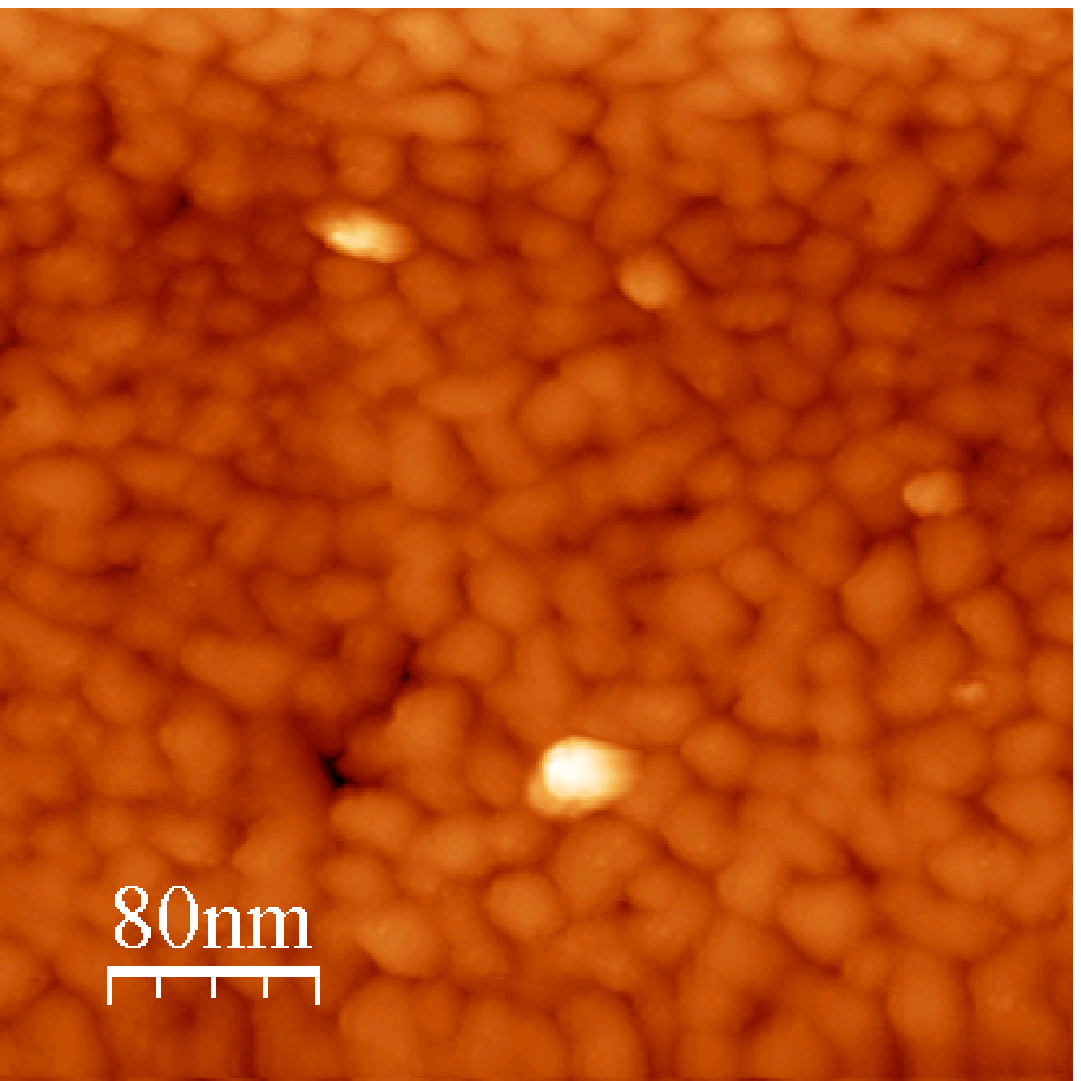}}
\subfigure{\includegraphics[width=0.5\textwidth]{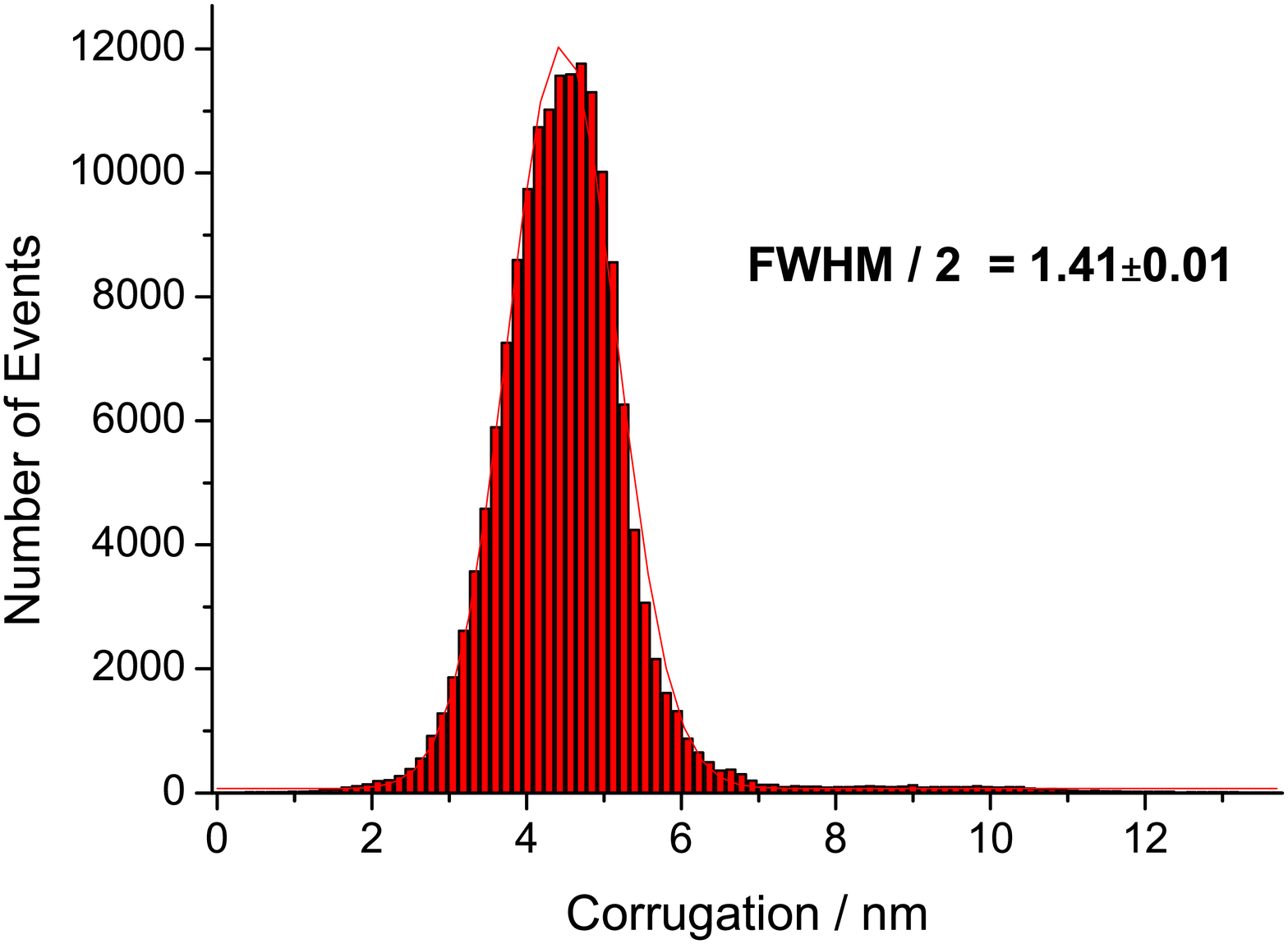}}
\hfill
\subfigure{\includegraphics[width=0.3\textwidth]{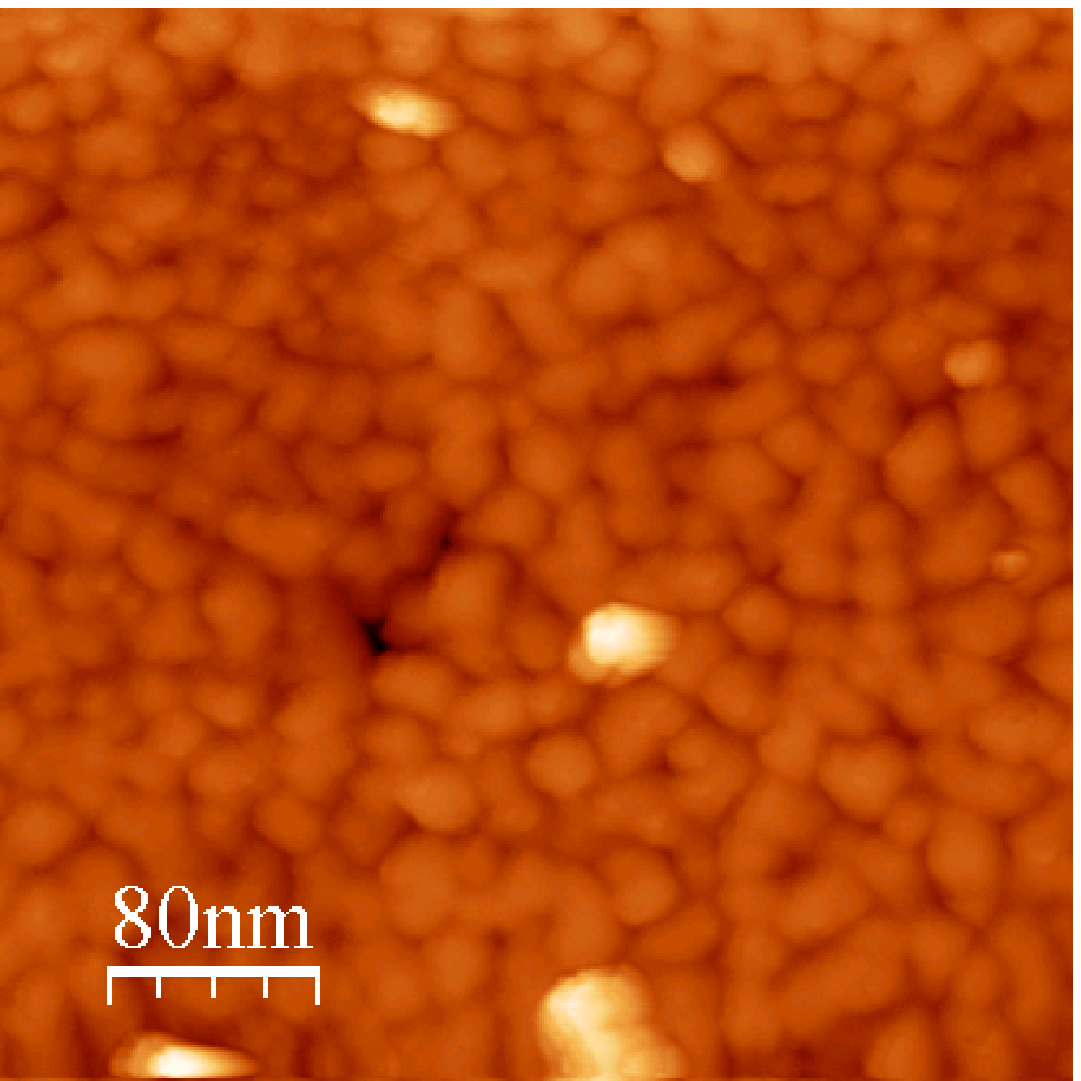}}
\subfigure{\includegraphics[width=0.5\textwidth]{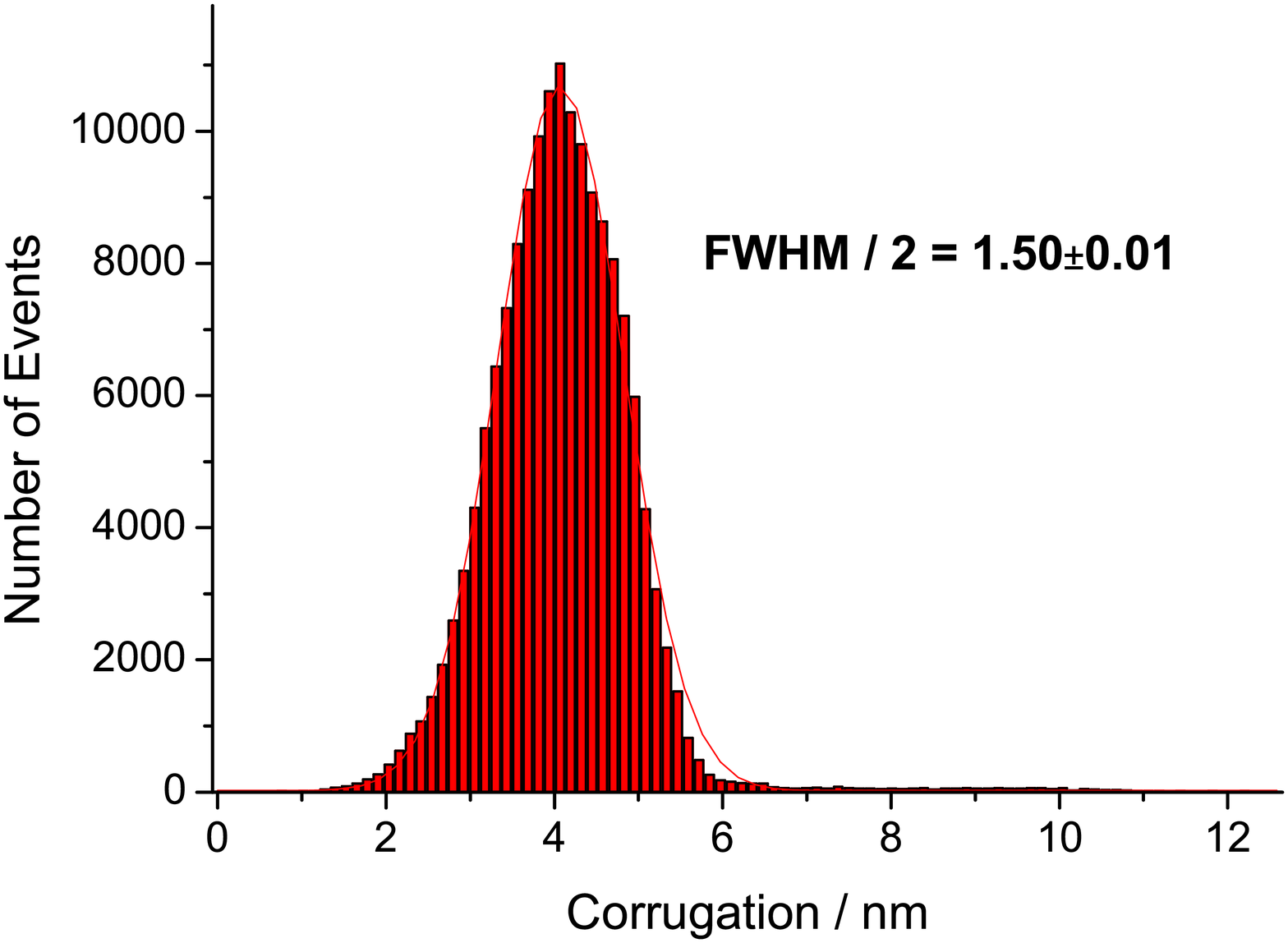}}
\hfill
\subfigure{\includegraphics[width=0.3\textwidth]{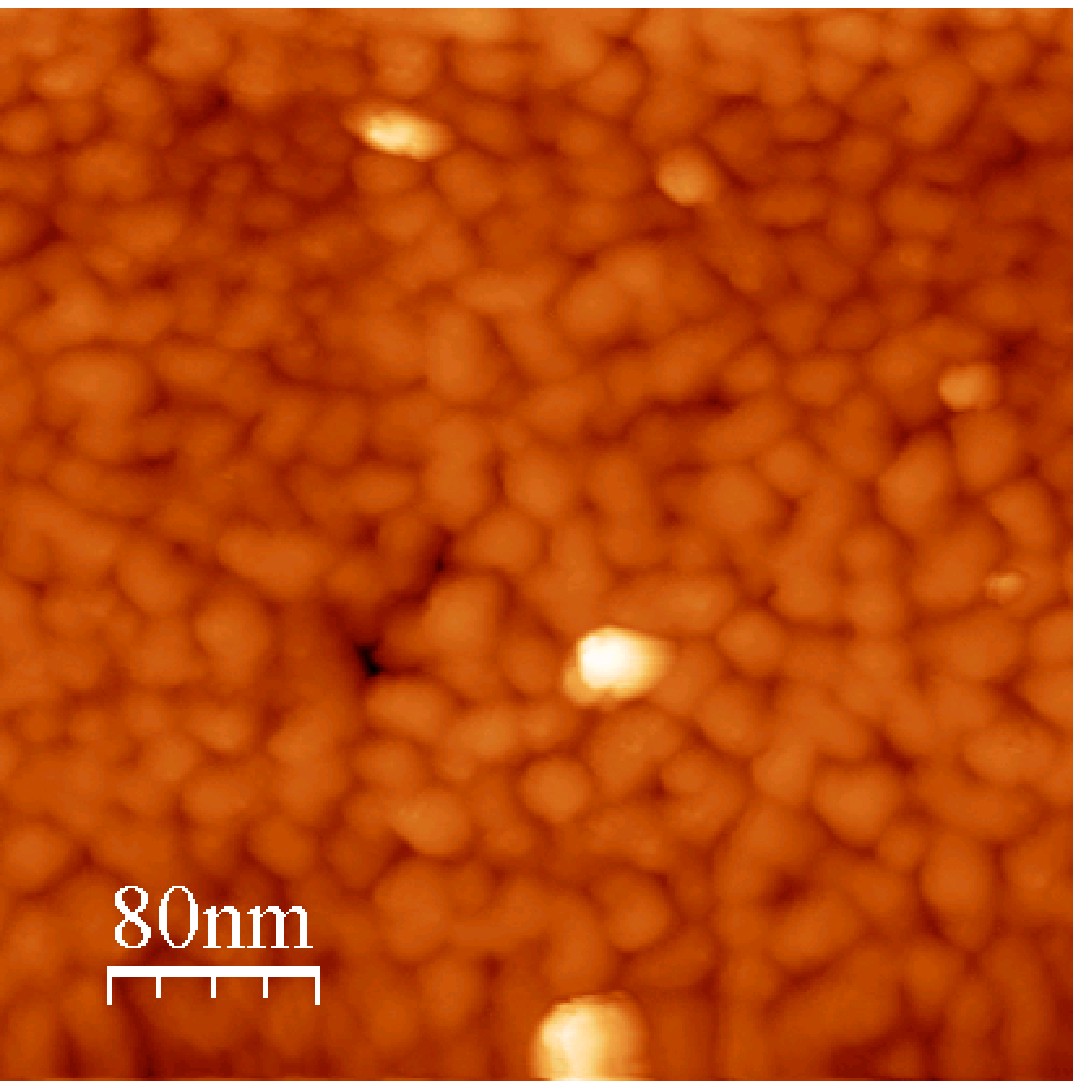}}
\caption[Corrugation histograms of Fe evaporated at $600~^\circ$C]{Corrugation histograms of
  Fe evaporated at $600~^\circ$C}
\label{fig:stmtemp600}
\end{flushleft}
\end{figure}

\subsection{Grain size analysis}
\label{sec:feaugrainsize}
In addition to an increase of the corrugation with higher substrate temperatures during
evaporation, it has become obvious that the average size of the grains of the iron
film has a temperature dependency as well. Looking at the STM images of the corrugation
analysis (Fig. \ref{fig:stmtemp100}-\ref{fig:stmtemp600}), it can be immediately
observed that grain sizes increase with higher temperatures. To analyze
grain sizes, we needed to find a solution which would yield reproducible
and comparable results. Simply performing a profile analysis with \emph{wsxm}
would not be able to yield exact values for corrugations as the grain sizes
of a single film vary statistically over the whole film area. Analyzing lattice
constants with \emph{wsxm} works best when the lattice has high symmetry which
is not the case for the iron films. A good method therefore was to calculate
the auto correlation (AC) of the image data and plot it against the grain size.
The result from this analysis is a plot which yields a kind of histogram
of the distances on the film and the shape of it allows to infer the long
distance order and the typical grain size of the film.

\medskip

We have written a small program in \emph{C} (\emph{autocorr.c}) which performs
the auto correlation, its source code can be found in the appendix
(see \ref{chapter:sourcecode}). It parses STM image data exported from \emph{wsxm}
as ASCII matrix input files, calculates the auto correlation and outputs
the result to the terminal. The result would be redirected into a file
which then could be read into \emph{Origin} to analyze the peaks.

\medskip

\begin{figure}
\begin{flushleft}
\subfigure[AC for 100$^\circ$C]{\includegraphics[width=0.45\textwidth]{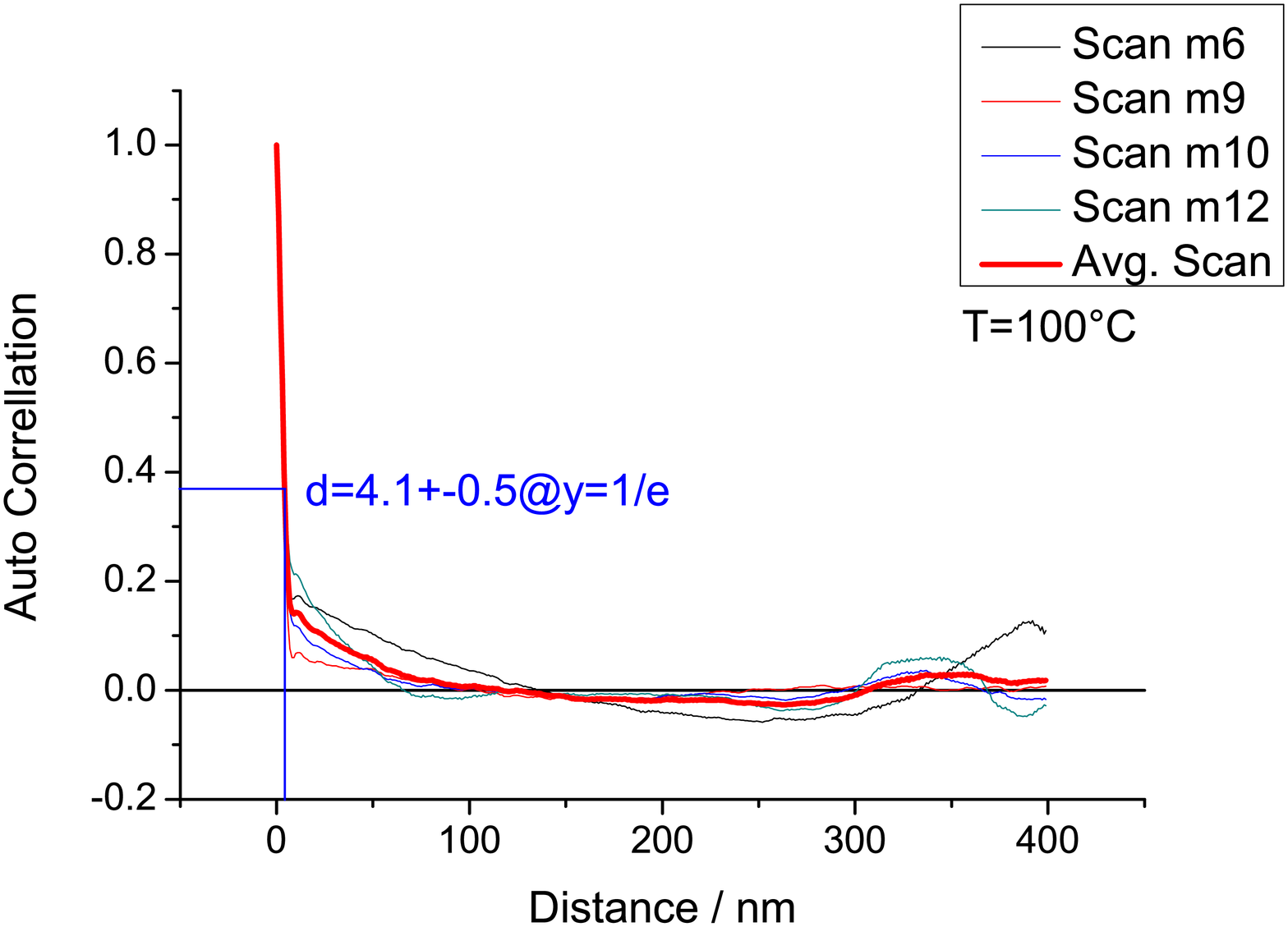}}
\hfill
\subfigure[AC for 200$^\circ$C]{\includegraphics[width=0.45\textwidth]{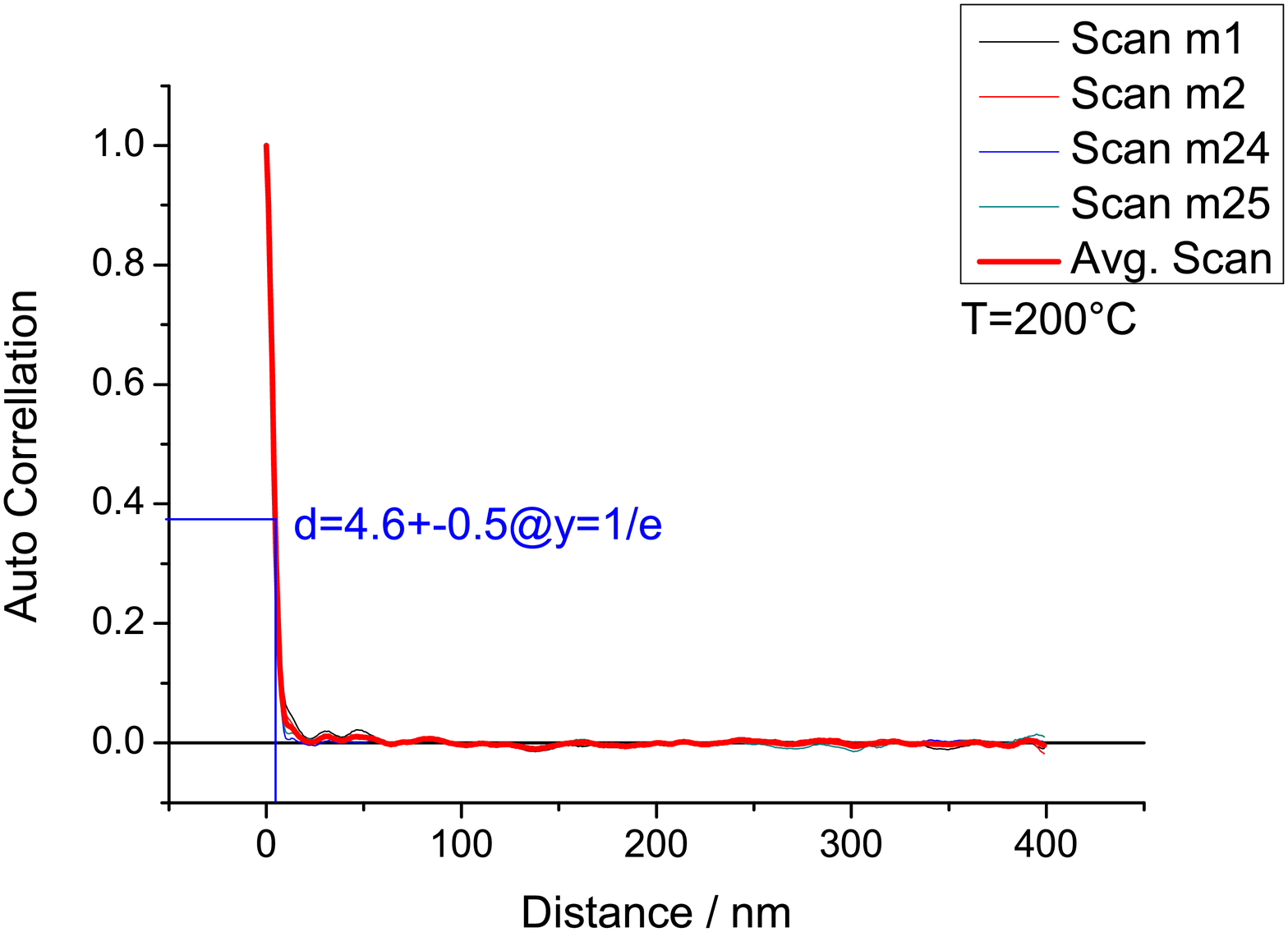}}
\subfigure[AC for 300$^\circ$C]{\includegraphics[width=0.45\textwidth]{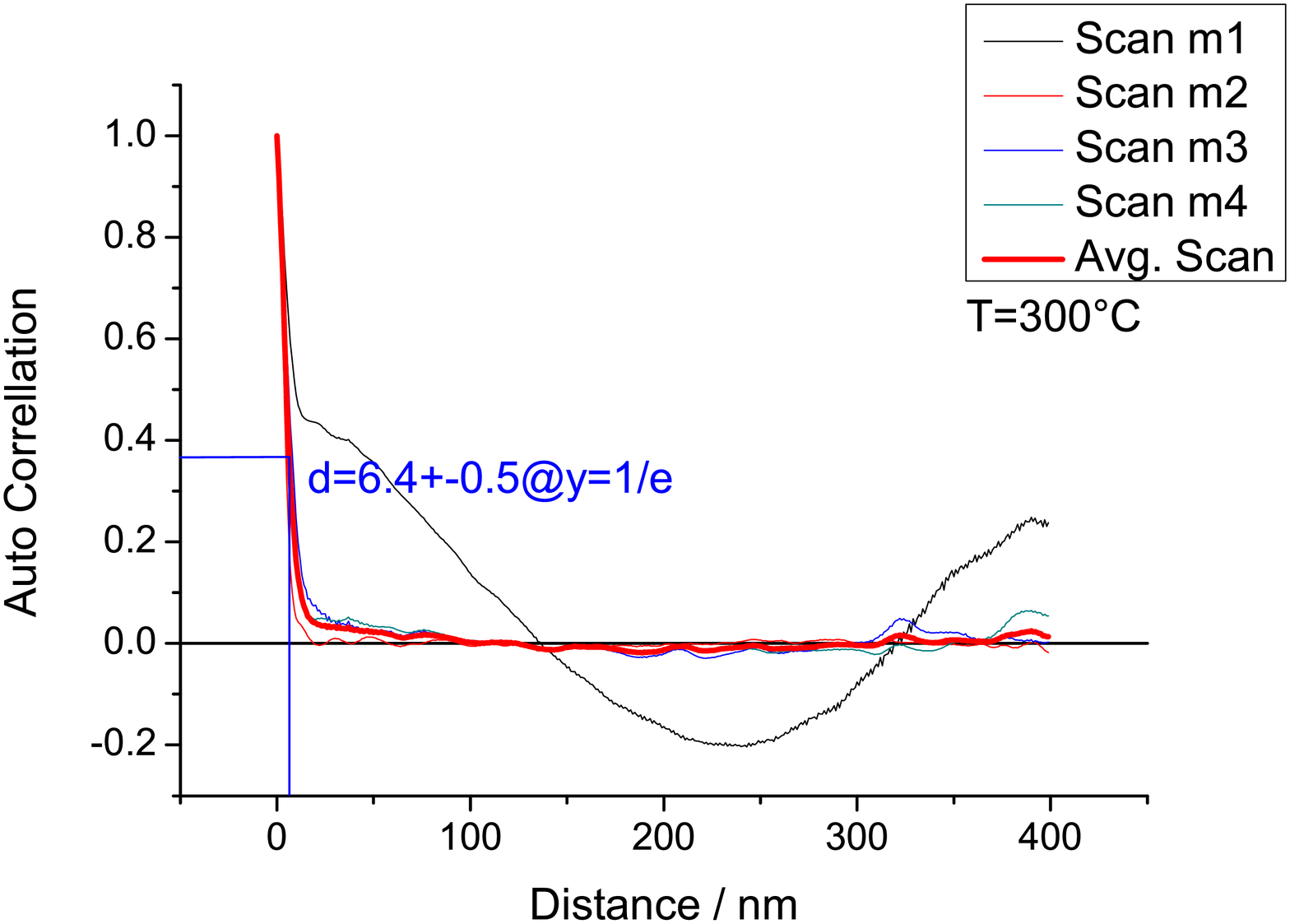}}
\hfill
\subfigure[AC for 400$^\circ$C]{\includegraphics[width=0.45\textwidth]{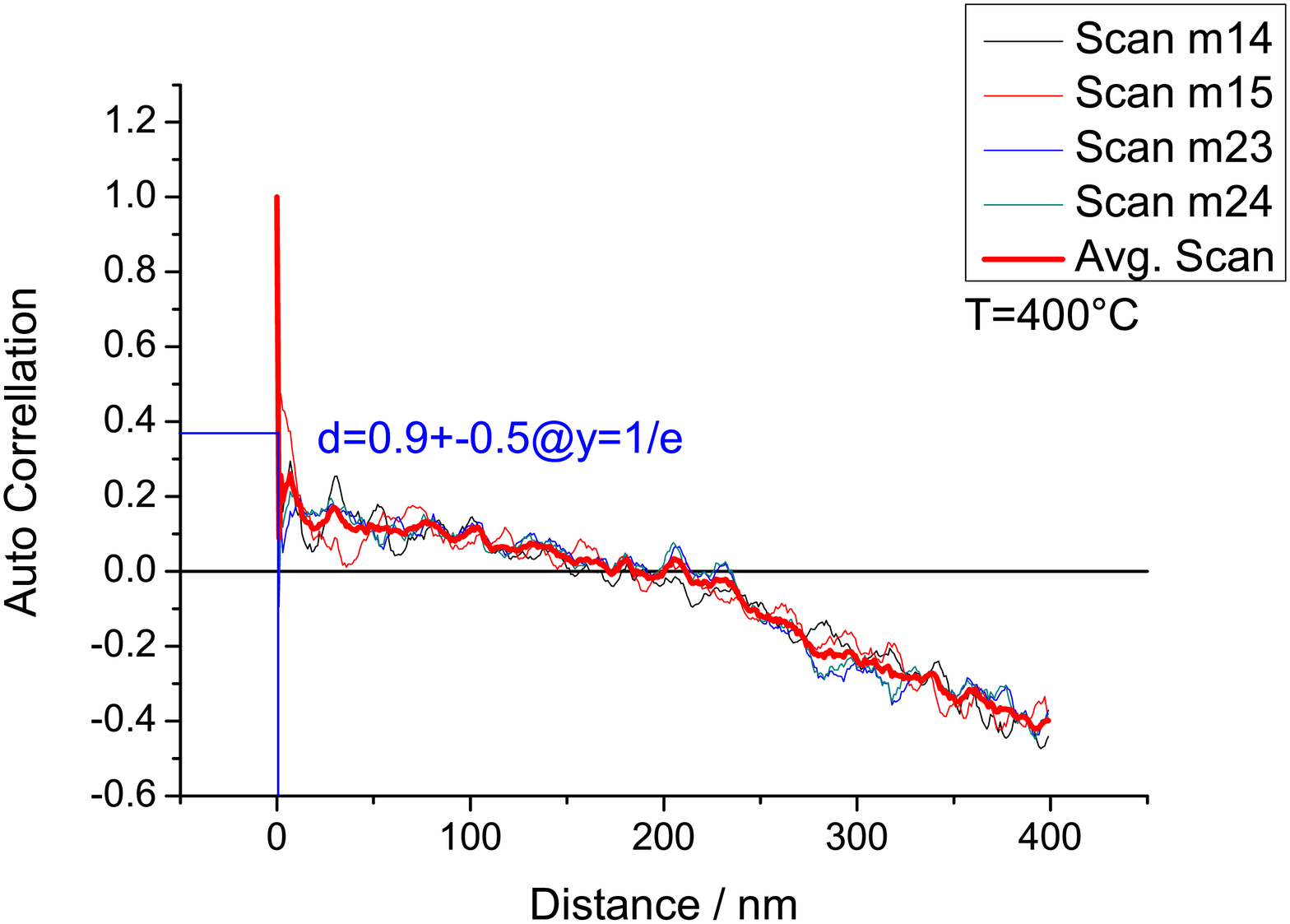}}
\subfigure[AC for 600$^\circ$C]{\includegraphics[width=0.45\textwidth]{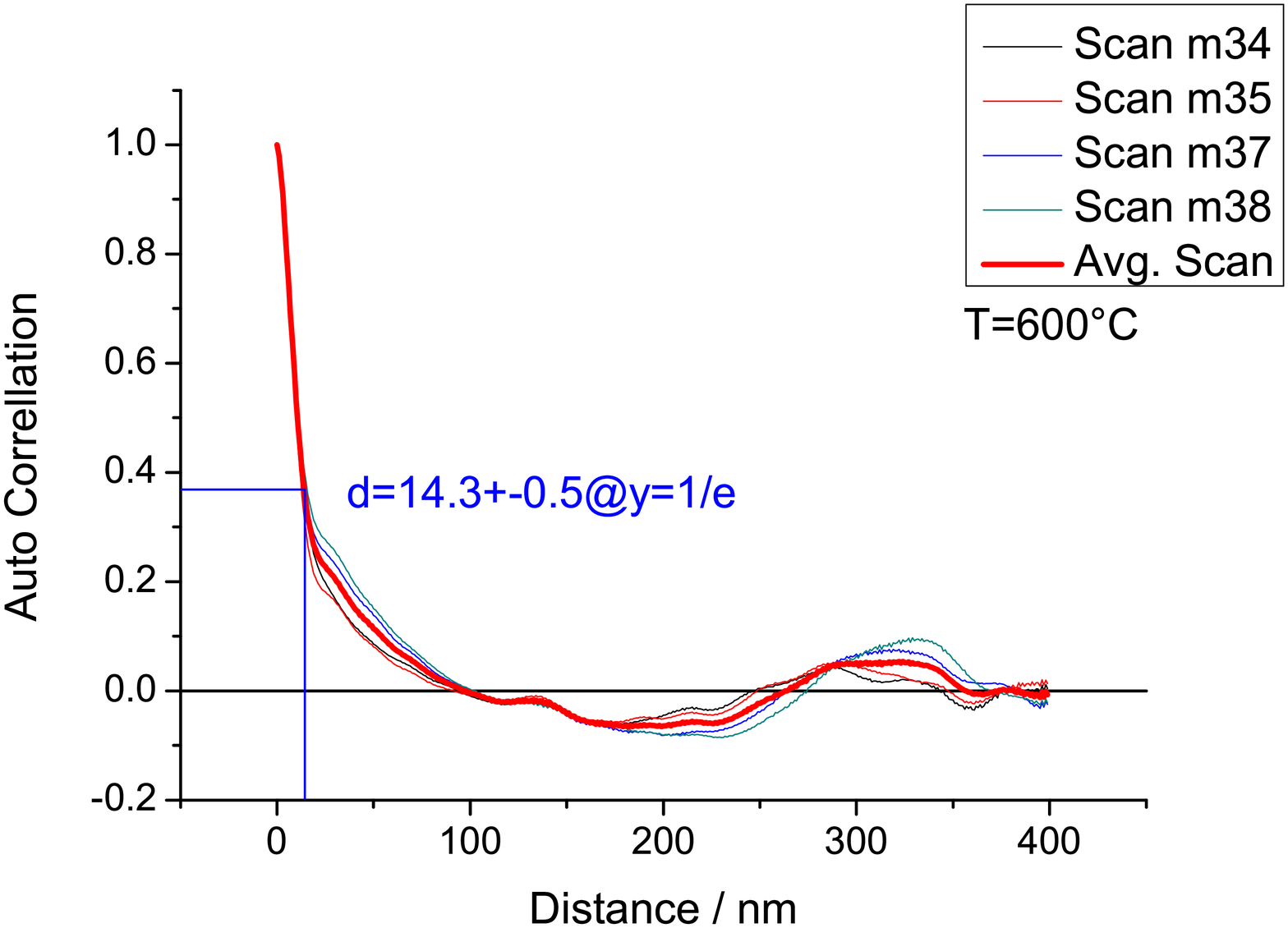}}
\hfill
\subfigure[Grain size against temperature]{\includegraphics[width=0.45\textwidth]{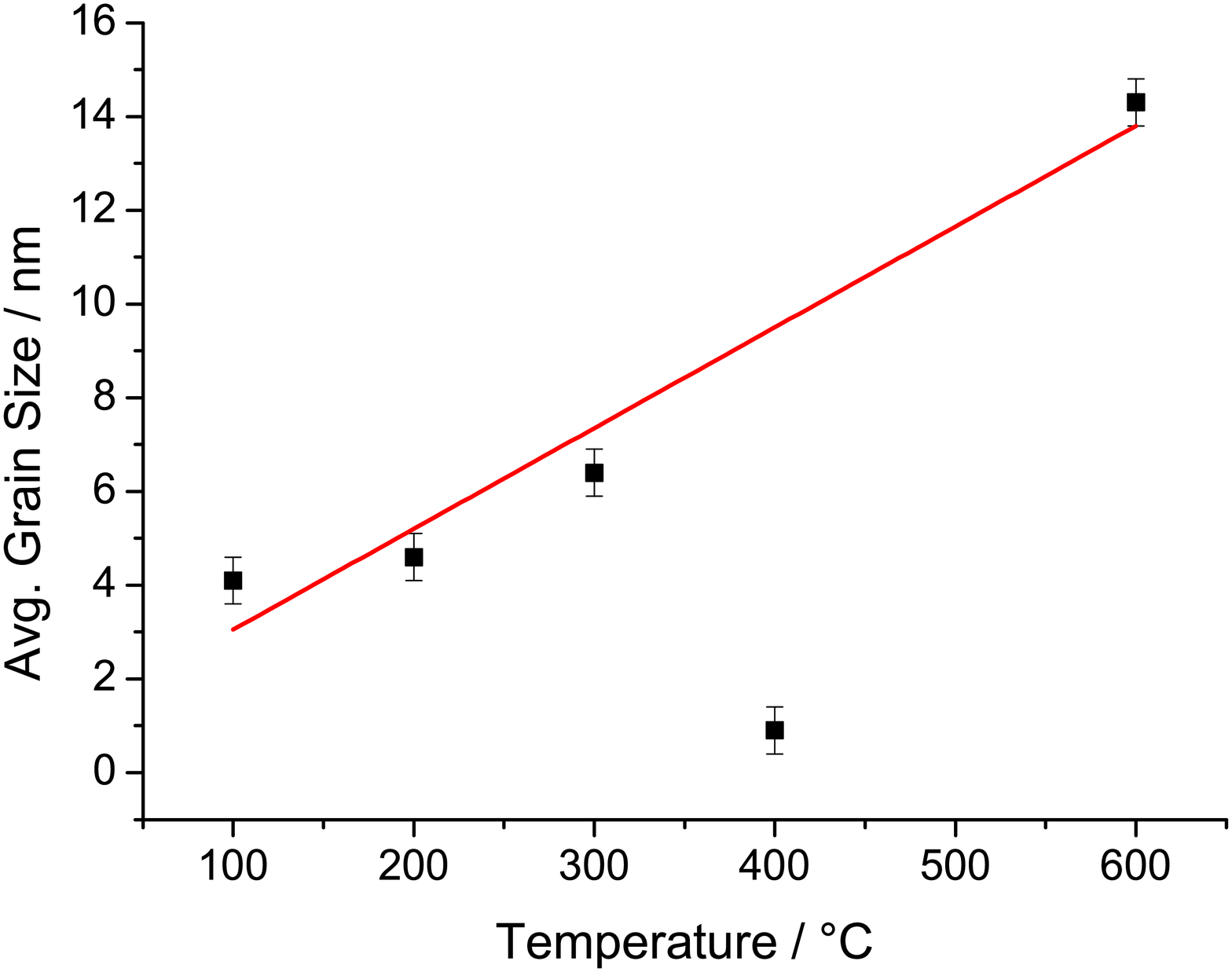}}
\caption[Auto correlation plots for different temperatures]{Auto correlation plots for different temperatures}
\label{fig:grainplots}
\end{flushleft}
\end{figure}

The input data (the STM images) is the same that was used for the corrugation analysis
in the previous section. We calculated and plotted the auto correlation
for each of the four images of each temperature and combined the four images
at one temperature into one plot. Then we plotted the mean value of the
auto correlation from the four images, disregarding outliers if any. The resulting
plots are shown in Fig. \ref{fig:grainplots}.

\medskip

To test our algorithm, we generated two model lattices (\emph{simple cube} (sc)
and \emph{hexagonal closed-packed} (hcp)) and analyzed the grain sizes there
with the auto correlation algorithm. Since we know the lattice constants
for our model lattices, we can cross-check the plots to find the corresponding
peaks.



\begin{figure}
\subfigure[Model for \emph{hcp}-lattice]{\includegraphics[width=0.5\textwidth]{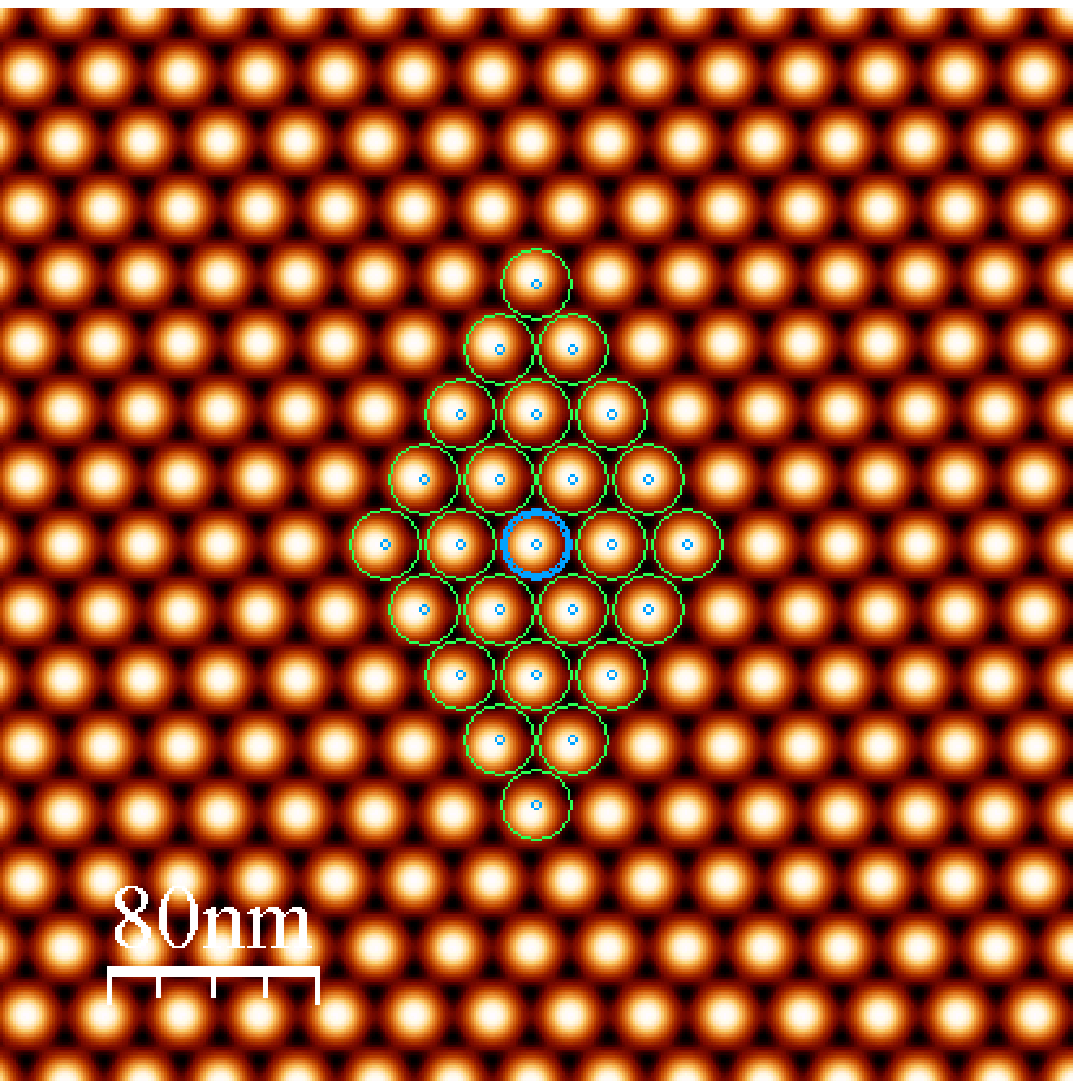}}
\subfigure[Model for \emph{sc}-lattice]{\includegraphics[width=0.5\textwidth]{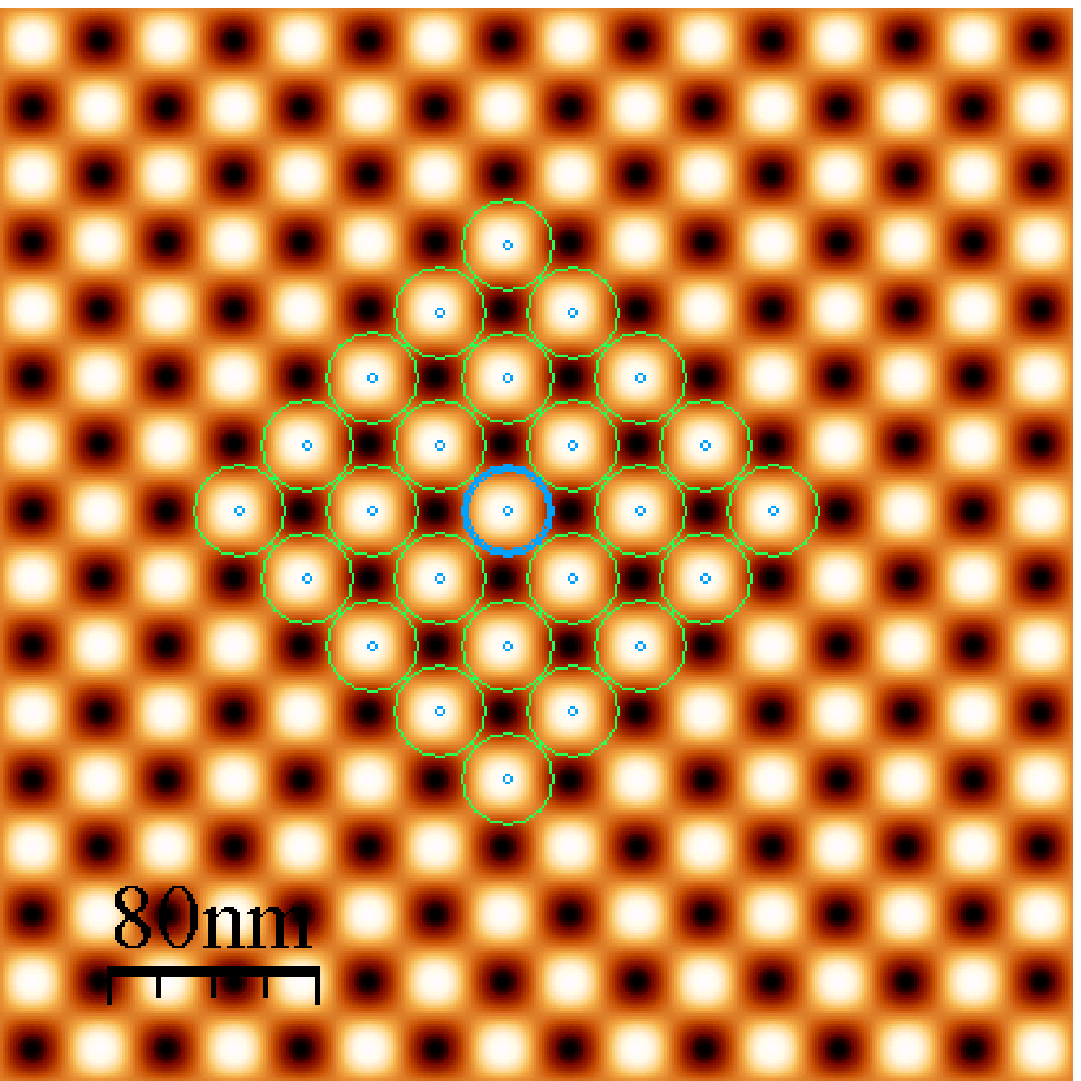}}
\caption[STM images of model lattices]{STM images of model lattices generated with \emph{Origin}. Both images were
  overlaid with a grid for lattice constant analysis. Since the surface symmetries of the surfaces are very
  high, the lattice constants could be determined very easily with this method.}
\label{fig:grain_modelimages}
\end{figure}

The model images could be easily analyzed with \emph{wsxm} since their lattices have a high
symmetry. To determine the lattice constant, the software provides the possibility to overlay
the image with a grid whose lattice constant and type (e.g. \emph{hexagonal}, \emph{sqaure},
\emph{7x7} and so on) has to be adjusted to match the underlying image. One can then immediately
read the lattice constant and type from the software. Figure \ref{fig:grain_modelimages} shows
the results.

\medskip

The lattice constants for the model images were determined as follows:

\begin{eqnarray}
  d_{hcp} & = & 28~\text{nm} \\
  d_{sc}  & = & 35~\text{nm}
\end{eqnarray}

where $d$ is the shortest distance to the next lattice point.

\medskip

\begin{figure}
\subfigure[AC plot for \emph{hcp}-lattice]{\includegraphics[width=0.5\textwidth]{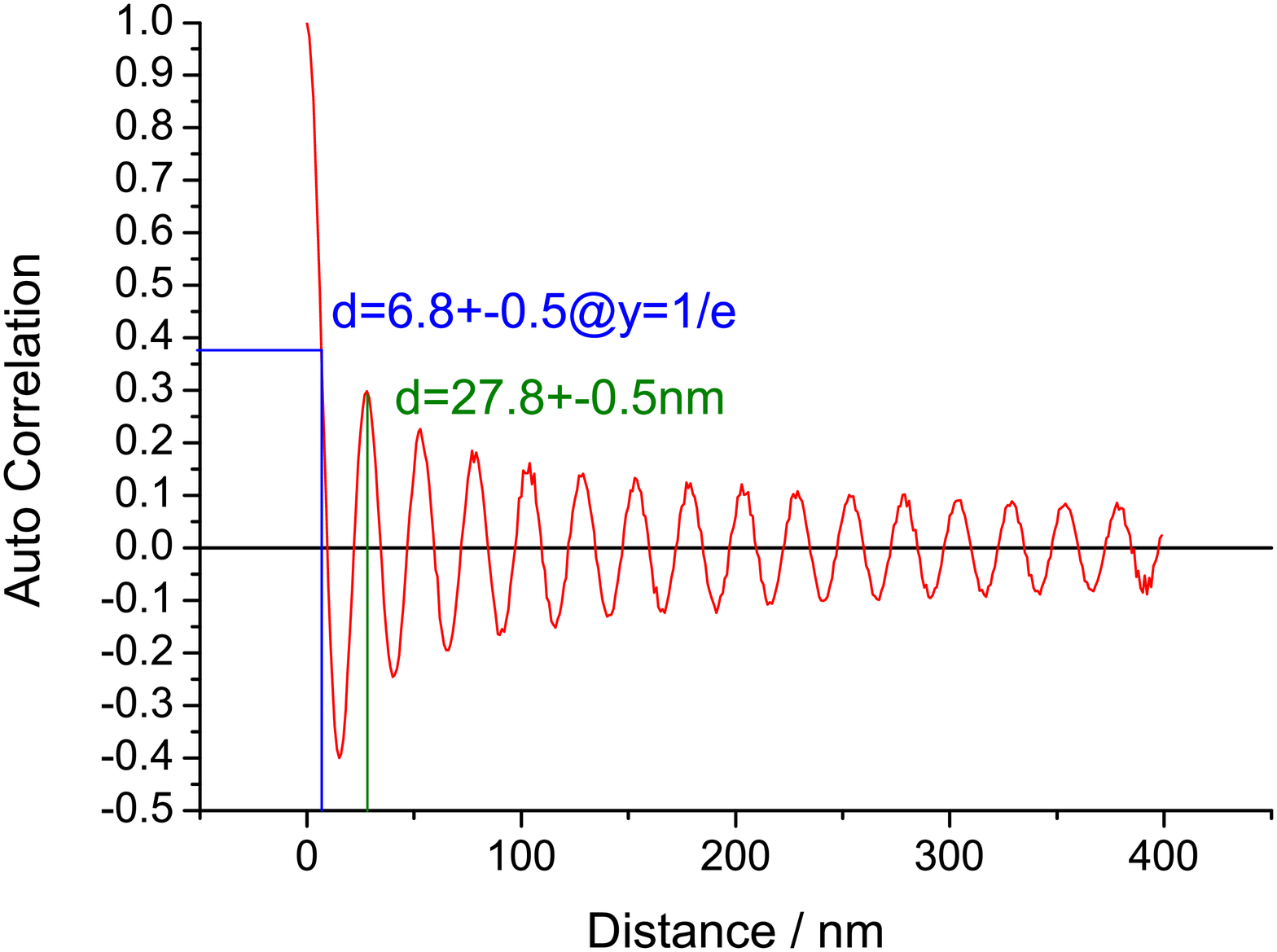}}
\subfigure[AC plot for \emph{sc}-lattice]{\includegraphics[width=0.5\textwidth]{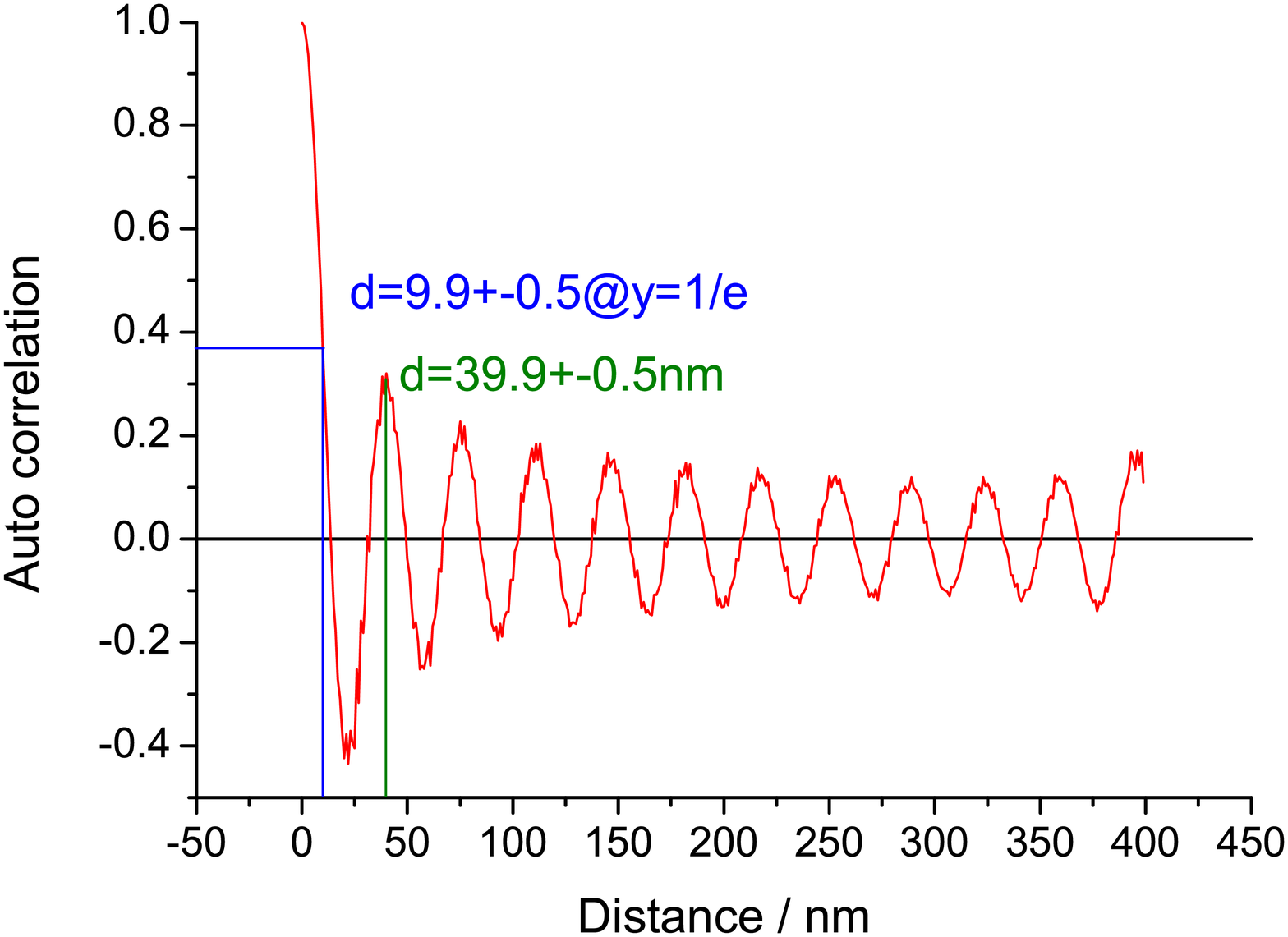}}
\caption[Auto correlation plots of model lattices]{Auto correlation plots for the STM model images generated with \emph{Origin}.
  The grain sizes can be read from the dominating distances found between atoms in the lattice. Every peak represents such
  a distance. The first peak reflects the distance of one atom to its next neighbors. It can also be read at $y = \frac{1}{e}$.}
\label{fig:grain_modelgraphs}
\end{figure}

Now, if we take a look at the auto correlation plots for the two model lattices in Figure \ref{fig:grain_modelgraphs}, we
can read the lattice constants from the first peaks of the oscillations:

\begin{eqnarray}
  d_{hcp} & = & 27.8\pm0.5~\text{nm} \\
  d_{sc}  & = & 39.9\pm0.5~\text{nm}
\end{eqnarray}

The lattice constant for the hcp-lattice read from the auto correlation analysis matches with the value determined with
lattice analysis with \emph{wsxm} within error range. For the sc-lattice, however, the lattice constant read from
the auto correlation analysis differs by a magnitude of the error range. The reason is that the auto correlation
analysis accumulates the distances to all next neighbors in the lattice. For hcp, the distance to all the next
neighboring lattice points are equidistant. For sc, however, one point has four neighbors with distance $d$
and four with distance $d \times \sqrt{2}$. Thus, we have to calculate the average distance for all eight
neighbors, which is:

\begin{equation}
  d_{sc}^{avg} = \frac{d_{sc} + d_{sc} \times \sqrt{2}}{2} = \frac{(35 + 35 \times \sqrt{2})~\text{nm}}{2} \approx 42.2~\text{nm}
\end{equation}

Thus the values lie within fivefold error intervals which is still acceptable. Moreover, the value for the lattice
constant can also be read at ordinate value $\frac{1}{e} \approx 0.37$, more precisely the axis of abscissae
at $y = \frac{1}{e}$ corresponds to $\frac{1}{4}$ of the lattice distance. This becomes helpful when the
auto correlation has very small oscillations to determine the lattice constant from the peak occurrences.

\medskip

Taking a look at the AC plots from the real STM image data in Figure \ref{fig:grainplots}, it's apparent that
the oscillations in the AC are too small to be accounted for. Thus, we looked at the values for the distances
at $y = \frac{1}{e}$ and plotted those over the temperature. The result is shown in Figure \ref{fig:grainplots}
(f). Likewise the corrugation, the grain size clearly has a proportional dependence on the substrate temperature
during evaporation in the sense, that basically the grain size grows with the temperature, ranging from
$16~\text{to}~56~\text{nm}$ ($4~\text{to}~14~\text{nm} \times 4$).

\medskip

Thus this grain size analysis also suggests growing the films at lowest possible substrate temperatures to
achieve smooth and flat films.

\subsection{Grain size comparison: Fe and Au}
\label{sec:feaugrainsizecomp}
\begin{figure}
  \begin{center}
  \includegraphics[width=0.5\textwidth]{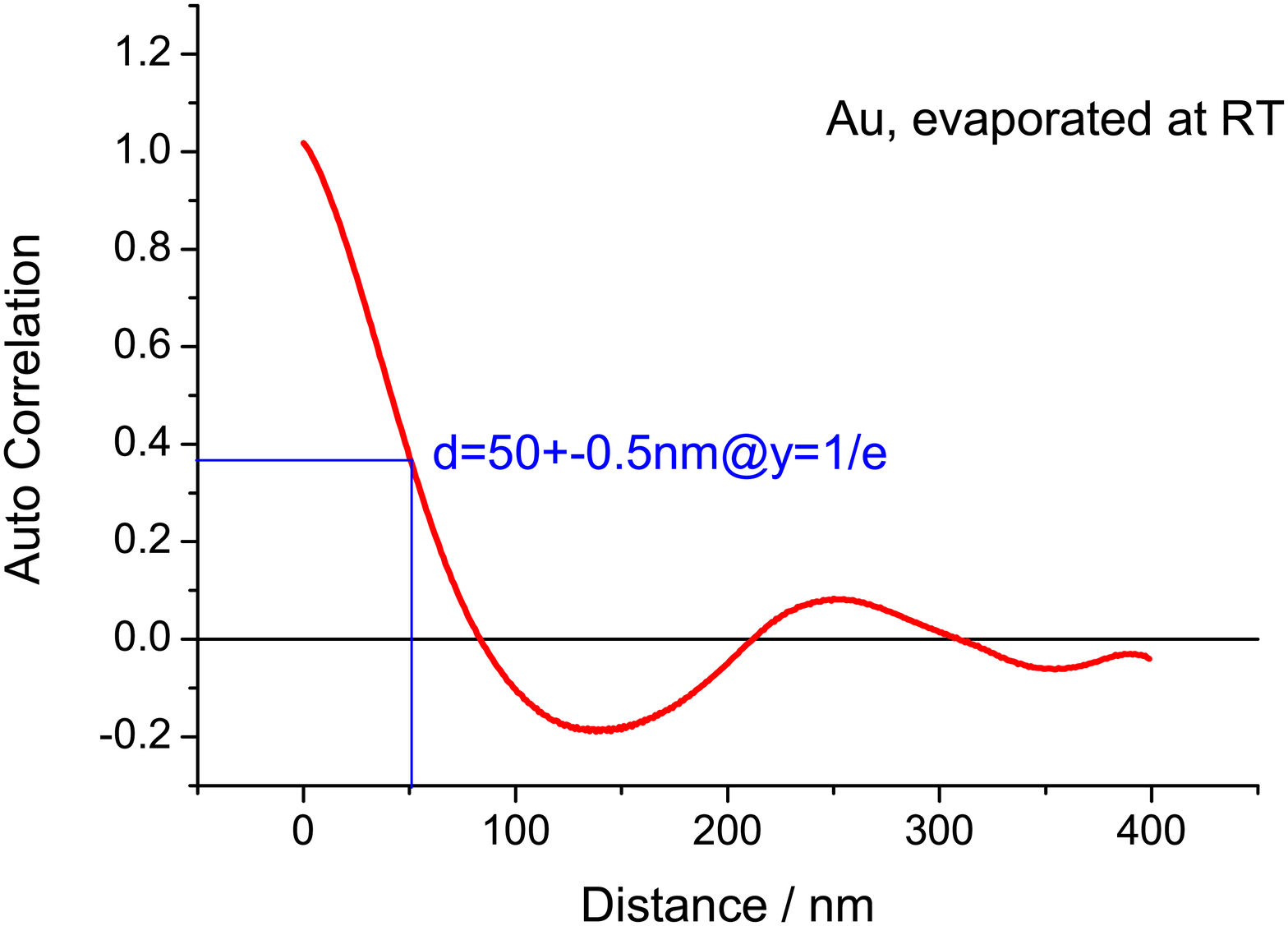}
  \end{center}
  \caption[Auto correlation plot for gold STM image]{Plot of the auto correlation for the STM image of gold. The
    grain size (more accurately $\frac{1}{4}$ of it) was read again at ordinate value $y = \frac{1}{e}$,
    around $200~\text{nm}$. As opposed to the AC analysis for iron, a smooth oscillation was observed.}
  \label{fig:goldgrainac}
\end{figure}

From the STM image analysis in the last chapter we know that the average grain size for our iron films
varies from $16~\text{to}46~\text{nm}$, depending on the temperature of the substrate during evaporation.
Now, let us compare the grain sizes of our iron and gold films as they were prepared with MBE. For this,
we let the auto correlation algorithm analyze the STM image of gold ($d \approx 50~\text{nm}$) from
Figure \ref{fig:stmfeau}. The resulting AC plot is shown in Figure \ref{fig:goldgrainac}, the typical
grain size is around $200~\text{nm}$.

\medskip

To conclude the grain sizes (evaporation at room temperature):

\begin{eqnarray}
  d_{Fe}  & \approx & 16~\text{nm} \\
  d_{Au}  & \approx & 200~\text{nm}
\end{eqnarray}

Since the images are only $400\times400~$nm in dimension and the blobs on the gold surface cover
half of the image (they are $200~$nm), heights are not Gaussian distributed as they are for iron
so we can not apply the method from Section \ref{sec:corrugationanalysis} to determine the corrugation
of the surface.

\subsection{Analysis Conclusion}

To conclude the quality analysis for Fe on MgO$(001)$ from RHEED and STM:

\begin{itemize}
  \item Fe grows as expected with low corrugation ($\approx 1~\text{nm}$) on MgO$(001)$
  \item layer-by-layer growth can be observed
  \item corrugation results from small islands, these can be seen as grains in the STM
  \item corrugation and grain size depend on the substrate temperature during evaporation
  \item the lower the substrate temperature, the lower the corrugation
  \item the lower the substrate temperature, the smaller the grain size
  \item Fe has a small grain size (about $20~\text{nm}$) while Au forms
        blobs of up to $200~\text{nm}$ lateral dimension
\end{itemize}








\section{Final sample parameters}

\begin{figure}
\subfigure[Sample 1]{\includegraphics[scale=0.3]{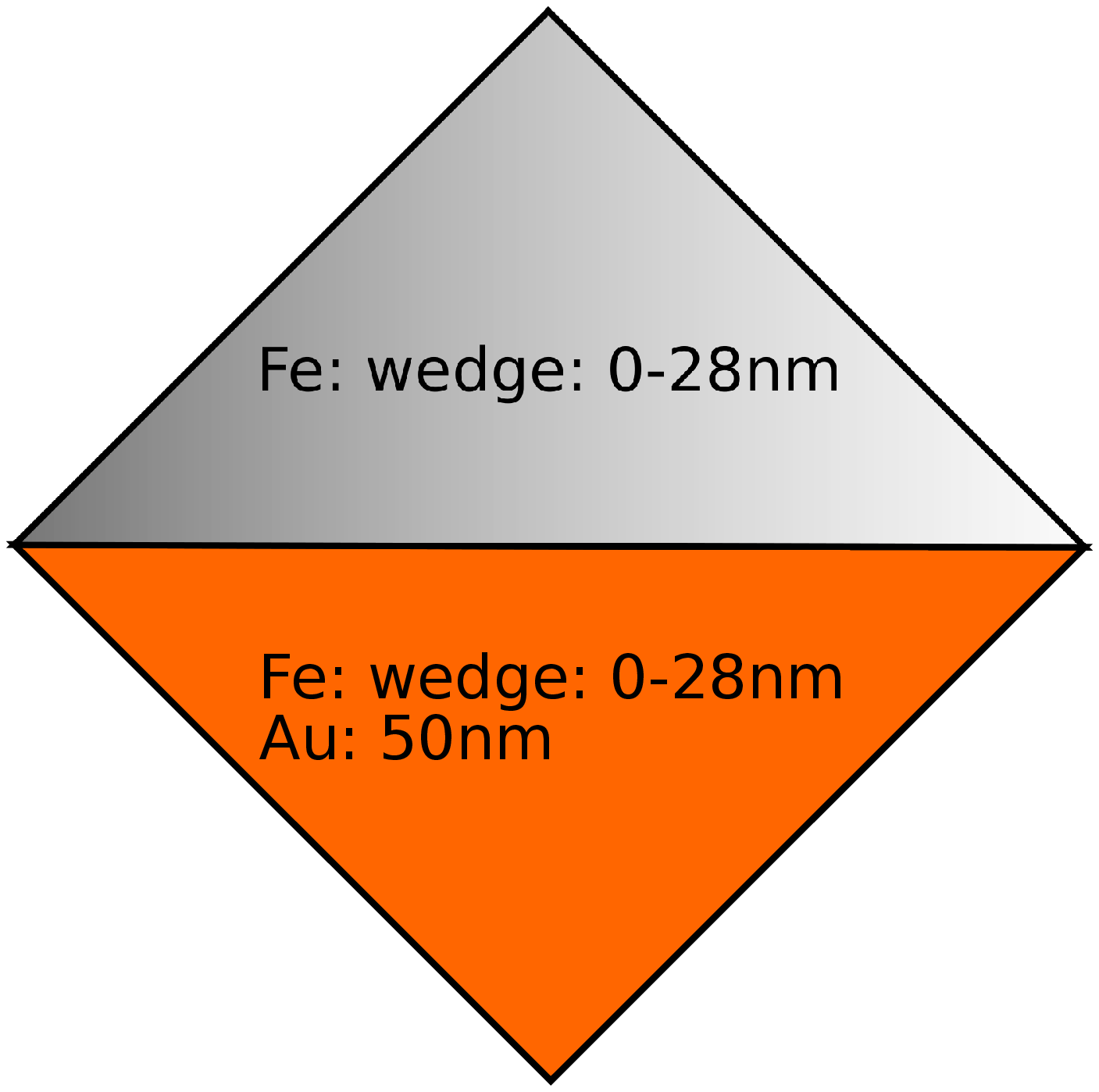}}
\subfigure[Sample 2]{\includegraphics[scale=0.3]{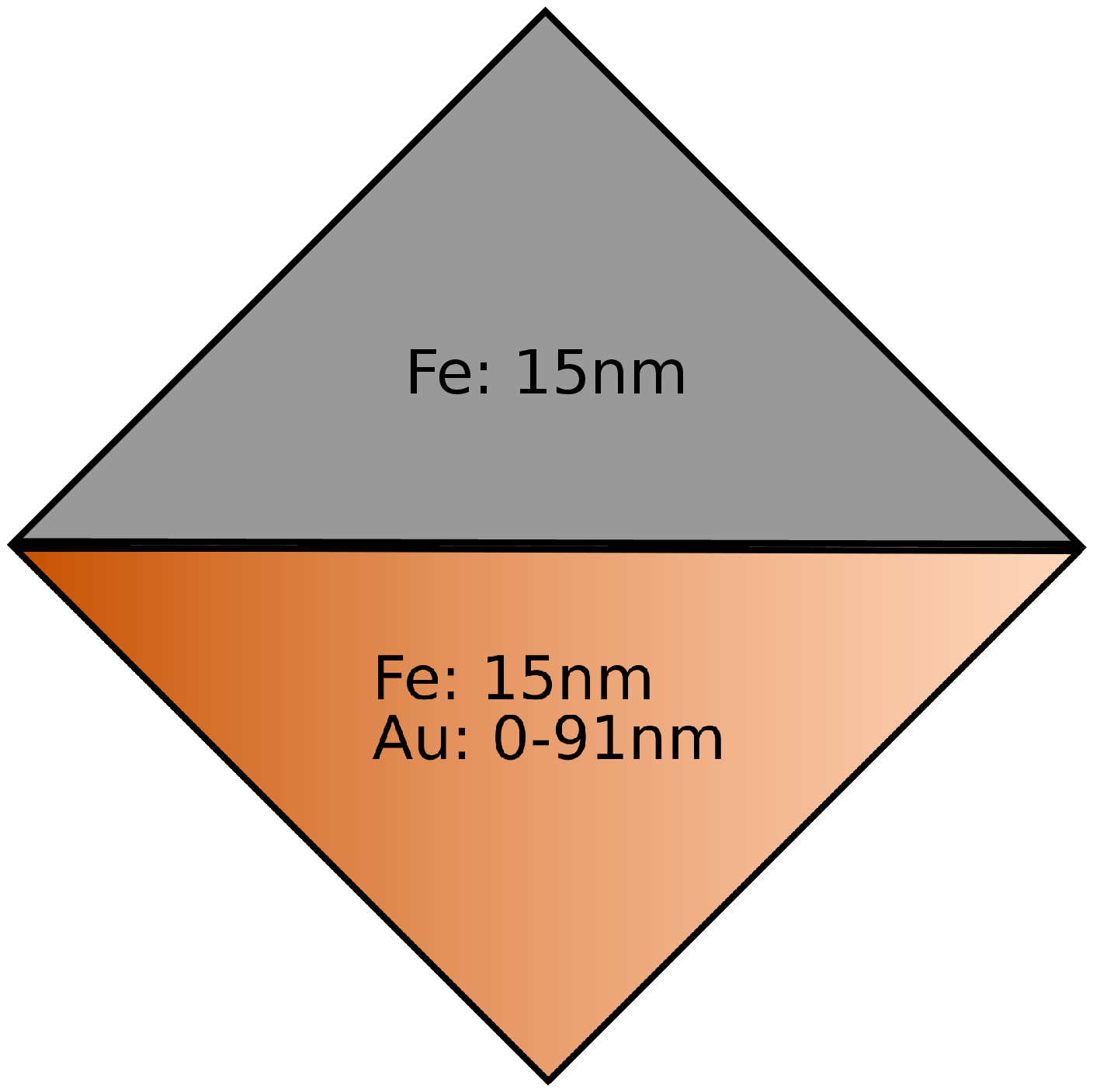}}
\subfigure[Sample 3]{\includegraphics[scale=0.3]{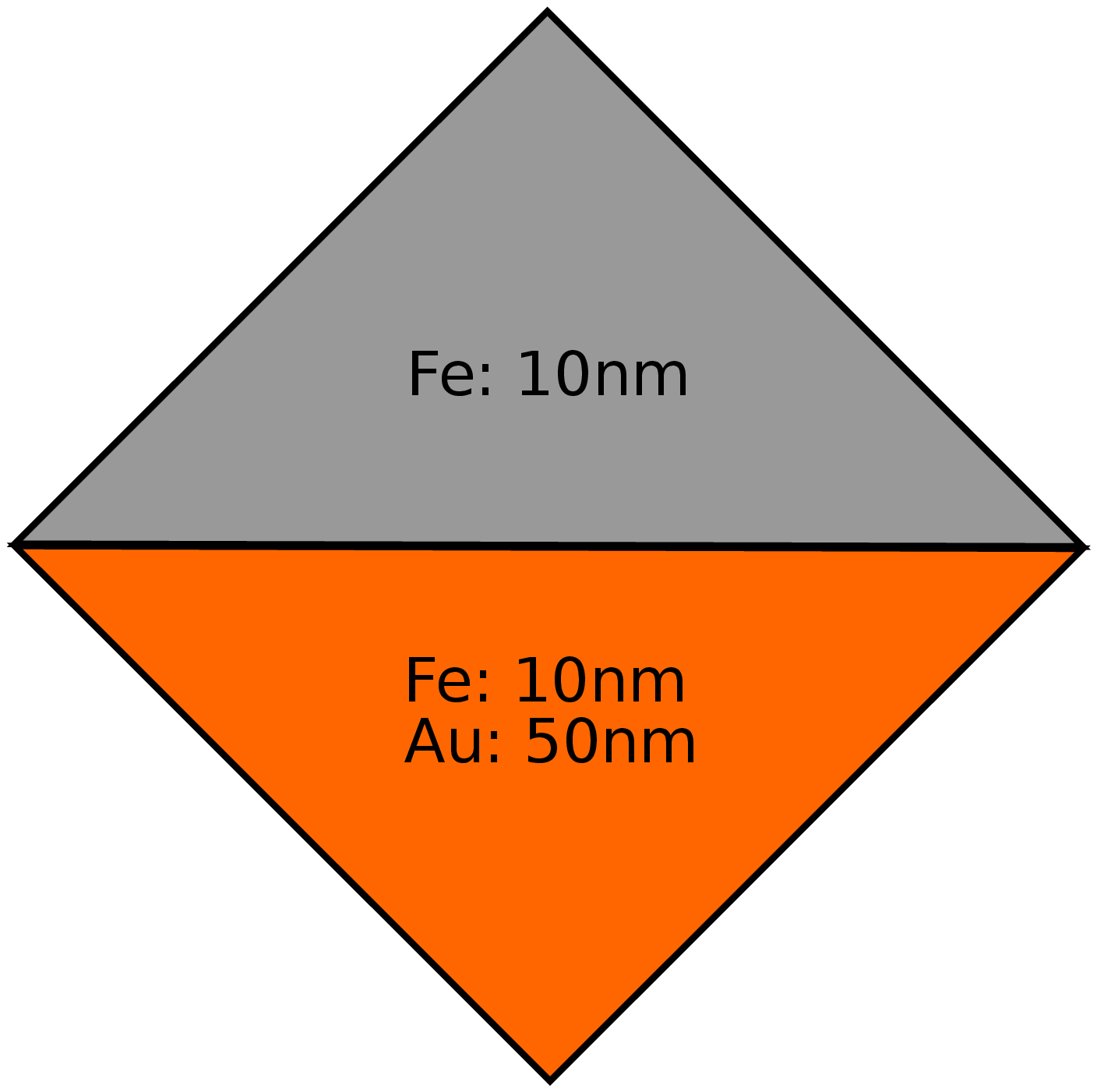}}
\subfigure[Samples 4, 5 and 6]{\includegraphics[scale=0.3]{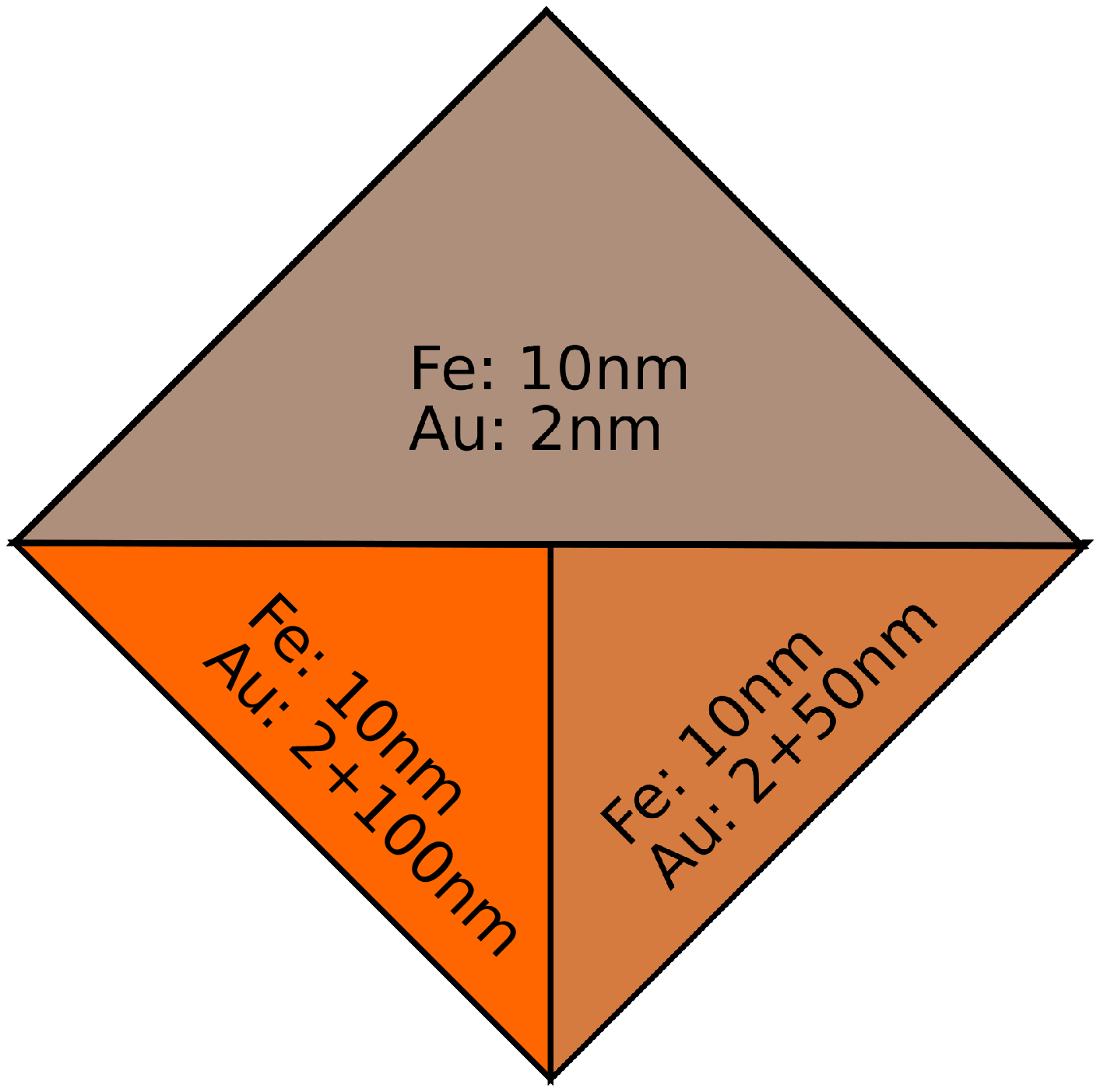}}
\subfigure[Sample 7]{\includegraphics[scale=0.3]{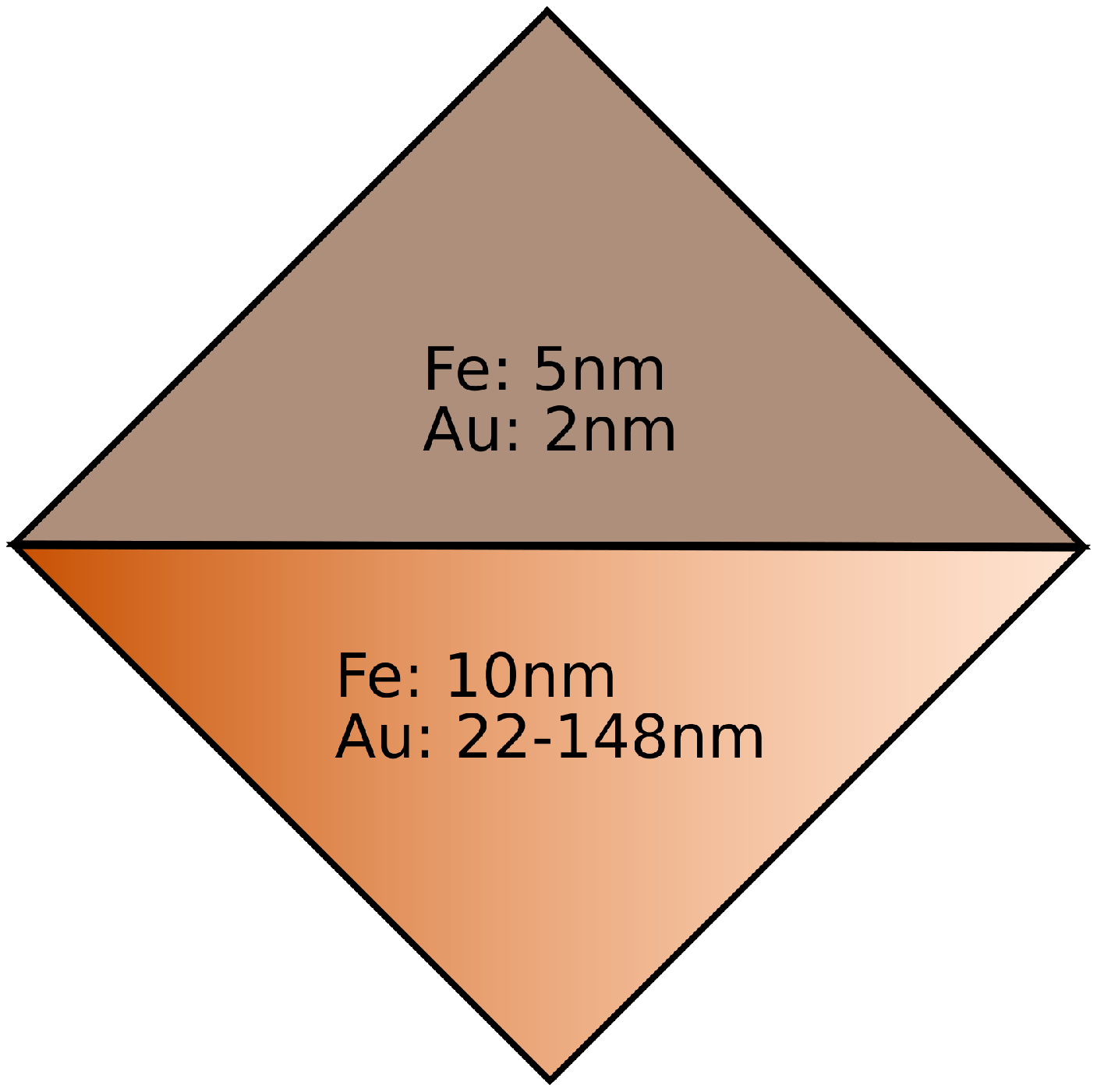}}
\subfigure[Sample 8]{\includegraphics[scale=0.3]{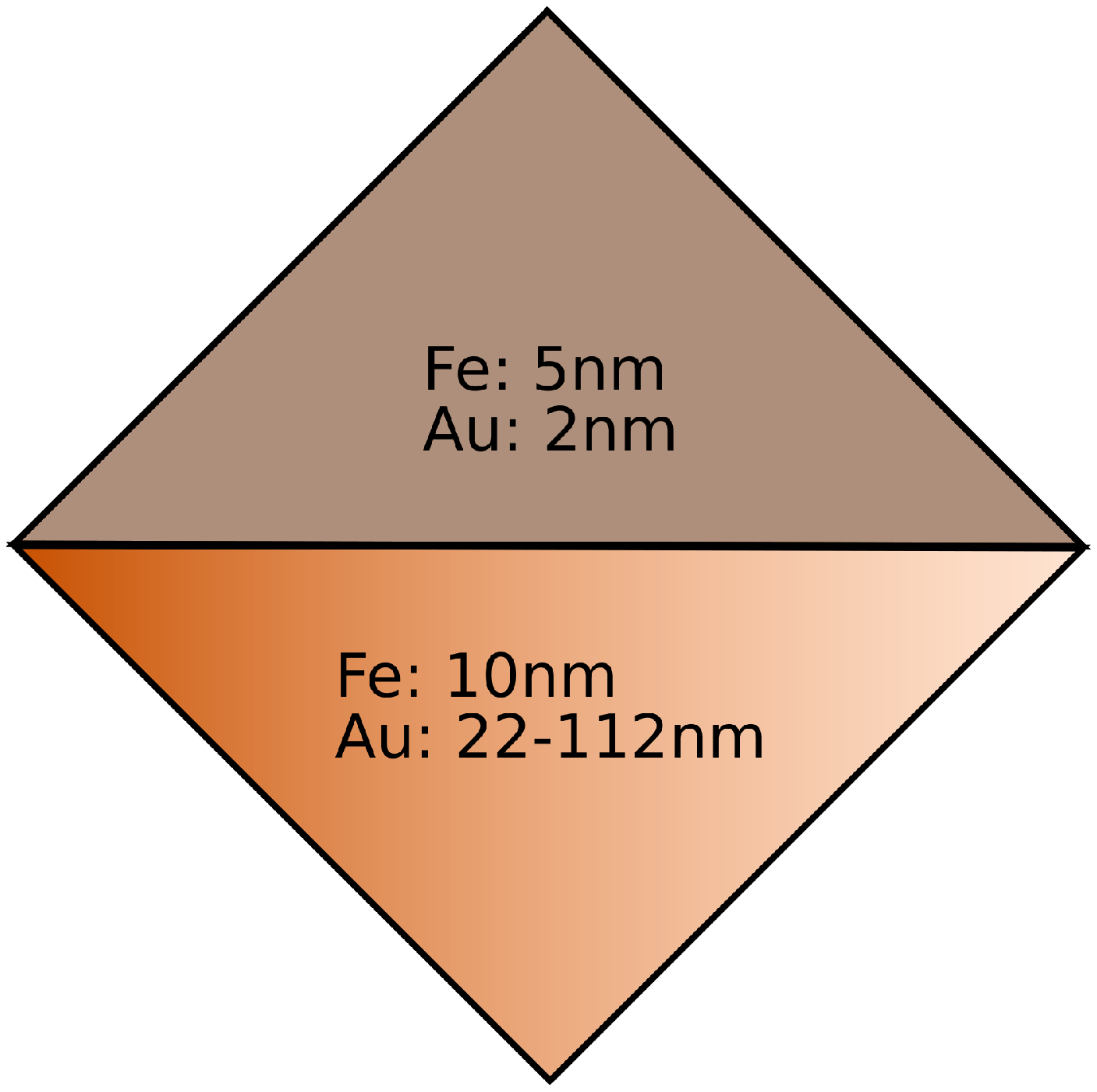}}
\subfigure[Sample 9]{\includegraphics[scale=0.3]{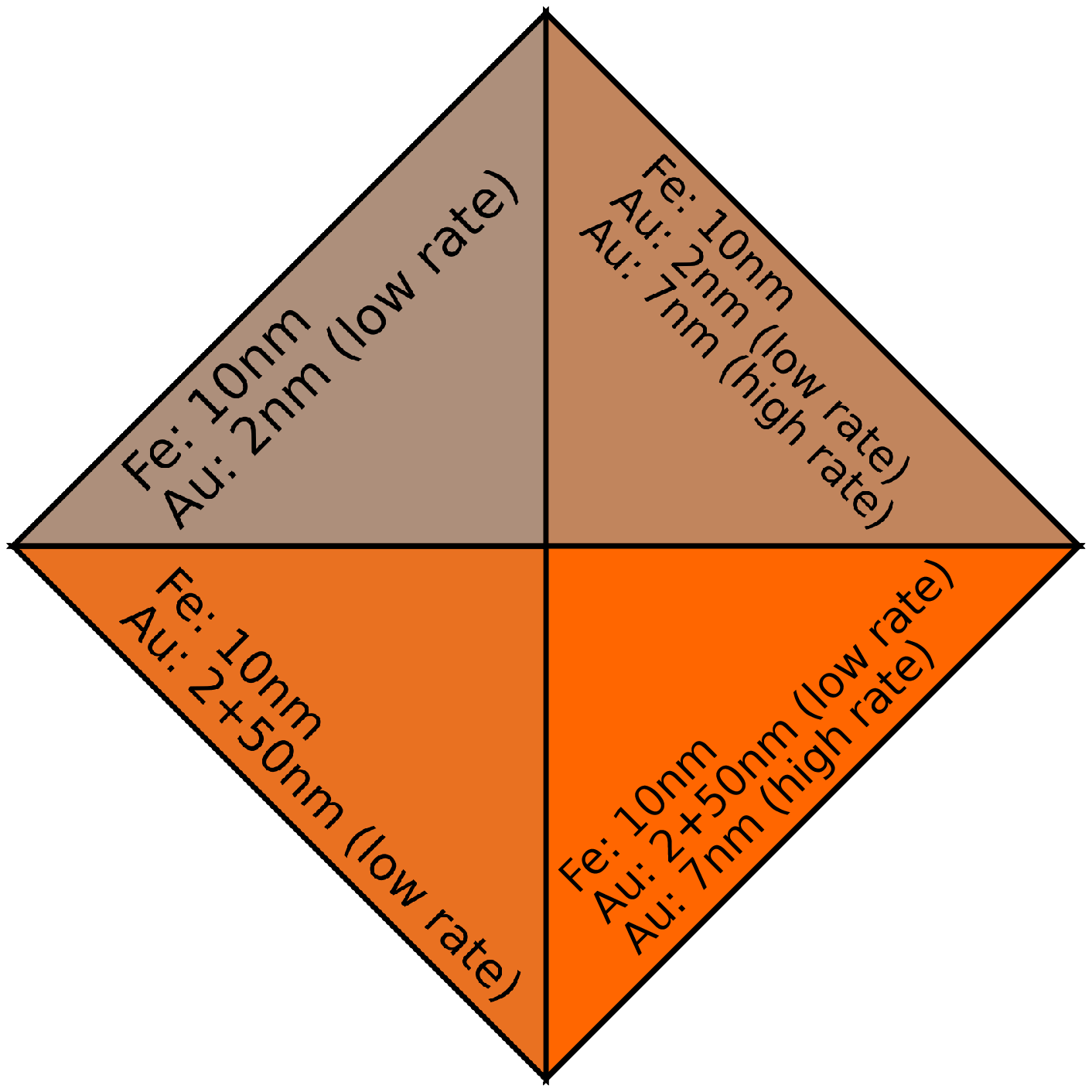}}
\caption[Illustrations of the samples used for SHG measurements]{Illustrations of the
  samples used for SHG measurements. All samples consist of $10 \times 10 \times 0.5~\text{mm}^3$
  MgO$(001)$ substrates on which Fe films of various thicknesses
  are evaporated over the whole substrate area. The Au films cover the Fe films with different
  geometries and thicknesses over the substrate area. See table \ref{tab:sampleparams}
  for detailed sample parameters.}
\label{fig:samples}
\end{figure}

The samples which were used for the measurements were evaporated with parameters
determined from the previous analysis (samples 1-3 were measured in the SHG setup
while we were still investigating with the sample preparation). That is, the
substrates were not to be heated during evaporation and the samples were cleaned,
outgassed and annealed prior evaporation. We produced wedges for the thickness
dependency analysis. Sample 9 was made to check whether the evaporation rate
has any influence on the roughness of the surface. The samples have associated
numbers which are referred to in the chapter on measurements and results.
This allows to crosscheck the evaporation parameters once two samples show
different results in the SHG experiments\footnote{This was actually the
reason why sample 9 was made. We had different results on samples with identical
film thicknesses but which were evaporated at slightly different rates.
Since the rate can play a role in the surface roughness we investigated whether
this was actually the reason for the different results.}. Table \ref{tab:sampleparams}
lists all samples and their parameters with date and sample number, figure
\ref{fig:samples} illustrates the samples.

\begin{table}[!h]
\Large
  \hfill
  \begin{sideways}
        \centering
                \begin{tabular}{|c|c|c|c|c|c|c|c|c|c|} \hline
                \textbf{No.} & \textbf{Type} & \textbf{Date}  & \textbf{d$_{Fe}$} & \textbf{Rate$_{Fe}$} & \textbf{d$_{Au}$} & \textbf{Rate$_{Au}$} & \textbf{$T_{evap}$} & \textbf{$T_{anneal} / t$} & \textbf{$T_{degas} / t$} \\
                     & & & nm  & \AA s$^{-1}$ & nm  & \AA s$^{-1}$ & $^{\circ}$C & $^{\circ}$C & $^{\circ}$C \\ \hline
                   1 & Fe wedge    & 10.4.2009 & 0-28 & 0.4  & 50      & 0.1     & 400 & 800  & 100-110 \\ \hline
                   2 & Au wedge    & 20.5.2009 & 15   & 0.05 & 0-91    & 0.1     & 400 & 1000 & 100-110 \\ \hline
                   3 & Au 50       & 22.5.2009 & 10   & 0.05 & 50      & 0.1     & 400 & 800  & 100-110 \\ \hline
                   4 & Au 50/100   & 1.7.2009  & 10   & 0.1  & 48 / 98 & 0.5/2.0 & RT  & 600  & 100-110 \\ \hline
                   5 & Au 50/100   & 7.7.2009  & 10   & 0.1  & 48 / 98 & 0.1     & RT  & 600  & 100-110 \\ \hline
                   6 & Au 50/100   & 9.7.2009  & 10   & 0.05 & 49 / 96 & 0.2     & RT  & 600  & 100-110 \\ \hline
                   7 & Au wedge    & 31.7.2009 & 5    & 0.05 & 22-148  & 0.3-0.4 & RT  & 600  & 100-110 \\ \hline
                   8 & Au wedge    & 5.8.2009  & 5    & 0.05 & 22-112  & 0.4     & RT  & 600  & 100-110 \\ \hline
                   9 & Au flat/rough & 2.9.2009  & 10   & 0.05 & 50 / 57 & 0.2/2.4 & RT  & 600  & 100-110 \\ \hline
                 \end{tabular}
      \end{sideways}
      \hspace*{\fill}
      \caption{Parameters of the samples prepared for SHG measurements.}
      \label{tab:sampleparams}
\end{table}

\chapter{Measurements and Results}

In the previous chapters we have outlined our motivation for this project
as well as the prerequisites which include the improvements to the
experimental setup as well the sample preparation. We now turn our
focus to the laser experiments performed. We have run
various measurements with the samples to investigate
electron and spin dynamics over different time scales as well as
different thicknesses for the non-magnetic spacer layer (gold).
Furthermore we used both non-linear (SHG) and linear optical effects
to be able to compare both techniques in regard to their suitability
for the investigation of electron and spin dynamics. The measurements
performed can be summarized as follows:
\begin{itemize}
  \item Hysteresis analysis: ferromagnetic properties of the iron layer
  \item Shorttime SHG: ballistic vs. diffusive transport
  \item Longtime SHG: electronic vs. phononic transport
  \item Wedge measurements: thickness dependency of transports
  \item MOKE measurements: alternative method to analyze spin dynamics
\end{itemize}

We have dedicated one section for each of the measurements. Prior to any series
of measurements, especially after a new sample has been inserted, the optical
setup has to be realigned. Since this involves a number of steps, we
describe this procedure in detail in the following section.

\section{Alignment procedure}

The laser cavity (see Sec. \ref{sec:femtosecondlaser}, Page \pageref{sec:femtosecondlaser})
is first adjusted for maximal output. For this, the end mirrors (\textbf{OC} and \textbf{HR}))
are adjusted for maximal output in \emph{cw} mode. The laser is put into
mode-locking and optimized for maximum output with the cavity dumper enabled
by adjusting the end mirrors, pick-out mirror, Ti:Sa crystal, prisms
(\textbf{PR1} and \textbf{PR2}) (see Fig. \ref{fig:laserscheme},
Page \pageref{fig:laserscheme}. The cavity is adjusted such that in the
spectrum of mode-locking regime the cw-component will be suppressed,
The control unit of the cavity dumper allows to set the delay, phase, dumping
rate and output power of the RF signal (see Sec. \ref{sec:femtosecondlaser}, Page
\pageref{sec:femtosecondlaser}, Subsection ``Cavity Dumping''). Once the laser has been
aligned for maximal and stable output, we performed the adjustment of the beam overlap.

\subsection{Beam overlap}
\begin{figure}
\begin{center}
  \includegraphics[width=0.7\textwidth]{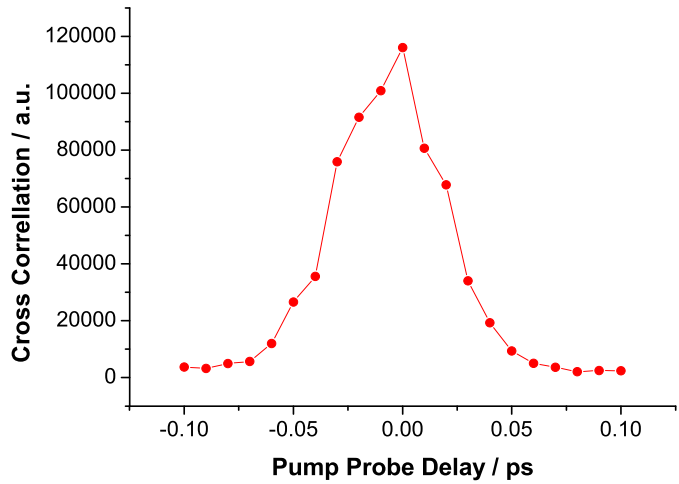}
\end{center}
 \caption[Typical shape of cross correlation]{The cross correlation is the intensity
   signal obtained from the photo multiplier while scanning the time delay between
   pump and probe. At the exact time of overlap, the cross correlation signal
   will be maximum which sets the zero delay between pump and probe.
 }
 \label{fig:crosscorr}
\end{figure}

The two beams (pump and probe) have to be in time and spatial overlap.
This means, that both beams will join on the sample with identical delay
and position. The overlap is conveyed by coherence effects which can be
detected with the linear detector (see Sec. \ref{sec:linearref}, Page \pageref{sec:linearref}).
To find the overlap, the following steps have to be performed. First
we focus both beams on the sample, then try to find a rough spatial
overlap of the two beams on the sample with the help of an optical
microscope. Since 50 or even 100 nm thick gold films won't let the
pump beam pass through in back pump configuration, we are using the
thin gold regions to perform this alignment. Then both the pump
and the probe can be seen as bright spots on the sample.
\emph{The higher the roughness of the films, the higher the scattering
and the better the visibility of the laser spots on the sample.} Once
the spatial overlap is found, we measure the \emph{cross correlation}
signal by scanning through the time delay (Fig. \ref{fig:crosscorr}).
We usually block the signal from the pump beam with the knife for this (see Figure
\ref{fig:experimentalsetup}, Page \pageref{fig:experimentalsetup}).
The zero delay is found by determining the center of the Gaussian.
Once the zero delay has been found, fine spatial adjustment is performed
by maximizing the intensity in the cross correlation signal. Further
improvement in the signal is achieved by proper adjustment of
the monochromator (alignment of beams into slit, fine tuning
of wave length). This means that the foci and the beam position are adjusted with the
lenses which focus the beam onto the sample (see Fig. \ref{fig:completesetup}, (f),
Page \pageref{fig:completesetup}). The previous time scan may be repeated to
verify the zero is still valid. It has shown that the cross correlation signal heavily depends on
the morphology of the sample. Samples which showed high scattering
(bright spots visible in the microscope) usually yielded very
irregularly shaped Gaussians which made it hard to find the zero
delay. Once the proper overlap has been found, the knife is closed
further to suppress also the cross correlation signal to
make sure we measure the probe SHG only. In case of high scattering
due to a rough sample surface, the blocking of the separate
beams becomes more difficult and we have to find a compromise
between blocking the cross correlation and letting the
probe pass. If the scattering of the sample is too high,
the sample cannot be used for measurements however.

\subsection{Laser-induced damage of the films }
During the pump-probe experiments we experienced several
problems with laser-induced damage. This means that the intensity
of the incident pump laser beam was so high that it
caused irreversible damage to the films. This damage
could either be due to \emph{heat accumulation} or
it could be a \emph{single shot effect} (high
intensity). To verify this, we ran two small
tests with \textbf{sample \#6}. First we
\textbf{reduced the repetition rate}
by adjusting the cavity dumper to find a damage
threshold. Within the range of adjustments, however,
we couldn't determine a sane threshold rate.
In the second test, we put an attenuator consisting
of a polarizer and a $\lambda$/2-plate to rotate
the polarization plane.
\begin{table}[htbp]
        \centering
                \begin{tabular}{|c|c|c|} \hline
                \textbf{Rel. Angle} & \textbf{Power} & Pulse Energy \\
                   $^\circ$ & mW & nJ/pulse\\ \hline
                   -5 & 37 & 25 \\
                   0  & 42 & 28 \\
                   10 & 42 & 28 \\
                   15 & 40 & 26 \\
                   20 & 34 & 22 \\
                   25 & 26 & 17 \\
                   30 & 18 & 12 \\
                   35 & 11 & 7  \\
                   40 & 5  & 3  \\ \hline
                \end{tabular}
                \caption{Calibration table: Power and pulse energy vs. angle of polarization filter}
                \label{tab:laserattentuator}
\end{table}
We first set the angle to \textbf{30}$^\circ$ which corresponds
to a beam power of approximately \textbf{18}~mW or \textbf{12}~nJ/pulse
(see Tab. \ref{tab:laserattentuator}) and positioned the laser beam
onto a new spot on the sample and observed the burning in
the microscope for several minutes. If there was any damage,
the spot from the pump beam would gain intensity.
For \textbf{30}$^\circ$ and \textbf{20}$^\circ$
we did not see any ``burning'', even after more than 5 minutes
of irradiation. We then changed the angle gradually from
\textbf{20}$^\circ$ towards \textbf{0}$^\circ$. Since burning
was visible at an angle of \textbf{14}$^\circ$, we ran
a second test starting at this angle. We repeated this
way until we found that at \textbf{17}$^\circ$, which
corresponds to approximately \textbf{37}~mW or \textbf{25} nJ/pulse,
the sample would not be subject to any visible burning. \emph{The laser-induced
damage is therefore a single shot effect and can be
avoided by reducing the beam intensity. This can also be
achieved by defocusing the pump beam on the sample
with the collimator lens. Subsequently we de-focused
the pump to reduce the intensity to below 90\% in the
following experiments.}

\subsection{Alignment of the photodiodes for MOKE and LR}
Once the setup has been properly aligned for SHG, the alignment
for the MOKE and linear reflectivity detectors is rather simple.
The diodes just have to be aligned for maximum signal. And in
case of the MOKE detector, the output signal of both diodes has
to be balanced.

\subsubsection{Alignment for MOKE}
\begin{figure}
\begin{center}
  \includegraphics[width=0.8\textwidth]{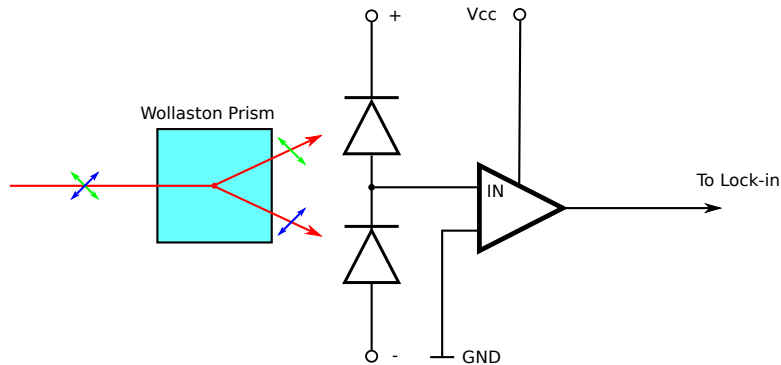}
\end{center}
 \caption[Setup used to measure MOKE]{Our setup to measure MOKE uses
    two photodiodes wired in a voltage divider configuration. In
    case of identical signal intensities at both diodes, the voltage
    at the input of the pre-amplifier will be zero. With the help of
    the Wollaston prism, the diodes measure perpendicular
    components of the polarization which will be equal if the
    diodes are balanced and there is no MOKE effect. If we have
    a MOKE signal, the voltage at the pre-amplifier will be
    non-zero and an electric current is measured, amplified
    and fed into the lock-in.
 }
 \label{fig:mokescheme}
\end{figure}

To measure the rotation of the polarization plane
from the MOKE effect, we use a setup of two
photodiodes which are preceded with a Wollaston
prism (see Fig. \ref{fig:mokescheme}). The
diodes are connected to a pre-amplifier which
feeds the lock-in. The prism separates the light
into two orthogonal, linearly polarized beams.
By measuring in perpendicular configuration,
we are able to determine the rotation angle by adding
the two components of the two vectors. For a proper
alignment, the output of both diodes has to be
balanced so that for a non-attenuated fundamental with
no external magnetic field applied to the sample, the
signal on both diodes is identical. In order to
balance the diodes, the detector casing is attached
to rotary holder which allows to set proper azimuth.

\subsubsection{Alignment for LR}
To measure linear reflectivity, we disconnected one of the
two photo diodes in the detector. Then the spot is focussed
and aligned onto the detector until the maximum output
signal is observed. In case the output signal is too high,
it will overdrive the pre-amplifier connected to the diodes.
We therefore used filters to attenuate the intensity to a proper
level so that we could use the full scale of detection
of the diode.

\section{Hysteresis analysis}
To verify the ferromagnetic properties of the iron layer, we have performed optical hysteresis analysis. We wanted
to make sure that the ferromagnetic properties would not be hampered by the adjacent gold layers. We therefore
measured at spots on the sample where the gold layer had different thicknesses. Since the non-ferromagnetic gold layer
suppresses the optical signal completely, measuring the hysteresis required the samples to be inserted into
the sample holder with their back side (from which pumping normally is performed) facing to the front to be able
to probe the ferromagnetic layer directly. Another motivation for the hysteresis measurements
was to find out what external field is necessary to saturate the magnetization, that is find
the critical external field at which the magnetization remains constant. The hysteresis was
measured in \emph{transversal geometry} of the external magnetic field. The current for the
magnetic coils was swept from \textbf{-2.5} to \textbf{2.5} Amperes which corresponds to
sweeping the external field from \textbf{-75} to \textbf{75} Gauss. We measured on
\textbf{sample \#6}.
\begin{figure}
  \subfigure[Spot 1]{\includegraphics[width=0.5\textwidth]{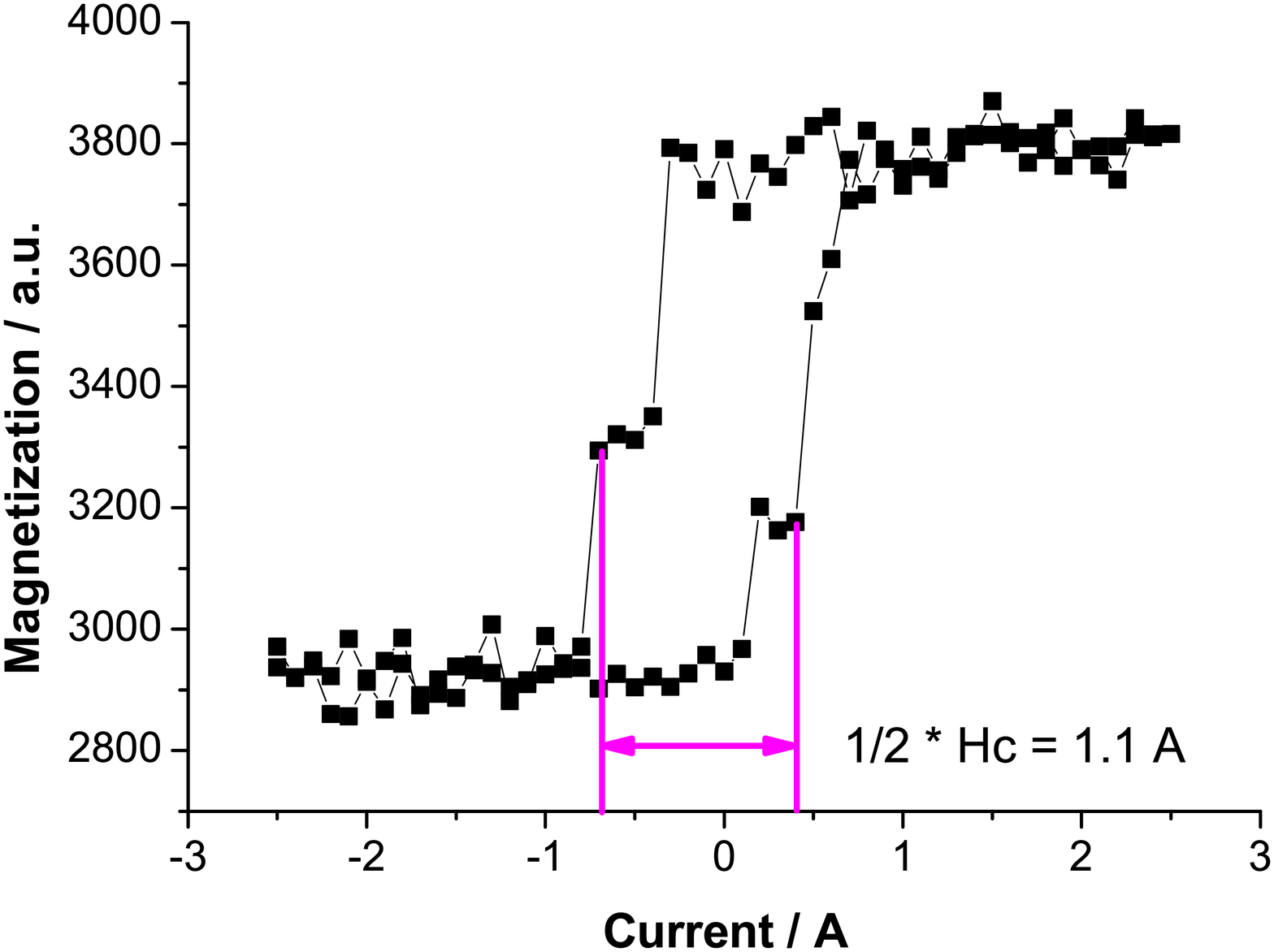}}
  \hfill
  \subfigure[Spot 2]{\includegraphics[width=0.5\textwidth]{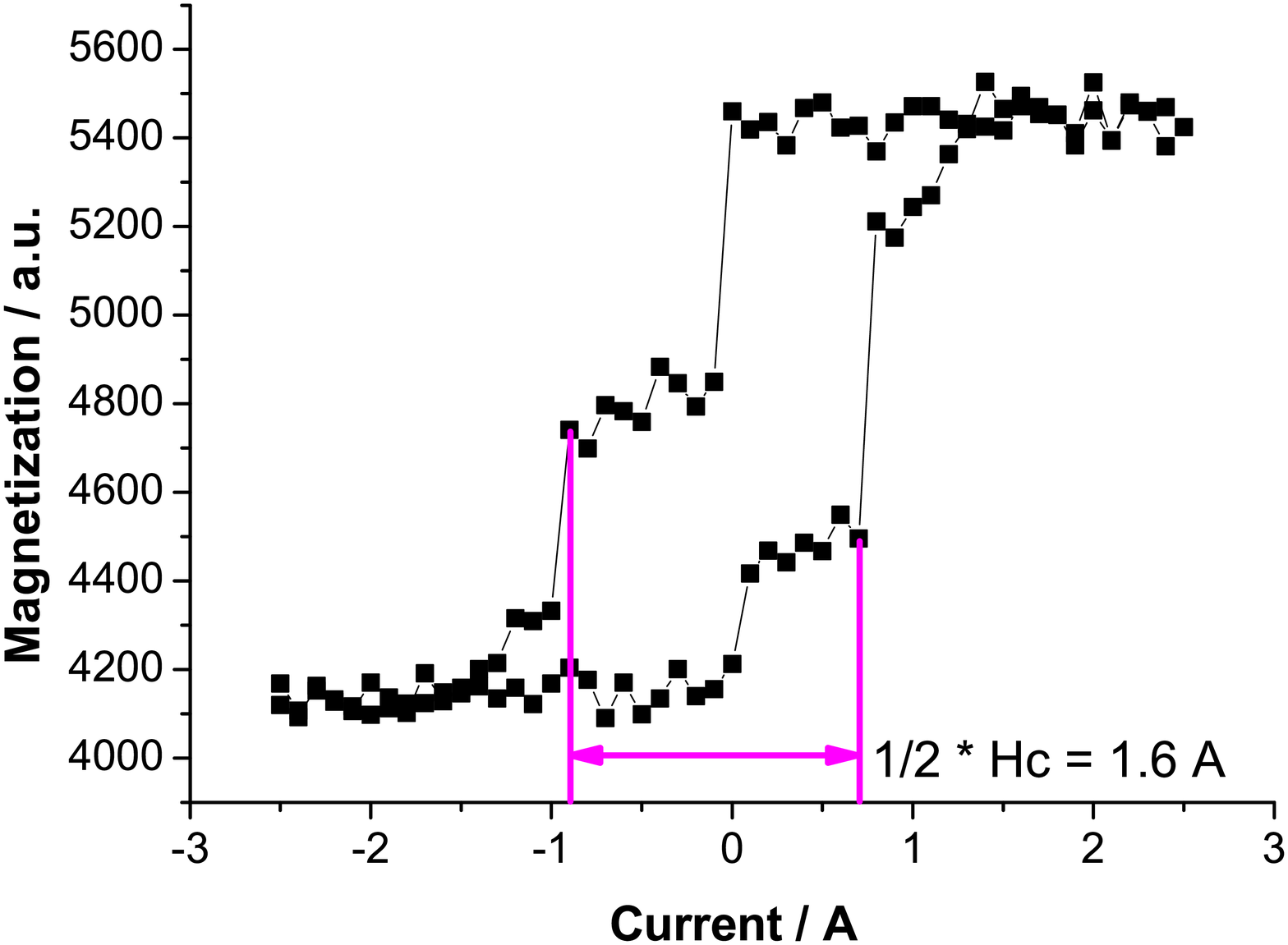}}
  \subfigure[Spot 3]{\includegraphics[width=0.5\textwidth]{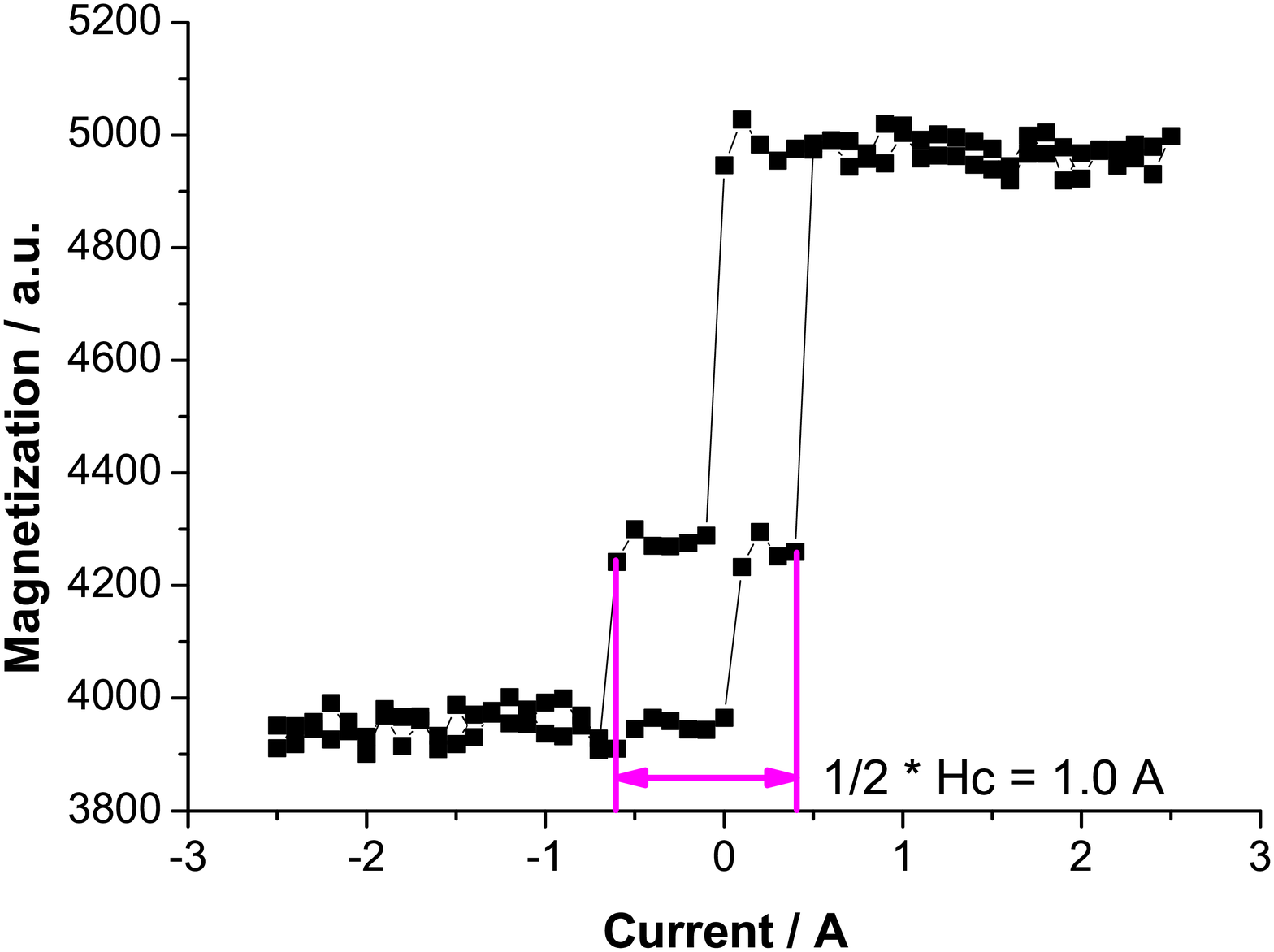}}
  \hfill
  \subfigure[Spot 4]{\includegraphics[width=0.5\textwidth]{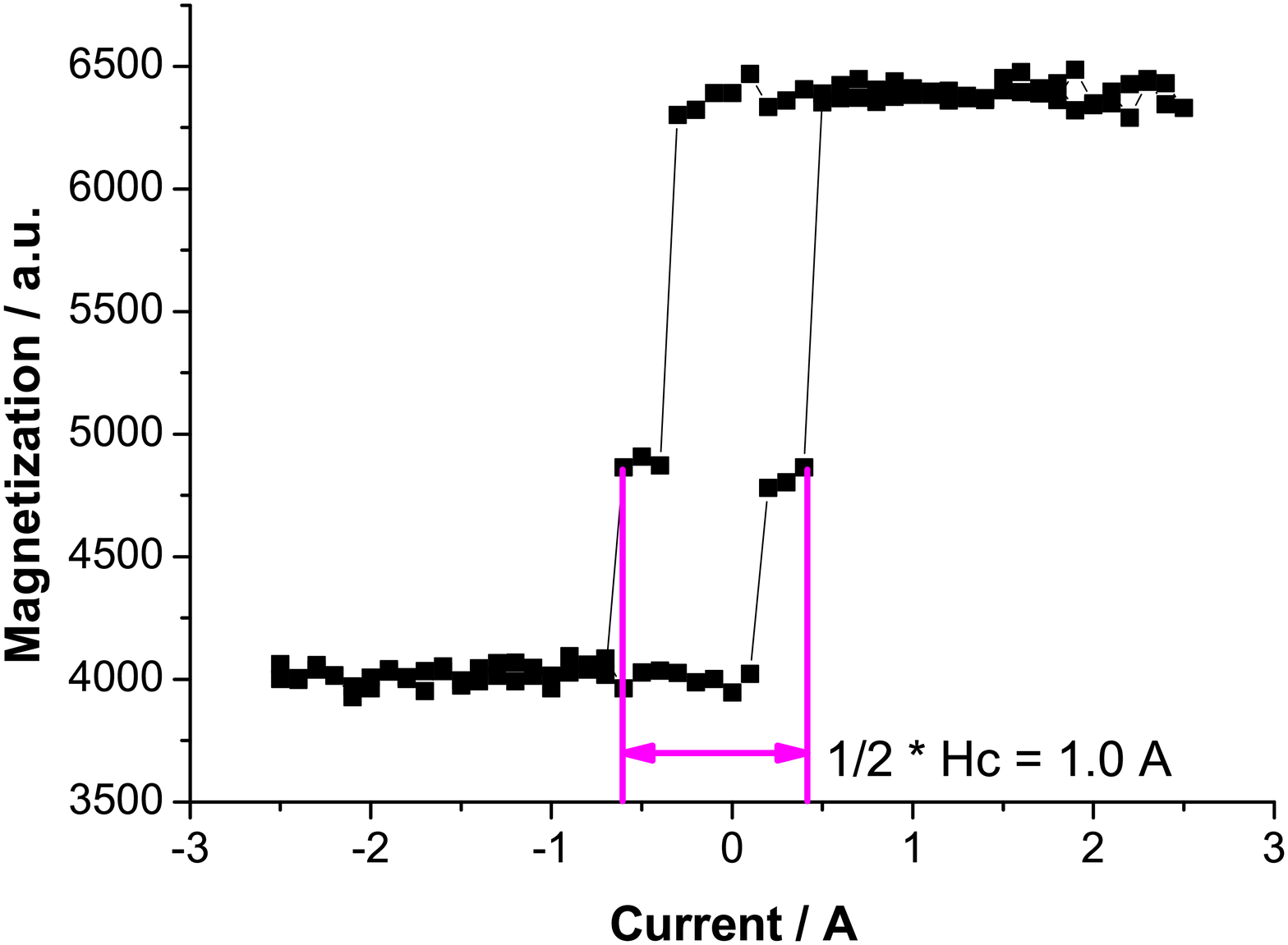}}
  \subfigure[Spot 5]{\includegraphics[width=0.5\textwidth]{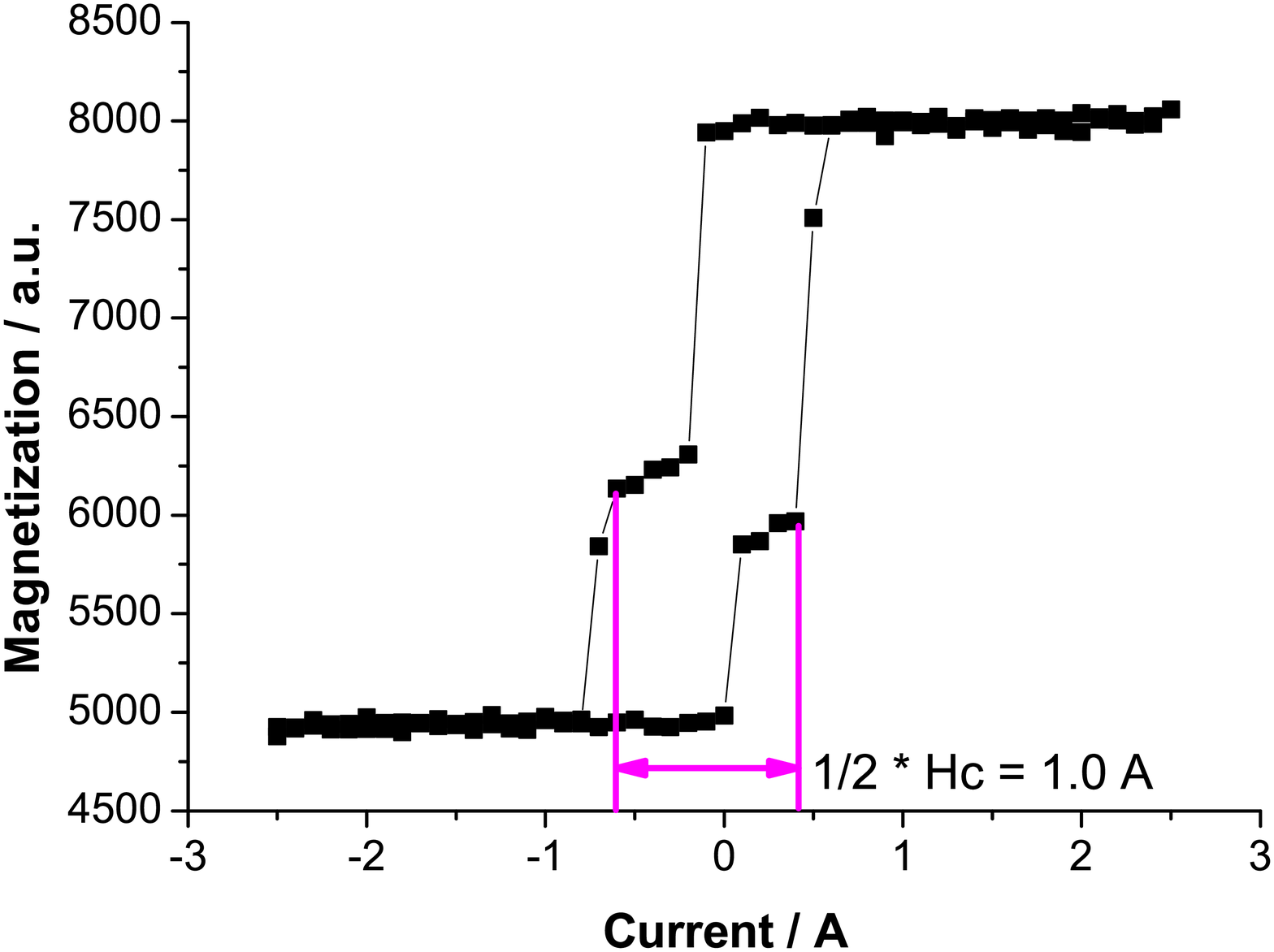}}
  \hfill
  \subfigure[Spotmap]{\includegraphics[width=0.5\textwidth]{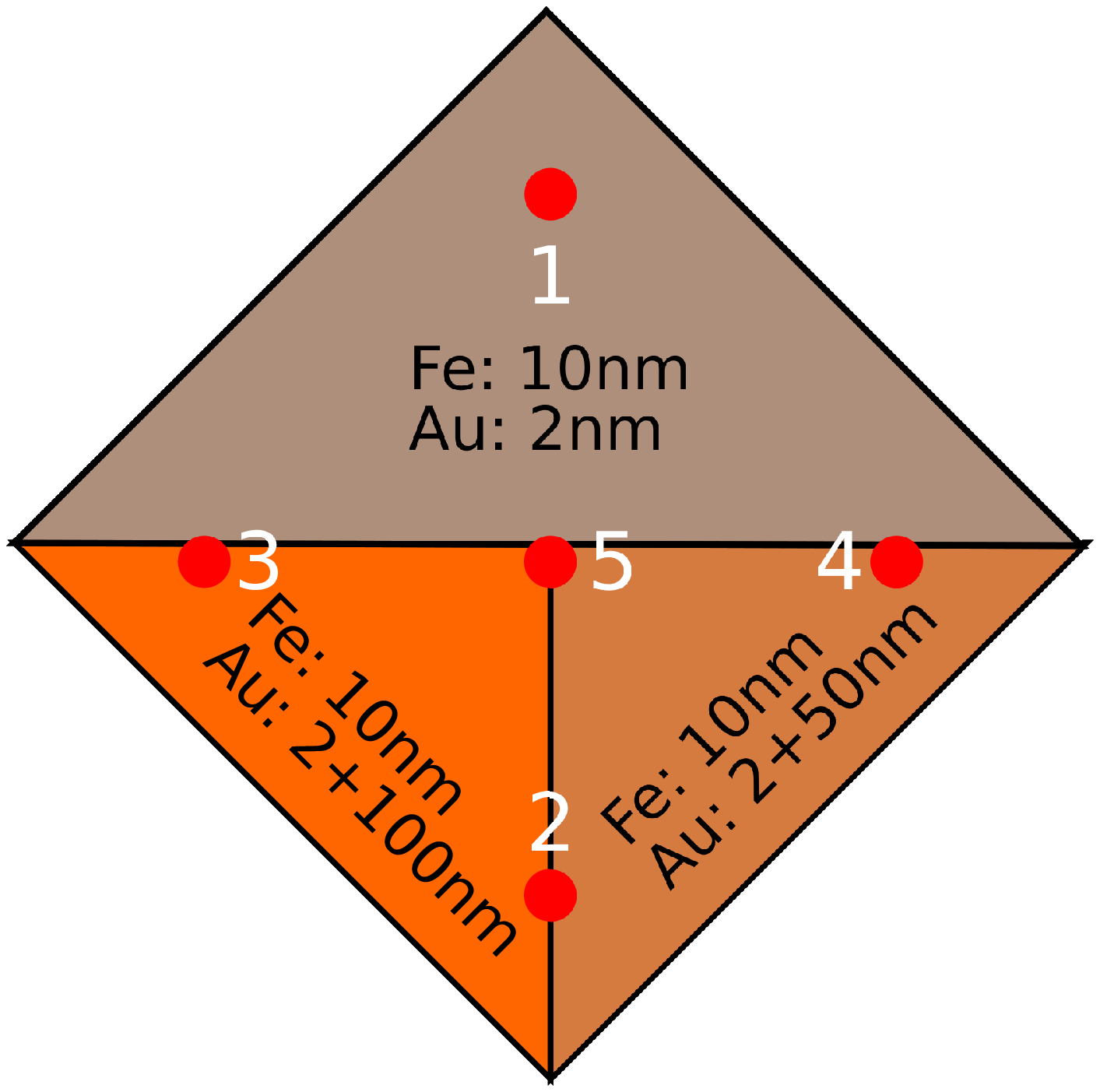}}
 \caption[Hysteresis analysis at different spots]{The Figures (a)-(e) show the hysteresis loops measured
   at different spots on the sample and therefore for different thicknesses of the gold layer. They have
   a more or less rectangular shape with some slight distortions. These distortions have also been found
   by Costa-Kr\"amer et al. in \cite{hernando_magnetization}. What is important for us, is that the
   hysteresis curves have equivalent values for the coercivity ($H_c \approx 0.5~\text{A} = 15~\text{Gauss}$).
   We can conclude that the gold layer does not hamper the ferromagnetic properties of the iron layer.
   Saturation of magnetization is found at around 1 Ampere.}
 \label{fig:hysteresisplots}
\end{figure}
Figure \ref{fig:hysteresisplots} shows the results of the hysteresis analysis. We have probed
at 5 different spots which lie below different thicknesses of the non-ferromagnetic gold
layer. The hysteresis loops are comparable for all spots and show rectangular shapes as also reported
by Costa-Kr\"amer et al. in \cite{hernando_magnetization} for single iron layers. We therefore conclude that
the ferromagnetic properties of the iron film are not influenced by the adjacent gold layer.
A saturation of the magnetization occurs at around \textbf{1 Ampere}, that is around
\textbf{30 Gauss}. Consequently we will sweep the current through the magnet between
\textbf{-1.5 A} and \textbf{1.5 A} in the following experiments.

\section{Shorttime: ballistic vs. diffusive transport}
The first task of the series of optical measurements of the core experiment was to reproduce the
previous results obtained within the scope of this project since we have
made modifications to the setup (new magnet and sample holder, rearrangement
of some components on the optical table). Furthermore, the new samples were produced
with a different type of evaporator and chamber. Thus we have produced samples with a 10 nm
iron layer and 50 and 100 nm gold layer using the shutter. For this series of measurements,
\textbf{sample \#5} was used (see Tab. \ref{tab:sampleparams}, Page \pageref{tab:sampleparams}).
The external magnetic field was driven to saturate magnetization in the ferromagnetic layer
($\approx$ \textbf{30 Gauss}). The time range for the pump-probe was set
from \textbf{-0.5} to \textbf{2.0} picoseconds.
\begin{figure}
  \subfigure[Pump-induced effects]{\includegraphics[width=0.5\textwidth]{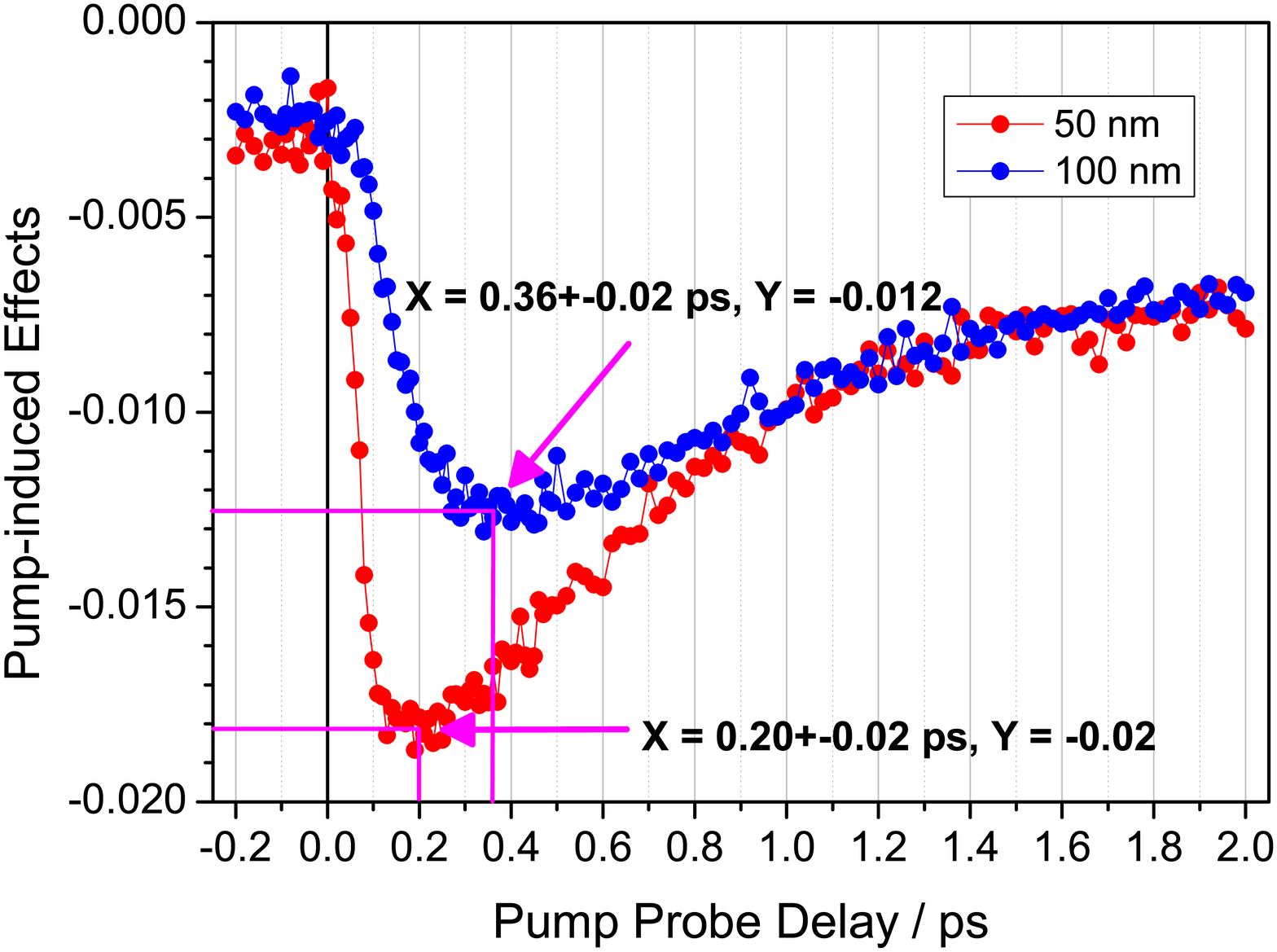}}
  \hfill
  \subfigure[Transient contrast]{\includegraphics[width=0.5\textwidth]{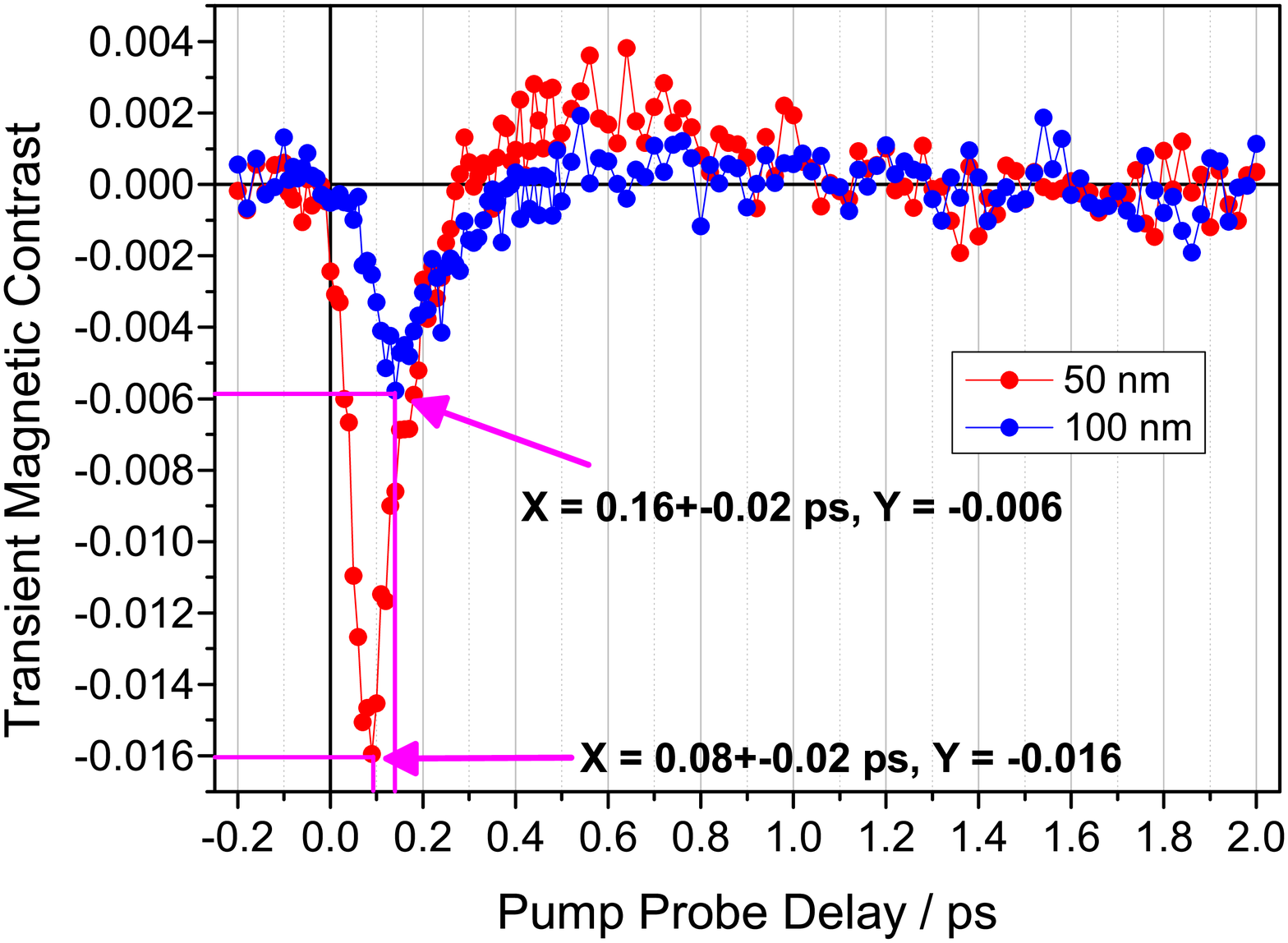}}
  \subfigure[Linear reflectivity]{\includegraphics[width=0.5\textwidth]{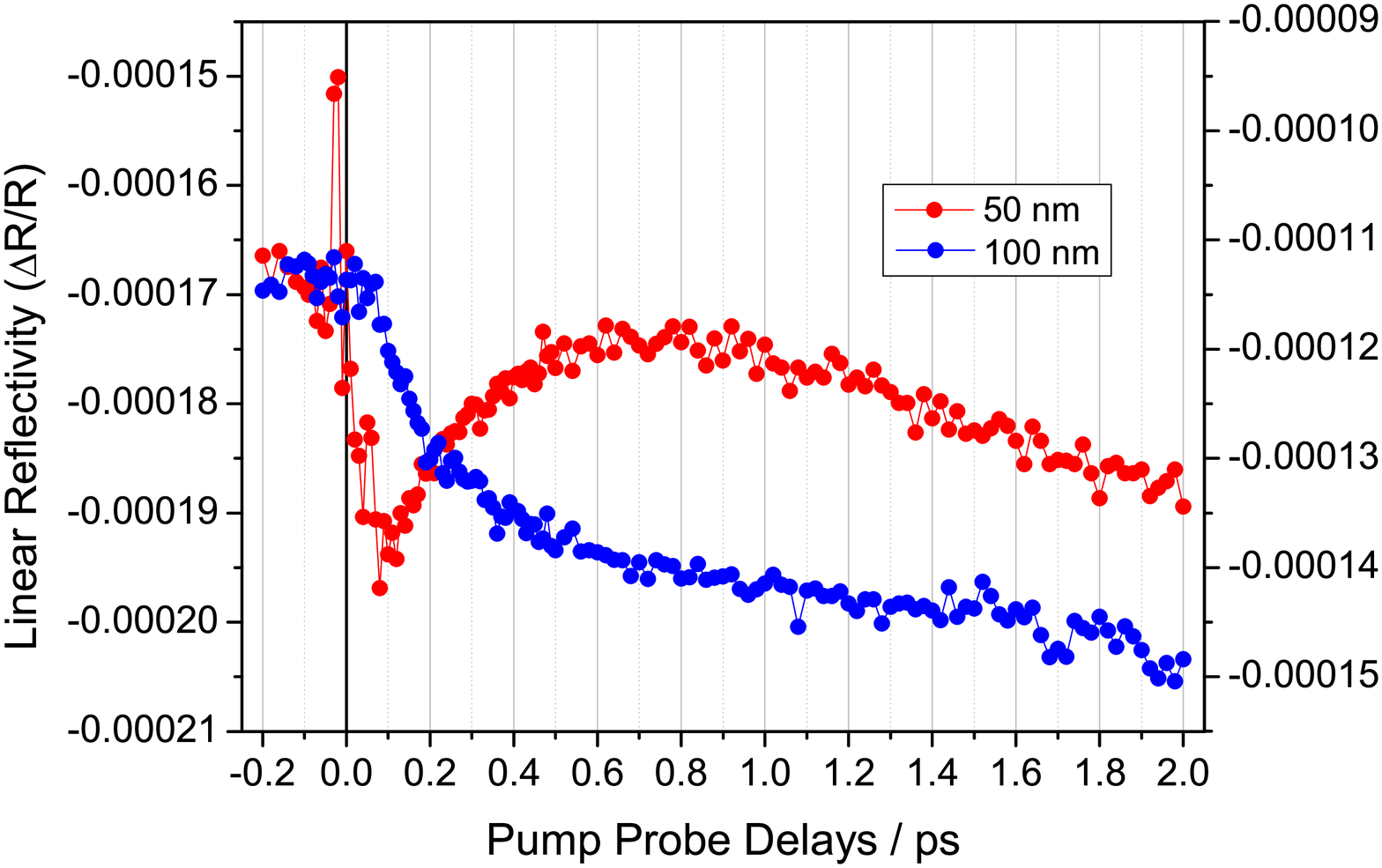}}
 \caption[Shorttime measurement result plots]{Result plots for the pump-induced effects (a), transient magnetic
   contrast (b) and the linear reflectivity (c) for \textbf{50} and \textbf{100 nm} gold layers. The ballistic carriers
   change the electron distribution versus energy at the gold surface which is detected in a change of SHG intensity. From the
   time scale we can read the average time that the carriers have travelled since excitation. Plot (a) does not
   distinguish between majority and minority carriers. This can be seen in the transient contrast plot (b) which
   is proportional to the surface magnetization. The signal is characterized by a short peak resulting
   from the ballistic, minority carriers (spin-down electrons and spin-up holes) and by an change
   of sign and overshoot caused by the majority carriers (spin-up electrons and spin-down holes)
   which travel diffusively (see text). The linear reflectivity show the bulk dynamics at the
   same time. It also show that the zero delay is set correctly for both thicknesses.}
 \label{fig:shorttimecomp}
\end{figure}
\subsection{Discussion of the results}
We have recorded three different signals during each run of the measurements:
\begin{itemize}
  \item $\Delta E_{even} (t)$: non-magnetic SHG (see Section \ref{sec:shgvsmshg}, Page \pageref{sec:shgvsmshg})
  \item $\rho (t)$: magnetic SHG
  \item linear reflectivity (see Section \ref{sec:linearref}, Page \pageref{sec:linearref})
\end{itemize}
\subsubsection{Non-magnetic SHG}
The non-magnetic SHG signal yields information about the electronic
transport from the moment of excitation ($t = 0~$ps). The signal is
distinguished by a strong peak shortly after delay which increases
steeply but recovers on a longer time scale. The peak results from a
sudden changes in the electron distribution versus energy at the surface
caused by hot carriers excited from the ferromagnetic layer reaching
the gold surface. The arriving of hot carriers results in a non-equilibrium
carrier distribution which differs from equilibrium Fermi distribution.
The charge density remains unchanged, however, since any surplus
charge from the ballistic carriers is compensated by carriers with
energies around $E_F$. The relaxation into energy
ground states takes place on a longer time scale than perturbation
from the carrier injection which explains the asymmetric shape of
the peak. The intensity of the peak and its position on the time
axis depend heavily on the thickness of the gold layer. For a
\textbf{50 nm} gold layer, the peak has its maximum
at $\approx$ \textbf{200$~\pm~$20} femtoseconds, the intensity reaches up
to \textbf{2 percent} of the total intensity. For \textbf{100 nm},
the peak is shifted to $\approx$ \textbf{360$~\pm~$20 fs} and has a reduced
relative intensity of around \textbf{1.25 percent}. The explanations
for both effects are obvious. A thicker gold layer means, of course,
that the path for the carriers to the surface is longer
and thus the longer they take to reach the gold surface. On the other
hand a thicker gold layer means that the carriers experience more
scattering within the crystal. If the gold layer is too thick,
more carriers will be subject to diffusive transport rather than
ballistic transport and the signal peak ``smears out''.
\subsubsection{Magnetic SHG}
\begin{figure}
\begin{center}
  \includegraphics[width=0.6\textwidth]{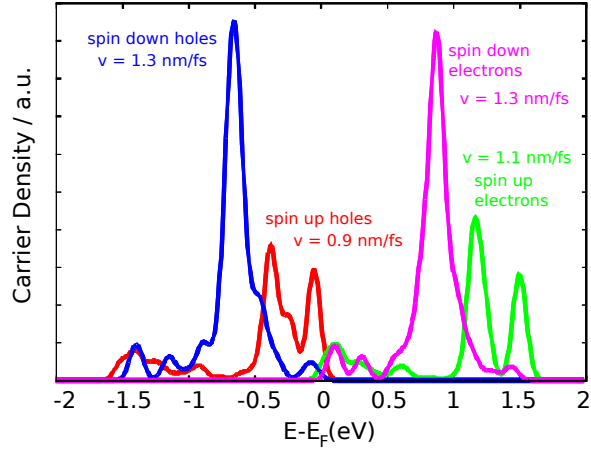}
\end{center}
 \caption[Excitation spectrum for iron]{Excitation spectrum for iron. There
   are four types of carriers which can be excited from iron: Spin-up
   electrons (majority carriers, since they increase the magnetic moment
   in Au), spin-down electrons (minority carriers, since they decrease the
   magnetic moment in Au), spin-up holes (minority carriers) and spin-down holes
   (majority carriers). The carriers are excited from different bands in
   the energy spectrum of iron and have therefore different velocities
   after excitation. The spectrum shows that the carrier densities
   for spin-down carriers are higher than the ones for spin-up
   carriers. Spectrum calculated by \emph{Tim Wehling} \cite{wehling_webpage}.
 }
 \label{fig:fespectrum}
\end{figure}

\begin{figure}
\begin{center}
  \includegraphics[height=1.3\textwidth]{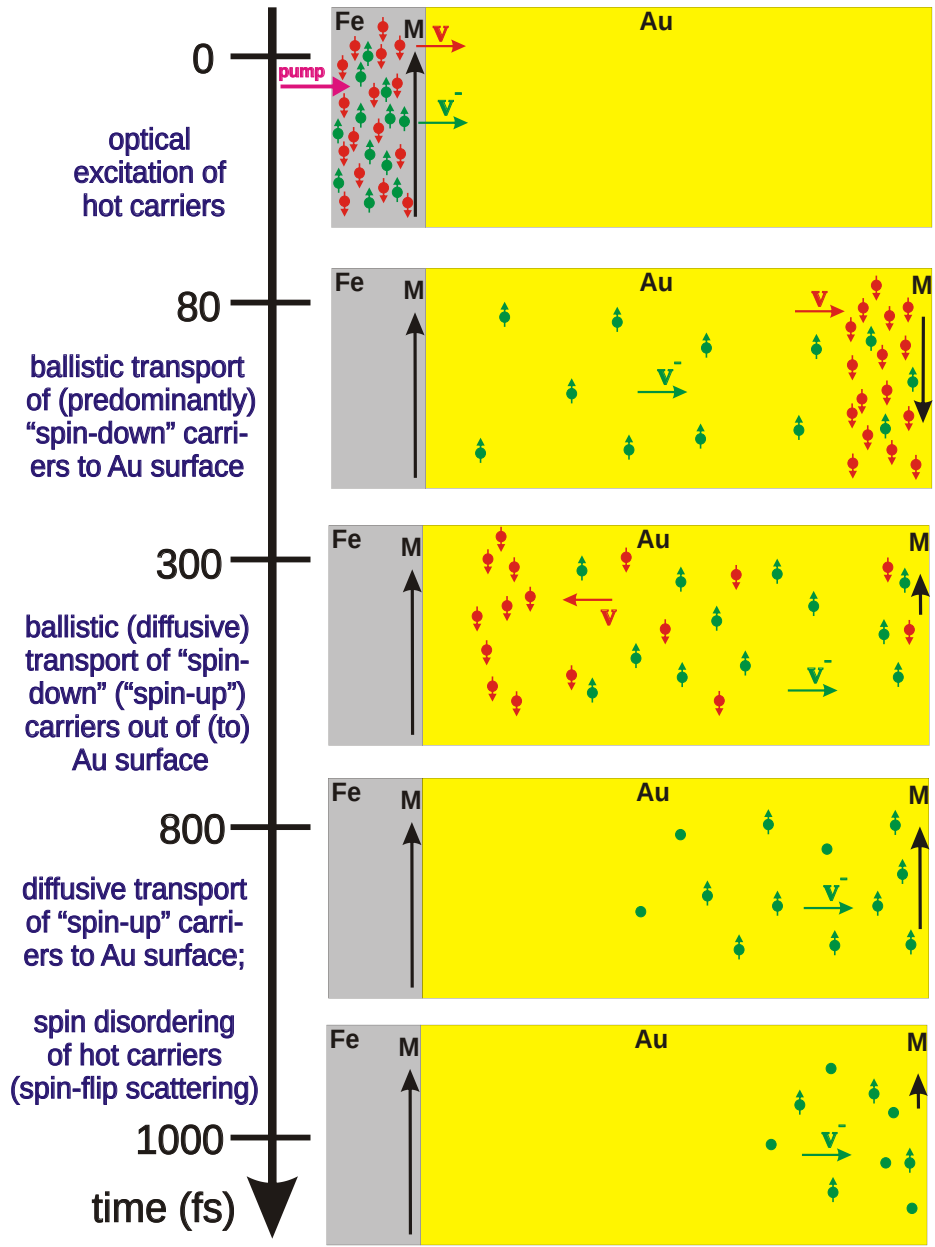}
\end{center}
\caption[Illustration of the electronic transport]{Illustration two types of
  the carrier transports within the gold layer after excitation. While
  the \textbf{ballistic} carriers (minority) reach the gold surface after only
  \textbf{80 fs}, the \textbf{diffusive} carriers (majority) take \textbf{300 fs}
  and more to reach the surface and to flip magnetization. After more
  than \textbf{1000 fs}, all electrons will have distributed within
  evenly and magnetization levels to zero.}
 \label{fig:timelapse}
\end{figure}

The results in the magnetic component show that there are clearly
two types of carrier transports possible. We have identified
these to be \textbf{ballistic} and \textbf{diffusive}. Ballistic
here means that the carriers travel through the gold layer
without scattering as a short packets. Due to this nature of
ballistic transport, they reach the gold surface
after a very short time upon excitation (\textbf{80$~\pm~$20 fs} and
\textbf{160$~\pm~$20 fs} for \textbf{50 nm} and \textbf{100 nm} of gold
respectively). The ballistic carriers can be identified as
the minority carriers (spin-down electrons and spin-up holes)
from the excitation spectrum for iron in Fig. \ref{fig:fespectrum},
calculated by \emph{Tim Wehling, University of Hamburg} \cite{wehling_webpage}.
This ``ballistic'' peak is compensated shortly thereafter by the majority carriers
(spin-up electrons and spin-down holes) which are transported \textbf{diffusively}.
The magnetization does not only compensate to zero but also an \textbf{overshoot}
is detected (thus these are the majority carriers).
The diffusive carriers from longer packets than the ballistic
carriers which is marked by a much broader peak of the
magnetization. Since they are also much slower than the
ballistic carriers (Fig. \ref{fig:fespectrum}),
the magnetization takes much longer until it recovers to zero. Figure
\ref{fig:timelapse} illustrates the two types of
transport of the minority and majority carriers and the
resulting magnetization.
\subsubsection{Linear reflectivity}
The linear reflectivity monitors the bulk dynamics in the gold layer. The
reflectivity for \textbf{50 nm} shows a similar curve like the
non-magnetic SHG component: a first pronounced peak followed by decay
indicates that there is again ballistic transport followed
by diffusive carriers. For \textbf{100 nm}, the peak from
the ballistic transport is no longer visible which arises
from the fact that the too many ballistic carriers have scattered
in the bulk to be detectable in the linear reflectivity signal.
The proper zero delay is also verified from this signal, as the
peaks from the coherence artifact appear at $t = 0~\text{ps}$.
\section{Longtime: electronic vs. phononic}
The first variation in the experimental scheme that we made within the context of this
thesis was that we were repeating the original experiment but on a much longer timescale.
In previous measurements, the timescales for the pump-probe delay usually ranged between
\textbf{-0.5} and \textbf{3.0} picoseconds. In this series of measurements we have extended
the range to up to \textbf{500} picoseconds. The motivation behind this is that we expect
more effects other than spin and electron dynamics on the longer timescale, namely the lattice
dynamics or phononic transport. Since the excited hot carriers have high velocities
high above the Fermi energy (around $v_F \approx 1~\text{nm/fs}$, see Fig. \ref{fig:fespectrum}),
they reach the gold surface after short delays. Lattice deformations propagate through the
crystal with the velocity of sound (for Au: $v_s = 3.24\cdot10^{-3}~\text{nm/fs}$). Since lattice deformations
do not change the equilibrium between majority and minority carriers, we do not expect any variations
in the magnetic channel. Again, \textbf{sample \#5} was used and the other parameters
are identical to the previous shorttime measurements.
\begin{figure}
\begin{sideways}
  \includegraphics[height=1.\textwidth]{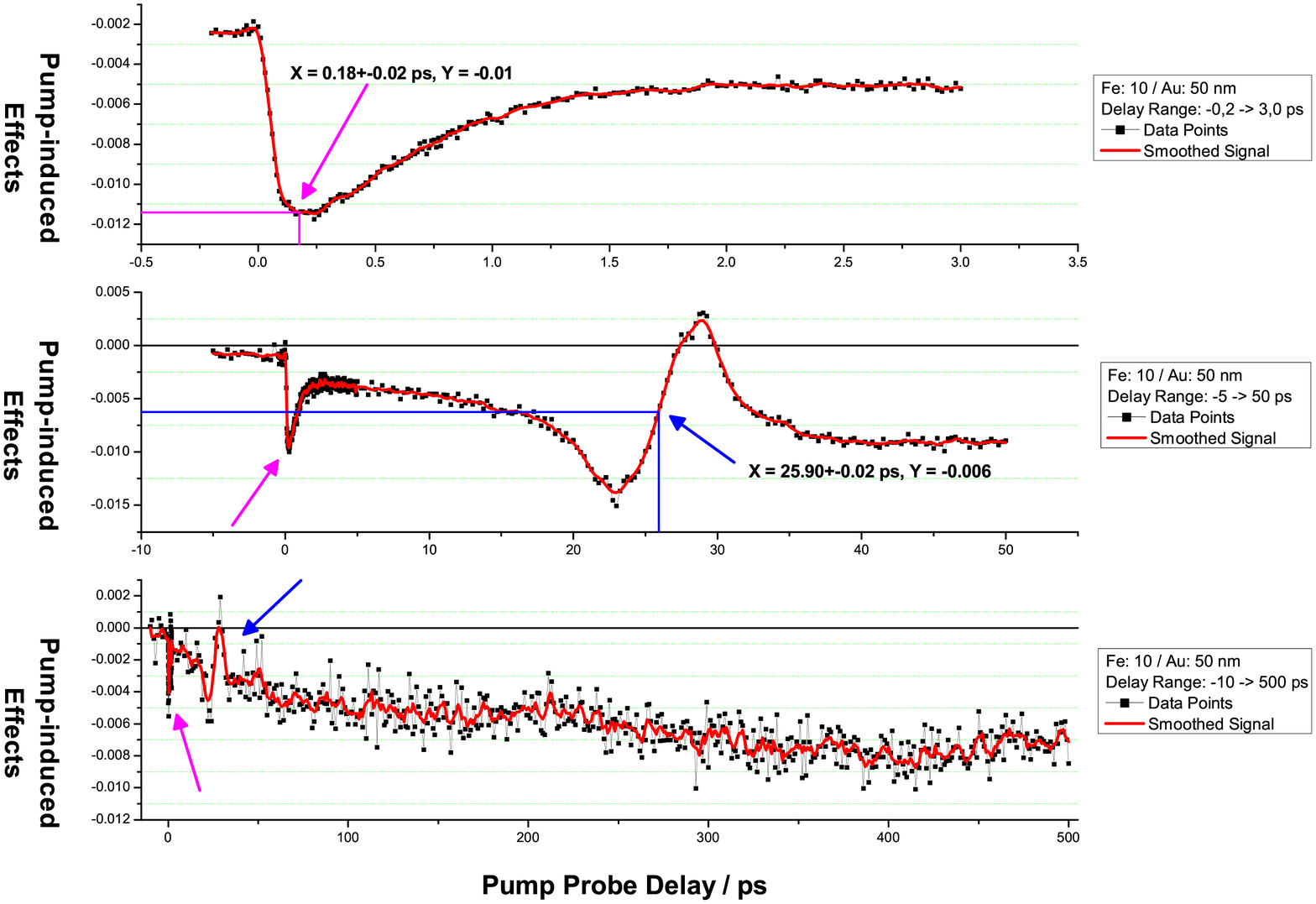}
\end{sideways}
\caption[Pump-induced effects on different time scales]{The non-magnetic part of the SHG signal
  shown on different time scales: \textbf{(a): -0.5-3.0 ps}, \textbf{(b): -0.5-50.0 ps}
  and \textbf{(c): -0.5-500.0 ps}. The curves in (a) are almost identical with the
  results from the shorttime measurements. However, the effect is less pronounced here
  (only around \textbf{1.2} percent compared to the \textbf{2.0} percent). The peak
  still arrives at around \textbf{180 fs}. Looking at the two longtime runs, a peculiarity
  can be distinguished at a delay of $\approx $\textbf{26 ps}. This results from
  lattice dynamics (see text).}
 \label{fig:longtimeevencomp}
\end{figure}

\begin{figure}
\begin{sideways}
  \includegraphics[height=1.\textwidth]{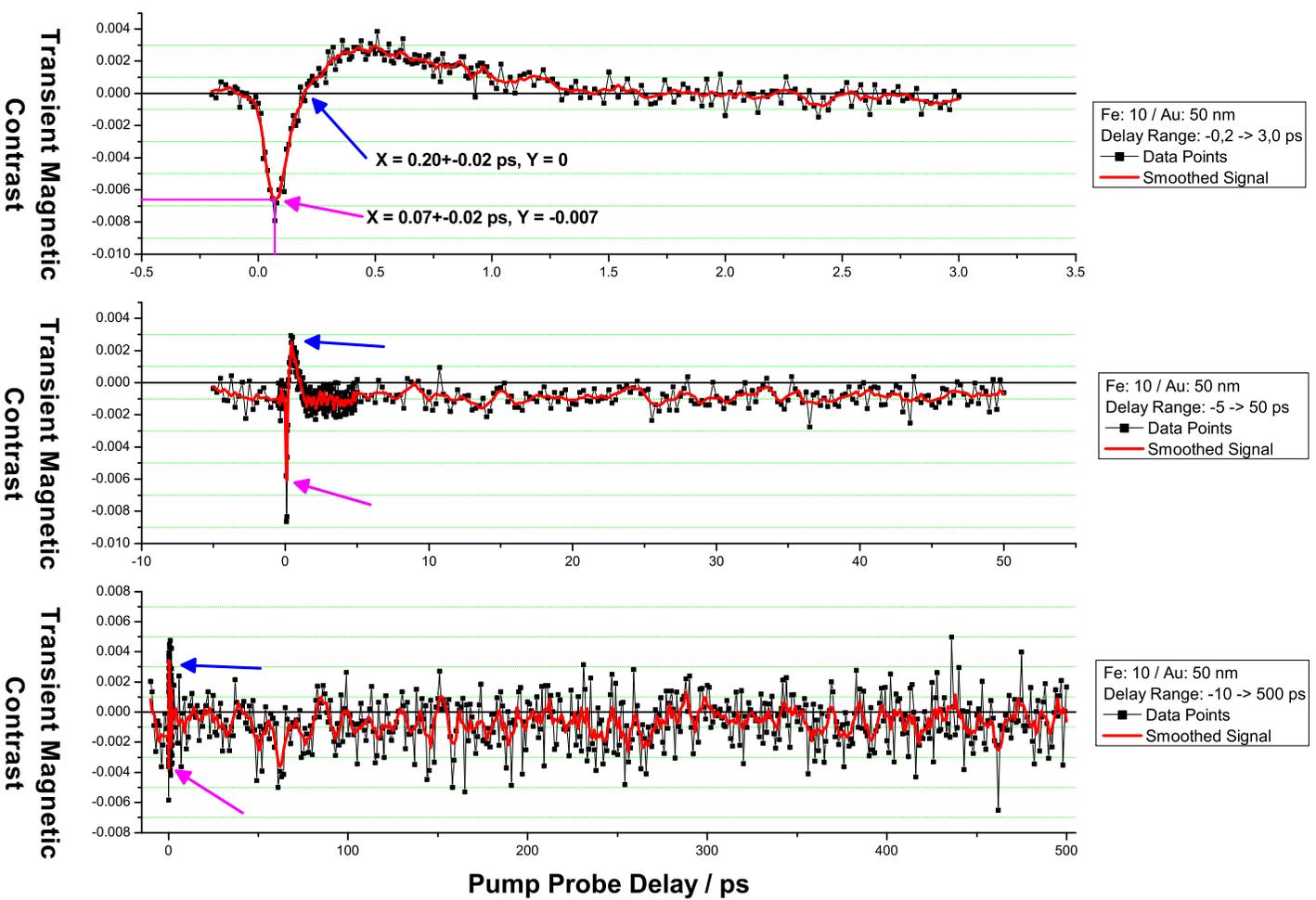}
\end{sideways}
\caption[Dynamic contrast on different time scales]{The magnetic component of the SHG
  signal does not show any further effects even on longer time scales. While we can
  see two additional peaks from the shock wave resulting from the lattice vibrations
  after excitation in the non-magnetic part, there are no further changes in the surface
  magnetization after the initial ballistic peak and the accompanying overshoot.}
 \label{fig:longtimeoddcomp}
\end{figure}

\begin{figure}
\begin{sideways}
  \includegraphics[height=1.\textwidth]{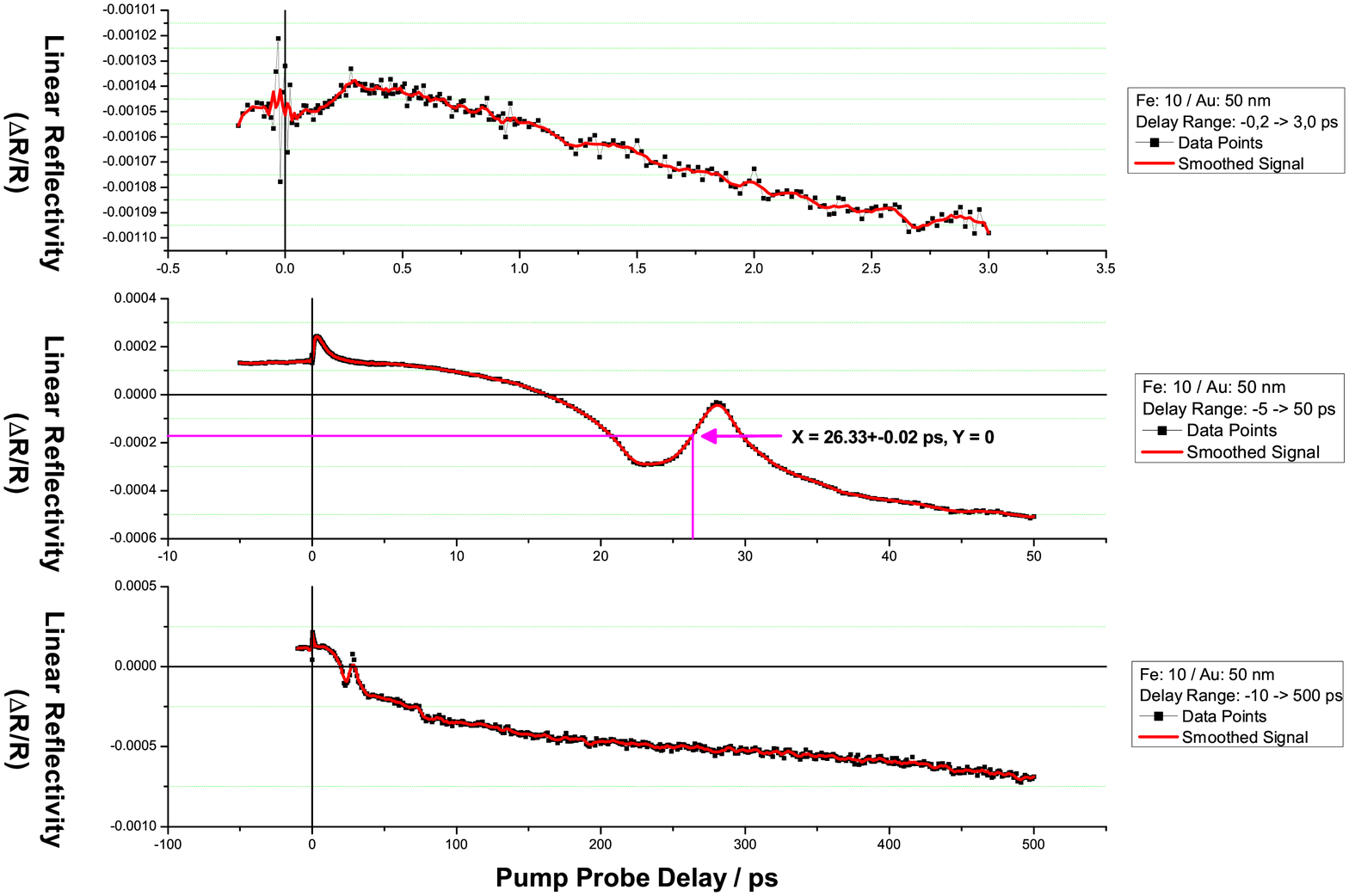}
\end{sideways}
\caption[Linear reflectivity on different time scales]{The linear reflectivity shows
  the same additional two peaks at $\approx $\textbf{26 ps} like they already appeared
  in the non-magnetic component of the SHG signal at the same time. Scanning on the
  longer time range of \textbf{500 ps} does not reveal any further peaks. However the signal should
  recover on even longer time scales. The coherence artifact at $t = 0~$ps indicates
  a proper set zero delay.}
 \label{fig:longtimelinearref}
\end{figure}
\subsection{Discussion of the results}
From the result plots in \ref{fig:longtimeevencomp}-\ref{fig:longtimelinearref}
we see what happens on longer time scales. A peculiarity arising from an acoustic
pulse can be distinguished at a delay of \textbf{26 picoseonds}. It is triggered by
a lattice deformation (expansion) in the iron after the energy absorbed by electrons
from the optical pulse is transferred to the lattice heat via electron-phonon
coupling (the typical time scale is on the order of 1 ps). From the sound velocity
in Au and the time delay for the peculiarity, we can calculate the film thickness:
\begin{eqnarray}
  s & = & v \cdot t \\
    & = & 3.24 \cdot 10^{-3}~\text{nm}\cdot\text{fs}^{-1} \cdot 26 \cdot 10^3~\text{fs} \nonumber \\
    & \approx & 84.24~\text{nm} \nonumber
\end{eqnarray}
This means that we have a deviation from the thickness determined by the
quartz micro balance (QMB) of $\approx$\textbf{70}\%. The balance
needs to be re-calibrated therefore.
The lattice vibrations or \emph{phonons} do not induce changes in the
spin equilibrium and therefore is no visible peak in the magnetic SHG component.
The measurement of the linear reflectivity is sensitive enough to detect
the acoustic pulse, too, as opposed to the initial ballistic peak which can only be
seen in the non-magnetic SHG measurement. The acoustic pulse
is detected in the linear reflectivity at the same delay as it is in the
non-magnetic SHG.
\section{Wedge: thickness dependence}
The idea of the wedge was to be able to conduct a systematic analysis of the thickness
dependence of the electron and spin dynamics. With this analysis, we would be able to
determine the velocity of the ballistic and diffusive electron and spin transport
and compare these velocities with the calculated values from theory. The production
of the wedges is discussed in Section \ref{sec:wedge-production}, Page \pageref{sec:wedge-production}
which also elaborates the motivation behind the wedge concept.
\textbf{Sample \#8} was used (see Tab. \ref{tab:sampleparams}, Page \pageref{tab:sampleparams})
and the external magnetic field was driven to saturate magnetization in the ferromagnetic layer
($\approx$ \textbf{30 Gauss}). The time range for the pump-probe was set from \textbf{-0.2}
to \textbf{3.0} picoseconds. The sample was moved with the micrometer screws of the
sample holder in order to measure at gold thicknesses of \textbf{50}, \textbf{60}
and \textbf{70} nanometers. To calibrate the displacement of the sample with
the current thickness of the wedge, we used the results from the evaporation
process of the wedges: we know from there the thickness for a certain displacement
of the shutter.
\begin{figure}
\begin{sideways}
  \includegraphics[height=1.\textwidth]{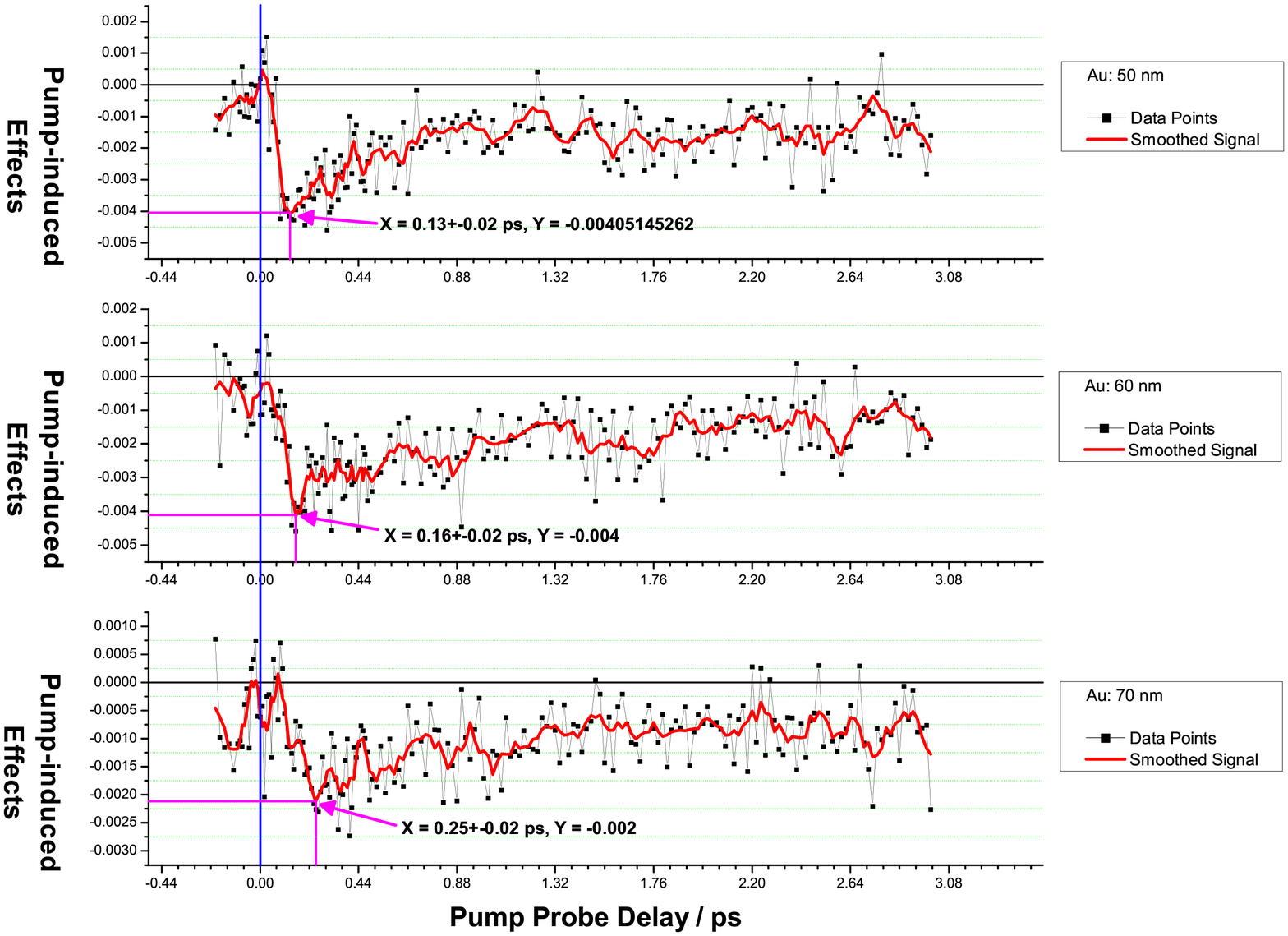}
\end{sideways}
\caption[Pump-induced effects for different thicknesses]{Pump-induced effects for different
thicknesses along the wedge. The peak with the local maximum is shifted from \textbf{130} (50 nm) over
\textbf{160} (60 nm) to \textbf{250} femtoseconds (70 nm) as one would expect for
increasing thicknesses. The results for 70 are very noisy, however, as alignment
was relatively poor during this measurement (zero delay, time/spatial overlap).}
 \label{fig:wedgeevencomp}
\end{figure}

\begin{figure}
\begin{sideways}
  \includegraphics[height=1.\textwidth]{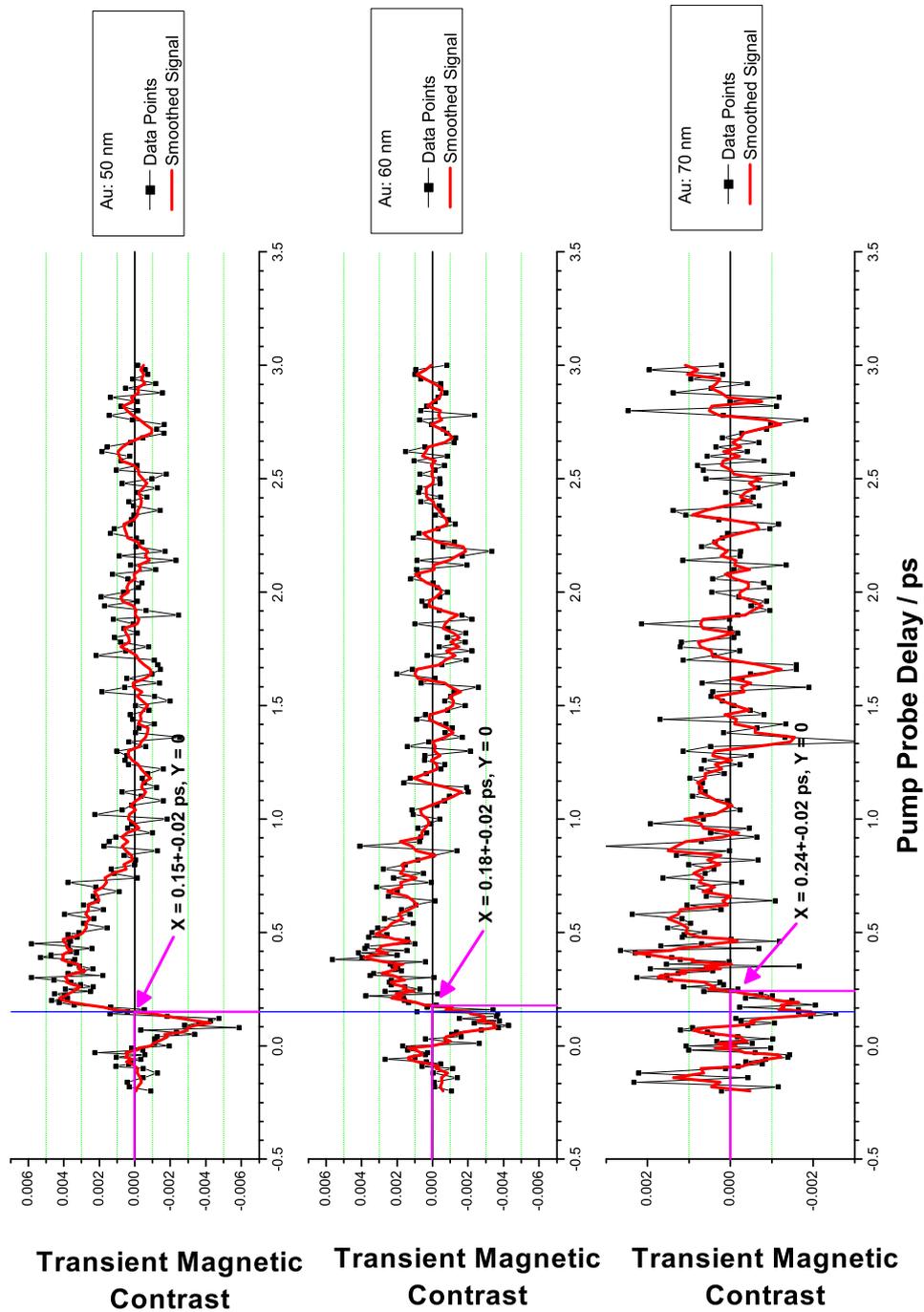}
\end{sideways}
\caption[Dynamic contrast for different thicknesses]{Dynamic contrast for different
  thicknesses along the wedge. We have used the zero crossing of the magnetization
  due to the overshoot from the diffusive electrons as a mean to determine the
  propagation time. The zero crossing is shifted from \textbf{150} (50 nm) over
  \textbf{180} (60 nm) to \textbf{240} femtoseconds (70 nm). Again, the signal
  for 70 nm is very noisy due to the poor alignment.}
 \label{fig:wedgeoddcomp}
\end{figure}

\begin{figure}
\begin{sideways}
  \includegraphics[height=1.\textwidth]{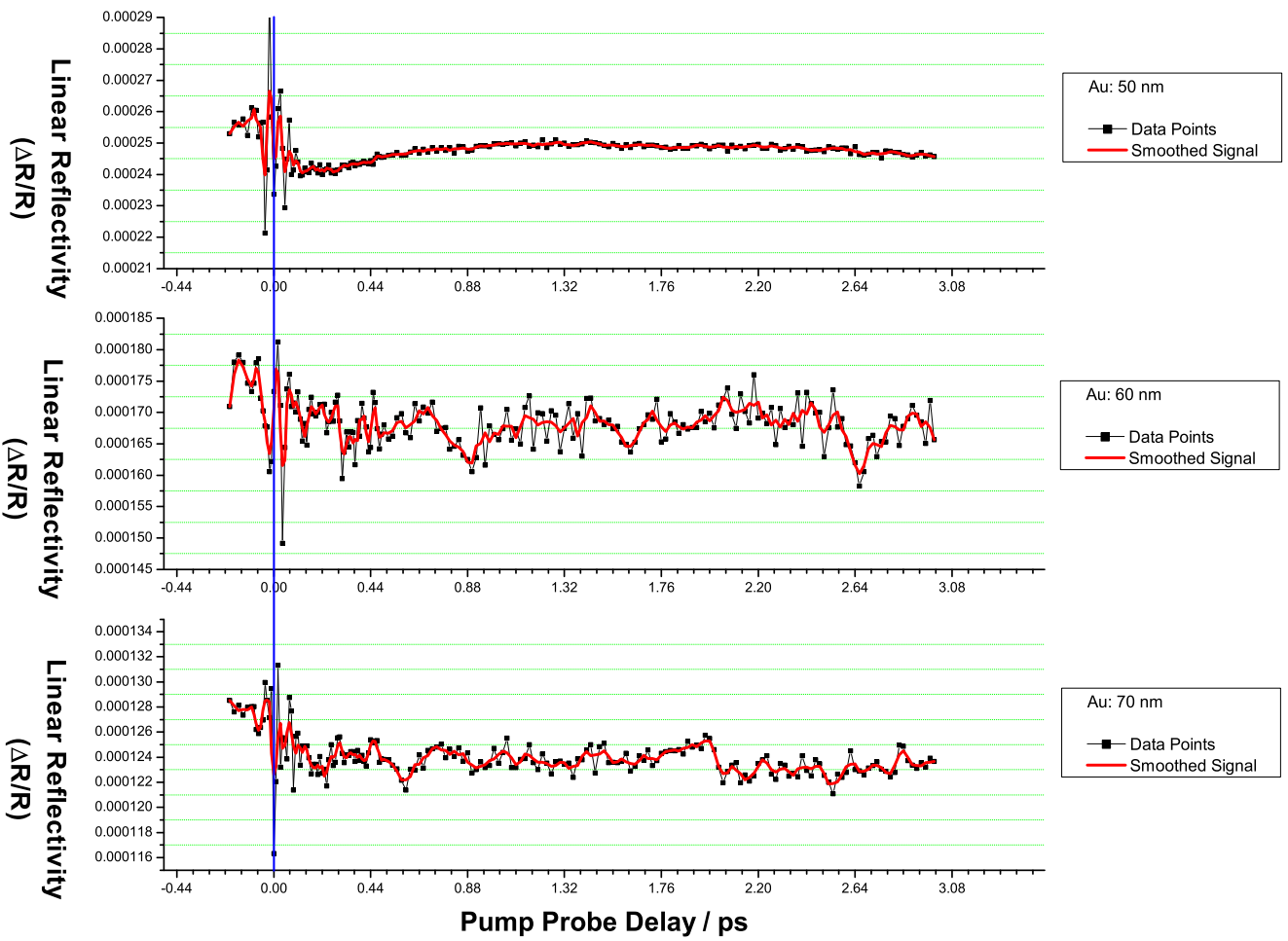}
\end{sideways}
\caption[Linear reflectivity for different thicknesses]{Linear reflectivity for different
  thicknesses along the wedge. The zero delay is definitely properly set for \textbf{50 nm}
  as the distinguishable coherence artifact lies at $t = 0~$fs while the rest of the
  signal is rather flat. For \textbf{60} and \textbf{70 nm}, however, we're picking
  up too much noise thus we cannot be sure that we are still at the proper zero delay
  since the coherence artifact might blur within the noise.}
 \label{fig:wedgelinearref}
\end{figure}
\subsection{Discussion of the results}
The wedge measurements did not yield the expected results. It was much more
difficult to maintain the alignment (time- and spatial overlap and zero delay)
while moving to higher thicknesses than we thought. The stability of the new sample
holder is still not sufficient so that an automated sweeping of the thickness
by translation of the sample along the wedge could be performed\footnote{Albeit
the fact that the setup still doesn't have a motorized stage but has to be
translated manually with the micrometer screws.}. For the first attempt
we have swept the thickness over a range of \textbf{20 nm} only, which
corresponded to a lateral translation of the sample of approximately
\textbf{1.62 mm} which was already enough impair the alignment such
that the results obtained should be considered with care. However,
the trend of the ballistic peak in non-magnetic SHG and the zero crossing
of the mSHG component shifting to later times is definitely there.
At least, for the thicknesses \textbf{50 nm} and \textbf{60 nm}
the shift of the characteristic data points that we have used
to determine the propagation time matches both in the magnetic
(Fig. \ref{fig:wedgeevencomp})and non-magnetic (Fig. \ref{fig:wedgeoddcomp})
signal: \textbf{30 fs}. The difference between \textbf{60 nm} and \textbf{70 nm}
is approximately \textbf{60 fs} for the non-magnetic and \textbf{90 fs}
for the magnetic component, so this data point should be rather discarded. Unfortunately,
we could not achieve a stable signal for even higher thicknesses so that we were
unable to measure any sensible signal. Thus, in order to perform
more reliable wedge measurements, the stability of the alignment
has to be improved further.
\section{Linear MOKE: alternative access to spin dynamics}
\begin{figure}
  \subfigure[MOKE vs mSHG, Au50]{\includegraphics[width=0.5\textwidth]{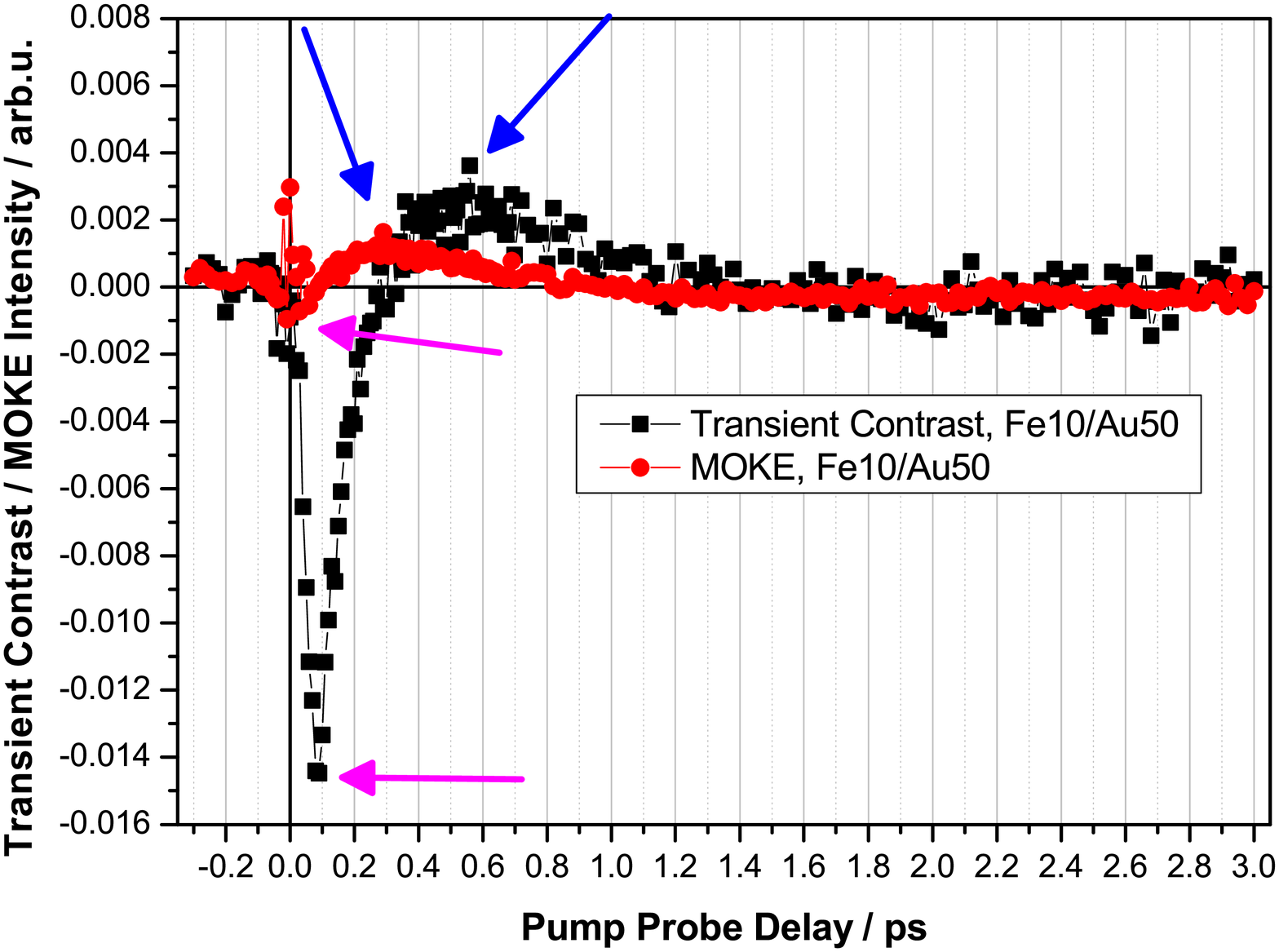}}
  \hfill
  \subfigure[MOKE vs mSHG, Au100]{\includegraphics[width=0.5\textwidth]{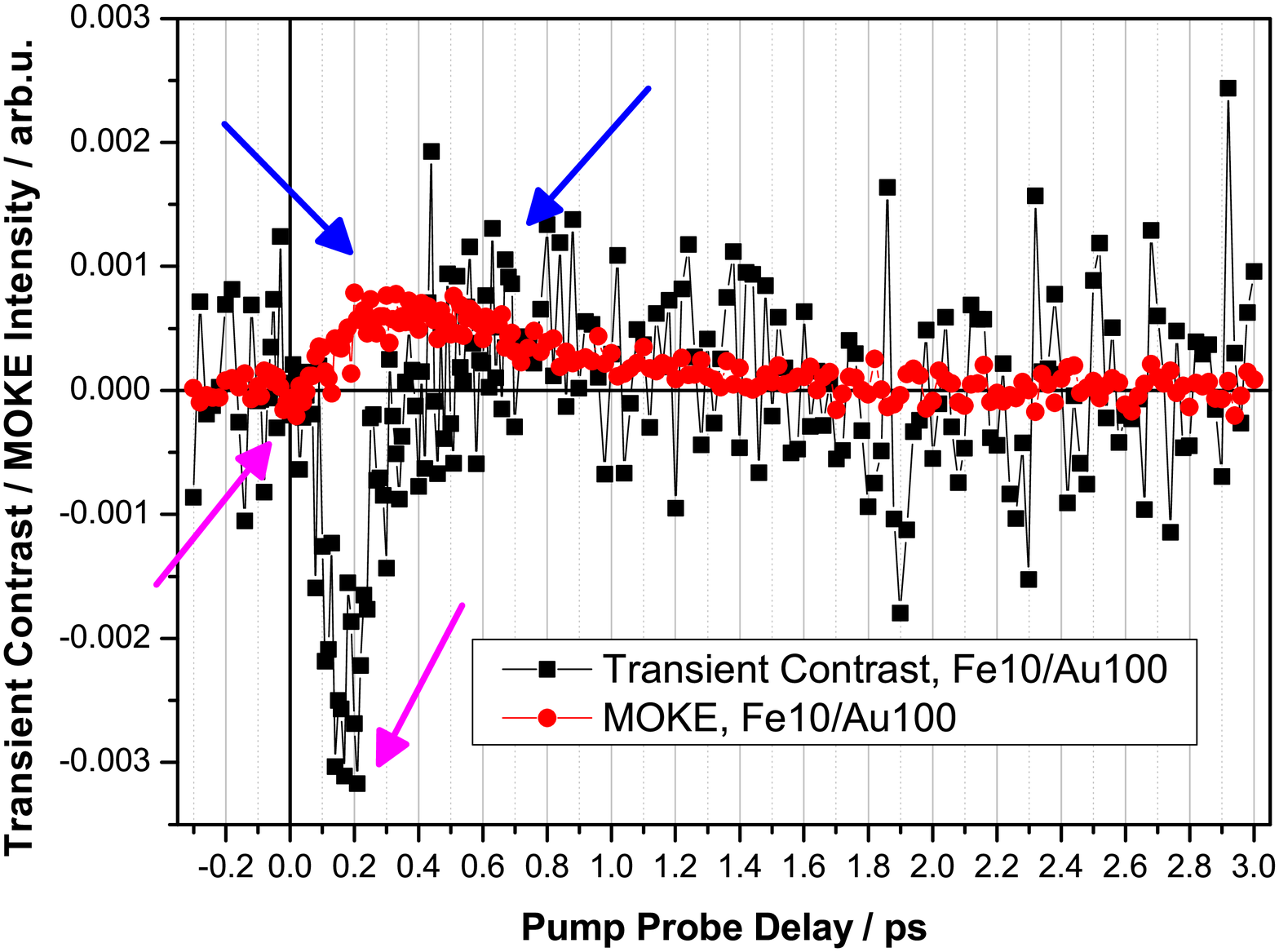}}
  \subfigure[MOKE, Au50 and Au100]{\includegraphics[width=0.5\textwidth]{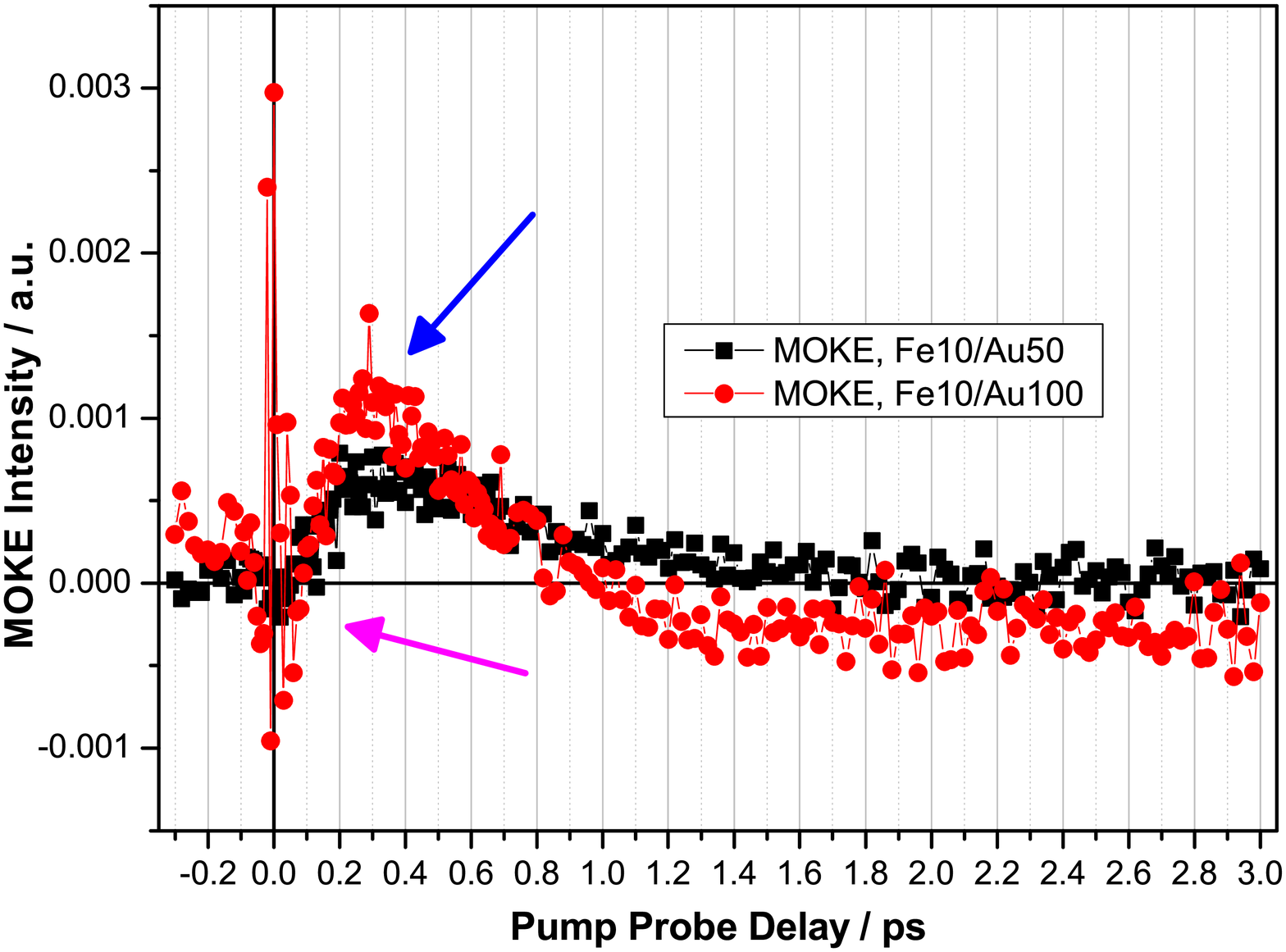}}
  \hfill
  \subfigure[Pump-induced effects, Au50]{\includegraphics[width=0.5\textwidth]{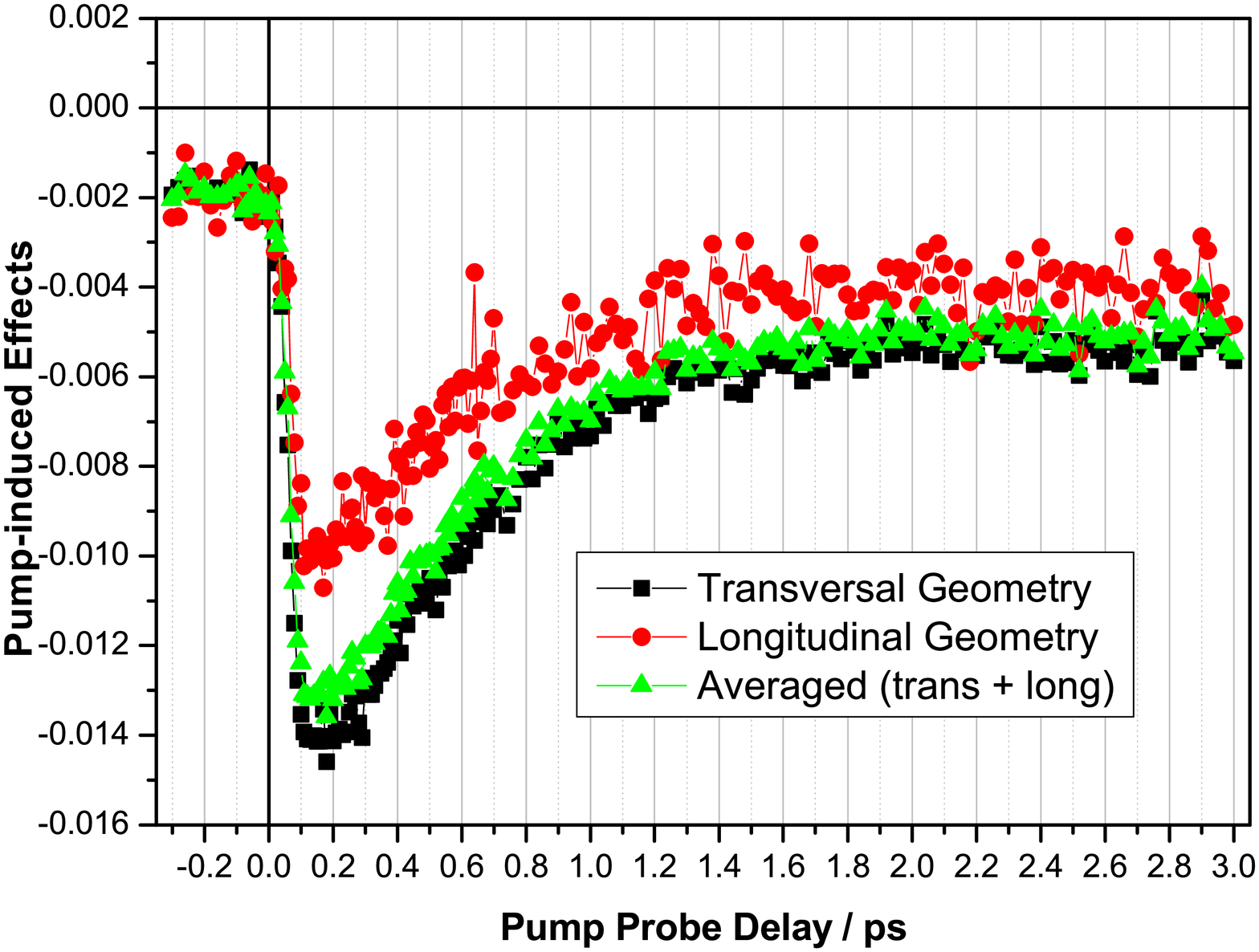}}
  \subfigure[Pump-induced effects, Au100]{\includegraphics[width=0.5\textwidth]{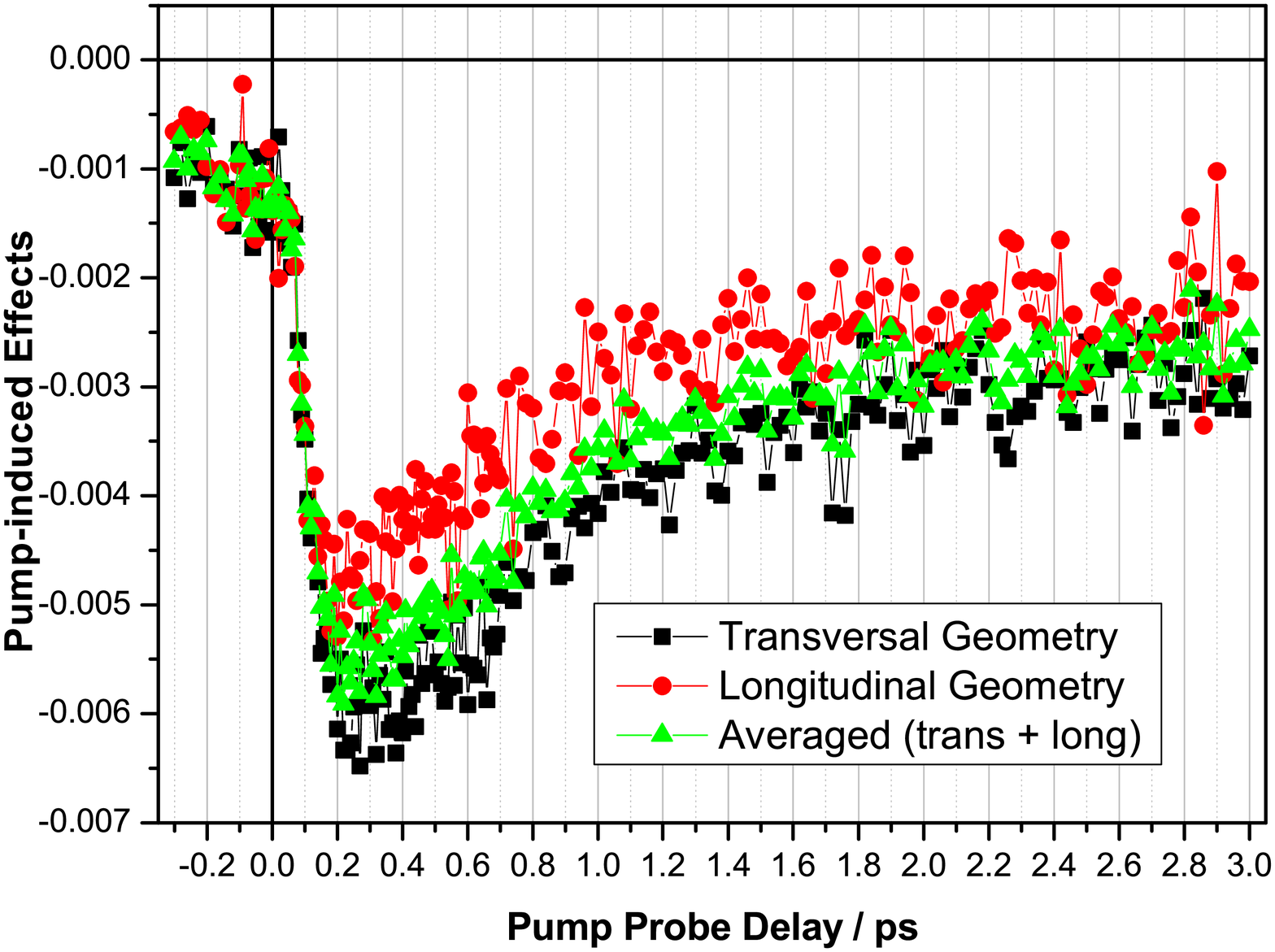}}
  \hfill
  \subfigure[Linear Reflectivity]{\includegraphics[width=0.5\textwidth]{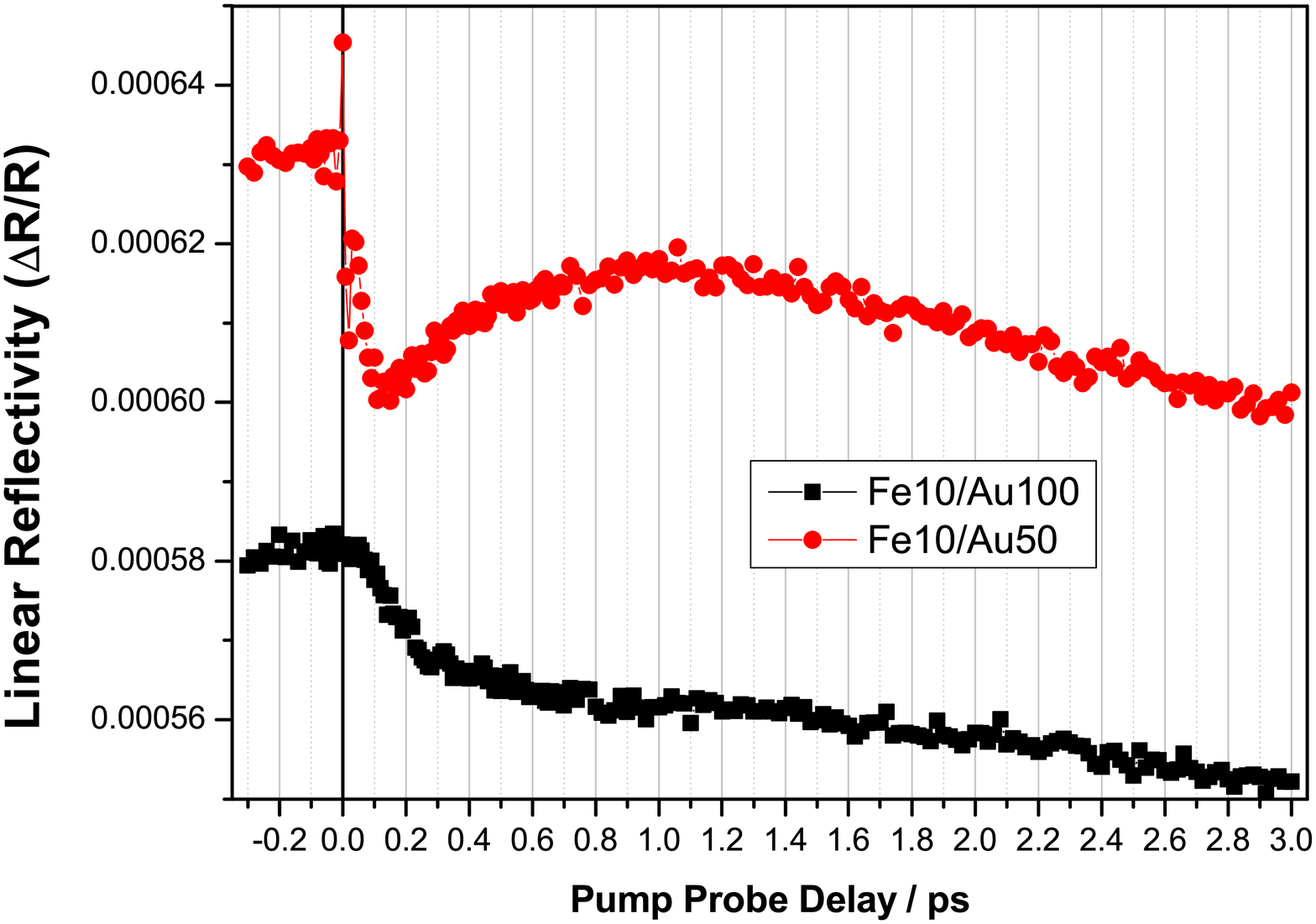}}
 \caption[Comparison of linear MOKE and SHG measurements]{Comparison of the results for MOKE and SHG measurements.
   The MOKE measurements were performed in \textbf{longitudinal geometry}. Comparing the MOKE measurements
   with the magnetic SHG measurements in \textbf{(a)} and \textbf{(b)} one can clearly see that the
   first peak (purple arrow) from the ballistic carriers is not visible in the MOKE signal while we can still observe
   the overshoot from the diffusive transport (blue arrow). The ballistic peak is missing both for \textbf{50} and
   \textbf{100 nm} scans \textbf{(c)}. Since the non-magnetic signal is independent from the external magnetic
   field, we can use these results to in \textbf{longitudinal} and \textbf{transversal} geometry of
   the external field are comparable \textbf{(d)}. The same applies for the linear reflectivity \textbf{(e)}.}
 \label{fig:mokecomp}
\end{figure}

Besides the measurements with the advanced SHG techniques, we have also performed conventional
linear MOKE measurements to be able to compare them with our method of choice, mSHG. The MOKE measurements
provide information on bulk spin dynamics analog to the linear reflectivity which yields
bulk electron dynamics. While we measured mSHG with the field applied in transversal
geometry (inner coils), for MOKE we chose the longitudinal configuration (outer coils).
This has to be kept in mind when comparing our mSHG and MOKE results.
However, since the non-magnetic part of the SHG signal can be measured in
both geometries, we had a mean to verify that the observed effect is comparable
in both configurations: If the geometry had any considerate influence on the SHG signal, it
should reflect in the non-magnetic component also. The plots for non-magnetic components can
be found in Figures \ref{fig:mokecomp} (d) and (e).
\subsection{Discussion of the results}
The bulk spin dynamics analyzed with MOKE in (\ref{fig:mokecomp} (a) and (b)) show different results when
comparing to the mSHG signal. The first striking difference is that there is no peak (purple arrows)
from the ballistic carriers, both for \textbf{50 nm} and \textbf{100 nm} gold layers. The overshoot from
the diffusive transport (blue arrow) is, however, detected. We haven't found yet any reasonable
explanation for this behavior but one can speculate: One possible scenario is that the different
penetration depths of the sensitivity (see Sections \ref{sec:moke}, Page \pageref{sec:moke} and
\ref{sec:linearvsnonlinear}, Page \pageref{sec:linearvsnonlinear}) might be related to that issue.
Since the ballistic carriers form short packets, the magnetization is concentrated within a
thin space layer whereas the majority carriers from diffusive transport occupy larger spaces
and thus their magnetization. And since MOKE is bulk-sensitive (as opposed to the surface
sensitivity of mSHG) it has larger response functions for magnetizations within large spatial
than for magnetizations within a surface or interface. The plot in  \ref{fig:mokecomp} (c)
compares the MOKE signal for different thicknesses of the gold layer, 50 and 100 nm.
The overshoots (blue arrow) for 50 and 100 nm and the ones from the measurement of the
magnetic component in \ref{fig:shorttimecomp} (b) are on comparable time scales,
however the overshoot for 50 nm has a higher intensity than the one for 100 nm in the
mSHG measurement while this is the opposite case for MOKE. Since the difference
between 50 and 100 nm is rather small, this might rather be an artifact than a
real effect. The ballistic peak would yield more information if it was detectable
with MOKE. The plot for linear reflectivity in \ref{fig:mokecomp} shows the same
dynamics as found for the shorttime measurements.

\chapter{Conclusion \& Discussion}
In the this work we have introduced a new concept to switch
magnetizations in ferromagnetic materials. Instead of using external magnetic
fields produced by coils or other ferromagnetic materials, we inject spin-polarized
carriers directly. These hot carriers are excited from a magnetized
ferromagnetic layer and thus form a spin-polarized current. The spin-polarized
current passes a non-ferromagnetic spacer layer over several
transport mechanisms which we have investigated in this work
and change, what has not been achieved yet, the magnetization
in a second ferromagnetic layer.

\medskip

The model systems that we have perceived for this setup are
made from single-crystalline, dual-side polished magnesium oxide substrates
(MgO (001)) on which metallic multi-layers are grown epitaxially
by means of evaporation in ultra-high vacuum environment. A
layer of 10-20 nanometers of iron serves both as the source for
spin-polarized electrons as well providing a buffer layer to improve
epitaxial growth of the overlaying gold layer. The gold layer, usually
50-100 nanometers in thickness, provides a non-ferromagnetic spacer
medium to separate a second iron layer. This is the target layer, whose
magnetization is altered by the spin-polarized carriers generated in
the source layer. Before the actual switching by spin injection can
be achieved, however, we have to perform a lot of preliminary research
to understand ultrafast spin dynamics, i.e. the mechanisms
of spin-polarzied carrier transport in the metallic multilayers.
For this, we have reduced our model system to a \emph{MgO(001)/Fe/Au}
bilayer setup thus neglecting the target ferromagnetic layer for now.

\medskip

To generate spin-polarized electrons in the source ferromagnetic layer,
an external magnetic field is first applied to drive the magnetization in
the specimen into saturation, then a femtosecond laser excites
``hot electrons'' by pumping the sample from the optically transparant
side of the MgO substrate. Due to the exchange-split bands in the source layer,
there are more electrons with one spin orientation than with the opposite orientation and the flow
of electrons becomes spin-polarized. After excitation the electrons
enter the gold spacer layer. Since the \emph{majority} and \emph{minority}
electrons have different energies, their velocities and transport
mechanism in Au are \emph{different}. Within the framework
of this project we were able to demonstrate that.

\medskip

The first focus of this thesis was put on the techniques of
producing the samples by means of molecular beam
epitaxy and investigating what factors have a direct
influence on the results of the optical experiments. In order
to achieve best possible sample qualities, we have varied
different evaporation parameters like substrate temperatures,
evaporation rates and annealing procedures. To conclude these,
we have found out that the films should be evaporated at
lowest possible substrate temperatures (usually room temperature),
lowest reasonable evaporation rates and the substrates themselves
need to be thoroughly cleansed and annealed to remove
as much contamination as possible and maintain high substrate
purity. Good quality films are desirable to reduce
the scattering of the laser beam at the sample surfaces
and electron scattering in both layers and at Fe/Au interfaces.

\medskip

The second focus of this thesis was to continue the investigations
performed in the core experiment, that is the SHG measurements.
Within the framework of this thesis we have extended the
time range of the pump-probe delay to observe electronic
and spin transport on a longer time scale after excitation.
In fact, we were able to monitor lattice dynamics of the gold
as acoustic waves resulting of lattice deformations triggered
by heat due to absorption of the laser pulse. There were,
however, no effects visible in the magnetic component of
the SHG probe signal.

\medskip

We have also performed conventional linear MOKE measurements
to monitor the bulk spin dynamics. The results were distinct
from the mSHG results. Like the measurement of the bulk
electron dynamics with linear reflectivity, we were able
to detect the majority carriers, but the sharp peak resulting
from the minority carriers (ballistic) was not visible at all.
We presume that the packets formed by the ballistic carriers
are too short to induce enough changes in the bulk magnetization
to be detected by MOKE. This would also explain, why the
overshoot from majority spins is still detectable.

\medskip

In an additional experiment, we wanted to determine the
velocities of both the ballistic and diffusive transport
mechanisms directly. For this, we created a gold wedge
in the evaporation process and installed the sample
into a displayable sample holder. Unfortunately, we
weren't yet able to design a sample holder that was
stable enough to maintain proper optical alignment
and beam overlap while moving the sample between the
measurements. Up to now, the experiments didn't yield
any truly usable results to determine the velocities.
We can say though, that the results show that there is
a trend that clearly suggests a shift of the peaks
towards longer times for higher thicknesses. We think,
that a more reliable sample holder with a motorized stage
will not only yield better alignment but also allow
a completely automated measurement for the wedge to
finally determine the electron velocities and
compare them with theory.

\medskip

Finally, we have realized that there is still a long
way to go and more caveats have to be ruled out until
we can perform the experiments which we have originally
perceived, namely the switching of magnetization by
injection of spin-polarized ``hot electrons''. However,
we have shown that such experiments are generally feasible
and we have already successfully tackled the first
obstacles to finally reach our goal.


\appendix
\cleardoublepage 
\phantomsection  
\addcontentsline{toc}{chapter}{Appendix} 

\chapter{Engineering Drawings}
\label{chapter:drawings}
The following engineering drawings were created using the open-source
CAD-software \emph{qcad}, version 2.2.0 (see \cite{ribbonsoft_qcad}).
All scales are in millimeters, screws used are in metric dimensions (``M x''). Material
is either high-graded steel or aluminium as indicated in the drawings. Shaded
areas indicate the metal parts; drilled holes or threads are indicated with solid fill. Large
blank areas indicate larger apertures. These drawings were actually used to handcraft
the parts or file them to the departments precision engineering group.

\vspace{3cm}


\begin{figure}[h!]
\begin{center}
  \includegraphics[width=1\textwidth]{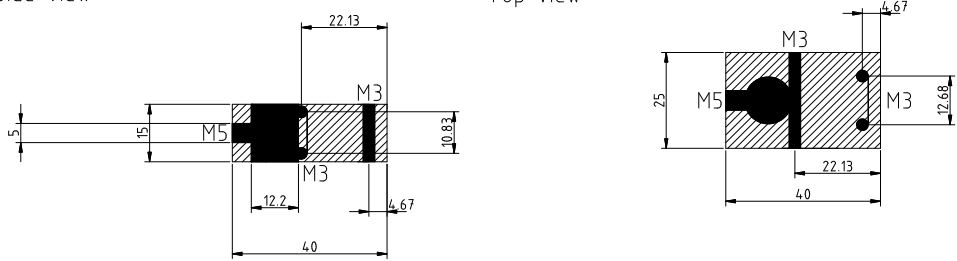}
\end{center}
 \caption[Helmholtz Coils - Mounting Block]{Helmholtz Coils - Mounting Block}
 \label{fig:helmholtz_mount}
\end{figure}

\begin{figure}
\begin{center}
  \includegraphics[width=1\textwidth]{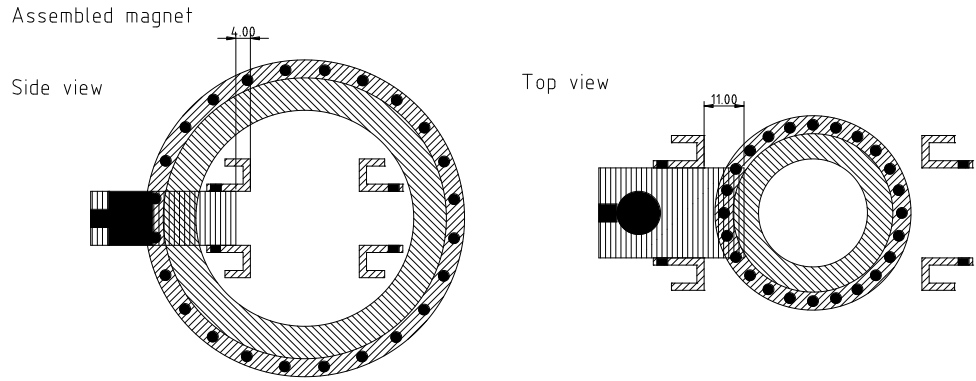}
\end{center}
 \caption[Helmholtz Coils - Assembled View]{Helmholtz Coils - Assembled View}
 \label{fig:helmholtz_all}
\end{figure}

\begin{figure}
\begin{center}
  \includegraphics[width=0.8\textwidth]{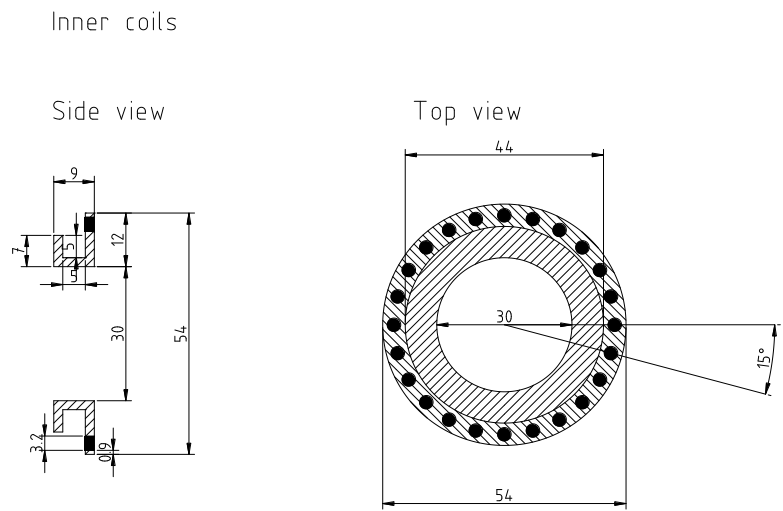}
\end{center}
 \caption[Helmholtz Coils - Inner Coils]{Helmholtz Coils - Inner Coils}
 \label{fig:helmholtz_inner}
\end{figure}

\begin{figure}
\begin{center}
  \includegraphics[width=0.65\textwidth]{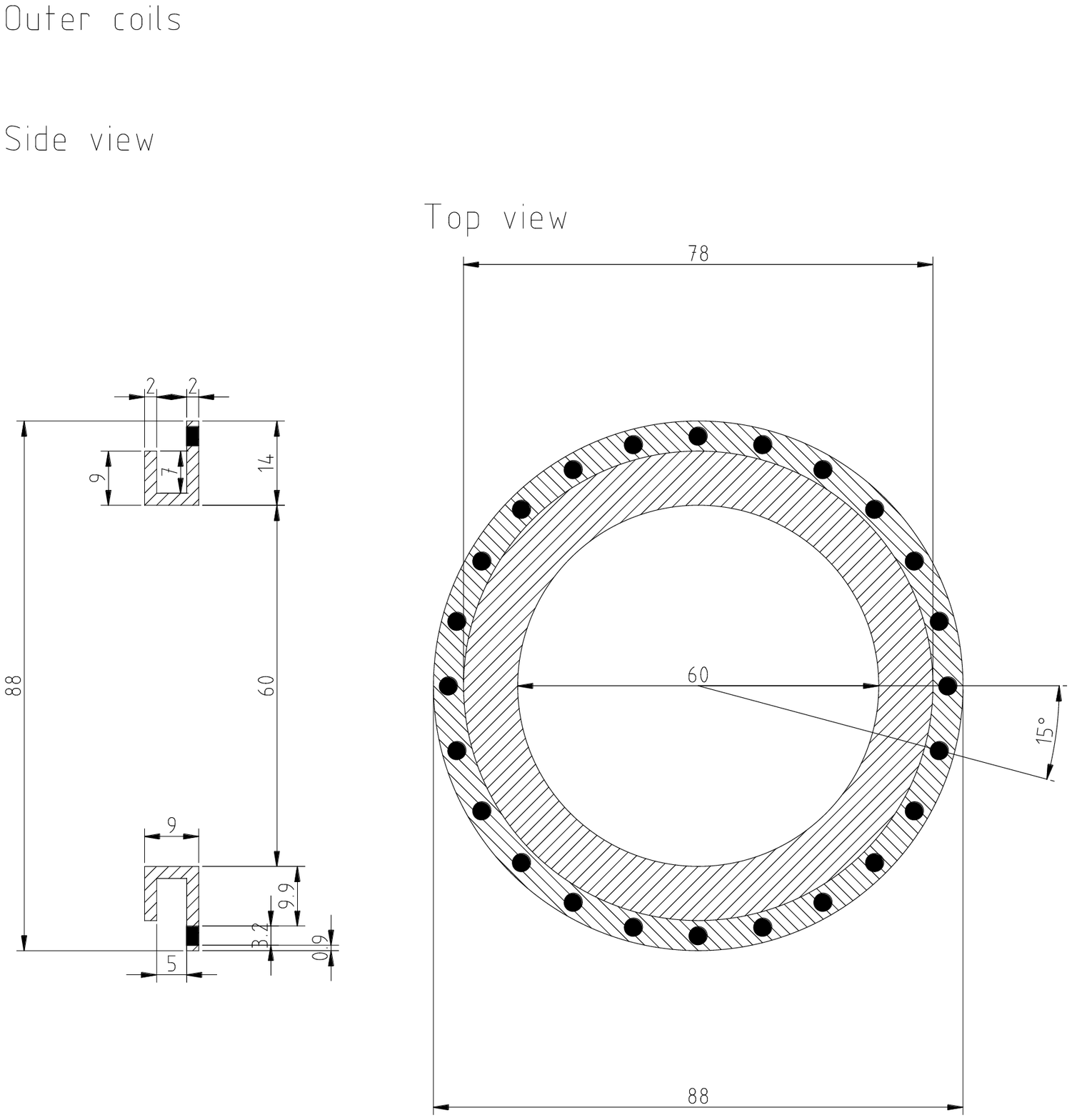}
\end{center}
 \caption[Helmholtz Coils - Outer Coils]{Helmholtz Coils - Outer Coils}
 \label{fig:helmholtz_outer}
\end{figure}

\begin{figure}
\begin{center}
  \includegraphics[width=0.6\textwidth]{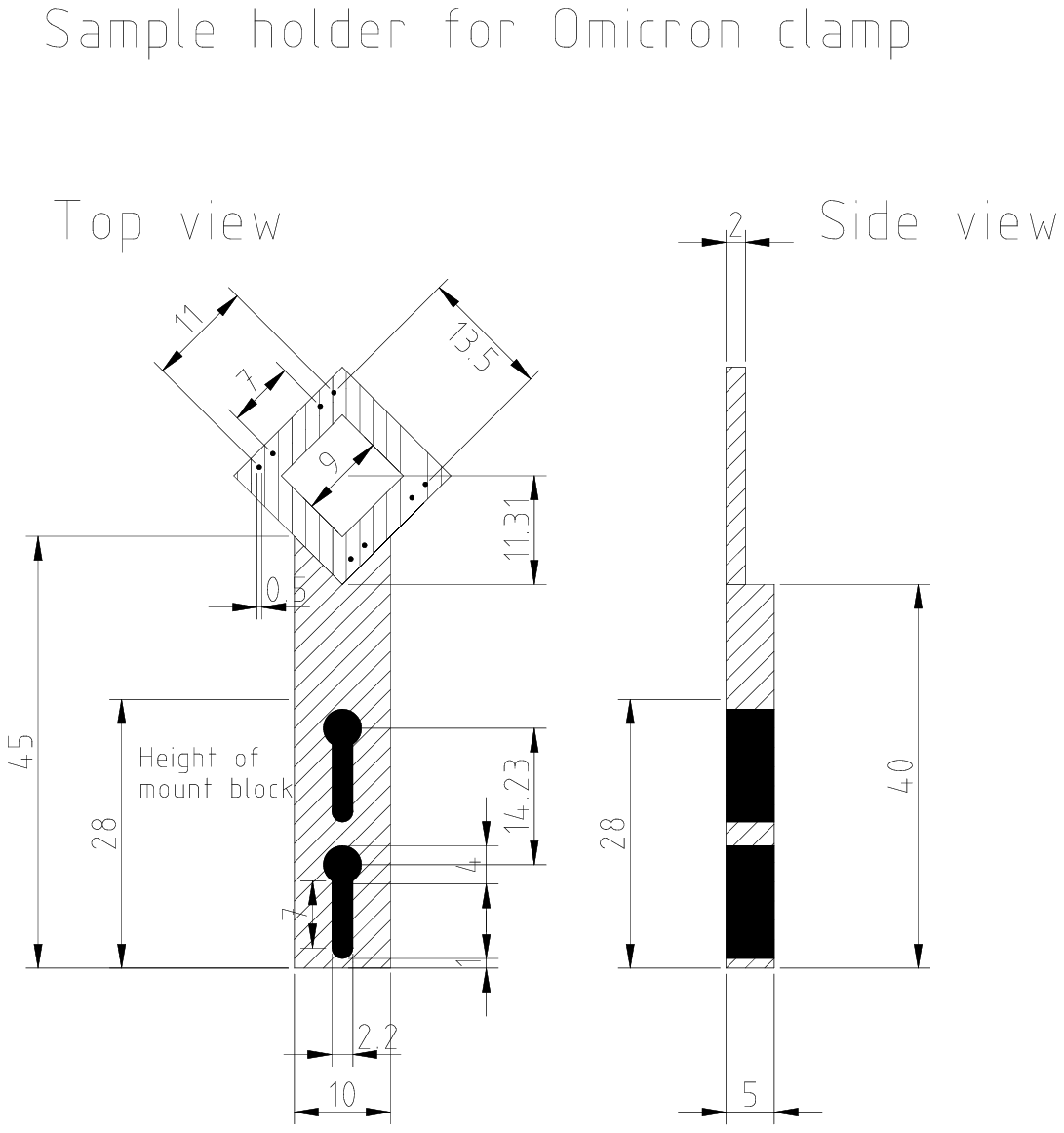}
\end{center}
  \caption[Sample holder for positioning the samples in the magnet]{Sample holder for positioning the samples in the magnet}
  \label{fig:sample_holder}
\end{figure}

\begin{figure}
\begin{center}
  \includegraphics[width=0.65\textwidth]{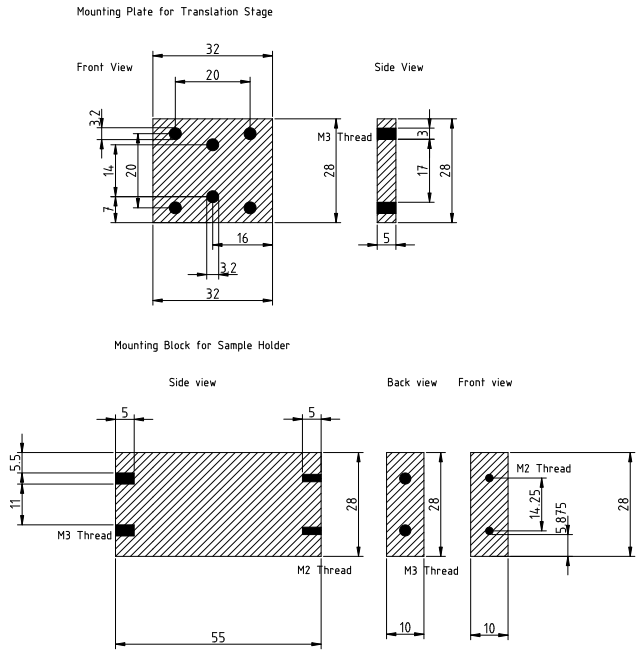}
\end{center}
 \caption[Sample Holder - Mount Assembly]{Sample Holder - Mount Assembly}
 \label{fig:samplemount}
\end{figure}

\begin{figure}
\begin{center}
  \includegraphics[width=0.65\textwidth]{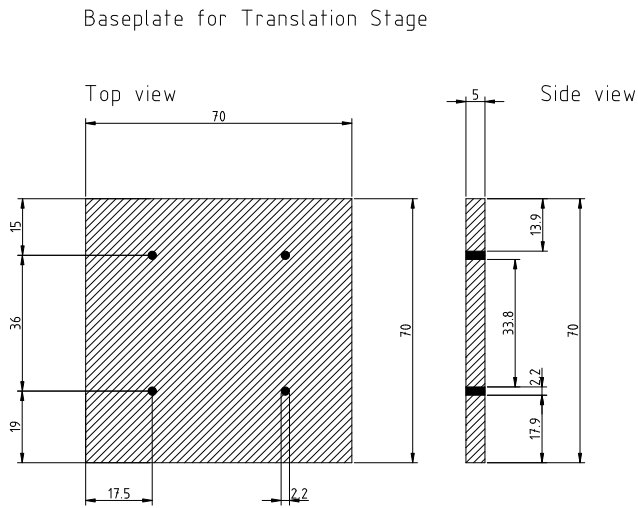}
\end{center}
 \caption[Sample Holder - Base plate for Translation Stage]{Sample Holder - Base plate for Translation Stage}
 \label{fig:baseplate}
\end{figure}


\begin{figure}
\begin{center}
  \includegraphics[width=0.65\textwidth]{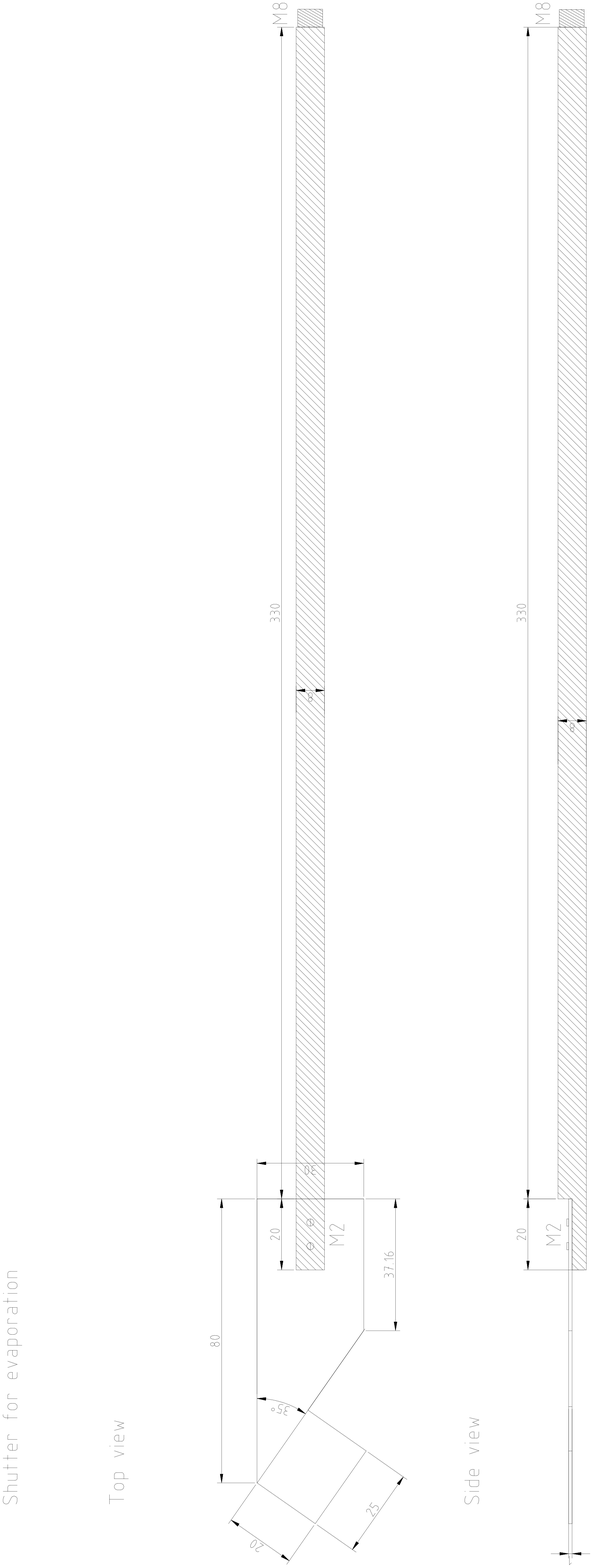}
\end{center}
 \caption[Shutter used for evaporation]{Shutter used for evaporation}
 \label{fig:shutter_complete}
\end{figure}


\chapter{Source code of autocorrelation program}
\label{chapter:sourcecode}
\lstset{
language=C,                     
basicstyle=\footnotesize,       
numbers=left,                   
numberstyle=\footnotesize,      
stepnumber=2,                   
numbersep=5pt,                  
showspaces=false,               
showstringspaces=false,         
showtabs=false,                 
frame=single,	                
tabsize=2,	                
captionpos=b,                   
breaklines=true,                
breakatwhitespace=false,        
}
\begin{lstlisting}
/* autocorr.c - program code for autocorrelation analysis of grain size of STM images */

#include <stdio.h>
#include <string.h>
#include <math.h>

/* Number of lines in data files exported from wsxm */
#define NLINES 160000

/* Number of points per line */
#define NPOINTS 400

/*
 * array data:
 * X - distance of point on X-axis
 * Y - distance of point on Y-axis
 * Z - elavation of point at x,y
 */
double x[NLINES], y[NLINES], z[NLINES];

/* parse - read in data file exported from wsxm
*          and store data into glocal arrays
*/
int parse(char * filename)
{
  int i;
  FILE* file;
  char buf[100];

  if ((file = fopen(filename, "r"))) {

    for (i = 0; i < 4; i++)
      fgets(buf, 100, file);

    for (i = 0; i < NLINES; i++) {
      fscanf (file ,"%lf %lf %lf", &x[i], &y[i], &z[i]);
    }
  }
}

/* autocorr - actual function to perform autocorrelation on image data */
int autocorr() {

  /* variables:
   *
   * ac - autocorrelation for each point
   * n - holds the amount occurrences for each distances
   * i,j - loop indices
   * s - mean value of z-elevation
   */

  double ac[NPOINTS];
  unsigned long long n[NPOINTS];
  int i, j, r;
  double s;

  s = 0;

  /* make sure buffers are zeroed */
  memset(ac, 0, sizeof(ac));
  memset(n, 0, sizeof(n));
  

  /* calculate mean */
  for (i = 0; i < NLINES; i++)
    s = s + z[i];
  
  s = s / NLINES;

  /* substract mean from z-data */
  for (i = 0; i < NLINES; i++)
    z[i] = z[i] - s;

  //   for (i = 0; i < NLINES; i++)
  //   z[i] = sin (2*M_PI*x[i]/50) * sin (2*M_PI*y[i]/50);

  for (i = 0; i < NLINES; i++)
    for (j = i; j < NLINES; j++) {
      /* calculate distance of points at i and j index */
      r = round(sqrt((x[i] - x[j])*(x[i] - x[j]) + (y[i] - y[j])*(y[i] - y[j])));
      /* count only distances which are smaller than a scan line */
      if (r < NPOINTS) {
	/* increase count of this specific distance */
	n[r] = n[r] + 1;
	/* calculate autocorrelation for this distance */
	ac[r] = ac[r] + z[i]*z[j];
      }
    }

  /* normalize autocorrelations */
  for (i = 0; i < NPOINTS; i++)
    ac[i] = ac[i]/n[i];

  /* output calculated AC data to terminal */
  for (i = 0; i < NPOINTS; i++)
    printf ("%d\t%f\n", i, ac[i]);
}

int main (int argc, char** argv) {


  char filename [256];

  /* case-matching for the amount of arguments specified: */
        
  switch (argc) {

    /* if argc == 1 only the programname was specified during invocation,
     * thus prompt for both operands/arguments
     */
  case 1:
    printf ("\nNo arguments specified: \n\nInput filename: ");
    scanf ("%s", &filename); /* scan input directly to integer */
    break;
    
  case 2:
    /* if we got argc == 2, then we have one argument specified and will
     * now try to parse the filename, if this fails, we proceed
     * to default: (no break) */
    
    if (sscanf (argv [1], "%s", &filename) == 1)
      break;
  default:
    
    /* something we cannot interpret was specified on command line, print help
     * message and exit with error
     */
    
    printf ("\nUsage: %s filename\n\n", argv [0], argv [0]);
    return -1; /* Return with EXIT_ERROR (= -1) */
  }

  parse (filename);
  autocorr();

  return 0;
  
}
\end{lstlisting}



\nocite{*}
\bibliographystyle{unsrt} 
\cleardoublepage 
\phantomsection  
\addcontentsline{toc}{chapter}{Bibliography} 
\bibliography{main}

\begin{thebibliography}{10}

\bibitem{hillebrands_spindynamics1}
B.Hillebrands and K.Ounadjela.
\newblock {\em Spin Dynamics in Confined Magnetic Structures I}.
\newblock Springer-Verlag Berlin, 2002.

\bibitem{hillebrands_spindynamics2}
B.Hillebrands and K.Ounadjela.
\newblock {\em Spin Dynamics in Confined Magnetic Structures II}.
\newblock Springer-Verlag Berlin, 2003.

\bibitem{hillebrands_spindynamics3}
B.Hillebrands and A.Thiaville.
\newblock {\em Spin Dynamics in Confined Magnetic Structures III}.
\newblock Springer-Verlag Berlin, 2006.

\bibitem{gruenberg_gmr}
{P. Gr{\"{u}}nberg, R. Schreiber, Y. Pang, M. B. Brodsky and H. Sowers}.
\newblock {Layered Magnetic Structures: Evidence for Antiferromagnetic Coupling
  of Fe Layers across Cr Interlayers}.
\newblock {\em Physical Review Letters}, 57:2442--2445, 1986.

\bibitem{baibich_gmr}
{M.N. Baibich, J.M. Broto, A. Fert, F. Nguyen Van Dau, and F. Petroff}.
\newblock Giant magnetoresistance of (001)fe/(001)cr magnetic superlattices.
\newblock {\em Physical Review Letters}, Volume 61, Number 21:2472--2475, 1988.

\bibitem{gao_perpendicular}
{Kai-Zhong Gao (Seagate Technology)}.
\newblock Magnetic thin films for perpendicular recording.
\newblock In {\em 2009 APS March Meeting - Monday–Friday, March 16–20,
  2009; Pittsburgh, Pennsylvania}, 2009.

\bibitem{tudosa_magnetization}
Ioan Tudosa.
\newblock {\em Magnetization dynamics using ultrashort magnetic field pulses}.
\newblock PhD thesis, Stanford University, 2005.

\bibitem{stoehr_mangnetism}
{Joachim St{\"{o}}hr, Hans Christoph Siegmann}.
\newblock {\em {Magnetism - From Fundamentals to Nanoscale Dynamics}}.
\newblock {Springer-Verlag Berlin}, 2006.

\bibitem{slonczewski_current}
J.~C. Slonczewski.
\newblock Current-driven excitation of magnetic multilayers.
\newblock {\em Journal of Magnetism and Magnetic Materials}, 159(1-2):L1--L7,
  June 1996.

\bibitem{ralph_spintransfer}
M.D.~Stiles D.C.~Ralph.
\newblock Spin transfer torques - current perspectives.
\newblock {\em Journal of Magnetism and Magnetic Materials}, 320:1190--1216,
  2008.

\bibitem{tsoi_excitation}
M.~Tsoi, A.~G.~M. Jansen, J.~Bass, W.-C. Chiang, M.~Seck, V.~Tsoi, and
  P.~Wyder.
\newblock Excitation of a magnetic multilayer by an electric current.
\newblock {\em Phys. Rev. Lett.}, 80(19):4281--, May 1998.

\bibitem{myers_current}
E.~B. Myers, D.~C. Ralph, J.~A. Katine, R.~Louie, and R.~A. Buhrman.
\newblock Current-induced switching of domains in magnetic multilayer devices.
\newblock {\em Science}, 285(5429):867--870, August 1999.

\bibitem{misochko_coherent}
O.~Misochko, M.~Lebedev, N.~Georgiev, and T.~Dekorsy.
\newblock Coherent phonons in ndba2cu3o7−x single crystals: Optical-response
  anisotropy and hysteretic behavior.
\newblock {\em Journal of Experimental and Theoretical Physics},
  98(2):341--347, February 2004.

\bibitem{zvezdin_magnetooptics}
A.K. Zvezdin and V.~A. Kotov.
\newblock {\em {Modern magnetooptics and magnetooptical materials}}.
\newblock Institute of Physics Publishing, 1997.

\bibitem{bennemann_nonlinear}
Karl~Heinz Bennemann.
\newblock {\em Nonlinear optics in metals}.
\newblock Oxford Science Publications, 1998.

\bibitem{shen_nonlinear}
Yuen-Ron Shen.
\newblock {\em {The Principles of Nonlinear Optics}}.
\newblock Wiley-Interscience, John Wiley and Sons, 1984.

\bibitem{radu_laser}
I.~Radu, G.~Woltersdorf, M.~Kiessling, A.~Melnikov, U.~Bovensiepen, J.-U.
  Thiele, and C.~H. Back.
\newblock Laser-induced magnetization dynamics of lanthanide-doped permalloy
  thin films.
\newblock {\em Phys. Rev. Lett.}, 102(11):117201--4, March 2009.

\bibitem{conrad_phdthesis}
{Uwe Conrad}.
\newblock {\em Statische und dynamische Untersuchungen {ul\-tra\-d\"un\-ner}
  Metallfilme mit optischer Freqeunzverdopplung und nichtlineare Mikroskopie}.
\newblock PhD thesis, Freie Universit\"at Berlin, 1999.

\bibitem{shen_shgintroduction}
Y.R. Shen.
\newblock Surface second harmonic generation: A new technique for surface
  studies.
\newblock {\em Annual Review of Materials Science}, 16:69--86, 1986.

\bibitem{lueth_surfaces}
{Hans L{\"{u}}th}.
\newblock {\em {Surfaces and Interfaces of Solid Materials}}.
\newblock {Springer-Verlag Berlin}, 1995.

\bibitem{lewitz_thesis}
{Bj\"orn Lewitz}.
\newblock {Aufbau und Inbetriebnahme eines polaren Kerrspektrometers zur
  Messung im Ultrahochvakuum}.
\newblock Master's thesis, {Freie Universit\"at Berlin}, 2007.

\bibitem{muehge_structural}
{Th. M{\"{u}}hge, A. Stierle, N. Metoki, H. Zabel, U. Pietsch}.
\newblock {Structural properties of high-quality sputtered Fe films on
  Al$_2$O$_3(11\overline{2}0)$ and MgO$(001)$ substrates}.
\newblock {\em Applied Physics A - Solids and Surfaces}, 59:659--665, 1994.

\bibitem{fahsold_epitaxial}
{G. Fahsold, G. K{\"o}nig, W. Theis, A. Lehmann, K.H. Rieder}.
\newblock {Epitaxial FeO films from ultrathin Fe on MgO(001) studied by He-atom
  scattering}.
\newblock {\em Applied Surface Science}, 137:224--235, 1999.

\bibitem{demokritov_morphology}
{M. Rickart, B.F.P Roos, T. Mewes, J. Jorzick, S.O. Demokritov, B.
  Hillebrands}.
\newblock {Morphology of epitaxial metallic layers on MgO substrates: influence
  of submonolayer carbon contamination}.
\newblock {\em Surface Science}, 495:68--76, 2001.

\bibitem{gruenberg_interlayer}
{P. Gr{\"{u}}nberg, S. Demokritov, A. Fuss, R. Schreiber, J. A. Wolf and S. T.
  Purcell}.
\newblock Interlayer exchange, magnetotransport and magnetic domains in fe/cr
  layered structures.
\newblock {\em Journal of Magnetism and Magnetic Materials},
  104-107:1734--1738, 1992.

\bibitem{krebs_properties}
B.~T.~Jonker J.~J.~Krebs and G.~A. Prinz.
\newblock Properties of fe single-crystal films grown on (100)gaas by
  molecular-beam epitaxy.
\newblock {\em J. Appl. Phys.}, 61:2596--2599, 1987.

\bibitem{oxford_evaporator}
{Oxford Applied Research}.
\newblock {\em Mini E-Beam Evaporator - Model EGCO4}.
\newblock Crawley Mill, Witney, Oxon, OX8 5TJ, United Kingdom, June 1999.

\bibitem{sauerbrey_qmb}
G{\"u}nther Sauerbrey.
\newblock Verwendung von schwingquarzen zur w{\"a}gung d{\"u}nner schichten und
  zur mikrow{\"a}gung.
\newblock {\em Zeitschrift f{\"u}r Physik}, 155:206--222, 1959.

\bibitem{maxtek_qmb}
{MAXTEK, INC.}
\newblock {\em {Maxtek Thickness Monitor - Model TM-350/400}}.
\newblock 5980 Lakeshore Drive, Cypress, CA 90630-3371 Tel: (714) 828-4200 Fax:
  (714) 828-4443 Email: sales@maxtekinc.com support@maxtekinc.com, 11 edition,
  August 2005.

\bibitem{blomqvist_structural}
{P. Blomqvist, R. W{\"a}ppling}.
\newblock {Structural properties of ultrathin bcc Co (111) layers}.
\newblock {\em Journal of Crystal Growth}, 252:120--127, 2001.

\bibitem{henzler_oberflaechenphysik}
W.G{\"o}pel M.Henzler.
\newblock {\em Oberfl{\"a}chenphysik des Festk{\"o}rpers}.
\newblock Teubner Verlag, 1994.

\bibitem{specs_safire}
{SPECS GmbH, Voltastrasse 5, 13355 Berlin, Germany}.
\newblock Safire 4 - a software for rheed analysis.

\bibitem{schatz_rheed}
{A. Schatz, S. Dunkhorst, S. Lingnau, U. von Hörsten and W. Keune}.
\newblock {RHEED intensity oscillations during epitaxial growth of fcc Fe on
  Cu(001)}.
\newblock {\em Surface Science}, 310:595--600, 1994.

\bibitem{schwabl_quanten}
Franz Schwabl.
\newblock {\em Quantenmechanik}.
\newblock Springer-Verlag Berlin, 1992.

\bibitem{omicron_userguide}
{Omicron Vakuumphysik GmbH}.
\newblock {\em {The UHV AFM/STM User's Guide}}.
\newblock Idsteiner Stra{\ss}e 78 D-65232 Taunusstein Germany, June 2000.

\bibitem{horcas_wsxm}
{I. Horcas, R. Fernandez, J.M. Gomez-Rodriguez, J. Colchero, J. Gomez-Herrero,
  and A.M.Baro}.
\newblock {WSXM: A software for scanning probe microscopy and a tool for
  nanotechnology}.
\newblock {\em {Review of Scientific Instruments}}, 78:013705, 2007.

\bibitem{thuermer_dynamic}
{K.Th\"urmer, R. Koch, M. Weber, and K.H. Rieder}.
\newblock {Dynamic Evolution of Pyramid Structures during Growth of Epitaxial
  Fe$(001)$ Films}.
\newblock {\em Physical Review Letters}, 75, 9:1767--1771, 1995.

\bibitem{hernando_magnetization}
{Costa-Kr\"amer, J. L. and Menendez, J. L. and Cebollada, A. and Briones, F.
  and Garcia, D. and Hernando, A.}
\newblock Magnetization reversal asymmetry in fe/mgo(001) thin films.
\newblock {\em Journal of Magnetism and Magnetic Materials}, 210(1-3):341--348,
  February 2000.

\bibitem{wehling_webpage}
Homepage of tim wehling, university of hamburg.

\bibitem{ribbonsoft_qcad}
Switzerland Ribbonsoft~GmbH, 8046~Zurich.
\newblock qcad, 2d computer aided design.

\bibitem{slonczewski_spinwaves}
{J.C. Slonczewski}.
\newblock {Excitation of spin waves by an electric current}.
\newblock {\em {Journal of Magnetism and Magnetic Materials}}, 195:261--268,
  1998.

\bibitem{jackson_electrodynamics}
{John David Jackson}.
\newblock {\em {Classical Electrodynamics}}.
\newblock {John Wiley \& Sohns New York}, 1962.

\bibitem{barnes_goldevaporation}
{Mark C. Barnes, Doh-Y. Kim and Nong M. Hwang}.
\newblock {The mechanism of gold deposition by thermal evaporation}.
\newblock {\em {Journal of Ceramic Processing Research}}, 1:45--52, 2000.

\bibitem{melnikov_shggd}
A.~Melnikov, O.~Krupin, U.~Bovensiepen, K.~Starke, M.~Wolf, and E.~Matthias.
\newblock Shg on ferromagnetic gd films: indication of surface-state effects.
\newblock {\em Applied Physics B: Lasers and Optics}, 74(7):723--727, May 2002.

\bibitem{miyazaki_tmr}
{T. Miyazaki and N. Tezuka}.
\newblock {Giant magnetic tunneling effect in Fe/Al$_2$O$_3$/Fe junction}.
\newblock {\em Journal of Magnetism and Magnetic Materials}, 139:231--234,
  1995.

\bibitem{weaver_opm}
J.H. Weaver and E.E.~Koch C.~Krafka, D.W.~Lynch.
\newblock {\em Optical Properties of Metals}, volume I and II.
\newblock Fachinformationszentrum Energie - Physik - Mathematik GmbH,
  Karlsruhe, Germany, 1981.

\end{thebibliography}

\cleardoublepage 
\phantomsection  
\addcontentsline{toc}{chapter}{Acknowledgments} 

\chapter*{Acknowledgements}
\chaptermark{Acknowledgements}

This work wouldn't have been possible without the help and support of
my friends and colleagues. I would therefore like to thank the following
people for their precious input.

\medskip

First of all, I would like to thank my supervisor \emph{Uwe Bovensiepen} for
introducing me into this topic and making this thesis possible. He
was always there to help with words and deeds and had a lot of good
ideas in cases some of the experiments didn't work.

\medskip

\emph{Alexey Melnikov} was the one whom I directly worked together with. He
is an absolute expert when it comes femtosecond lasers and electron dynamics
and the project covered by this thesis was mainly his idea. Also, Alexey
was the one who helped me shaping the text of this thesis and who had an
infinite amount of patience reading and correcting my texts. Thank you,
Alexey. Without you, this thesis would never have been possible.

\medskip

I would like to thank \emph{Michael Karcher} who was my lab-mate in the
MBE laboratory at Prof. Fumagalli's workgroup. Michael knows almost the entire experimental
setup in the Fumagalli workgroup by heart and he was just an invaluable help during
all the experiments for sample preparation. I have learned so incredibly
many things from Michael and still keep learning from him. Thanks Michael!

\medskip

My friend \emph{Daniel Sachse} for helping me with \LaTeX , helping with inkscape
to make the drawings and giving to good tips on designing the magnet. Also,
Daniel was always there to support me mentally, also together with his
enchanting wife \emph{Lili}. He kept on driving me all the time.

\medskip

\emph{Anja Diesing} helped me with some of the CAD drawings, she is a real
talent when it comes to making things visual on the computer. Thanks Anja.

\medskip

\emph{Axel Luchterhand} assisted me in the workshop when building the magnet.

\medskip

The following people haven't helped me directly but I still would like to thank
them for their support: \emph{Wolf Dieter Woidt}, \emph{Jonas Hoffmann},
\emph{Milian Wolff}, \emph{Marco Starace}, \emph{Jens Dreger}, \emph{Philipp Neuser},
\emph{Prof. Robert Schrader} and many more I forgot to mention. Thank you,
everybody!

\medskip

Last but not least, I'm especially thankful to my parents and my sister Anna
and my brother Philippe. Without their restless support and help I probably would have never
got so far and being able to earn a degree at a university. Danke \emph{Mama}, \emph{Papa},
\emph{Philippe} und \emph{Anna}!

\end{document}